UNIVERSITÉ DE GENÈVE
Section de Physique
Département de Physique Théorique

FACULTÉ DES SCIENCES
Professeur Antonio RIOTTO

# Signals from the Early Universe:

## Black Holes, Gravitational Waves and Particle Physics

THÈSE

présentée à la Faculté des sciences de l'Université de Genève
pour obtenir le grade de Docteur ès sciences, mention physique

par

## Valerio De Luca

de

Pescara (Italie)

Thèse N° – 5665

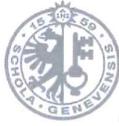

UNIVERSITÉ
DE GENÈVE
FACULTÉ DES SCIENCES

DOCTORAT ÈS SCIENCES, MENTION PHYSIQUE

## Thèse de Monsieur Valerio DE LUCA

intitulée :

## «Signals from the Early Universe: Black Holes, Gravitational Waves and Particle Physics»

La Faculté des sciences, sur le préavis de Monsieur A. W. RIOTTO, professeur ordinaire et directeur de thèse (Département de physique théorique), Monsieur F. RIVA, professeur assistant (Département de physique théorique), Monsieur G. BALLESTEROS, professeur (Instituto de fisica teorica, Universidad Autónoma de Madrid, Madrid, Spain), Monsieur H. VEERMÄE, professeur (National Institute of chemical physics and biophysics, Tallin, Estonia), autorise l'impression de la présente thèse, sans exprimer d'opinion sur les propositions qui y sont énoncées.

Genève, le 5 juillet 2022

Thèse  - 5665 -

Le Doyen

# Abstract

We dedicate this thesis to the study of signatures coming from the primordial epochs of the universe. We will focus in particular on Primordial Black Holes (PBHs), which may be formed from perturbations generated during inflation and might comprise a fraction of the dark matter in the universe. In the first part of the thesis, we will address the PBH properties at the time of formation, that are their masses, spins and abundance, and investigate the generation of Gravitational Wave (GW) signals during their production. In the second part, we will describe the PBHs evolution across the cosmic history due to their assemble in binaries, phases of baryonic mass accretion and clustering effects. We will then discuss GW signatures coming from their coalescence, compare these predictions with present GW data detected by the LIGO/Virgo Collaboration (LVC) and assess the role of future GW experiments like 3G detectors and LISA in discovering these objects. Finally, in the third part, we will investigate some aspects of the interplay between black holes and fundamental physics in the early universe, focusing on the role of GWs to shed light on their properties.

# Résumé

L'objectif de cette thèse a été d'étudier des phénomènes qui auraient pu se produire au début de l'univers, en se concentrant en particulier sur la formation des trous noirs primordiaux (PBH), leur liens avec la physique fondamentale, et de décrire les nombreuses façons dont les observations des les ondes gravitationnelles (GW) fournissent des aperçus fascinants liés à la physique de l'univers primordial.

Dans la partie I, nous avons étudié la génération des PBH à partir de l'effondrement gravitationnel des perturbations cosmologiques produites pendant l'ère inflationniste. Nous avons vu comment les propriétés des perturbations qui s'effondrent sont cruciales pour déterminer les caractéristiques des PBH résultants. Ensuite, nous avons étudié les signatures associées à la formation des PBH, en nous concentrant sur le fond d'ondes gravitationnelles stochastiques généré à partir des mêmes perturbations. Nous avons montré que le spectre GW correspondant peut se situer dans les courbes de sensibilité des expériences comme NANOGrav ou comme LISA, en fournissant des informations sur la connexion entre les PBH et la matière noire dans l'univers.

Dans la partie II, nous avons étudié l'évolution des PBH à travers l'histoire cosmique. Nous montrons que les PBH peuvent s'assembler dans des systèmes binaires, et que ces binaires peuvent connaître une phase d'accrétion de masse baryonique. De plus, nous avons vu que, si les PBH constituent une grande partie de la matière noire de l'univers, après l'égalité matière-rayonnement, ils formeront des amas, dont l'effet principal serait d'affecter leurs taux de fusion. Dans la même partie de la thèse, nous avons comparé le modèle PBH avec les données actuelles détectées par la collaboration LIGO/Virgo/KAGRA, en constatant que le scénario dans lequel les PBH expliquent tous les événements observés est compatible avec les contraintes actuelles sur l'abondance des PBH. Ensuite, nous avons étudié les perspectives de futurs détecteurs GW comme Einstein Telescope, Cosmic Explorer et LISA pour découvrir ces objets compacts.

La troisième partie est consacrée à décrire le liens entre les PBH et la physique fondamentale au cours des premiers stades de l'univers. En particulier, nous avons étudié certaines propriétés des transitions de phase du premier ordre, que peuvent généré PBH. Nous avons également décrit le rôle des PBH dans la catalyse des processus de violation du nombre de baryons via les transitions de sphaléron du modèle standard. Enfin, nous avons étudié la capacité de ET et de LISA à fournir des informations sur les propriétés des champs scalaires condensés dans des halos autour des BH à travers les nombres de marée de Love.

Dans cette direction, des nombreux travaux sont encore nécessaires pour comprendre le rôle des PBH dans l'évolution de l'univers et leur lien avec la physique fondamentale. Cependant, le grand nombre d'événements GW qui seraient détectés dans un avenir proche par les expériences GW actuelles et futures représentent une voie prometteuse vers une meilleure compréhension des nombreux phénomènes et signaux provenant de l'univers primordial.

# Members of the Jury

# List of publications

## Papers appearing in this thesis

## Other journal papers

## White papers and collaboration papers

# Acknowledgments

Inizio questi ringraziamenti con il mio sensei, *Toni Riotto*. Mi sento estremamente fortunato ad averlo avuto come mentore in questi quattro anni, durante i quali è riuscito a trasmettermi tutta la passione e la dedizione nel fare ricerca in fisica, inseguendo anche le più piccole curiosità. Custodirò gelosamente tutte le illuminanti discussioni avute nel suo ufficio su dei ritagli di carta e le belle chiacchierate da grandi amici davanti al una pizza.

La seconda persona che voglio ringraziare è *Paolo Pani*, per la sua disponibilità e il prezioso supporto durante il periodo di visita trascorso a Roma. A special thank goes to *Alex Kehagias*, for his contagious cheerfulness and for our unforgettable dinners full of jokes. I am grateful to *Vincent Desjacques* and *Marco Peloso* for their support over these years, and *Guillermo Ballesteros*, *Francesco Riva*, *Hardi Veermäe* for being members of the jury.

I want to thank the Cosmology and Astroparticle physics group at UNIGE for such a stimulating place to do research. Thank you also to all my colleagues in the gravity group at Sapienza, with whom I also had the pleasure of playing football together. I am particularly indebted to *Andrea, Enrico, Ilia, Susanna, Swetha* for precious collaborations. During this PhD I had the great opportunity to attend the GGI school in Florence. I thank all the people I met there, especially *Ariane*, for our destructive trips and wrong skiing experiences.

Passo ora ai ringraziamenti personali partendo da Ginevra. I want to thank my roommates in the apartment 5A for their support along these years, *Felipe, Mark, Salman, Srinivas* and especially *Kuo* for our tennis matches. Ringrazio tutti gli amici conosciuti in questi anni per avermi fatto sentire a casa, *Anna, Azadeh, Carl, Davide R., Diego, Eleonora, Francesco I., Francesco L., Julia, Laura, Maëlle, Marc, Pietro*. In particolare, voglio ringraziare *Lula* per le fancy serate e bruciature al lago, il mozzo abruzzese *Lorenzo* per le chiacchierate alle sei del mattino, *Davide* per le incredibili abbuffate e discussioni filosofiche, e la tigre *Giulia* per la sua instancabile simpatia e i quotidiani aperitivi a suon di spritz. Infine ringrazio il mio grande amico *Gabriele* per le vittorie e le tranvate condivise insieme, grazie al quale tutto è stato più bello.

La seconda tappa è Pisa. Voglio fare dei ringraziamenti presidenziali a tutti i miei amici di "FeC" e "No no grazie a lei" che, nonostante la distanza, mi sono stati vicini durante questi anni, *Alessandro, Bruno, Budrus, Chessa, Elisa, Giulio, Laura, Maria, Massimo, Nicola, Pierp, Peppo, Pippo, Simon, Sofia, Zip*. Un ringraziamento speciale va a *Pierino*, per la nostra infinita amicizia e ai cuccioli giapponesi *Mr.P* ed *Erika*, con i quali il mio cammino si intreccia continuamente e spero continui a farlo per sempre.

La terza ed ultima tappa è casa mia, Pescara. Parto con il ringraziare i miei zii acquisiti *Emilio* e *Nicola*, per le loro traversate in macchina carichi di pasta e sughi al pomodoro. Voglio ringraziare tutte le mie bimbe, *Carolina, Davide, Eugenio, Giacomo*, il pane orizzontale *Giada, Gianmarco*,

*Giulia, Massi, Lorenzo, Lozzio, Paolo, Pierluigi, Rebecca, Vittorio*, ed in particolare *Laura* per le nostre gite nella capitale ed *Alessandra* per il suo incessante entusiasmo per i buchi neri che evaporano. Infine ringrazio *Luca*, per esserci da sempre.

Siamo giunti alla fine, e poche righe non potranno mai bastare per esprimere la mia gratitudine verso la mia famiglia. Ringrazio mia nonna *Dina* e mia zia *Carla* per il loro affetto. Sono eternamente debitore a mia sorella *Francesca* e ai miei genitori *Claudia* e *Nicola* per aver creduto in me in ogni circostanza e per il loro infinito supporto, senza il quale questo traguardo non sarebbe stato possibile. Infine, dedico questa tesi a Re *Scar* per essere entrato a zampe tese nella mia vita.

# Contents

# Chapter 1

# Introduction

One of the most fascinating properties of our universe is that, on large scales, it appears to be homogeneous and isotropic. These features pointed towards the foundation of the Cosmological Principle, which states that all spatial positions and directions in the universe are essentially equivalent, and one of its most direct consequence, based on the Einstein's theory of general relativity, is the observed expansion of the Universe [38–40].

In this Introduction we describe the main features of the Big-Bang model and the evolution of the hot thermal universe. Then we review the revolutionary theory of inflation and describe the formation of PBHs during this epoch. Finally we introduce GWs as a probe to investigate the properties of these objects.

## 1.1 Big-Bang cosmology

In this section we want to briefly summarise the main aspects of the Big-Bang model, whose formulation began in the 1940s and it is based on the proposal that the early universe was once very hot and dense, and has subsequently expanded and cooled to its present state [41, 42]. Direct consequences of this model are the presence of the Cosmic Microwave Background (CMB) [43], that is a relic background radiation with a temperature of the order of a few Kelvin [44, 45], and the Big-Bang nucleosynthesis, thanks to which heavier elements like hydrogen could get formed during the evolution of the universe. Many uncertainties however reside in the initial conditions, whose best solution is provided by the inflationary cosmology.

For standard texts on the Big-Bang cosmological model see, e.g., Refs. [46–50].

### 1.1.1   ΛCDM

By assuming homogeneity and isotropy in space, the most generic metric that is consistent with this set of symmetries is the Friedmann-Robertson-Walker (FRW) metric, whose expression is given by

$$ds^2 = dt^2 + a^2(t) \left( \frac{dr^2}{1 - kr^2} + r^2 d\theta^2 + r^2 \sin^2\theta d\phi^2 \right). \tag{1.1}$$

It depends only on two cosmological parameters, that are the spatial curvature $k$ and the scale factor $a$. The first one can take only the discrete values $+1, -1, 0$, which correspond to closed, open or spatially flat geometries, while the second determines distances in comoving coordinates.

From the Einstein's equations

$$R_{\mu\nu} - \frac{1}{2} g_{\mu\nu} R = 8\pi G T_{\mu\nu} + \Lambda g_{\mu\nu}, \tag{1.2}$$

where $G$ is the Newton's gravitational constant and $\Lambda$ is the cosmological constant, one can write down the Friedmann equations

$$H^2 = \left( \frac{\dot{a}}{a} \right)^2 = \frac{8\pi G}{3} \rho - \frac{k}{a^2} + \frac{\Lambda}{3},$$
$$\frac{\ddot{a}}{a} = \frac{\Lambda}{3} - \frac{4\pi G}{3} (\rho + 3p). \tag{1.3}$$

Here $H$ denotes the Hubble parameter, which represents the expansion rate of the universe and whose present value is given by $H_0 = 100\, h\, \mathrm{km}\, s^{-1}\, \mathrm{Mpc}^{-1}$ with $h = 0.73 \pm 0.03$, as reported by the Planck collaboration [51].[1] We have assumed that the matter content of the universe is in the form of a perfect fluid, for which the stress tensor $T_{\mu\nu}$ takes the form

$$T_{00} = \rho, \qquad T_{ij} = p g_{ij}, \tag{1.4}$$

in terms of the energy density $\rho$ and pressure $p$. The conservation of the energy momentum tensor then translates into the equation

$$\dot{\rho} = -3H(\rho + p). \tag{1.5}$$

By assuming the equation of state $p = w\rho$, it can be easily shown that the previous equation gives the general behaviour $\rho \propto a^{-3(1+w)}$. One can then consider different types of matter. For a gas of radiation (relativistic matter), one has $w = 1/3$ and

$$\text{radiation}: \qquad \rho_r \propto a^{-4}, \qquad a \propto t^{1/2}, \qquad H = 1/2t. \tag{1.6}$$

---

[1]Increasing attention in the literature has been focusing on the "Hubble tension", that is a discrepancy in the value of the Hubble parameter between the measurements obtained using the cosmic microwave background [51, 52] and local measurements from supernovae [53, 54] and lensing time delays [55, 56], with the latter suggesting a higher value.

For non relativistic matter $w = 0$ and one has

$$\text{matter}: \qquad \rho_m \propto a^{-3}, \qquad a \propto t^{2/3}, \qquad H = 2/3t. \tag{1.7}$$

For a universe dominated by vacuum energy, for example acting like a positive cosmological constant $\Lambda$ with $w = -1$, the spacetime is called de Sitter space, and one has

$$\text{vacuum}: \qquad \rho_v = \text{const.}, \qquad a = \sinh^{2/3}(\sqrt{3\Lambda}t/2), \qquad H^2 = \frac{8\pi G}{3}\Lambda. \tag{1.8}$$

By introducing the critical energy density $\rho_c = 3H^2/8\pi G$ and the ratio $\Omega = \rho/\rho_c$, the Friedmann equation can be recast as

$$\frac{k}{H^2 a^2} = \Omega - 1. \tag{1.9}$$

This implies that for a closed universe with $k = 1$, $\Omega > 1$, for an open universe with $k = -1$, $\Omega < 1$ and for a flat one with $k = 0$, $\Omega = 1$. Our universe is found to be basically flat, and the decomposition of its total energy density into the various components we have outlined above is given by $\Omega_r h^2 = 0.02242 \pm 0.00014$, $\Omega_m = 0.3111 \pm 0.0056$ and $\Omega_v = 0.6889 \pm 0.0056$ [51], with the vacuum energy dominating the current energy budget.

## 1.1.2 A history of the hot thermal universe

We now discuss the main stages of the evolution of the universe. From the FRW solutions of the Einstein's equations, one can see that the scale factor gets monotonically smaller in the past, which implies that the universe was much smaller in size and that objects were closer together, compressed and hot. It is therefore likely that the universe underwent a phase dominated by radiation.

During this radiation-dominated phase, thermal equilibrium was established in the matter content thanks to rapid particle interactions with respect to the expansion rate of the universe, which takes the form $H \sim T^2/M_p$, where $T$ is the temperature of the thermal bath and $M_p = 1/8\pi G$ is the reduced Planck mass. During this epoch, the age of the universe can be related to its temperature as

$$t = \sqrt{\frac{90}{32\pi^3 G N(T)}} T^{-2}. \tag{1.10}$$

At very high temperatures, massive particles are pair produced and constitute part of the thermal bath. In this case the total energy density take the form

$$\rho = \left( \sum_B g_B + \frac{7}{8} \sum_F g_F \right) \frac{\pi^2}{30} T^4 \equiv \frac{\pi^2}{30} N(T) T^4 \tag{1.11}$$

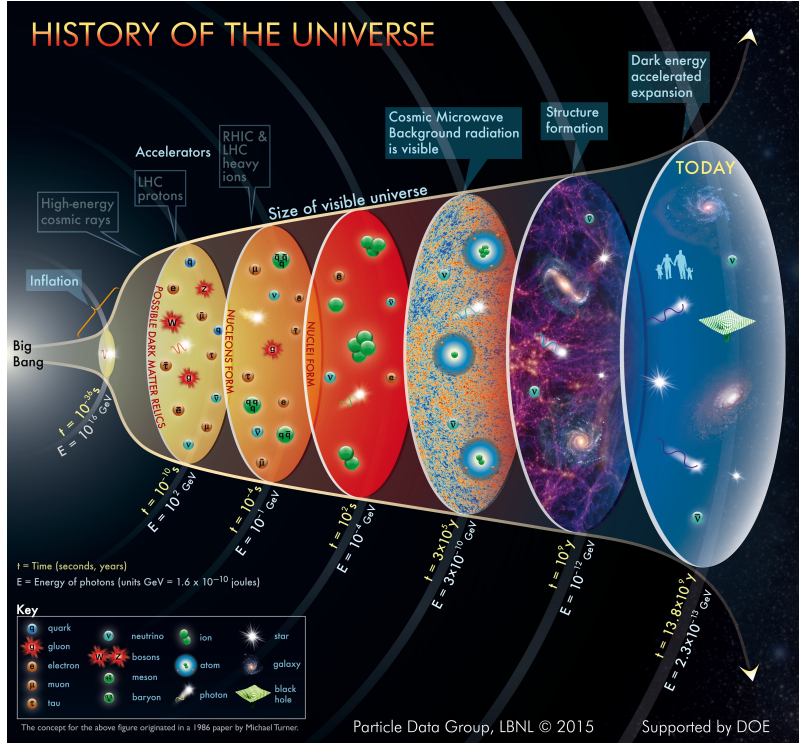

**Figure 1.1**: Pictorial representation of the history of the universe (Source & Credit: Particle Data Group at Lawrence Berkeley National Lab).

in terms of the number of degrees of freedom of each boson (B) and fermion (F) with mass $m_i \ll T$. $N(T)$ denotes the effective number of degrees of freedom, whose value changes as the temperature increases to account for new particle states, and depends on the underlying model of particle physics. A given particle species should be considered if it experiences interactions with a rate $\Gamma_i \gtrsim H$, that on average implies at least one interaction occurred over the lifetime of the universe. When this condition is not satisfied anymore, then the species will decouple from the thermal bath.

The standard picture of the evolution of the universe proceeds as follows. At temperatures larger than the QCD phase transition, quarks and gluons behave like free particles. This implies that at this stage they will contribute to the effective number of degrees of freedom, along with photons and leptons. Then, after the QCD phase transition, they will end up in a hadronic phase and decouple from the thermal bath. At temperatures smaller than the pion mass, the only light species are electrons and neutrinos, with the antineutrons annihilating with neutrons and antiprotons with protons, giving rise to a small net baryon number. At this stage, protons and neutrons are in equilibrium thanks to interactions with neutrinos. When the latter decouple from the thermal bath, many neutrons will pair with protons to form stable light elements, whose abundance is mainly determined by the baryon-to-photon ratio $\eta \equiv n_B/n_\gamma \simeq 6 \cdot 10^{-10}$. After the universe temperature cools down to about 0.1 eV, electrons and nuclei start to combine to form neutral atoms, in a phase which is called recombination. During this phase, the universe become essentially transparent to photons, which stream freely and characterise the CMB radiation, with a temperature about $\approx 2.7\,\mathrm{K}$. Furthermore, matter and radiation starts to have comparable densities, which then become equal at the matter-radiation equality. Then matter takes the lead, and dominates the energy density up to the present time. A pictorial representation of the evolution of the universe is shown in Fig. 1.1.

Even though this is the basic picture, there are many puzzles which have to be further investigated. First of all, most of the energy density is in two forms which we cannot observe directly. One is Dark Matter (DM), whose existence was postulated from the realisation that the expected rotational velocity of the galaxies rotation curves does not agree with the observed distribution of stars and galaxies [57–59]. Its contribution to the total matter energy density is estimated to be around $\Omega_m \approx 0.3$, while the contribution from nucleosynthesis and CMB to baryons is found to be much smaller, $\Omega_b \approx 0.05$. The second one is dark energy, which is quite possibly a cosmological constant and dominates the present energy budget [51].

Furthermore, the universe appears to be populated with an excess of particles over antiparticles. To produce a net baryon asymmetry one needs to satisfy the Sakharov conditions [60], according to which it is necessary that the underlying microphysics violates baryon number and CP, and that there is a departure from thermal equilibrium. Famous examples of baryogenesis are the ones through heavy particles decay [61, 62], electroweak baryogenesis, which involve the presence of non-perturbative interactions of sphalerons [63], and lepto-baryogenesis [64]. Moreover, it is very likely that the universe has undergone phase transitions [65, 66], which can occur as first-order with a tunneling process through a potential barrier, or as second-order with the driving field evolving from one state to another. These transitions can leave detectable signatures like GWs and formation of BHs, which we will discuss in chapter 6 of this thesis.

Finally, homogeneity, flatness and the putative absence of topological objects induce some puzzles which are solved by inflation, that we will discuss in the next section.

### 1.1.3   Inflation

The paradigm of inflation was made to solve some problems associated to the initial conditions of Big Bang cosmologies. The first problem is homogeneity: by looking at the temperature anisotropies in the CMB, which are of the order of $10^{-5}$, one would expect that two points in space had been in causal contact before recombination. However in the standard radiation-dominated era, such a causal contact cannot be achieved. The problem therefore stands in how these two points managed to have almost identical temperatures. The second problem is flatness: a flat universe today would imply a universe which was even flatter in the past, giving rise to a fine tuning problem. Finally, there is the monopole problem: these are objects predicted in grand unified theories which, unless their density were many orders of magnitude smaller than the baryon one, would have a present energy density far greater than the observed one.

A possible solution to these problems is inflation, that is a phase of extremely rapid expansion of the universe [67–71]. The most standard scenario is slow-roll inflation, where the expansion is driven by a scalar field $\phi$, called the inflaton, with a potential $V(\phi)$ that is slowly varying for some field range. In particular, a classical, potential-dominated scalar field cosmology can drive a de Sitter expansion, with the slow time-evolution of the energy density responsible for breaking the exact $O(1, 3)$ symmetry of a 4-D de Sitter spacetime down to a FRW space, where the scalar field plays the role of cosmic time coordinate.

The equation of motion of the scalar field reads

$$\ddot{\phi} + 3H\dot{\phi} + V'(\phi) = 0, \tag{1.12}$$

where its evolution is driven by the potential gradient $V'(\phi)$ and is subject to Hubble damping. If the field is slowly rolling the potential, one can neglect $\ddot{\phi}$, such that the equation of motion reduces to [72]

$$3H\dot{\phi} = -V'(\phi). \tag{1.13}$$

The Hubble expansion is then dominated by the potential energy

$$H^2 \simeq \frac{1}{3M_p^2} V(\phi), \tag{1.14}$$

which is then responsible for an almost exact de Sitter expansion. This result occurs only if the slow-roll condition is satisfied, which is usually expressed in terms of the slow-roll parameters

$$\epsilon = \frac{1}{2} M_p^2 \left( \frac{V'}{V} \right)^2 \ll 1, \qquad |\eta| = M_p^2 \left| \frac{V''}{V} \right| \ll 1. \tag{1.15}$$

The inflaton equation of motion can be integrated to get the time it takes the field to traverse the plateau of the potential. This time can be related to the number of e-folds

$$N = \frac{1}{M_p^2} \int \mathrm{d}\phi \frac{V(\phi)}{V'(\phi)}, \tag{1.16}$$

which has to be larger than about 60 to solve the flatness, horizon and monopole problems.

Even though slow-roll inflation can lead to an exponentially large universe, the energy density is locked in the inflaton potential energy, which at a certain point has to be converted to particles and thermalised to recover the standard hot Big-Bang cosmology. This process is usually called reheating [73, 74], which occurs when the inflaton reaches the minimum of its potential. At that stage, the potential can be approximated to be quadratic, such that the inflaton satisfies the equation of motion

$$\ddot{\phi} + 3H\dot{\phi} + m^2\phi = 0, \tag{1.17}$$

in terms of its mass $m$, which has an oscillatory solution $\phi = \phi_0 \cos(mt)$ as $m \gg H$. This solution describes a coherent state of particles with energy density $\rho = m^2\phi_0^2/2$, zero momentum and no pressure, which therefore behaves like pressureless dust, that is then diluted by the expansion of the universe.

The inflaton condensate can lose energy through perturbative decays due to interaction terms in its lagrangian, characterised by a finite decay rate $\Gamma$. When $\Gamma \approx H$, the condensate decays and gives rise to the Standard Model particles, whose interactions may bring them in thermal equilibrium before the conventional hot Big-Bang cosmology. This sets therefore an upper bound on the reheating temperature given by [74]

$$T_{\mathrm{rh}} = 0.8 g_*^{-1/4} m^{1/2} (\Gamma M_p)^{1/4}, \tag{1.18}$$

in terms of the effective number of degrees of freedom $g_*$.

Even though inflation was initially discussed to solve the problem of initial conditions for our homogeneous and isotropic universe, it was soon realised that it also offers a simple mechanism to generate inhomogeneities, that are crucial to explain the formation of large-scale structure [75–80]. In particular, to have inflation, we need a field and a metric which are uniform over a size comparable to the Hubble horizon at that epoch; however, the presence of quantum fluctuations breaks this spatial symmetry. Since during inflation $m \ll H$, then we can treat the inflaton as a massless free field in de Sitter space, which then experiences fluctuations of the order of $\delta\phi \sim H/2\pi$. The equation of motion for these perturbations reads

$$\delta\ddot{\phi} + 3H\delta\dot{\phi} + \frac{k^2}{a^2}\delta\phi = 0, \tag{1.19}$$

in terms of their comoving momentum $k$. Since during a de Sitter phase the scale factor grows exponentially, as soon as the fluctuations pass out the horizon during inflation, $k/a \sim H$, the system is overdamped and they are frozen. Then, they re-enter the horizon during the subsequent eras and start to grow, until they become non linear and account for the observed structures in the universe.

Since the scalar field and scalar metric perturbations transform under temporal gauge transformations of the form $t \to t + \delta t(t, x^i)$, it is customarily to introduce a gauge invariant quantity called the curvature perturbation $\zeta$, which remains constant on super-Hubble scales for adiabatic density perturbations both during and after inflation [79, 81]. In the flat slicing, it is related to the inflaton perturbations $\delta\phi$ as

$$\zeta = H\frac{\delta\phi}{\dot{\phi}}. \tag{1.20}$$

One can write down the power spectrum of the primordial curvature perturbations on super-Hubble scales produced in single-field inflation as

$$\mathcal{P}_\zeta(k) = \frac{1}{2\epsilon M_p^2} \left( \frac{H}{2\pi} \right)^2,\tag{1.21}$$

at the time when the perturbation with momentum $k$ exit the Hubble horizon. Its amplitude value $A_0$ is constrained by the Planck 2018 temperature and polarization measurements to be $\ln(10^{10} A_0) = 3.044 \pm 0.0014$ on scales ranging from $0.008 h^{-1}\,\mathrm{Mpc}^{-1} \le k \le 0.1 h\,\mathrm{Mpc}^{-1}$ [51].

Other observables of inflation are the spectral index $n_s$, which indicates deviations from perfect scale invariance of the curvature power spectrum and is measured to be around $n_s \simeq 0.9668 \pm 0.0037$ [51], and the tensor-to-scalar ratio $r$, which measures the strength of the tensor modes and has an upper bound of $r < 0.063$ at 95% C.L. on scales $k = 0.002\,\mathrm{Mpc}^{-1}$ from Planck 2018 and BICEP2/Keck Array 2015 data [82].

Before concluding this section we stress that the curvature power spectrum is currently unconstrained on scales which are much smaller than those probed by CMB. Therefore, there could be some features on those scales giving rise to signatures which we can possibly measure and use to set constraints on the power spectrum. One of these signatures is the generation of PBHs, which is the topic of the next section and of many chapters of this thesis.

## 1.2 Introduction on primordial black holes

Following the conclusions of the previous section, we are going to introduce the topic of PBHs. The interested reader may find many details in recent reviews on this topic [83–86].

The study of PBHs goes back to the 1960s and 70s [87], where it was proposed that sizeable inhomogeneities in the early universe can directly undergo gravitational collapse to form BHs [88]. The dynamics of such collapse is different from the one experienced by stars, that give rise to astrophysical BHs with a mass larger than the Chandrasekhar value $\approx M_\odot$. Indeed, the strong gravitational force inside these highly compressed radiation or matter overdensities allows for the formation of BHs which can span a range of masses from stellar/super-massive values to much lighter ones, in principle as light as the Planck mass [89]. After the inflationary cosmology took the stage, the formation and properties of PBHs, such as their masses and abundance, had been studied in the context of inflation models. At the same time, observational information about PBHs (including their non-detection) may provide important clues to constrain early universe scenarios.

It was then realised that PBHs are a potential dark matter candidate. In particular, if they form before the matter-radiation equality, on cosmological scales they behave like a cold and collisionless fluid which experiences only gravitational interactions. At the same time, on galactic and smaller scales its granularity can have observable consequences, which can be used to constrain their formation mechanisms. Due to Hawking evaporation, PBHs which are lighter than $M_{\rm PBH} \lesssim 5 \cdot 10^{14}$g are evaporated by the cosmic age [90]. Nevertheless, the high energy particles which they have emitted in the process may play a role in Big Bang nucleosynthesis, and thus could be used to set stringent constraints on the PBH abundance in that mass range [84]. On the other hand, those which have a larger mass have a lifetime longer than the age of the universe and can therefore comprise a fraction of the dark matter [91–93].

In the recent years the interest in PBHs has drastically increased after the discovery of GWs emitted from the coalescence of tens of Solar mass BHs by the LVC in 2016 [94]. In particular, it was pointed out that PBH binaries could have been formed by accidental encounters of PBHs in dense environments in the late universe or due to the presence of tidal perturbations in the early universe [95–97]. This highlighted the importance of GWs to probe the parameter region of PBHs, which is very difficult with electromagnetic waves.

In the following we briefly summarise one of the most standard scenario for the formation of PBHs in the early universe, and then outline the present observational constraints on their abundance with respect to the dark matter one.

### 1.2.1   Some preliminaries

The formation of PBHs in the very early universe may occur within different scenarios. The most standard one involves the gravitational collapse of sizeable overdensities generated during inflation [98–123], which will be the main focus of this thesis. Others models include PBH generation during an early matter era [124–128], from phase transitions [129–131], collisions of bubbles [132–135], collapse of domain walls [136–142], standard model Higgs instability [143, 144], and many more.

The scenario we consider is described as follows. In the previous section we have introduced inflation as a period of accelerated expansion in the early Universe, able to generate primordial perturbations via quantum fluctuations of scalar fields. Stringent constraints on the power spectrum of the curvature perturbation $\zeta$ have been set by CMB and large scale structure data. However, the current constraints on the amplitude of the power spectrum on small scales are fairly weak [145]. In particular, the COBE/FIRAS constraints on CMB spectral distortions require $\mathcal{P}_\zeta(k) \lesssim 10^{-4}$ for scales $k \approx (10 - 10^4)\text{Mpc}^{-1}$ [146, 147] and Pulsar Timing Array (PTA) limits on GWs require $\mathcal{P}_\zeta(k) \lesssim 10^{-2}$ for scales $k \approx (10^6 - 10^7)\text{Mpc}^{-1}$ [148–150]. This implies that on such small scales, an enhancement of the power spectrum with respect to its value at CMB scales is possible.

We therefore assume that there is such an enhancement on small scales, which can be achieved for example by modifying the standard slow roll paradigm close to the end of inflation [101] (we will return to this point later in the thesis). These perturbations, that during inflation are stretched on super-horizon scales, are then transferred to the radiation fluid during the reheating phase. As the Hubble horizon grows during the radiation-dominated phase, at a certain epoch it becomes comparable to the characteristic scales of the perturbations. As horizon crossing occurs, gravity becomes active and the overdense region starts contracting. If these regions are dense enough, they can collapse and give rise to a PBH. Due to the presence of radiation pressure which can rapidly disperse the overdensity peaks, the dynamics is quite fast, and occurs only if these overdensities are sizable enough at the time of horizon crossing.

The characteristic PBH mass is basically determined by the mass $M_H$ enclosed in the horizon at the time of formation, i.e. $M_{\mathrm{PBH}} = \gamma M_H$, in terms of an efficiency factor $\gamma$, whose value was estimated analytically to be of the order of $\gamma \simeq 0.2$ [92]. The horizon mass can be related to the corresponding momentum scale $k$ of the collapsing perturbation at the time of horizon crossing, giving the relation [97]

$$M_{\mathrm{PBH}} \simeq 30 M_\odot \left( \frac{\gamma}{0.2} \right) \left( \frac{g_*(t_{\mathrm{form}})}{106.75} \right)^{-1/6} \left( \frac{k}{2.9 \cdot 10^5 \mathrm{Mpc}^{-1}} \right)^{-2}, \qquad (1.22)$$

in terms of the effective number of degrees of freedom at the formation time, $g_*(t_{\mathrm{form}})$. One can therefore appreciate that perturbations with larger wavelengths, that are those which re-enter the Hubble horizon later, give rise to heavier PBHs. From this equation one finds also the hierarchy between the observable scales by CMB observations and, for example, a $30 M_\odot$-PBHs in terms of the number of e-folds [97]

$$N = \ln \left( \frac{k_{\mathrm{PBH}}}{k_{\mathrm{CMB}}} \right) = \ln \left( \frac{2.9 \cdot 10^5 \mathrm{Mpc}^{-1}}{0.002 \mathrm{Mpc}^{-1}} \right) \sim 20. \qquad (1.23)$$

By comparing this number with the one measured from when the present horizon scale exits the Hubble horizon to the end of inflation, that is typically around 60, one concludes that to form such PBHs we need a mechanism which can amplify the perturbations during inflation.

The abundance of PBHs in the universe is typically expressed in terms of the ratio between their energy density and the one of the DM as

$$f_{\mathrm{PBH}} = \frac{\Omega_{\mathrm{PBH}}}{\Omega_{\mathrm{DM}}}, \qquad (1.24)$$

where $\Omega_{\mathrm{DM}} h^2 = 0.120 \pm 0.001$, as currently constrained by the Planck measurements [151]. Depending on the considered formation model, one predicts specific properties of the collapsing perturbations, thus affecting the resulting masses and abundance of the PBH population.

In the next section we are going to discuss the observational constraints on this parameter to understand the viable regions where PBHs can still comprise the dark matter in the universe.

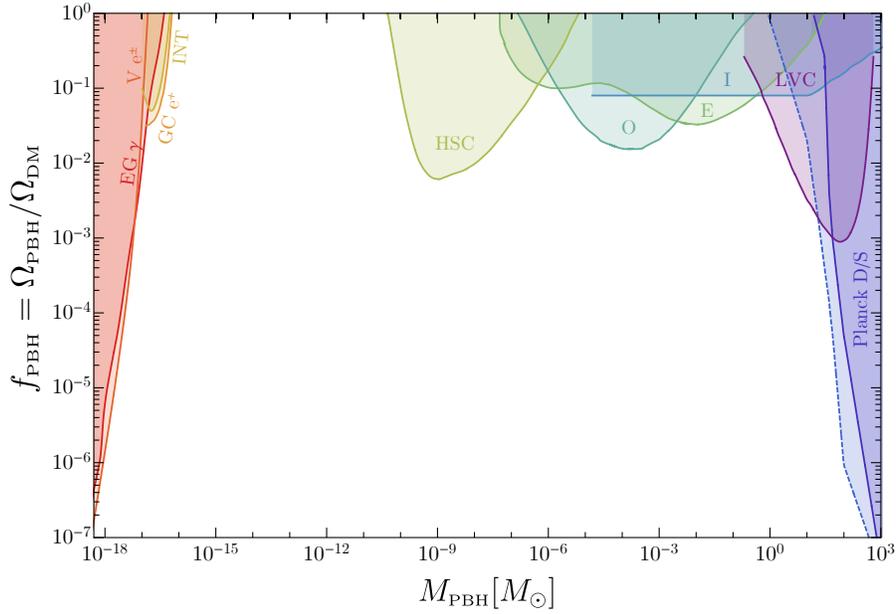

**Figure 1.2**: Plot of some of the observational constraints on the PBH abundance normalised to the dark matter one, for a monochromatic population with mass $M_{\rm PBH}$. For a more comprehensive plot of the constraints see, for example, Ref. [84].

## 1.2.2 Constraints on the PBH abundance

In this section we summarise some of the most relevant constraints on the PBH abundance in the universe. We stress that each constraint is derived based on a specific set of assumptions, and may be relaxed or strengthened by modifying the formation mechanism. We restrict our attention to PBHs with masses $M_{\rm PBH} \lesssim 10^3 M_\odot$. The interested reader can have a look at Ref. [84] for the most recent and complete review on the topic.

PBHs with masses smaller than $10^{-18}$ have completed their Hawking evaporation by the present day. On the other hand, the emission from slightly heavier PBHs can be used to set constraints on their abundance. In particular, limits can be derived from the production of extra-galactic gamma-rays (EG $\gamma$) [152], from positron annihilations in the Galactic Center (GC $e^+$) [153], from gamma-ray observations by INTEGRAL (INT) [154] and from $e^\pm$ observations by Voyager 1 (V $e^\pm$) [155]. See also Refs. [156–161] for additional constraints in this mass range.

In principle, PBHs with mass from $10^{-16} M_\odot \leq M_{\rm PBH} \leq 10^{-11} M_\odot$ can be constrained by the consequences of their capture or transient through stars. For example, a PBH passing through a star induces energy loss by dynamical friction, with the PBH consequently captured by the star; then, it will sink in the center of the star, accreting matter and potentially destroying it. Alternatively, their transit in white dwarfs leads to localised heating and potentially cause a runaway explosion. However, both constraints derived by these mechanisms have been revised in recent works [162, 163], leaving the asteroid region free from observational bounds.

When PBHs with mass in the range $5 \cdot 10^{-10} M_\odot \leq M_{\mathrm{PBH}} \leq 10 M_\odot$ cross the line of sight of a star, a temporary amplification of its flux occurs, giving rise to a microlensing phenomenon. The duration of such event is proportional to $M_{\mathrm{PBH}}^{1/2}$, and therefore the constrained range of masses depends on the cadence of the microlensing survey. The absence of observations of these events can then be used to set constraints on the PBH abundance in the corresponding mass range. Indeed, limits have been deduced from microlensing searches by Subaru Hyper Supreme Camera (HSC) [164, 165], from searches of massive compact halo objects (MACHOs) towards the Large Magellanic Clouds (EROS, E) [166, 167], from fast transient events near the critical curves of massive clusters of galaxies (Icarus, I) [168], and from stars observations in the Galactic bulge by the Optical Gravitational Lensing Experiment (Ogle, O) [169].

PBHs with mass larger than the solar mass can experience a phase of baryonic mass accretion in the early universe. This phase is usually accompanied by the emission of ionizing radiation, which may impact onto the CMB temperature and polarization anisotropies. Measurements from Planck can then be used to set constraints on the PBH abundance [170, 171]. Due to the several uncertainties in the physics of accretion, one can set two different constraints depending on the geometry of the accretion flow; those are the disk and spherical geometry (Planck D/S respectively). However, since the relevant electromagnetic emission takes place in the redshift range $300 \lesssim z \lesssim 600$, the spherical model (Planck S) is expected to be more accurate, as thin accretion disks could form only at much smaller redshift [172]. Furthermore, these constraints take into account the catalysing effect of an early dark matter halo formed around isolated PBHs, as we will discuss later in this thesis. Finally, since the relevant emission takes place at high redshifts, its physics is independent of uncertainties in the accretion model due to the onset of structure formation.

Other constraints can be set from the comparison between the emission at late times of electromagnetic signals from PBHs accreting gas from the interstellar medium, and observations of galactic radio and X-ray isolated sources (XRay) [173, 174] and X-ray binaries (XRayB) [175]. Furthermore, interactions of PBHs with the interstellar medium can induce Dwarf Galaxy Heating (DGH), which can be constrained using data from Leo T dwarf galaxy observations [176].

Finally, one can set constraints from GW observations. In particular, LVC measurements can be used to set an upper bound on the PBH abundance to not overproduce GW signals [177–179]. Furthermore, in the scenario in which PBHs form from the collapse of density perturbations without non-Gaussianities [180–182], the null observation in the 11-yr dataset of the NANOGrav experiment, searching for a stochastic GW background in the frequency range around $f \simeq 1 \, \mathrm{yr}^{-1}$, can be used to constrain the GWs induced at second order by the same curvature perturbations responsible for PBH formation [150].

Some of the most relevant constraints on the PBH abundance for a population with a monochromatic mass function are shown in Fig. 1.2. One can appreciate that there are no stringent constraints in the asteroid mass range. This implies that PBHs with those masses can comprise the totality of the DM, while in other regions they can represent only a fraction of it.

One of the crucial point is therefore to find possible signatures which may help in prove or disprove the possibility that PBHs comprise the totality or a fraction of the dark matter. One of the most concrete avenues is provided by GWs, which we introduce in the next section.

## 1.3 Introduction on gravitational waves

The first observation of GWs opened the era of GW astronomy. In particular, the detection of coalescing binary BHs represents a unique opportunity to get information of the universe which could never be obtained from electromagnetic signals, and this certainly includes the possibility to detect PBHs.

Aside from the primordial channel, which we describe in details in the next chapters of this thesis, there are also astrophysical channels which may give rise to heavy stellar-mass binary BHs: those are the isolated field channel, where two stars in an isolated binary collapse to BHs, and the dynamical formation scenario, in which BHs in dense stellar environments give rise to binaries, that are eventually ejected by three-body interactions, see Refs [183, 184] for reviews. Since in principle both primordial and astrophysical channels may be consistent with present GW observations, it is crucial to investigate their main characteristics to discriminate between those scenarios and pin down the correct ones. These studies are already in progress for what concerns present data detected by the LIGO/Virgo/KAGRA collaboration (LVKC). Moreover, with the advent of future GW experiments both on ground and in space, the sensitivity to GW signals will greatly improve and many more merger events will be detected. By those experiments, we will obtain much information of BH binaries such as mass distribution, space distribution, redshift distribution and spin distribution, which will definitely help us in discriminating between different scenarios.

Furthermore, GWs could be unique opportunities to probe the strong gravity regime and investigate the putative role of fundamental physics in GWs emission. This, for example, includes the presence of scalar fields modifying the environment of binary coalescences or the emission of gravitational signals from phenomena which occur in the very early universe, like phase transitions, cosmic strings and many others (see Refs. [185–187] for reviews).

In the following we will summarise the main aspects of the coalescence of binary systems and the consequent emission of gravitational radiation [188]. GWs can be described as propagating deformations of the spacetime geometry. Considering a system of two BHs moving in a circular orbit, one can solve the Einstein's equations to find the corresponding emitted quadrupole radiation. In Fourier space, the general expression for the GW signal for the binary inspiral phase, after performing an average over the sky locations, is given by [189] (we set $G = c = 1$ for simplicity)

$$h(f) = \mathcal{A} f^{-7/6} e^{i\psi(f)}. \tag{1.25}$$

Here $f$ denotes the corresponding GW frequency and $\mathcal{A}$ is the GW amplitude

$$\mathcal{A} = \frac{1}{\sqrt{30}\pi^{2/3}} \frac{\mathcal{M}^{5/6}}{D_L}, \tag{1.26}$$

in terms of the luminosity distance $D_L$ of the source and the binary chirp mass

$$\mathcal{M} = \nu^{3/5} M_{\rm tot} = \frac{(m_1 m_2)^{3/5}}{(m_1 + m_2)^{1/5}}, \tag{1.27}$$

which depends on the symmetric mass ratio $\nu = \eta_1 \eta_2 = m_1 m_2 / M_{\rm tot}^2$ and the total mass of the binary $M_{\rm tot} = m_1 + m_2$, both expressed in terms of the individual masses $m_i$.

The dependence of the binary parameters is captured in the GW phase $\psi$, which is usually expressed as the sum of three different contributions in a post-Newtonian (PN) expansion. The first contribution is the point-particle contribution up to 3.5PN order [189–192]. It depends on the mass ratio of the binary system $q = m_2 / m_1 \leq 1$ (assuming the primary component to be heavier than the secondary one), and on their spins $\vec{S}_1$ and $\vec{S}_2$. They enter in the combinations

$$\chi_{\rm eff} = \frac{\vec{S}_1/m_1 + \vec{S}_2/m_2}{m_1 + m_2} \cdot \hat{L}, \tag{1.28}$$

which is called the effective spin of the binary system and describes the orbital projection of the individual spins onto the direction of the binary initial angular momentum $\vec{L}$, and the precession spin parameter $\chi_{\rm p}$, that denotes the spin components perpendicular to the binary angular momentum responsible for the precession of the orbital plane. The presence of spinning objects impacts onto the amount of GWs fluxes emitted by the binary, since larger spins lead to longer inspiral phases. The second contribution is tidal heating, which depends on the energy absorbed at the BH horizon, and introduces a 2.5PN(4PN) correction for spinning (nonspinning) binaries relative to the leading term [193, 194]. Finally, at the 5PN and 6PN orders, there is the contribution from tidal deformations, which depends on the parameter [195, 196]

$$\Lambda = \left(264 - \frac{288}{\eta_1}\right) \frac{\lambda_2^{(1)}}{M_{\rm tot}^5} + (1 \leftrightarrow 2), \tag{1.29}$$

in terms of the quadrupolar tidal Love number, $\lambda_2^{(A)} = 2m_A^5 k_2^{(A)}/3$, of the $A$-th body, that describes the tidal deformations of a compact object under the presence of external tidal fields, like a companion in a binary system. We will come back to the role of tidal effects in GW observations in the last chapter of this thesis.

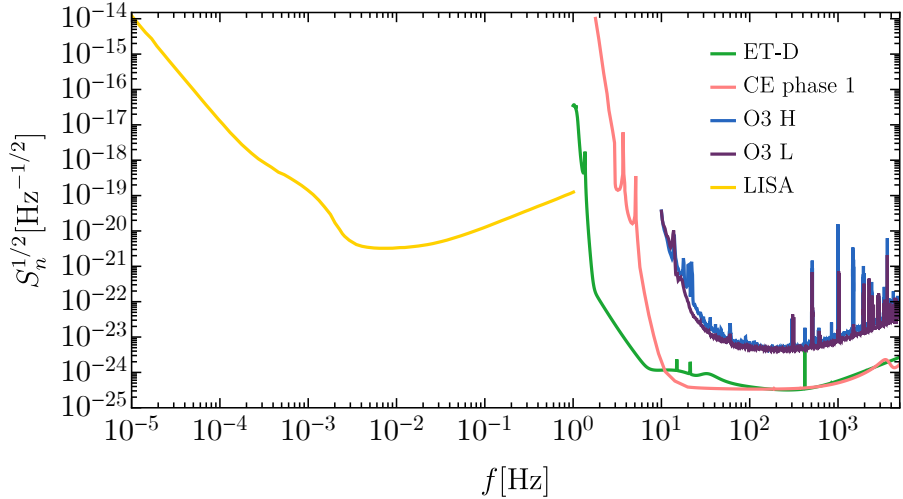

**Figure 1.3**: Noise sensitivity curves for the LIGO Hanford (H) and Livingston (L) detector during the O3 run [197], the 3G detectors Einstein Telescope (ET) at design sensitivity from [198] and Cosmic Explorer (CE) during phase 1 from [199], and the space-based LISA experiment [200]. Figure taken from Ref. [5].

This description holds for the inspiral phase at radial distances larger than the Innermost Stable Circular Orbit (ISCO), after which the dynamics is dominated by strong-field effects, and the two objects plunge toward each other. The corresponding frequency is given by [188]

$$f_{\text{ISCO}} = \frac{1}{6\sqrt{6}(2\pi)} \frac{1}{M_{\text{tot}}}. \tag{1.30}$$

At larger frequencies the binary evolves in the merger phase, where the two bodies give rise to a remnant, whose oscillations to reach its stationary configuration are described in the ringdown phase.

This implies that the properties of the binary system manifest into a different GW signal and frequencies. This motivates the realization of several GW experiments, which are able to probe different parameter spaces. For example, the LIGO and Virgo experiments are sensitive to frequencies around $f \simeq (10 - 10^3)$Hz. On the other hand, Pulsar Timing Array (PTA) experiments like Parkes Pulsar Timing Array (PPTA) [201] and the North American NanoHertz Observatory for Gravitational Waves (NANOGrav) [202] can look for GWs with frequencies close to the nHz. For frequencies around the mHz, GWs will be detected by the space-based Laser Interferometer Space Antenna (LISA) [203]. Finally, in the range around $10^2$ Hz, third generation GW detectors Einstein Telescope (ET) [204] and Cosmic Explorer (CE) [205] are expected to overcome the sensitivity of the LVKC experiments. Some of the corresponding detector sensitivities are shown in Fig. 1.3.

It is exciting that PBHs could give rise to GW signals which can be detected at present and future GW experiments. We dedicate Part I of this thesis to discuss the formation of PBHs and the associated gravitational signatures. Part II is devolved to the evolution of PBHs across the cosmic history, and the corresponding GW signals coming from their mergers. Finally, Part III is dedicated to the interplay between BHs and fundamental physics, and the role of GWs in sheding lights on their connection.

# Part I

# Signatures from primordial black holes: formation

# Chapter 2

# Properties of primordial black holes at formation

This chapter is devoted to the discussion of the properties of PBHs at the time of formation, in the standard scenario where enhanced curvature perturbations, generated during inflation on very small scales and transferred to the radiation fluid at reheating, may gravitationally collapse at horizon re-entry and give rise to a PBH.

We will discuss about the initial properties of PBHs, such as their initial masses, which would be of the order of the mass contained in the cosmological horizon at the time of re-entry, their abundance, which depends on the statistics of the collapsing perturbations, and their initial spins, which is found to be very small due to the mainly spherical nature of the collapsing perturbation peaks and the small timescales of collapse because of the presence of the radiation pressure.

## 2.1   On the abundance of PBHs

The computation of the PBH abundance follows some step. Once the probability distribution function of the density fluctuations is known, that depends on the inflationary mechanism which give rise to such fluctuations, one can determine the threshold to collapse thanks to dedicated general relativity simulations. Then, one can evaluate the probability to have a density contrast larger than the threshold to compute the initial PBH abundance. This section is therefore dedicated to the estimation of the threshold and statistics of the collapsing perturbations to determine the PBH abundance.

### 2.1.1 Threshold for collapse

The current understanding of PBH formation relies on dedicated general relativity numerical simulations [206–209] which show that a cosmological perturbation may collapse and give rise to a PBH if its amplitude is larger than a certain threshold, $\delta > \delta_c$. The first estimate of this quantity was based on the Jeans length argument in Newtonian gravity [92], from which $\delta_c \sim c_s^2 = 1/3$, in terms of the sound speed of the cosmological radiation fluid. Then, generalising the Jeans length argument with the theory of General Relativity, one gets a value $\delta_c \approx 0.4$ for a Universe dominated by radiation [210], which represents a lower bound since it does not account for the non-linear effects due to pressure gradients, which require a relativistic numerical simulations. Finally, recent studies showed that the value of the threshold depends on the shape of the collapsing perturbations, giving rise to the range $0.4 \leq \delta_c \leq 2/3$ [211, 212].

In this section, following Ref. [11], we will provide a prescription to compute the threshold for PBH formation from the shape of the power spectrum of the collapsing curvature perturbations, which for simplicity we assume to be Gaussian. At the end of the section we will also show how to generalise the procedure to non-Gaussian perturbations, showing some of the results of Ref. [1].

**PBH formation: initial conditions**

In this section we are going to consider the standard formation scenario in which PBHs form from the collapse of sizable density perturbations after they re-enter the cosmological horizon during the radiation-dominated era. Those perturbations are peaks of the density field which are generated during inflation and transferred to the radiation fluid during the reheating phase.

On superhorizon scales, one can assume those extreme peaks to be spherically symmetric [213], and the local region of the Universe characterised by such perturbations to be described by the metric

$$ds^2 = -dt^2 + a^2(t)\left[\frac{dr^2}{1 - K(r)r^2} + r^2 d\Omega^2\right] = -dt^2 + a^2(t)e^{2\zeta(\hat{r})}\left[d\hat{r}^2 + \hat{r}^2 d\Omega^2\right], \qquad (2.1)$$

in terms of the scale factor $a(t)$ and the conserved comoving curvature perturbations $K(r)$ and $\zeta(\hat{r})$ on superhorizon scales, which approach zero at large distances where the Universe is taken to be unperturbed and spatially flat. From the radial and angular parts one gets the correspondence

$$\begin{cases} r = \hat{r}e^{\zeta(\hat{r})}, \\ \dfrac{dr}{\sqrt{1 - K(r)r^2}} = e^{\zeta(\hat{r})}d\hat{r}. \end{cases} \qquad (2.2)$$

The choice between the curvature perturbations $K(r)$ or $\zeta(\hat{r})$ is related to the choice of "spatial gauge" of the comoving radial coordinate. In particular, the Lagrangian coordinate $\hat{r}$ describes the perturbed region as a local separated FRW universe, with $\zeta(\hat{r})$ modifying the local expansion, while $K(r)$ is defined with respect to the background FRW solution ($K = 0$). One can relate them to get

$$K(r)r^2 = -\hat{r}\zeta'(\hat{r})\left[2 + \hat{r}\zeta'(\hat{r})\right]. \tag{2.3}$$

On super-Hubble scales, where the comoving curvature profile is time independent, one can adopt the gradient expansion approach [207, 214–216], according to which the time dependent variables can be expanded as a series of the small ratio between the Hubble radius and the perturbation length scale, $\epsilon \ll 1$. Under this assumption pressure gradients are small and they do not dramatically affect the evolution of the perturbations [217]. In this expansion, the energy density profile can be written as [211, 218]

$$\delta(r) \equiv \frac{\delta\rho}{\rho_b} = \frac{\rho(r,t) - \rho_b(t)}{\rho_b(t)} = \frac{1}{a^2H^2}\frac{3(1+w)}{5+3w}\frac{[K(r)\,r^3]'}{3r^2} = -\frac{1}{a^2H^2}\frac{4(1+w)}{5+3w}e^{-5\zeta(\hat{r})/2}\nabla^2 e^{\zeta(\hat{r})/2}, \tag{2.4}$$

in terms of the Hubble parameter $H(t) = \dot{a}(t)/a(t)$ and the mean background energy density $\rho_b$. The parameter $w$ relates the total (isotropic) pressure $p$ to the total energy density $\rho$ in the fluid equation of state $p = w\rho$, which we will assume to be equal to $w = 1/3$, being focused on the formation of PBHs in a radiation dominated Universe.

An important quantity which enters in the criterion for the formation of a PBH is the compaction function [207, 211], which is defined as the ratio between the areal radius $R(r,t)$ and the difference $\delta M(r,t)$ between the Misner-Sharp mass within a sphere of radial radius and the background mass $M_b(r,t) = 4\pi\rho_b(r,t)R^3(r,t)/3$, calculated with respect to a spatially flat FRW metric, as

$$\mathcal{C} \equiv 2\frac{\delta M(r,t)}{R(r,t)}. \tag{2.5}$$

On superhorizon scales, the compaction function is found to be time independent, and is related to the curvature perturbations as

$$\mathcal{C} = \frac{2}{3}K(r)r^2 = -\frac{2}{3}\hat{r}\zeta'(\hat{r})\left[2 + \hat{r}\zeta'(\hat{r})\right]. \tag{2.6}$$

The comoving radius $r_m$ at which the compaction function reaches its maximum

$$\mathcal{C}'(r_m) = 0 \qquad \rightarrow \qquad \zeta'(\hat{r}_m) + \hat{r}_m\zeta''(\hat{r}_m) = 0, \tag{2.7}$$

describes the comoving length scale of the perturbation, such that the gradient expansion parameter $\epsilon$ can be defined as

$$\epsilon \equiv \frac{R_H(t)}{R_b(r_m,t)} = \frac{1}{aHr_m} = \frac{1}{aH\hat{r}_m e^{\zeta(\hat{r}_m)}}, \tag{2.8}$$

where $R_H = 1/H$ is the cosmological horizon and $R_b(r, t) = a(t)r$ is the background component of the areal radius.

The perturbation amplitude can then be defined as the energy density mass excess within the scale $r_m$, measured at the cosmological horizon crossing time $t_H$ when $\epsilon = 1$ ($aHr_m = 1$). Even though, in principle, the applicability of the gradient expansion in this regime is not very accurate and the horizon crossing time is obtained as a linear extrapolation, one can still define a criterion to measure consistently the amplitude of different perturbations. This is done by studying how the threshold for PBH varies assuming different initial curvature profiles [211].

The amplitude of the perturbation is then defined as the smoothed density contrast, averaged over a spherical volume of radius $R_m$, evaluated at the horizon crossing time $t_H$

$$\delta_m \equiv \frac{4\pi}{V_{R_m}} \int_0^{R_m} \frac{\delta\rho}{\rho_b} R^2 \mathrm{d}R = \frac{3}{r_m^3} \int_0^{r_m} \frac{\delta\rho}{\rho_b} r^2 \mathrm{d}r \,, \tag{2.9}$$

where $V_{R_m} = 4\pi R_m^3/3$ and where we have neglected higher order terms in $\epsilon$. One can then show that the volume averaged density contrast evaluated at $r_m$ is equal to the compaction function at the same point, $\delta_m = \mathcal{C}(r_m)$, and that

$$\delta_m = 3\frac{\delta\rho}{\rho_b}(r_m, t_H)\,. \tag{2.10}$$

The criterion for PBH formation then states that a PBH may form if the perturbation amplitude $\delta_m$ is larger than a threshold $\delta_c$, which depends on the shape of the energy density profile and is bounded in the range $2/5 \leq \delta_c \leq 2/3$. The smoothed density contrast can be rewritten in terms of the quantity $\Phi \equiv -\hat{r}\zeta'(\hat{r})$ as

$$\delta_m = \frac{4}{3}\Phi_m \left(1 - \frac{1}{2}\Phi_m\right)\,, \tag{2.11}$$

where $\Phi_m = \Phi(\hat{r}_m)$, such that the corresponding threshold for $\Phi$ is bounded by $0.37 \lesssim \Phi_c \leq 1$. The quantity $\Phi_m$ measures the amplitude of the perturbation in terms of the local curvature, while the quantity $\delta_m$ is related to the global geometry, and it is in turn related to the compactness of the region of radius $r_m$.

The PBHs which are formed from this dynamics would be characterised by a mass spectrum described by the scaling law of critical collapse [211]

$$M_{\mathrm{PBH}} = \kappa(\delta_m - \delta_c)^\gamma M_H\,, \tag{2.12}$$

as a function of the horizon mass $M_H$ at the horizon crossing time, the scaling coefficient $\gamma \simeq 0.36$ in a radiation dominated fluid, and the parameter $\kappa$ which depends on the particular profile of $\delta\rho/\rho_b$. Numerical simulations have shown that $1 \lesssim \kappa \lesssim 10$, and that this scaling relation holds for constant $\gamma$ when $\delta_m - \delta_c \lesssim 10^{-2}$.

**The shape parameter**

The dependence of the PBH threshold on the shape of the cosmological perturbations can be captured by the dimensionless parameter [211, 212]

$$\alpha = -\frac{\mathcal{C}''(r_m)r_m^2}{4\mathcal{C}(r_m)}, \tag{2.13}$$

in terms of the width of the peak of the compaction function. Given that the volume average of the compaction function $\mathcal{C}(r)$ in a comoving radius $r_m$

$$\bar{\mathcal{C}}(r_m) = \frac{3}{r_m^3} \int_0^{r_m} \mathcal{C}(r)\, r^2 \mathrm{d}r\,, \tag{2.14}$$

has a nearly constant value $\bar{\mathcal{C}}_c \simeq 2/5$ at the threshold for PBH formation, one can derive an analytic expression to compute the threshold $\delta_c$ in terms of the shape parameter $\alpha$ as [212]

$$\delta_c \simeq \frac{4}{15} e^{-1/\alpha} \frac{\alpha^{1-5/2\alpha}}{\Gamma\left(\frac{5}{2\alpha}\right) - \Gamma\left(\frac{5}{2\alpha}, \frac{1}{\alpha}\right)}, \tag{2.15}$$

where $\Gamma$ identifies the special Gamma-functions. The peak amplitude of the overdensity field, corresponding to its value at the centre of symmetry which we assume to be at $r = 0$, is related to $\delta_m$ by $\delta\rho_0/\rho_b = e^{1/\alpha}\delta_m$, such that

$$\left(\frac{\delta\rho_0}{\rho_b}\right)_c \simeq \frac{4}{15} \frac{\alpha^{1-5/2\alpha}}{\Gamma\left(\frac{5}{2\alpha}\right) - \Gamma\left(\frac{5}{2\alpha}, \frac{1}{\alpha}\right)}. \tag{2.16}$$

The shape parameter $\alpha$ describes the main properties of the profile in the region $0 < r \lesssim r_m$, where PBHs are generated, and any additional parameter would describe only secondary modification of the tail at $r \gtrsim r_m$, with few percent deviations from the value of $\delta_c$ computed as above.

The shape of the perturbation is not correlated with its amplitude when the shape is measured in the comoving coordinate $r$-gauge, while a correlation arises when measured in the $\hat{r}$-gauge. In particular, one gets

$$\mathcal{C}''(r_m) = \frac{1}{e^{2\zeta(\hat{r}_m)} \left[1 + \hat{r}_m \zeta(\hat{r}_m)\right]^2} \mathcal{C}''(\hat{r}_m), \tag{2.17}$$

such that the shape parameter becomes

$$\alpha = -\frac{\mathcal{C}''(\hat{r}_m)\hat{r}_m^2}{4\mathcal{C}(\hat{r}_m)\left[1 - \frac{3}{2}\mathcal{C}(\hat{r}_m)\right]}, \tag{2.18}$$

where the peak of the compaction function does not cancel out with the peak of its second derivative. Expressing the compaction function in terms of $\Phi$ gives

$$\alpha = -\frac{\Phi_m''\, \hat{r}_m^2}{4\Phi_m \left(1 - \frac{1}{2}\Phi_m\right)(1 - \Phi_m)}, \tag{2.19}$$

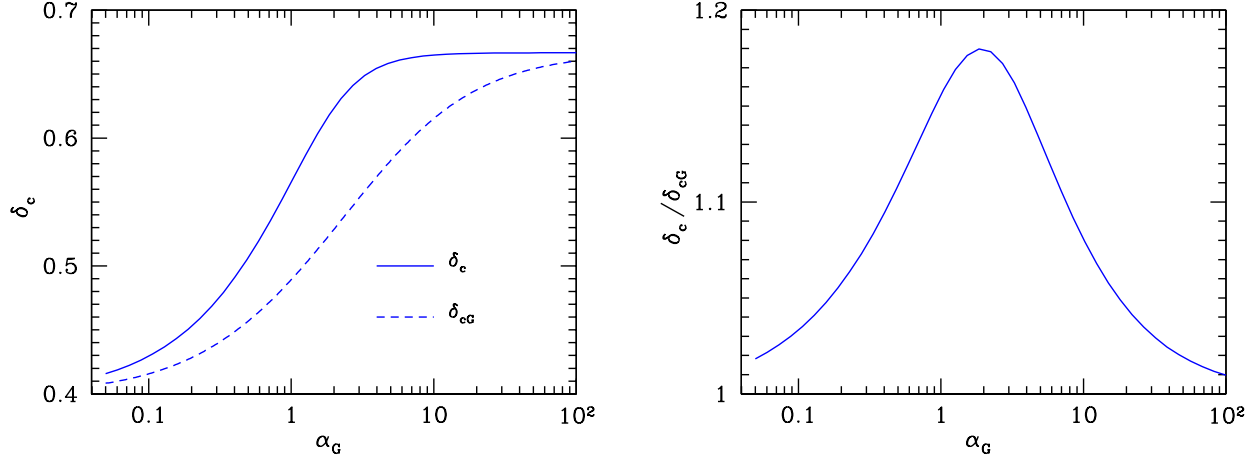

**Figure 2.1**: Behavior of the linear $\delta_{c,\mathrm{G}} = \delta_c(\alpha_{\mathrm{G}})$ and non linear $\delta_c$ averaged thresholds in terms of the Gaussian shape parameter $\alpha_{\mathrm{G}}$. Figure taken from Ref. [11].

showing the correlation between the values $\Phi_m''$ and $\Phi_m$, when varying the amplitude of the perturbation, for a fixed value of $\alpha$. For example, in the limit of $\Phi_m \to 1$ one finds $\alpha \to \infty$, which implies that $\Phi_m'' \to -\infty$ as $\mathcal{C}''(r_m) < 0$ for any positive peak of the compaction function.

An explicit example of correlation between the amplitude and the shape of the curvature perturbation profile $\zeta(\hat{r})$ is obtained assuming the benchmark profile [219]

$$\zeta(\hat{r}) = \mathcal{B} \exp\left[-\left(\frac{\hat{r}}{\hat{r}_m}\right)^{2\beta}\right], \tag{2.20}$$

which can be inserted into Eq. (2.19) to give

$$\alpha = \frac{\beta^2}{(1 - \beta\zeta(\hat{r}_m))(1 - 2\beta\zeta(\hat{r}_m))}. \tag{2.21}$$

In the linear approximation $\mathcal{B} \ll 1 \Rightarrow \beta\zeta(\hat{r}_m) \ll 1$, such that $\alpha \simeq \beta^2$. This implies that, for a fixed value of $\alpha$, the corresponding value of $\beta$ is fixed and there is no correlation between the shape and the amplitude. On the other hand, for a perturbation amplitude of the order of the threshold $\delta_c$, one has that $\mathcal{B} \sim 1$ and a non negligible correlation. For example, for a typical Mexican-hat shape, $\alpha = 1$, the threshold for collapse would be $\delta_c \simeq 0.5$, while for a value of $\beta = 1$, corresponding to a Mexican-hat shape only in the linear approximation, the threshold is $\delta_c \simeq 0.55$ [219].

In the simplified assumption of Gaussian curvature perturbation $\zeta$, also $\Phi_m$ and $\Phi_m''\hat{r}_m^2$ obey Gaussian statistics. One can therefore rewrite the shape parameter as

$$\alpha = \frac{\alpha_{\mathrm{G}}}{\left(1 - \frac{1}{2}\Phi_m\right)\left(1 - \Phi_m\right)}, \tag{2.22}$$

where

$$\alpha_{\mathrm{G}} = -\frac{\Phi_m'' \, \hat{r}_m^2}{4\Phi_m} \tag{2.23}$$

is the Gaussian shape parameter obtained in the linear approximation ($\Phi_m \ll 1$), that does not depend on the amplitude of $\Phi_m$ since $\Phi_m'' \propto \Phi_m$. When the value of $\Phi_m$ is non linear, the term $(1 - \Phi_m)(1 - \Phi_m/2)$ gives a non negligible contribution to $\alpha$ with respect $\alpha_{\mathrm{G}}$. In the non linear limit, therefore, $\alpha$ would depend on both the statistics of $\Phi_m''$ and the amplitude $\Phi_m$.

By implementing the relation between $\Phi_m$ and $\delta_c$ using (2.11), one gets

$$\Phi_m = 1 - \sqrt{1 - \frac{3}{2}\delta_c}, \tag{2.24}$$

such that inserting this equation combined with Eq. (2.15) into Eq. (2.22) gives

$$F(\alpha)\left[1 + F(\alpha)\right]\alpha = 2\alpha_{\mathrm{G}}, \tag{2.25}$$

where

$$F(\alpha) = \sqrt{1 - \frac{2}{5}e^{-1/\alpha}\frac{\alpha^{1-5/2\alpha}}{\Gamma\left(\frac{5}{2\alpha}\right) - \Gamma\left(\frac{5}{2\alpha}, \frac{1}{\alpha}\right)}}. \tag{2.26}$$

Eq. (2.25) can be solved numerically to compute the value of $\alpha$ in terms of $\alpha_{\mathrm{G}}$. By inserting this value into Eq. (2.15), one can determine the value of $\delta_c$ as a function of $\alpha_{\mathrm{G}}$.

In the left panel of Fig. 2.1 we compare the behavior of the averaged thresholds $\delta_{c,\mathrm{G}} = \delta_c(\alpha_{\mathrm{G}})$ and $\delta_c(\alpha(\alpha_{\mathrm{G}}))$ in terms of the linear parameter $\alpha_{\mathrm{G}}$, while in the right panel we show their ratio. One can easily appreciate the correction of $\delta_c$ due to the modification of the shape with respect to the one obtained in the Gaussian approximation, due to the non linear effects coming from the solution of Eq. (2.25). Given that at the boundaries $F(0) \to 1$ ($F(\infty) = 1$), there is no correction to the Gaussian case and $\delta_c = \delta_{c,\mathrm{G}}$ in the limits $\alpha \to 0$ ($\alpha \to \infty$).

**The average value of $\delta_c$**

In the previous section we have seen how to compute the averaged value of the threshold $\delta_c$ from the shape parameter $\alpha$. We now want to determine the proper value of $\alpha$ to be used starting from a given cosmological power spectrum, which carries the information of the formation scenario under consideration.

Assuming the comoving curvature perturbation $\zeta$ to be a Gaussian random field, the first step is to determine the value of $\alpha_{\mathrm{G}}$ from the power spectrum $P_\zeta(k, \eta)$ defined as

$$P_\zeta(k, \eta) = \frac{2\pi^2}{k^3}\mathcal{P}_\zeta(k)T^2(k, \eta), \tag{2.27}$$

evaluated at the proper time $\eta$ when $\hat{r}_m \gg r_H$, where $r_H = 1/aH$ is the comoving horizon radius. The power spectrum is expressed in terms of its dimensionless form $\mathcal{P}_\zeta(k)$, and the linear transfer function $T(k, \eta)$

$$T(k, \eta) = 3 \frac{\sin\left(k\eta/\sqrt{3}\right) - \left(k\eta/\sqrt{3}\right)\cos\left(k\eta/\sqrt{3}\right)}{\left(k\eta/\sqrt{3}\right)^3}, \tag{2.28}$$

which has the effect of smoothing out the subhorizon modes and acts like pressure gradients during the collapse. To ensure that the modes collapsing within $r_H$ do not affect the collapse on the larger scale $\hat{r}_m$, one can smooth out at scales larger than $\hat{r}_m \simeq 10\, r_H$, in order to be consistent with the gradient expansion approach used to specify the initial conditions of the numerical simulations. The details on this topic have been discussed in Ref. [220], where it was proven that introducing the transfer function on superhorizon scales avoids the need of using a window function on the perturbation scale $\hat{r}_m$.[1]

The scale at which the compaction function is maximum, $\hat{r}_m$, can be determined in terms of the power spectrum assuming Gaussian peak theory to write $\zeta(\hat{r})$

$$\zeta(\hat{r}) = \zeta_0 \int dk\, k^2 \frac{\sin(k\hat{r})}{k\hat{r}} P_\zeta(k, \eta), \tag{2.29}$$

and solving for $\Phi'(\hat{r}_m) = 0$, to get

$$\int dk\, k^2 \left[ (k^2\hat{r}_m^2 - 1) \frac{\sin(k\hat{r}_m)}{k\hat{r}_m} + \cos(k\hat{r}_m) \right] P_\zeta(k, \eta) = 0. \tag{2.30}$$

Given its complexity, the integral has usually to be solved numerically once the expression of $P_\zeta$ is known.

The Gaussian shape parameter $\alpha_{\rm G}$ can then be estimated from the average profile of $\zeta(\hat{r})$ of Eq. (2.29) as

$$\alpha_{\rm G} = \frac{1}{2} - \frac{\hat{r}_m^2}{4} \frac{\zeta'''(\hat{r}_m)}{\zeta'(\hat{r}_m)}, \tag{2.31}$$

where we have used the constraint relation $\Phi'(\hat{r}_m) = 0$ to get

$$\hat{r}_m^2 \Phi''_m = \hat{r}_m [2\zeta'(\hat{r}_m) - \hat{r}_m^2 \zeta'''(\hat{r}_m)]. \tag{2.32}$$

Inserting the explicit expression of the curvature profile one then obtains

$$\alpha_{\rm G} = -\frac{1}{4} \left[ 1 + \hat{r}_m \frac{\int dk\, k^4 \cos(k\hat{r}_m) P_\zeta(k, \eta)}{\int dk\, k^3 \sin(k\hat{r}_m) P_\zeta(k, \eta)} \right], \tag{2.33}$$

which shows that both $\alpha_{\rm G}$ and $\alpha$ vary by changing the shape of the cosmological power spectrum. From their value one can calculate the corresponding values of $\delta_{\rm c,G}$ and $\delta_c$ using the results of the previous section, and therefore get a relation between the threshold and the shape of the cosmological power spectrum $P_\zeta$.

---

[1] We stress that the use of a window function would introduce corrections in the computation of the threshold, even though the latter can be reduced if the same window function is adopted to evaluate the perturbation variance [221].

We can now apply this prescription to compute the threshold assuming different shapes of the curvature perturbation power spectrum.

- *Peaked Power Spectrum*: the shape of a monochromatic power spectrum with support in a single comoving scale $k_*$ and amplitude $\mathcal{P}_0$ is given by

$$\mathcal{P}_\zeta(k) = \mathcal{P}_0 k_* \delta_D(k - k_*)\,, \tag{2.34}$$

where $\delta_D$ is the Dirac-delta distribution. The scale at which the compaction function is maximum is determined by Eq. (2.30) to be $k_* \hat{r}_m \simeq 2.74$. This implies that, in the linear approximation, $\delta_{c,G} \simeq 0.51$ [222] and, including the effect of non linearities of Eq. (2.25), one gets $\alpha \simeq 6.33$ and $\delta_c \simeq 0.59$. The same value would be found assuming the average profile of $\zeta(r)$ for a monochromatic power spectrum, characterised by a sync function, to set up the initial conditions for the numerical simulations [223]. This property lies on the consistency in using peak theory in $\zeta$ or in the density contrast $\delta\rho/\rho_b$ for highly peaked power spectra [21].

- *Broad Power Spectrum*: a broad and flat power spectrum of the curvature perturbations may take the parameterized form [19, 224]

$$\mathcal{P}_\zeta(k) = \mathcal{P}_0 \Theta(k - k_{\min}) \Theta(k_{\max} - k)\,, \quad k_{\max} \gg k_{\min}. \tag{2.35}$$

In this case one has $k_{\max} \hat{r}_m \simeq 4.49$, which gives $\alpha_G \simeq 0.9$ and $\delta_{c,G} \simeq 0.48$, while the corresponding values including the non linear effects are $\alpha \simeq 3.14$ and $\delta_c \simeq 0.56$.

- *Gaussian Power Spectrum*: a Gaussian shape of the curvature power spectrum is given by

$$\mathcal{P}_\zeta(k) = \mathcal{P}_0 \exp\left[-(k - k_*)^2/2\sigma^2\right]\,, \tag{2.36}$$

in terms of the central reference scale $k_*$ and width $\sigma$. The relation between the perturbation length scale $\hat{r}_m$ and $k_*$ is shown in the left panel of Fig. 2.2. In the limit $\sigma \to 0$ the result approaches the one obtained for a monochromatic power spectrum, while for broader power spectra the expected overdensity length scale multiplied by $k_*$ decreases. This results from the fact that multiple modes participate to the collapse for broader shapes, resulting in a narrower curvature profile and a flatter compaction function. Finally, given that in this limit pressure gradients are reduced and facilitate the collapse, the corresponding threshold decreases for larger values of $\sigma$, as one can see from the right panel of the same figure. Moreover, as non-linearities are taken into account, the critical threshold $\delta_c$ reaches larger values than that for $\delta_{c,G}$.

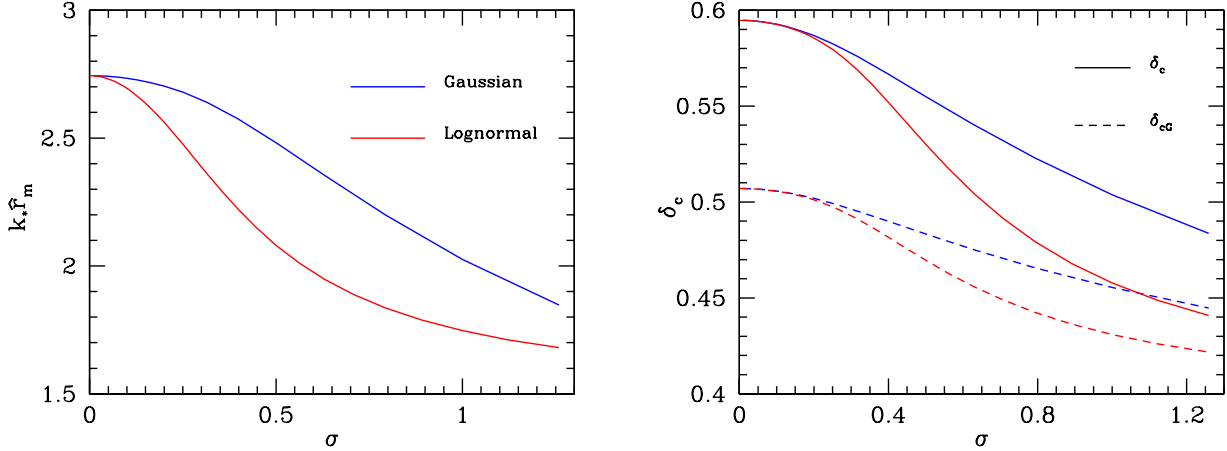

**Figure 2.2**: *Left:* Position of the peak of the compaction function $\hat{r}_m$ for a gaussian and lognormal shape of the curvature power spectrum. *Right:* Average threshold for collapse for both $\delta_{c,\mathrm{G}}(\alpha_{\mathrm{G}})$ and $\delta_c$ for the same power spectra. Figure taken from Ref. [11].

- *Lognormal Power Spectrum*: a lognormal shape for the curvature power spectrum is given by

$$\mathcal{P}_\zeta(k) = \mathcal{P}_0 \exp\left[-\ln^2\left(k/k_*\right)/2\sigma^2\right],\qquad(2.37)$$

as a function of the width $\sigma$ and central scale $k_*$, whose corresponding overdensity length scale and threshold are shown in Fig. 2.2, and are characterised by a similar behavior than that found for the Gaussian power spectrum. Given that $\sigma$ identifies the width of the power spectrum in logarithmic space, the results of assuming larger values of $\sigma$ for the relative change of $k_*\hat{r}_m$ and $\delta_c$ are amplified if compared to the Gaussian case.

These examples show the relevance of non-linear effects in determining the threshold to collapse and provide an application of the procedure to compute this value assuming Gaussian curvature perturbations.

### Extension to the non-Gaussian case

The results obtained above were based on the assumption of Gaussian curvature perturbation $\zeta$. However, if the latter is characterised by the presence of some non-Gaussianity, then the estimate of the threshold would be different, see Ref. [223] for a work along this direction.

The threshold of the smoothed density contrast $\delta_c$ depends on the (mean) profile of the curvature perturbation $\zeta(r)$ [222], which implies that the knowledge of the joint two-point probability of the curvature field $P[\zeta(\vec{x}_1), \zeta(\vec{x}_2)]$ is necessary to estimate it. Following Refs. [1, 223], we provide the main steps to perform this calculation (for simplicity from now on we avoid the use of the hat in the radial dependence of the curvature field). Let us start with the conditional probability to have, at a distance $r$, a given value $\zeta(r)$

$$P(\zeta(r)|\zeta_0 > \nu\sigma) = \frac{\Big\langle \delta_D(\zeta(\vec{x}) - \zeta(r))\theta(\zeta_0 - \nu\sigma)\Big\rangle}{\Big\langle \theta(\zeta_0 - \nu\sigma)\Big\rangle}, \tag{2.38}$$

where $\sigma$ identifies the square root of the variance of the curvature perturbation, $\theta$ is the Heaviside step-function and we have defined the threshold as $\nu\sigma$. The mean curvature perturbation then takes the form

$$\begin{aligned}
\overline{\zeta}(r) = \langle \zeta(r)|\zeta_0 > \nu\sigma\rangle &= \int_{-\infty}^{\infty} d\zeta(r)\,\zeta(r)\frac{P(\zeta(r),\zeta_0 > \nu\sigma)}{P(\zeta_0 > \nu\sigma)} \\
&= \frac{1}{P(\zeta_0 > \nu\sigma)}\int_{-\infty}^{\infty} d\zeta(r)\,\zeta(r)\Big\langle \delta_D(\zeta(\vec{x}) - \zeta(r))\theta(\zeta_0 - \nu\sigma)\Big\rangle \\
&= \frac{1}{P(\zeta_0 > \nu\sigma)}\int_{-\infty}^{\infty} d\zeta(r)\,\zeta(r)\int [D\zeta(\vec{x})]P[\zeta(\vec{x})]\delta_D(\zeta(\vec{x}) - \zeta(r))\theta(\zeta_0 - \nu\sigma) \\
&= \frac{1}{P(\zeta_0 > \nu\sigma)}\int [D\zeta(\vec{x})]P[\zeta(\vec{x})]\,\zeta(\vec{x})\,\theta(\zeta_0 - \nu\sigma) = \frac{\Big\langle \zeta(\vec{x})\,\theta(\zeta_0 - \nu\sigma)\Big\rangle}{\Big\langle \theta(\zeta_0 - \nu\sigma)\Big\rangle}, \tag{2.39}
\end{aligned}$$

where $P[\zeta(\vec{x})]$ is the probability distribution of the curvature field $\zeta$. By introducing the partition function

$$Z[J] = \int [\mathcal{D}\zeta(\vec{y})]\,P[\zeta(\vec{y})]\exp\left[ i\int d^3y J(\vec{y})\zeta(\vec{y})\right], \tag{2.40}$$

as a function of the arbitrary external source $J$, one gets

$$\Big\langle \zeta(\vec{x}_1)\,\theta(\zeta(\vec{x}_2) - \nu\sigma)\Big\rangle = \frac{\sigma}{(2\pi)^2}\int_{-\infty}^{\infty} da_1\,a_1 \int_{\nu}^{\infty} da_2 \int_{-\infty}^{\infty} d\phi_1 \int_{-\infty}^{\infty} d\phi_2\, e^{-i\sigma(\phi_1 a_1 + \phi_2 a_2)}Z[J], \tag{2.41}$$

with

$$J(\vec{x}) = \sum_{i=1}^{2} J_i(\vec{x}_i, \vec{x}), \qquad J_i(\vec{x}_i, \vec{x}) = \phi_i\delta_D(\vec{x}_i - \vec{x}). \tag{2.42}$$

By employing the standard expansion of $\ln Z[J]$

$$\begin{aligned}
\ln Z[J] &= \sum_{n=2}^{\infty}\frac{i^n}{n!}\int d^3y_1 \cdots \int d^3y_n \sum_{i_1=1}^{2}\cdots\sum_{i_n=1}^{2} J_{i_1}(\vec{y}_1, \vec{x}_1)\cdots J_{i_n}(\vec{y}_n, \vec{x}_n)\xi^{(n)}(\vec{y}_1, \cdots, \vec{y}_n) \\
&= \sum_{n=2}^{\infty}\frac{i^n}{n!}\sum_{m=0}^{n}\binom{n}{m}\phi_1^m\phi_2^{n-m}\xi_{[m,n-m]}^{(n)}, \tag{2.43}
\end{aligned}$$

in terms of the curvature perturbation connected correlation functions

$$\xi^{(n)}(\vec{x}_1, \cdots, \vec{x}_n) = \int [\mathcal{D}\zeta(\vec{x})]\, P[\zeta(\vec{x})]\, \zeta(\vec{x}_1) \cdots \zeta(\vec{x}_n), \tag{2.44}$$

and

$$\xi^{(n)}_{[m,n-m]} = \xi^{(n)}\big(\underbrace{\vec{x}_1, \cdots, \vec{x}_1}_{m\text{-times}}, \underbrace{\vec{x}_2, \cdots, \vec{x}_2}_{(n-m)\text{-times}}\big), \tag{2.45}$$

one obtains

$$\begin{aligned}
\Big\langle \zeta(\vec{x}_1)\,\theta(\zeta(\vec{x}_2) - \nu\sigma) \Big\rangle = {}& (2\pi)^{-2}\sigma \int_{-\infty}^{\infty} \mathrm{d}a_1\, a_1 \int_{\nu}^{\infty} \mathrm{d}a_2 \int_{-\infty}^{\infty} \mathrm{d}\phi_1 \int_{-\infty}^{\infty} \mathrm{d}\phi_2 \\
& \times \exp\left\{ \sum_{n=2}^{\infty} \frac{(-1)^n}{n!} \sum_{m=0}^{n}{}' \binom{n}{m} \xi^{(n)}_{[m,n-m]} \frac{\partial^m}{\partial a_1^m} \frac{\partial^{n-m}}{\partial a_2^{n-m}} \right\} \\
& \times \exp\left\{ \left( -\frac{1}{2}\sigma^2(\phi_1^2 + \phi_2^2) - i\sigma(\phi_1 a_1 + \phi_2 a_2) \right) \right\}.
\end{aligned} \tag{2.46}$$

The prime on the sum indicates that it has to be performed by omitting the terms containing $\phi_1^2$ and $\phi_2^2$. Performing the integration over the variables $\phi_1$ and $\phi_2$

$$\begin{aligned}
\Big\langle \zeta(\vec{x}_1)\,\theta(\zeta(\vec{x}_2) - \nu\sigma) \Big\rangle = {}& (2\pi)^{-1}\sigma \int_{-\infty}^{\infty} \mathrm{d}a_1\, a_1 \int_{\nu}^{\infty} \mathrm{d}a_2 \\
& \times \exp\left\{ \sum_{n=2}^{\infty} \frac{(-1)^n}{n!} \sum_{m=0}^{n}{}' \binom{n}{m} \frac{1}{\sigma^n} \xi^{(n)}_{[m,n-m]} \frac{\partial^m}{\partial a_1^m} \frac{\partial^{n-m}}{\partial a_2^{n-m}} \right\} \exp\left( -\frac{1}{2}(a_1^2 + a_2^2) \right),
\end{aligned} \tag{2.47}$$

one gets

$$\begin{aligned}
\Big\langle \zeta(\vec{x}_1)\,\theta(\zeta(\vec{x}_2) - \nu\sigma) \Big\rangle = {}& \frac{\sigma}{\sqrt{2\pi}} e^{-\nu^2/2} \left( \frac{\xi^{(2)}(\vec{x}_1, \vec{x}_2)}{\sigma^2} + \frac{\nu}{2\sigma^3} \xi^{(3)}(\vec{x}_1, \vec{x}_2, \vec{x}_2) \right. \\
& \left. + \frac{\nu^2 - 1}{6\sigma^4} \xi^{(4)}(\vec{x}_1, \vec{x}_2, \vec{x}_2, \vec{x}_2) + \cdots \right).
\end{aligned} \tag{2.48}$$

The connected term of Eq. (2.48) is then given by

$$\Big\langle \zeta(\vec{x}_1)\,\theta(\zeta(\vec{x}_2) - \nu\sigma) \Big\rangle = \frac{\sigma}{\sqrt{2\pi}} e^{-\nu^2/2} \sum_{m=0}^{\infty} \frac{1}{2^{m/2}(m+1)!} \frac{1}{\sigma^{(m+2)}} \xi^{(m+2)}(\vec{x}_1, \vec{x}_2, \cdots, \vec{x}_2) H_m\left( \frac{\nu}{\sqrt{2}} \right), \tag{2.49}$$

in terms of the Hermite polynomials $H_m(x)$. Furthermore, the one-point non-Gaussian threshold probability for $\nu \gg 1$ is provided by the expression

$$\Big\langle \theta(\zeta_0 - \nu\sigma) \Big\rangle \approx \frac{e^{-\nu^2/2}}{\sqrt{2\pi}\nu} \exp\left( \sum_{n=3}^{\infty} (\nu/\sigma)^n \xi^{(n)}(0)/n! \right), \tag{2.50}$$

where $\xi^{(n)}(0)$ identify the $n$-point correlators computed at the same point. The mean curvature profile at distance $r$ from the origin and for large values of the threshold is given by $(|\vec{x}_2 - \vec{x}_1| = r)$

$$\overline{\zeta}(r) = \nu \left[ \frac{\xi^{(2)}(r)}{\sigma} + \frac{\nu}{2\sigma^2} \xi^{(3)}(\vec{x}_1, \vec{x}_2, \vec{x}_2) + \frac{\nu^2}{6\sigma^3} \xi^{(4)}(\vec{x}_1, \vec{x}_2, \vec{x}_2, \vec{x}_2) + \cdots \right]$$
$$\times \exp \left( - \sum_{n=3}^{\infty} (\nu/\sigma)^n \xi^{(n)}(0)/n! \right),  \tag{2.51}$$

and provides the starting point to compute the smoothed density contrast threshold as shown in the previous section. This expression makes manifest how at least the bivariate joint probability of the curvature perturbation $P[\zeta(\vec{x}_1), \zeta(\vec{x}_2)]$, necessary to calculate the connected $n$-point correlators evaluated in two different points, is needed to properly estimate the threshold to collapse.

A simple picture to understand this result follows. Consider a peak of the curvature perturbation, located at the spatial position $\vec{x} = 0$ where the curvature perturbation has a value $\zeta_0$ (which is a stochastic variable). Then, rotating the coordinate axes to be aligned with the principal axes of length $\lambda_i$ ($i = 1, 2, 3$) of the constant-curvature perturbation ellipsoid, and performing a Taylor expansion up to second-order gives [213]

$$\zeta(r) \simeq \zeta_0 - \frac{1}{2} \sum_{i=1}^{3} \lambda_i x_i^2, \qquad r^2 = \sum_{i=1}^{3} x_i^2.  \tag{2.52}$$

One then finds

$$\Phi_m = -r_m \zeta'(r_m) \simeq 2 \left[ \zeta_0 - \zeta(r_m) \right],  \tag{2.53}$$

which shows that the statistics of the smoothed density contrast, computed in a volume of radius $r_m$, requires the knowledge of the curvature perturbation correlations in two different spatial points.

### 2.1.2 Computation of the abundance

In this section we are going to discuss the computation of the PBH abundance, once the value for the threshold for collapse $\delta_c$ is known. As we have seen in the previous section, its value depends on the characteristics of the collapsing perturbations. In the following we will show how different statistics of the collapsing perturbations result in a different estimate of the PBH formation probability. We will report some of the results shown in Refs [1, 21], where the interested reader can find additional details.

**Gaussian statistics: threshold vs peak theory**

As we mentioned above, in the early radiation-dominated universe, PBHs may form from highly overdense regions where the density contrast becomes of order unity and eventually stops their expansion, leading to their gravitational collapse when they re-enter the horizon. In particular, one has to compute the volume average of the energy density contrast within the scale $r_m$, at which the compaction function is maximum, measured at the horizon crossing time $aHr_m = 1$ and shown in Eq. (2.9). This represents the crucial quantity to determine the abundance of PBHs, that can be simplified as

$$\delta_m = \delta_l - \frac{3}{8}\delta_l^2, \qquad \delta_l = -\frac{4}{3}r_m\zeta'(r_m) = \frac{4}{3}\Phi_m. \tag{2.54}$$

It is manifest that only gradients of the comoving curvature perturbation $\zeta$ may affect the formation of PBHs, given that on super-horizon scales one can always perform a coordinate transformation which shifts it and does not affect the physics. This stresses why the smoothed density contrast, and not the curvature perturbation, should be used in the criterion for the formation of PBHs.

The corresponding abundance of PBHs is then obtained by integrating the probability distribution function of the smoothed density contrast from the threshold $\delta_c$ on

$$\beta = \int_{\delta_c} P(\delta_m) d\delta_m. \tag{2.55}$$

Assuming that the curvature perturbation follows a Gaussian distribution (see next section for extensions), and considering only the first-order term in $\zeta$ in Eq. (2.4), i.e.

$$\delta(r) = -\frac{4}{9a^2H^2}\nabla^2\zeta(r), \tag{2.56}$$

then also the smoothed density contrast obeys Gaussian statistics. Its distribution is then determined by the overdensity variance

$$\sigma_\delta^2 = \int \frac{d^3k}{(2\pi)^3} W^2(k, r_H) P_\delta(k), \tag{2.57}$$

in terms of the window function $W(k, r_H)$ (which we assume to be a top-hat in real space) and the power spectrum of the overdensity in the linear approximation

$$P_\delta(k, t) = \frac{16}{81}\frac{k^4}{a^4H^4}P_\zeta(k), \tag{2.58}$$

as a function of the curvature one $P_\zeta$.

One then distinguishes between two different approaches. The first one is threshold statistics, where the PBH mass fraction $\beta$ comes about when identifying PBHs with regions whose overdensity is above a given threshold. In this case the mass fraction at formation time becomes

$$\beta^{\text{th}} \simeq 2\sqrt{\frac{1}{2\pi\nu_c^2}}e^{-\nu_c^2/2}, \qquad \nu_c = \frac{\delta_c}{\sigma_\delta}, \tag{2.59}$$

where the overall factor 2 is introduced to account for the cloud-in-cloud problem [225].

Alternatively, one can identify PBHs as local maxima of the overdensity field and apply the formalism of peak theory [213] to compute their initial mass fraction. In this case one has [226]

$$\beta^{\text{pk}} \simeq \frac{1}{3\pi} \left( \frac{\langle k^2 \rangle}{3} \right)^{3/2} r_H^3 \left( \nu_{\text{pk}}^2 - 1 \right) e^{-\nu_{\text{pk}}^2/2} \quad \text{with} \quad \langle k^2 \rangle = \frac{1}{\sigma_\delta^2} \int \frac{\mathrm{d}^3 k}{(2\pi)^3} \, k^2 \, P_\delta(k), \tag{2.60}$$

and [222]

$$\nu_{\text{pk}} = \frac{\delta_{\text{pk}}^{\text{c}}}{\sigma_\delta}, \tag{2.61}$$

where $\delta_{\text{pk}}^{\text{c}}$ identifies the critical value of the overdensity at the center of the peak above which the initial perturbation may eventually collapse into a PBH [211, 222]. At the Gaussian level, peak theory gives a PBH abundance which is systematically larger than the one provided by threshold statistics [226]. Furthermore, we stress that a correspondence between peaks in the curvature perturbation and peaks in the density contrast is guaranteed only at linear level [219].

If one goes beyond the linear term in $\zeta$, i.e. accounting for the full non-linear relation between the density contrast and the curvature perturbation, then one can make use of the conservation of the probability to write

$$P(\delta_l)\mathrm{d}\delta_l = P(\delta_m)\mathrm{d}\delta_m, \tag{2.62}$$

such that the linear smoothed density contrast $\delta_l$ is the crucial parameter which we have to determine the probability distribution of, with a corresponding threshold given by

$$\delta_{l,c} = \frac{4}{3} \left( 1 - \sqrt{1 - \frac{3}{2}\delta_c} \right). \tag{2.63}$$

The corresponding computation has been investigated in Refs. [21, 219] assuming a Gaussian curvature perturbation, and we leave the interested reader to those references for details.

These results show that, since PBHs are generated through very large and rare fluctuations, even a tiny change in the overdensity variance is exponentially amplified in the PBH mass fraction, and that the latter strongly depends on any possible non-Gaussianity [106, 227–239]. This implies that non-Gaussianities need to be accounted in the computation of the PBH abundance (an example is provided by the presence of primordial local non-Gaussianity in the comoving curvature perturbation, which can significantly modify the PBH number density through mode coupling [113, 115, 181, 240–242]).

## Non-Gaussian statistics

In this section we are going to generalise the previous results to non-Gaussian curvature perturbations, adopting the path integral technique used in the context of large scale structure [243–246]. In particular, we will outline the dependence of the probability of the smoothed density contrast in terms of higher-order correlators of the curvature perturbation field, see Ref. [1] for details.

As we have seen above, the crucial quantity to compute the probability of is the smoothed linear density contrast $\delta_l$. The joint probability that the linear field $\delta_l(\vec{x})$ at a spatial point $\vec{x}$ attains a value in the range between $\delta_l$ and $\delta_l + \mathrm{d}\delta_l$ is given adopting the path integral approach as

$$P(\delta_l) = \int [\mathcal{D}\zeta(\vec{x})] \, P[\zeta(\vec{x})] \, \delta_D[\delta_l(\vec{x}) - \delta_l]. \tag{2.64}$$

Being

$$\delta_l = -\frac{4}{3} r_m \zeta'(r_m) = -\frac{4}{9a^2 H^2} \frac{3}{R^3} \int \mathrm{d}r \, r^2 \, \theta(R-r)\nabla^2\zeta \bigg|_{R=r_m}, \tag{2.65}$$

we can compute the more general integral

$$P(\alpha) = \int [\mathcal{D}\zeta(\vec{x})] \, P[\zeta(\vec{x})] \, \delta_D[\nabla^2\zeta_R(\vec{x}) - \alpha], \tag{2.66}$$

as a function of the quantity

$$\nabla^2\zeta_R(\vec{x}) \equiv \int \mathrm{d}^3 y \, W_R(|\vec{x}-\vec{y}|)\nabla^2\zeta(\vec{y}) = \frac{3}{R^3} \int \mathrm{d}r \, r^2 \, \theta(R-r)\nabla^2\zeta, \tag{2.67}$$

choosing a top-hat window function in real space $W_R(|\vec{x}-\vec{y}|) = \theta(R-|\vec{x}-\vec{y}|)/V_R$ with smoothing scale $R$ and $|\vec{x}-\vec{y}| = r$. The angular integral in the last step has been calculated assuming isotropy. The functional integration in Eq. (2.66) can then be computed using the integral representation of the Dirac-delta function

$$\delta_D(x) = \int_{-\infty}^{\infty} \frac{\mathrm{d}\lambda}{2\pi} e^{i\lambda x}, \tag{2.68}$$

and the definition of the partition function shown in Eq. (2.40), such that

$$P(\alpha) = \int_{-\infty}^{\infty} \frac{\mathrm{d}\lambda}{2\pi} \int [\mathcal{D}\zeta(\vec{x})] \, P[\zeta(\vec{x})] \exp\left[-i\lambda\alpha + i\lambda \int \mathrm{d}^3 y W_R(|\vec{x}-\vec{y}|)\nabla^2\zeta(\vec{y})\right]. \tag{2.69}$$

One therefore obtains

$$P(\alpha) = \int_{-\infty}^{\infty} \frac{\mathrm{d}\lambda}{2\pi} e^{-i\lambda\alpha} Z[\widetilde{J}], \qquad \widetilde{J}(\vec{y}) = \lambda\nabla_x^2 W_R(|\vec{x}-\vec{y}|)\big|_{\vec{x}=0}. \tag{2.70}$$

The probability $P(\alpha)$ of forming a PBH can then be expressed in terms of the curvature perturbation connected correlation functions shown in Eq. (2.44), by expanding the logarithm of the partition function as

$$\ln Z[\widetilde{J}] = \sum_{n=2}^{\infty} \frac{i^n}{n!} \int \mathrm{d}^3 x_1 \cdots \mathrm{d}^3 x_n \, \xi^{(n)}(\vec{x}_1, \cdots, \vec{x}_n)\widetilde{J}(\vec{x}_1)\cdots\widetilde{J}(\vec{x}_n) = \sum_{n=2}^{\infty} \frac{(i\lambda)^n}{n!}\xi_n(R), \tag{2.71}$$

where

$$\xi_n(R) \equiv \int \left[ \prod_{a=1}^{n} \mathrm{d}^3 x_a \, \nabla_x^2 W_R(|\vec{x}_a - \vec{x}|) \big|_{\vec{x}=0} \right] \xi^{(n)}(\vec{x}_1, \cdots, \vec{x}_n). \tag{2.72}$$

The probability then takes the form

$$P(\alpha) = \frac{1}{\sqrt{2\pi}} \exp \left[ \sum_{n=2} \frac{(-1)^n}{n!} \xi_n(R) \left( \frac{\partial}{\partial \alpha} \right)^n \right] \exp \left[ -\alpha^2/2 \right]. \tag{2.73}$$

For a top-hat window function

$$V_R \nabla_x^2 W_R(|\vec{x}_a - \vec{x}|) |_{\vec{x}=0} = -2 \frac{\delta_D(R - |\vec{x}_a|)}{|\vec{x}_a|} - \frac{\partial \delta_D(R - |\vec{x}_a|)}{\partial |\vec{x}_a|}, \tag{2.74}$$

such that Eq. (2.72) can be recast as (labelling $|\vec{x}_a| = r_a$)

$$\xi_n(R) = -\left( \frac{4\pi}{V_R} \right)^n \int \left[ \prod_{a=1}^{n} \mathrm{d}r_a r_a^2 \left( \frac{2}{r_a} \delta_D(R - r_a) + \frac{\partial \delta_D(R - r_a)}{\partial r_a} \right) \right] \int \frac{\mathrm{d}\Omega_1}{4\pi} \cdots \frac{\mathrm{d}\Omega_n}{4\pi} \xi^{(n)}(\vec{x}_1, \cdots, \vec{x}_n)$$

$$\equiv -\left( \frac{4\pi}{V_R} \right)^n \int \prod_{a=1}^{n} \mathrm{d}r_a r_a^2 \left( \frac{2}{r_a} \delta_D(R - r_a) + \frac{\partial \delta_D(R - r_a)}{\partial r_a} \right) \xi_{\mathrm{av}}^{(n)}(r_1, \cdots, r_n). \tag{2.75}$$

One finally gets the expression

$$\xi_n(R) = \left( \frac{3}{R} \right)^n \prod_{a=1}^{n} \frac{\partial}{\partial r_a} \xi_{\mathrm{av}}^{(n)}(r_1, \cdots, r_n) \bigg|_{r_a = R}. \tag{2.76}$$

From this result one deduces that, in order to properly compute the abundance of PBHs, the knowledge of the $n$-point correlation functions of the curvature field is needed, which can be only determined once the joint probability of the field in different spatial points is known, $P[\zeta(\vec{x}_1), \cdots, \zeta(\vec{x}_n)]$. In particular, realising that the tail of the probability distribution of the curvature perturbation is not Gaussian but, e.g. an exponential function, is not enough to properly determine the formation probability. Indeed, while for a Gaussian-distributed variable the joint probability is a multivariate normal distribution, for other functions the joint probabilities are not unique. Finally, we stress that the knowledge of the first few correlation functions would not be enough to properly estimate the formation probability, as it was shown that a truncation in the perturbative expansion may not converge, as discussed in [21].

In order to overcome this problem, a possible solution could be to know the expression of the non-linear curvature perturbation in terms of a Gaussian component. In this case, one can calculate the joint probability of the curvature perturbation from the multivariate normal distribution of the linear Gaussian component, and then compute the PBH formation probability. In the next section we will show an explicit example where this procedure can be performed.

### 2.1.3 Formation from ultra-slow-roll

In this section we are going to consider one of the most standard scenario for the formation of PBHs, that is through a phase of ultra-slow-roll within single-field models of inflation, in which the curvature perturbation $\zeta$ is boosted from its value $\sim 10^{-9}$ at large scales to $\sim 10^{-2}$ on small scales [247, 248].

Following the discussion of the previous sections, it is clear that one of the main problems in computing the PBH abundance is related to the presence of rare large fluctuations, and it follows that the PBH formation probability strongly depends on changes in the tail of the perturbation distribution and to primordial or intrinsic non-Gaussianities [21, 219, 239, 249, 250].[2] Following Ref. [6], we will provide the exact probability for PBH formation in the ultra-slow-roll inflationary scenario accounting for the effect of non linearities and non-Gaussianities which enter in the model.

**The ultra-slow-roll model**

The ultra-slow-roll model of single-field inflation, also denoted non-attractor scenario, is a mechanism to boost the comoving curvature perturbation during inflation and it is based on the assumption that during its evolution the inflaton field $\phi$ moves through a flat region of its potential $V(\phi)$. The corresponding equation of motion of the inflaton field then reads

$$\phi''(N) + 3\phi'(N) = 0, \ \frac{\mathrm{d}V}{\mathrm{d}\phi} \simeq 0, \tag{2.77}$$

where the prime denotes differentiation with respect to the number of e-folds $N$. Its solution is given by

$$\phi(N) = \phi_e + \frac{\pi_e}{3}\left(1 - e^{-3N}\right), \ \pi(N) \equiv \phi'(N) = \pi_e e^{-3N}, \tag{2.78}$$

in terms of the value of the inflaton field at the end of the ultra-slow-roll phase $\phi_e$, occurring at $N = 0$, such that $N < 0$. One can immediately notice that the slow-roll conditions are badly violated.

To compute the curvature perturbation in this scenario one can adopt the $\delta N$ formalism [255], according to which, on superhorizon scales, each spatial point of the universe is characterised by an independent evolution and the latter is well approximated by the evolution of an unperturbed universe. This implies that, being the scalar field fluctuations quantised on the flat slices, one gets $\zeta = -\delta N$. This approach is particularly useful when one deals with non-linear perturbations. One

---

[2]Several works have been investigating the role of quantum diffusion on the computation of the PBH formation probability in the context of single field models of inflation [251–254], which however we do not discuss in this thesis.

immediately finds that [238, 248]

$$\zeta = -\delta N = -\frac{1}{3}\ln\left(1 + \frac{\delta\pi_e}{\overline{\pi}_e}\right), \tag{2.79}$$

where the overlines correspond to the background values. By using the relation for the inflaton velocity

$$\pi_e = 3\left[\phi(N) - \phi_e\right] + \pi(N), \tag{2.80}$$

one gets, up to irrelevant constants,

$$\zeta = -\frac{1}{3}\ln\left(1 + 3\frac{\delta\phi}{\overline{\pi}_e}\right). \tag{2.81}$$

This non-linear and simple expression highlights the enhancement of the curvature perturbation because of the exponential decrease of the inflaton velocity during the ultra-slow-roll phase.

Given that the dynamics of the inflaton perturbations $\delta\phi$ is the one of a massless perturbation in de Sitter, to a very good approximation they follow a Gaussian distribution

$$P(\delta\phi) = \frac{1}{\sqrt{2\pi}\,\sigma_{\delta\phi}}e^{-(\delta\phi)^2/2\sigma_{\delta\phi}^2}, \tag{2.82}$$

in terms of the variance

$$\sigma_{\delta\phi}^2 = \int \mathrm{d}\ln k\, \mathcal{P}_{\delta\phi}(k) \tag{2.83}$$

and the inflaton perturbation power spectrum $\mathcal{P}_{\delta\phi}(k)$. Conservation of probability then implies [1]

$$P(\zeta) = P[\delta\phi(\zeta)]\left|\frac{\mathrm{d}\delta\phi}{\mathrm{d}\zeta}\right| = \frac{\overline{\pi}_e}{\sqrt{2\pi}\sigma_{\delta\phi}}\exp\left[-\frac{1}{2\sigma_{\delta\phi}^2}\left(\frac{\overline{\pi}_e^2}{9}\left(e^{-3\zeta}-1\right)^2\right) - 3\zeta\right], \tag{2.84}$$

from which it is manifest the non-Gaussian nature of the curvature perturbation due to the non-linear mapping between $\delta\phi$ and $\zeta$. In the limit of small fluctuations $\zeta \ll 1$ one recovers a Gaussian distribution, while for large values of the curvature perturbation one gets

$$P(\zeta) \simeq \frac{\overline{\pi}_e}{\sqrt{2\pi}\,\sigma_{\delta\phi}}e^{-3\zeta}, \tag{2.85}$$

which shows that its probability has a non-Gaussian exponential tail. The same result was also found in Refs. [251, 254, 256] using the stochastic approach, under the realistic limit of large field displacement during the non-attractor phase in units of the Hubble rate. It is also valid in the limit of rapid transition into the subsequent slow-roll phase [248] and for a constant potential.

**The exact formation probability**

As discussed in the previous section, the crucial quantity of which we have to compute the probability of is the smoothed density contrast at the horizon crossing, which takes the non-linear expression shown in Eq. (2.54). In the following we will make the simplifying assumption of monochromatic curvature perturbation power spectrum, for which the critical threshold reads $\delta_c \simeq 0.59$ [11, 211].

From the non-linear relation between the curvature perturbation and the inflaton fluctuations one gets

$$\delta_l = -\frac{4}{3}r_m \zeta'(r_m) = \frac{4}{3}r_m \frac{\delta\phi'(r_m)/\overline{\pi}_e}{1 + 3\,\delta\phi(r_m)/\overline{\pi}_e}. \tag{2.86}$$

Given that $\delta_l$ is the ratio of two normally distributed and uncorrelated random variables

$$X = \frac{4}{3}r_m \frac{\delta\phi'(r_m)}{\overline{\pi}_e},$$
$$Y = 1 + 3\frac{\delta\phi(r_m)}{\overline{\pi}_e}, \tag{2.87}$$

one can write its probability as the ratio distribution

$$P(Z) \equiv P(X/Y) = \int \mathrm{d}Y'\,|Y'|P(ZY',Y'),$$
$$P(X,Y) = \frac{1}{2\pi\sqrt{\det C}}\exp\left(-\vec{V}^T C^{-1}\vec{V}/2\right),$$
$$\vec{V}^T = (X, Y-1),\ C = \mathrm{diag}\left(\sigma_X^2, \sigma_Y^2\right), \tag{2.88}$$

in terms of the effective variances

$$\sigma_X^2 = \frac{16r_m^2}{9\overline{\pi}_e^2}\int \mathrm{d}\ln k\, k^2\, \mathcal{P}_{\delta\phi}(k), \qquad \sigma_Y^2 = \frac{9}{\overline{\pi}_e^2}\sigma_{\delta\phi}^2, \tag{2.89}$$

and using isotropy argument to deduce that the cross correlation between $X$ and $Y$ is zero. We finally obtain

$$P(\delta_l) = \frac{e^{-1/2\sigma_Y^2}}{2\pi(\sigma_X^2 + \sigma_Y^2\delta_l^2)^{3/2}}\sigma_X \cdot \left[2\sigma_Y\sqrt{\sigma_X^2 + \sigma_Y^2\delta_l^2}\right.$$
$$\left. + \sqrt{2\pi}\sigma_X e^{\sigma_X^2/(2\sigma_X^2\sigma_Y^2 + 2\sigma_Y^4\delta_l^2)}\mathrm{Erf}\left(\frac{\sigma_X}{\sqrt{2}\sigma_Y\sqrt{\sigma_X^2 + \sigma_Y^2\delta_l^2}}\right)\right]. \tag{2.90}$$

This exact expression reduces to a simple Gaussian distribution in the limit of small inflaton perturbations $\sigma_Y\delta_l \ll 1$,

$$P(\delta_l) \simeq \frac{e^{-\delta_l^2/2\sigma_X^2}}{\sqrt{2\pi}\sigma_X}, \tag{2.91}$$

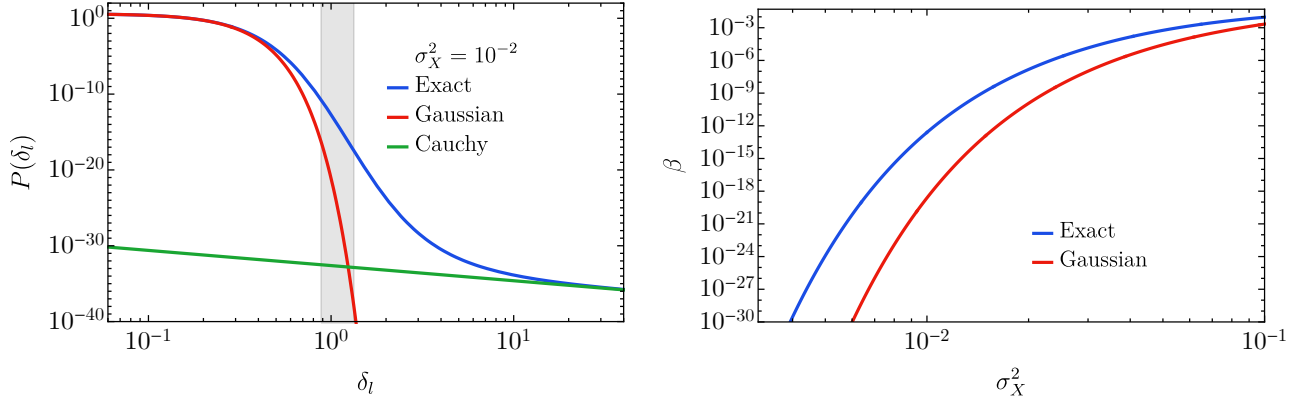

**Figure 2.3**: *Left:* Probability distribution of the linear density contrast for fixed variance $\sigma_X^2$ and for a monochromatic power spectrum of inflaton fluctuations. The blue line denotes the exact result, while the red and green lines indicate the limit for small and large linear density perturbations $\delta_l$, respectively. The vertical gray band indicates the relevant range for PBH formation. *Right:* The corresponding PBH mass fraction in terms of the variance $\sigma_X^2$ for the choices of Gaussian and exact probability distribution of the density field. Figure taken from Ref. [6].

while for large density contrast fluctuations $\sigma_Y \delta_l \gg 1$ we get

$$P(\delta_l) \simeq e^{-1/2\sigma_Y^2} \frac{\sigma_X}{\pi \sigma_Y \delta_l^2}, \tag{2.92}$$

that describes a Cauchy probability distribution.

In Fig. 2.3 we show the full probability for the choice of monochromatic power spectrum of the inflaton fluctuations $\mathcal{P}_{\delta\phi}(k) = Ak_\star \delta_D(k - k_\star)$ peaked at a momentum scale $k_\star$ such that $r_m k_\star \simeq 3$ [211]. For this choice one has $\sigma_Y^2 \simeq (9/16)\sigma_X^2$. From the figure one concludes that, in the relevant range of values of $\delta_l$, the probability is neither a Gaussian nor a Cauchy distribution.

From the probability function, one can compute the PBH mass fraction (taking into account that $\delta_m < 2/3$ [211])

$$\beta(M_H) = \int_{\delta_c}^{2/3} \mathrm{d}\delta_m \frac{M_{\text{PBH}}}{M_H} P(\delta_m) = \int_{\delta_{l,c}}^{4/3} \mathrm{d}\delta_l \frac{M_{\text{PBH}}}{M_H} P(\delta_l), \tag{2.93}$$

where we include the relation between the horizon mass $M_H$ at formation and the PBH mass for overdensities close to the critical threshold as shown in Eq. (2.12) [219, 257–259]. The result is plotted in the right panel of Fig. 2.3 for both the Gaussian limit and the exact behaviour of the density probability function. One therefore concludes that the final abundance is enhanced compared to the Gaussian case by various orders of magnitude.

## 2.2   The PBH initial spin

In this section we are going to compute the initial spin of PBHs born out of the collapse of large overdensities during the radiation-dominated era, closely following the results shown in Ref. [22]. The same topic has been investigated also by Refs. [260–262].

Given that PBHs originate from local maxima of the radiation density field, we will use the standard results of peak theory [213], and the fact that peak profiles of the density field may deviate from spherical symmetry, to show that the coupling of zeroth-order anisotropies of the collapsing object with surrounding tidal gravitational fields at first-order can generate a torque on the collapsing peak, which will result into a spinning PBH. This approach is similar to the "tidal-torque theory" used in the large-scale structure literature [263–267] to compute the spin of halos from the collapse of ellipsoidal perturbations under gravity [268–272], assuming a Lagrangian sphere to describe the collapsing object [264, 273]. For PBHs, the shape of the Lagrangian region is imprinted when the perturbation's characteristic length scale is super-horizon.

In the homogeneous ellipsoid approximation, we will see that the initial PBH spin does not vanish at first-order in perturbation theory as long as the lengths of the principal axes of the inertia tensor are different and the off-diagonal components of the velocity shear do not vanish. Even though these conditions are not satisfied on super-horizon scales, a torque is generated at the horizon re-entry until the perturbation decouples from the background, i.e. until turnaround. Given the fast timescales of the collapse in a radiation-dominated era, a small non-zero spin is generated. After turnaround, we assume that the angular momentum remains constant, as it happens for non-relativistic dust [264, 265, 273], even though a dedicated numerical simulation has to be performed to investigate this issue. We will show that the PBH Kerr parameter takes typical values of (in this section we use the notation $a_{\rm s}$ to indicate the Kerr parameter $\chi = S/GM^2$ to be consistent with Ref. [22])

$$a_{\rm s} = \frac{1}{2\pi} \sigma_\delta \sqrt{1-\gamma^2} \sim 10^{-2} \sqrt{1-\gamma^2}, \tag{2.94}$$

in terms of the overdensity variance at horizon crossing $\sigma_\delta^2$ and the shape parameter of the curvature power spectrum $\gamma$, which tends to unity for a monochromatic power spectrum (in this limit being the velocity shear aligned with the inertia tensor).

### 2.2.1   The angular momentum of PBHs as local density maxima

In general relativity the angular momentum, being a conserved quantity originated from rotational invariance, may be defined using the formalism introduced by Komar [274], which consists of taking flux integrals of the derivative of the rotational Killing vector $\phi^\nu$ over closed two-surfaces surrounding the matter sources. The amplitude of the Komar angular momentum on a given time-slicing $\Sigma$ is then defined to be [275]

$$S(\Sigma) = \frac{1}{16\pi G}\int_{\partial\Sigma} \mathrm{d}S_{\mu\nu}D^\mu\phi^\nu = \int_\Sigma \mathrm{d}S_\mu J^\mu(\phi), \qquad (2.95)$$

where in the second equality we have introduced $J^\mu(\phi) = T^\mu_\nu \phi^\nu - T\phi^\mu/2$ in terms of the energy-momentum tensor. For a relativistic perfect fluid one gets

$$S(\Sigma) = \int_\Sigma \mathrm{d}V\, T^0_\mu \phi^\mu = \frac{4}{3}\int_\Sigma \mathrm{d}V \rho\, \vec{v}\cdot\vec{\phi}, \qquad (2.96)$$

in terms of the velocity field $\vec{v}$ and the density field $\rho$. Given that spinning PBHs may originate from the collapse of local maxima of the overdensity field, one can expand around the peak located at $\vec{x}_{\mathrm{pk}}$ to obtain

$$S_i = \frac{4}{3}a^4(\eta)\epsilon_{ijk}\int_{V_e} \mathrm{d}^3 x\, \sqrt{\gamma}\,\rho(\vec{x},\eta)(x-x_{\mathrm{pk}})^j(v-v_{\mathrm{pk}})^k, \qquad (2.97)$$

where we have adopted conformal time $\eta$, $\vec{x}$ is a comoving coordinate and we have introduced the determinant of the induced metric $\gamma$. In the following we will focus on the computation of the PBH spin at first-order in perturbation theory, given that the velocity is a first-order quantity. The 3-dimensional volume $V_e$ over which we perform the integral can be described using the triaxial ellipsoid approximation given that it can be significantly aspherical. To do so, we use the results of peak theory [213] and expand the overdensity field around the peak as

$$\delta(\vec{x}) = \frac{\delta\rho(\vec{x})}{\bar{\rho}} \simeq \delta_{\mathrm{pk}} + \frac{1}{2}\zeta_{ij}(x-x_{\mathrm{pk}})^i(x-x_{\mathrm{pk}})^j > f\delta_{\mathrm{pk}}, \qquad \zeta_{ij} = \frac{\partial^2\delta}{\partial x^i \partial x^j}\Big|_{\mathrm{pk}}, \qquad (2.98)$$

where the first gradient term is zero given that we are considering maxima of the overdensity, and we have introduced the parameter $f$ to indicate the threshold above which the matter near the peak can collapse into a PBH. Performing a rotation of the coordinate axes to be aligned with the principal axes of length $\lambda_i$ of the constant-overdensity ellipsoids gives

$$\delta \simeq \delta_{\mathrm{pk}} - \frac{1}{2}\sigma_\zeta \sum_{i=1}^3 \lambda_i(x^i - x^i_{\mathrm{pk}})^2. \qquad (2.99)$$

in terms of the characteristic root-mean-square variance $\sigma_\zeta$ of the components of $\zeta_{ij}$. The boundaries of the integration volume $V_e$ can then be found by solving

$$2\frac{\sigma_\delta}{\sigma_\zeta}(1-f)\nu = \sum_{i=1}^3 \lambda_i(x^i - x^i_{\mathrm{pk}})^2, \qquad (2.100)$$

where we have focused on the ellipsoidal surface $\delta = f\delta_{\mathrm{pk}}$ and introduced the peak height $\nu = \delta_{\mathrm{pk}}/\sigma_\delta$ in units of the rms overdensity $\sigma_\delta$. The principal semi-axes of such an ellipsoid are therefore

$$a_i^2 = 2\frac{\sigma_\delta}{\sigma_\zeta}\frac{(1-f)}{\lambda_i}\nu, \qquad (2.101)$$

which reduce in the large $\nu$ limit to $a_i^2 \sim 6\sigma_\delta^2/\sigma_\times^2$, in terms of the characteristic cross-correlation $\sigma_\times^2$ between $\delta$ and $\zeta_{ij}$, being [213]

$$\lambda_i = \frac{\gamma\nu}{3}\left(1 + \epsilon_i\right), \;\; \epsilon_i = \mathcal{O}\left(\frac{1}{\gamma\nu}\right) \;\; \text{and} \;\; \gamma = \frac{\sigma_\times^2}{\sigma_\delta\sigma_\zeta}. \tag{2.102}$$

The difference $|\lambda_i - \lambda_j|$ is of order unity because the "ellipticity" $\epsilon_i$ scales like $1/(\gamma\nu)$.

The velocity field can also be expanded around the peak

$$(v - v_{\text{pk}})^k = v_l^k(x - x_{\text{pk}})^l, \qquad v_l^k = \left.\frac{\partial v^k}{\partial x^l}\right|_{\text{pk}}, \tag{2.103}$$

to get

$$S_i = \frac{4}{3}a^4(\eta)\epsilon_{ijk}\overline{\rho}_{\text{rad}}(\eta)v_l^k \int_{V_e} \text{d}^3x(x - x_{\text{pk}})^j(x - x_{\text{pk}})^l. \tag{2.104}$$

where we have expanded at zeroth-order the density field and defined the average density of the universe $\overline{\rho}_{\text{rad}}(\eta)$. Even though the volume $V_e$ is gradually deformed by tidal forces as the perturbation collapses to a BH, its deformation becomes significant only after the turnaround. Before this time, we can therefore approximate it as time independent and bounded by the isodensity surface $\delta = f\delta_{\text{pk}}$. The integral then becomes

$$S_i = \frac{4}{3}a^4(\eta)\epsilon_{ijk}\overline{\rho}_{\text{rad}}(\eta)g_v(\eta)\widetilde{v}_l^k \int_{V_e} \text{d}^3x \; x^j x^l, \tag{2.105}$$

where we have chosen $\vec{x}_{\text{pk}} = 0$ and introduced the normalised velocity shear $\widetilde{v}_{kl}$ by factorising out the time-dependence of $v_l^k$ as

$$v_l^k(\eta) = g_v(\eta)\widetilde{v}_l^k. \tag{2.106}$$

Perform a change of coordinates to ellipsoidal ones as

$$\int_{V_e} \text{d}^3x = a_1a_2a_3 \int_0^1 r^2\text{d}r \int_0^{2\pi} \text{d}\phi \int_0^\pi \text{d}\theta \sin\theta, \tag{2.107}$$

where $x_1 = a_1r\cos\phi\sin\theta$, $x_2 = a_2r\sin\phi\sin\theta$ and $x_3 = a_3r\cos\theta$, one gets

$$\int_{V_e} \text{d}^3x \; x^j x^l = \frac{4\pi}{15}a_1a_2a_3 \begin{bmatrix} a_1^2 & 0 & 0 \\ 0 & a_2^2 & 0 \\ 0 & 0 & a_3^2 \end{bmatrix}_{jl}, \tag{2.108}$$

from which one deduces [266]

$$\vec{S}^{(1)} = \frac{16\pi}{45}a^4(\eta)\overline{\rho}_{\text{rad}}(\eta)g_v(\eta)a_1a_2a_3([a_2^2 - a_3^2]\widetilde{v}_{23}, [a_3^2 - a_1^2]\widetilde{v}_{13}, [a_1^2 - a_2^2]\widetilde{v}_{12}). \tag{2.109}$$

This equation highlights that the spin does not vanish at first-order only for different lengths of the semi-axes and non-zero values of the off-diagonal components of the velocity shear, misaligned with the inertia tensor. Using Eq. (2.101), it can be recast as

$$\vec{S}^{(1)} = \left[ \frac{4}{3} a^4(\eta) \overline{\rho}_{\mathrm{rad}}(\eta) g_v(\eta) (1-f)^{5/2} R_*^5 \right] \frac{16\sqrt{2}\pi}{135\sqrt{3}} \left( \frac{\nu}{\gamma} \right)^{\frac{5}{2}} \frac{1}{\sqrt{\Lambda}} \left( -\alpha_1 \widetilde{v}_{23}, \alpha_2 \widetilde{v}_{13}, -\alpha_3 \widetilde{v}_{12} \right), \quad (2.110)$$

where we have introduced $\Lambda = \lambda_1 \lambda_2 \lambda_3$, $R_* = \sqrt{3}\sigma_\times / \sigma_\zeta$ and

$$\alpha_1 = \frac{1}{\lambda_3} - \frac{1}{\lambda_2}, \quad \alpha_2 = \frac{1}{\lambda_3} - \frac{1}{\lambda_1}, \quad \alpha_3 = \frac{1}{\lambda_2} - \frac{1}{\lambda_1}, \quad (2.111)$$

such that $\alpha_i \geq 0$, with $\alpha_2 \geq \alpha_1, \alpha_3$ due to the ordering choice $\lambda_1 > \lambda_2 > \lambda_3$. The magnitude of the dimensionless angular momentum can finally be written as

$$S^{(1)}(\eta) = S_{\mathrm{ref}}(\eta) s_{\mathrm{e}}^{(1)}, \quad (2.112)$$

where the reference spin

$$S_{\mathrm{ref}}(\eta) = \frac{4}{3} a^4(\eta) \overline{\rho}_{\mathrm{rad}}(\eta) g_v(\eta) R_*^5 (1-f)^{5/2}, \quad (2.113)$$

identifies the term inside the square brackets and can be identified as the time-dependent part of the spin, being the same for all peaks. The shape and the height of the individual peak are accounted in the remaining term. Given that $\alpha_i \sim |\lambda_j - \lambda_k| / \lambda_j \lambda_k \sim 1/(\gamma\nu)^2$, $s_{\mathrm{e}}^{(1)}$ is a quantity of order $\mathcal{O}(1)$, since the only first order quantity appearing in this calculation is the velocity shear $v_l^k$.

## 2.2.2 The PBH initial Kerr parameter

In this subsection we will discuss the two relevant regimes for the collapse of density perturbations, which are those corresponding to super-horizon and sub-horizon scales, and then provide an estimate for the value of the PBH spin at formation time.

**Before horizon crossing**

During the radiation-dominated era PBHs may be formed from the collapse of sizeable overdense regions. Before the time of collapse, the comoving sizes of such regions are larger than the horizon length and one can apply the separate universe approach [218] for which observables can be expanded at leading order in spatial gradients. At this stage, one should fix the slicing and the threading of the spacetime manifold, and the most appropriate choice is provided by the constant mean curvature slicing (CMC), which is often adopted to perform numerical relativity simulations for PBH formation [207]. In this slicing the overdensity can be written in the non-perturbative expression [218]

$$\delta(\vec{x}, \eta) = -\frac{4}{3} e^{-5\mathcal{R}(\vec{x})/2} \frac{\nabla^2 e^{\mathcal{R}(\vec{x})/2}}{\mathcal{H}^2}, \tag{2.114}$$

in terms of the comoving curvature perturbation $\mathcal{R}(\vec{x})$ and the Hubble parameter in conformal time $\mathcal{H}$. The overall coefficient is $3/2$ times larger than the one used in the comoving slicing (CG), and also the PBH threshold corresponds to a larger value $\delta_{\text{cmc}}^{\text{c}} = 3\delta_{\text{CG}}^{\text{c}}/2$. In the following we assume the reference value $\delta_{\text{CG}}^{\text{c}} = 0.45$, corresponding to $\delta_{\text{cmc}}^{\text{c}} = 0.675$ [218], neglecting the shape dependence of the threshold [211, 222, 249].

As the universe expands, the density contrast grows and regions where it reaches unity may eventually stop expanding and collapse, with comoving scales of the order of the horizon scale. Even though at this point the gradient expansion breaks down, one can still deduce a criterion for PBH formation, see Ref. [226] for related nonlinear numerical studies. One finds that the angular momentum keeps growing until the system turns around and torques are decreased.

In the threading where the perturbed $_{0i}$ component of the metric vanishes, the velocity is given, in the long wavelength regime, by [218]

$$v^i(\vec{x}, \eta) = \frac{1}{12\mathcal{H}} \partial^i \delta(\vec{x}, \eta), \tag{2.115}$$

which scales like $a^3$ on superhorizon scales. Given its proportionality to gradients of the overdensity field, one can immediately conclude that the off-diagonal components of the matrix $\tilde{v}_{ij}$ are vanishing and therefore that, before horizon-crossing, the PBH angular momentum $\vec{S}^{(1)}$ is zero. Physically, this implies that the principal axes of the inertia tensor are aligned with the velocity shear $v_l^k$, and no spin can be generated, independently from asphericities. This conclusion holds also for any particular realization of a density peak, regardless of the probability density of the random variables $\delta$ and $v^i$.

**After horizon crossing**

When the perturbation's characteristic wavelength becomes smaller than the Hubble horizon scale, the relation between the velocity and the overdensity changes. This implies that the spin can increase due to linear tidal torques until the perturbation decouples from the background. From the free-fall and sound crossing timescales one deduces the Jeans criterion, for which a gravitational instability occurs for perturbations with comoving wavenumber

$$\frac{k}{\mathcal{H}} < \sqrt{\frac{2\pi}{3}} \frac{(1 + \delta_{\text{pk}})^{1/2}}{c_{\text{S}}} \approx \frac{2}{c_{\text{S}}}, \tag{2.116}$$

in terms of the sound speed $c_{\text{S}} \simeq 1/\sqrt{3}$ in a radiation dominated-era and assuming $\delta_{\text{pk}} \sim 1$. When this happens, the perturbations decouple from the background around horizon crossing and eventually form PBHs (otherwise the presence of radiation pressure would stabilise them). In the following we will assume that turnaround occurs around the horizon crossing time, such that one can estimate the amount of angular momentum acquired by the perturbation through linear tidal-torque.

Given that we are considering PBH formation during a radiation dominated-era, we can focus on perturbations with characteristic comoving scales at horizon crossing $k_{\rm H} \gg k_{\rm eq}$, where $k_{\rm eq}$ identifies the mode re-entering the Hubble horizon at matter-radiation equality, Given that the mean free path is very short, one can ignore any anisotropic stress and assume the Bardeen potentials to be equal $\Phi = \Psi$. Under these assumptions, the radiation overdensity is given by [276]

$$\delta(\vec{k}, \eta) \simeq -6\Phi(\vec{k}, 0) \cos(kc_{\rm s}\eta) + 4\Phi(\vec{k}, \eta), \tag{2.117}$$

where

$$\Phi(\vec{k}, \eta) = 3\Phi(\vec{k}, 0) \, \frac{\sin(kc_{\rm s}\eta) - (kc_{\rm s}\eta)\cos(kc_{\rm s}\eta)}{(kc_{\rm s}\eta)^3}. \tag{2.118}$$

The corresponding Fourier modes of the radiation velocity field are instead

$$v^i(\vec{k}, \eta) = i\frac{9}{2}\frac{k^i}{k}\Phi(\vec{k}, 0) \, c_{\rm s} \sin(kc_{\rm s}\eta), \tag{2.119}$$

such that the velocity shear at horizon crossing time $\eta_{\rm H}$ is given by

$$v^j_i(\vec{k}, \eta_{\rm H}) = ik_i v^j(\vec{k}, \eta_{\rm H}) = -\frac{9}{2}\frac{k_i k^j}{k}\Phi(\vec{k}, 0)c_{\rm s}\sin(k/k_{\rm s}), \tag{2.120}$$

in terms of $k_{\rm s} \equiv k_{\rm H}/c_{\rm s}$. In the CMC gauge, the density perturbation and the velocity shear take the final form

$$\delta_{\rm cmc}(\vec{k}, \eta_{\rm H}) = -6\Phi(\vec{k}, 0)\frac{[2\,(3c_{\rm s}^2 + 1) + (k/k_{\rm s})^2]\cos{(k/k_{\rm s})} - 2\,(3c_{\rm s}^2 + 1)\,(k_{\rm s}/k)\sin{(k/k_{\rm s})}}{6c_{\rm s}^2 + (k/k_{\rm s})^2},$$

$$v^j_{{\rm cmc}i}(\vec{k}, \eta_{\rm H}) = -\frac{9}{2}\Phi(\vec{k}, 0)\frac{k_i k^j}{k}c_{\rm s}\left(\frac{k}{k_{\rm s}}\right)^2\frac{\sin{(k/k_{\rm s})}}{6c_{\rm s}^2 + (k/k_{\rm s})^2}. \tag{2.121}$$

One can now introduce the normalised density $\nu$, shear $\widetilde{v}_{ij}$ and Hessian $\zeta_{ij}$ to keep the analogy with Ref. [266] as

$$\sigma_{\delta_{\rm cmc}}\nu(\vec{x}, \eta_{\rm H}) = \frac{V}{(2\pi)^3}\int {\rm d}^3k\,\left(\frac{k}{k_{\rm s}}\right)^2 T_{\delta}(k, \eta_{\rm H})\,\Phi(\vec{k}, 0)\,W(k)\,e^{i\vec{k}\cdot\vec{x}},$$

$$\zeta_{{\rm cmc}ij}(\vec{x}, \eta_{\rm H}) = -\frac{V}{(2\pi)^3}\int {\rm d}^3k\,k_i k_j\,\delta_{\rm cmc}(\vec{k}, \eta_{\rm H})\,W(k)\,e^{i\vec{k}\cdot\vec{x}},$$

$$v^j_{{\rm cmc}i}(\vec{x}, \eta_{\rm H}) = -k_{\rm H}\frac{V}{(2\pi)^3}\int {\rm d}^3k\,\frac{k_i k^j}{k^2}\frac{T_v(k, \eta_{\rm H})}{T_{\delta}(k, \eta_{\rm H})}\,\delta_{\rm cmc}(\vec{k}, \eta_{\rm H})\,W(k)\,e^{i\vec{k}\cdot\vec{x}} \equiv g_v(\eta_{\rm H})\widetilde{v}^j_{{\rm cmc}i}(\vec{x}, \eta_{\rm H}), \tag{2.122}$$

in terms of the spherically symmetric window function $W(k)$ with smoothing scale $R = 1/k_{\mathrm{H}}$ and transfer functions

$$T_\delta(k, \eta_{\mathrm{H}}) = -6 \left(\frac{k_{\mathrm{S}}}{k}\right)^2 \frac{[2\,(3c_{\mathrm{S}}^2 + 1) + (k/k_{\mathrm{S}})^2]\cos{(k/k_{\mathrm{S}})} - 2\,(3c_{\mathrm{S}}^2 + 1)\,(k_{\mathrm{S}}/k)\sin{(k/k_{\mathrm{S}})}}{6c_{\mathrm{S}}^2 + (k/k_{\mathrm{S}})^2},$$

$$T_v(k, \eta_{\mathrm{H}}) = \frac{9}{2} \left(\frac{k}{k_{\mathrm{S}}}\right) \frac{\sin{(k/k_{\mathrm{S}})}}{6c_{\mathrm{S}}^2 + (k/k_{\mathrm{S}})^2}. \tag{2.123}$$

As one can appreciate from Eq. (2.122), in the sub-horizon regime the velocity shear is not proportional to gradients of the density contrast. From the spectral moments

$$\sigma_j^2 \equiv \frac{V}{2\pi^2} \int \mathrm{d}k\, k^{2+2j} \left|\delta_{\mathrm{cmc}}(\vec{k}, \eta_{\mathrm{H}})\right|^2 W^2(k), \tag{2.124}$$

in terms of a Fourier volume $V$, one can furthermore define the root-mean-square variances $\sigma_{\delta_{\mathrm{cmc}}} = \sigma_0$, $\sigma_{\zeta_{\mathrm{cmc}}} = \sigma_2$ and $\sigma_{\times_{\mathrm{cmc}}} = \sigma_1$. Finally, the time dependence of the velocity shear is captured by

$$g_v^2(\eta_{\mathrm{H}}) = k_{\mathrm{H}}^2 \frac{V}{2\pi^2} \int \mathrm{d}k\, k^2 \frac{T_v^2(k, \eta_{\mathrm{H}})}{T_\delta^2(k, \eta_{\mathrm{H}})} \left|\delta_{\mathrm{cmc}}(\vec{k}, \eta_{\mathrm{H}})\right|^2 W^2(k) \sim \frac{T_v^2(k_{\mathrm{H}}, \eta_{\mathrm{H}})}{T_\delta^2(k_{\mathrm{H}}, \eta_{\mathrm{H}})} k_{\mathrm{H}}^2 \sigma_{\delta_{\mathrm{cmc}}}^2(\eta_{\mathrm{H}}), \tag{2.125}$$

where in the last step we have approximated the contribution of the transfer functions to a constant given that the power spectrum is peaked at $k = k_{\mathrm{H}}$, with the numerical value to be $T_v(k_{\mathrm{H}}, \eta_{\mathrm{H}})/T_\delta(k_{\mathrm{H}}, \eta_{\mathrm{H}}) \sim 0.5$. Given that in this regime $\widetilde{v}_{\mathrm{cmc}ij}$ does not vanish and is not aligned with the Hessian of $\delta_{\mathrm{cmc}}$, some spin is generated at linear order, and the corresponding reference spin at turnaround is finally given by

$$S_{\mathrm{ref}}(\eta_{\mathrm{H}}) = \frac{4}{3} a^4(\eta_{\mathrm{H}}) \overline{\rho}_{\mathrm{rad}}(\eta_{\mathrm{H}}) g_v(\eta_{\mathrm{H}}) R_*^5 (1-f)^{5/2}. \tag{2.126}$$

**Estimate of the natal Kerr parameter**

At the horizon crossing time $\eta_{\mathrm{H}}$, the PBH Kerr parameter is given by

$$a_{\mathrm{s}} = \frac{S}{GM^2} = \frac{S_{\mathrm{ref}}(\eta_{\mathrm{H}})}{GM^2} s_{\mathrm{e}} \equiv A(\eta_{\mathrm{H}}) s_{\mathrm{e}}, \tag{2.127}$$

where

$$A(\eta_{\mathrm{H}}) = \frac{4}{3} \frac{a^4(\eta_{\mathrm{H}}) g_v(\eta_{\mathrm{H}}) \overline{\rho}_{\mathrm{rad}}(\eta_{\mathrm{H}}) R_*^5 (1-f)^{5/2}}{GM^2}. \tag{2.128}$$

The threshold parameter $f$ can be estimated by making the simplifying assumption of a power spectrum peaked at the scale $R_* \sim \sqrt{3}k_{\mathrm{H}}^{-1} \sim \sqrt{3}\mathcal{H}^{-1}(\eta_{\mathrm{H}})$ [211], such that in the CMC gauge $(1-f) \sim 1/3$. After some simplifications, one can deduce an estimate for the natal Kerr parameter as

$$a_{\mathrm{s}} = A(\eta_{\mathrm{H}}) s_{\mathrm{e}} \sim \left[\frac{1}{2\pi} \sigma_{\delta_{\mathrm{cmc}}}(\eta_{\mathrm{H}})\right] s_{\mathrm{e}}. \tag{2.129}$$

Assuming $\sigma_{\delta_{\text{cmc}}}(\eta_{\text{H}}) = \delta^{\text{c}}_{\text{cmc}}/\nu \sim 0.08$ for the indicative value $\nu = 8$, one gets a numerical estimate of $a_{\text{s}} \sim 10^{-2} s_{\text{e}}$.

The next parameter to be estimated is $s_{\text{e}}$, whose distribution dictates the probability distribution of the Kerr parameter at formation. One expects its value to be of the order of (for some $i \neq j$)

$$s_{\text{e}} \simeq \frac{16\sqrt{2}\pi}{135\sqrt{3}\Lambda}\nu^{5/2}\widetilde{v}_{ij}\alpha_i = \mathcal{O}(1) \cdot \sqrt{1 - \gamma^2}, \tag{2.130}$$

where we have introduced the scaling $\sqrt{1 - \gamma^2}$ of the velocity shear, as we will see in the following, such that the Kerr parameter is estimated to be

$$a_{\text{s}} \sim 10^{-2}\sqrt{1 - \gamma^2}. \tag{2.131}$$

In the limit of very narrow power spectra, for which $\gamma$ tends to unity, the spin vanishes. This happens because of the alignment between the velocity shear and the principal axes of the mass ellipsoid, due to the correlation with the inertia tensor. In the limit $\gamma = 1$ this alignment is total and no spin can be generated.

## 2.2.3 On the statistics of local maxima for the PBH spin

In this subsection we estimate the probability distribution of the parameter $s_{\text{e}}$ from the distribution of the relevant quantities entering in its definition, and then deduce the Kerr parameter initial distribution.

Starting from the sixteen variables (we drop the label CMC from now on)

$$\delta, \quad \zeta_i = \frac{\partial\delta}{\partial x_i}, \quad \zeta_{ij} = \frac{\partial^2\delta}{\partial x_i \partial x_j}, \quad v_i^j = \frac{\partial v^j}{\partial x^i}, \tag{2.132}$$

where the last two matrices are symmetric, we can define a vector $V$ of these sixteen components and write down its joint distribution as [266]

$$f(V_i)\mathrm{d}^{16}V_i = \frac{1}{(2\pi)^8|\mathbf{M}|^{1/2}}e^{-\frac{1}{2}(V_i - \langle V_i\rangle)\mathbf{M}^{-1}_{ij}(V_j - \langle V_j\rangle)}\mathrm{d}^{16}V_i, \tag{2.133}$$

where the covariance matrix $\mathbf{M}$ is given by

$$\mathbf{M}_{ij} = \langle(V_i - \langle V_i\rangle)(V_j - \langle V_j\rangle)\rangle. \tag{2.134}$$

We can rearrange the variables in the dimensionless form

$$\nu = \delta/\sigma_\delta,$$

$$x = -(\zeta_{11} + \zeta_{22} + \zeta_{33})/\sigma_\zeta, \quad y = -\frac{1}{2}(\zeta_{11} - \zeta_{33})/\sigma_\zeta, \quad z = -\frac{1}{2}(\zeta_{11} - 2\zeta_{22} + \zeta_{33})/\sigma_\zeta,$$

$$v_A = -(\widetilde{v}_{11} + \widetilde{v}_{22} + \widetilde{v}_{33}), \quad v_B = -\frac{1}{2}(\widetilde{v}_{11} - \widetilde{v}_{33}), \quad v_C = -\frac{1}{2}(\widetilde{v}_{11} - 2\widetilde{v}_{22} + \widetilde{v}_{33}),$$

$$w_1 = \widetilde{v}_{23}, \quad w_2 = \widetilde{v}_{13}, \quad w_3 = \widetilde{v}_{12},$$

$$\widetilde{\zeta}_1 = \zeta_1/\sigma_\times, \quad \widetilde{\zeta}_2 = \zeta_2/\sigma_\times, \quad \widetilde{\zeta}_3 = \zeta_3/\sigma_\times,$$

$$\widetilde{\zeta}_{12} = \zeta_{12}/\sigma_\zeta, \quad \widetilde{\zeta}_{13} = \zeta_{13}/\sigma_\zeta, \quad \widetilde{\zeta}_{23} = \zeta_{23}/\sigma_\zeta,$$

$$(2.135)$$

such that their non-zero correlators are given by (we set the expectation values to $\langle V_i \rangle = 0$)

$$\langle x^2 \rangle = \langle \nu^2 \rangle = \langle v_A^2 \rangle = 3\langle \widetilde{\zeta}_1^2 \rangle = 15\langle \widetilde{\zeta}_{12}^2 \rangle = 15\langle w_3^2 \rangle = \langle v_A \nu \rangle = ... = 1,$$

$$\langle x\nu \rangle = \langle x v_A \rangle = 5\langle v_C z \rangle = 15\langle v_B y \rangle = 15\langle \widetilde{\zeta}_{12} w_3 \rangle = ... = \gamma,$$

$$\langle z^2 \rangle = 3\langle y^2 \rangle = \langle v_C^2 \rangle = 3\langle v_B^2 \rangle = 1/5.$$

$$(2.136)$$

Notice also that the velocity shear is totally aligned with the inertia tensor in the limit $\gamma \to 1$. Given that $\langle v_A \nu \rangle^2 = \langle v_A^2 \rangle \langle \nu^2 \rangle = 1$, the variables $v_A$ and $\nu$ are correlated, and we can therefore reduce the number of independent variables to fifteen, by dropping $v_A$. By changing variables from the six entries of $\zeta_{ij}$ to the three eigenvalues plus the three Euler angles of the matrix $-\zeta_{ij}/\sigma_\zeta$, describing the length and orientation of its principal axes respectively, one can introduce the quantities

$$x = \lambda_1 + \lambda_2 + \lambda_3, \quad y = \frac{1}{2}(\lambda_1 - \lambda_3), \quad z = \frac{1}{2}(\lambda_1 - 2\lambda_2 + \lambda_3), \qquad (2.137)$$

and then integrate over the angles using the system's independence from them, leaving only twelve independent variables. The corresponding distribution is given by

$$f(\nu, \widetilde{\zeta}_i, \lambda_i, v_B, v_C, w_i) = \left( \frac{5^5 3^{11/2}}{2(2\pi)^{11/2}} \Gamma^3 \right) e^{-Q_2} |(\lambda_1 - \lambda_2)(\lambda_2 - \lambda_3)(\lambda_1 - \lambda_3)|, \qquad (2.138)$$

in terms of $\Gamma = 1/(1 - \gamma^2)$ and

$$2Q_2 = \Gamma \nu^2 - 2\gamma \Gamma x\nu + \Gamma x^2 + 15\Gamma y^2 - 30\gamma \Gamma y v_B + 15\Gamma v_B^2 + 5\Gamma z^2 - 10\gamma \Gamma z v_C$$
$$+ 5\Gamma v_C^2 + 15\Gamma(w_1^2 + w_2^2 + w_3^2) + 3(\widetilde{\zeta}_1^2 + \widetilde{\zeta}_2^2 + \widetilde{\zeta}_3^2). \qquad (2.139)$$

By Taylor expanding in the region around the peak

$$\zeta_i = \frac{\partial \delta}{\partial x_i} = \left( \frac{\partial^2 \delta}{\partial x_i \partial x_j} \right) \bigg|_{\text{pk}} (x - x_{\text{pk}})^j = \zeta_{ij}|_{\text{pk}}(x - x_{\text{pk}})^j, \qquad (2.140)$$

such that

$$\mathrm{d}^3 \widetilde{\zeta}_i = \left( \frac{\sigma_\zeta}{\sigma_\times} \right)^3 |\lambda_1 \lambda_2 \lambda_3| \mathrm{d}^3 x_i, \qquad (2.141)$$

and integrating the distribution over the three variables $\widetilde{\zeta}_i$, one has only nine variables. The resulting distribution gives the comoving number density of peaks in the element $\mathrm{d}\nu \mathrm{d}v_B \mathrm{d}v_C \mathrm{d}^3 w_i$, and integrating over the unconstrained variables $v_B$ and $v_C$ gives finally

$$N_{\mathrm{pk}}(\nu, \lambda_i, w_i)\mathrm{d}\nu \mathrm{d}^3\lambda_i \mathrm{d}^3 w_i = \frac{B}{R_*^3} e^{-Q_4} F(\lambda_i)\mathrm{d}\nu \mathrm{d}^3\lambda_i \mathrm{d}^3 w_i, \tag{2.142}$$

where

$$F(\lambda_i) = \frac{27}{2}\lambda_1\lambda_2\lambda_3(\lambda_1 - \lambda_2)(\lambda_2 - \lambda_3)(\lambda_1 - \lambda_3), \quad B = \frac{5^4 3^{9/2}}{2^{11/2}\pi^{9/2}}\Gamma^2, \tag{2.143}$$

and

$$2Q_4 = \nu^2 + \Gamma(x - x_*)^2 + 15y^2 + 5z^2 + 15\Gamma w^2, \tag{2.144}$$

with $x_* = \gamma\nu$, $w^2 = w_1^2 + w_2^2 + w_3^2$ identifies the modulus of a polar coordinate system, with angles $\theta = \arccos(u)$ and $\phi$, that we will adopt in the following. At this point we can rewrite the comoving number density of peaks

$$N_{\mathrm{pk}}(\nu, \lambda_i, w_i)\mathrm{d}\nu \mathrm{d}^3\lambda_i \mathrm{d}^3 w_i = N_{\mathrm{pk}}(\nu, \lambda_i, w, u, \phi)w^2 \mathrm{d}w \mathrm{d}u \mathrm{d}\phi \mathrm{d}^3\lambda_i \mathrm{d}\nu \frac{\mathrm{d}u}{\mathrm{d}s_\mathrm{e}}\mathrm{d}s_\mathrm{e}, \tag{2.145}$$

as a function of $s_\mathrm{e}$ through Eq. (2.110)

$$s_\mathrm{e} = \frac{2^{9/2}\pi\nu^{5/2}w}{5 \times 3^{7/2}\gamma^{5/2}\sqrt{\lambda_1\lambda_2\lambda_3}}\sqrt{\beta^2 + (\alpha_3^2 - \beta^2)u^2}, \tag{2.146}$$

where

$$\beta^2(\lambda_i, \phi) = \alpha_1^2 \cos^2\phi + \alpha_2^2\sin^2\phi. \tag{2.147}$$

Integrating over the variables $w$, $\phi$ and $\lambda_i$, with the integration over $\phi$ replaced by the integration over $\beta$, gives the distribution of peaks of given height $\nu$ and spin $s_\mathrm{e}$

$$N_{\mathrm{pk}}(\nu, s_\mathrm{e}) = \frac{4Cs_\mathrm{e}}{R_*^3\nu^5}\int_0^\infty \mathrm{d}\lambda_1 \int_0^{\lambda_1} \mathrm{d}\lambda_2 \int_0^{\lambda_2} \mathrm{d}\lambda_3 \int_{\alpha_1}^{\alpha_2} \mathrm{d}\beta \frac{e^{-Q_5}F(\lambda_i)\Lambda T(\alpha_3, \beta, s_\mathrm{e}, \nu)}{\sqrt{|(\alpha_1^2 - \beta^2)(\alpha_2^2 - \beta^2)(\alpha_3^2 - \beta^2)|}}, \tag{2.148}$$

where

$$\Lambda = \lambda_1\lambda_2\lambda_3, \quad C = \frac{3^{11}5^{11/2}\gamma^5\Gamma^{3/2}}{2^{13}\pi^{13/2}}, \quad 2Q_5 = \nu^2 + \Gamma(x - x_*)^2 + 15y^2 + 5z^2, \tag{2.149}$$

and

$$T = \Theta(\alpha_3^2 - \beta^2)e^{\frac{-15\Gamma w_3^2}{2}}D(X) + \Theta(\beta^2 - \alpha_3^2)\frac{\sqrt{\pi}}{2}e^{\frac{-15\Gamma w_\beta^2}{2}}\mathrm{erf}(X). \tag{2.150}$$

The argument of the Dawson's integral $D(X)$ is

$$X = \sqrt{\frac{15}{2}\Gamma|w_\beta^2 - w_3^2|}, \text{ with } w_3 = \frac{\sqrt{\Lambda}s_\mathrm{e}}{K\nu^{5/2}\alpha_3}, \quad w_\beta = \frac{\sqrt{\Lambda}s_\mathrm{e}}{K\nu^{5/2}\beta}, \quad K = \frac{2^{9/2}\pi}{5 \times 3^{7/2}\gamma^{5/2}}. \tag{2.151}$$

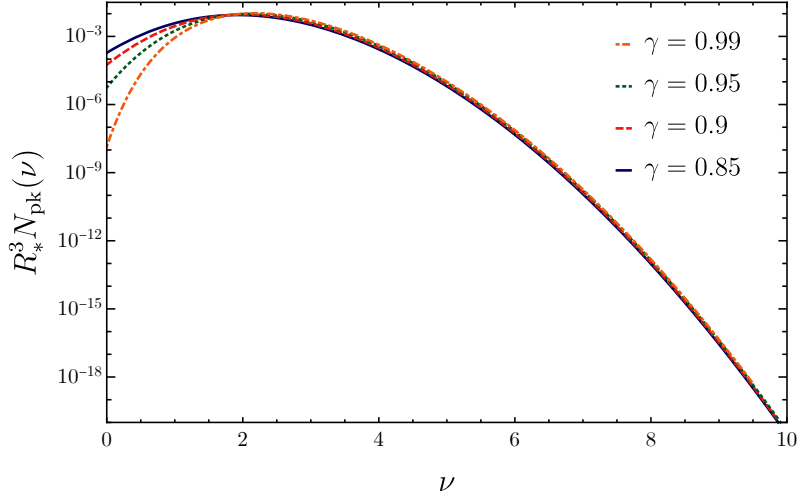

**Figure 2.4**: Rescaled comoving peak density for different values of $\gamma$. Around the interesting value for PBH formation $\nu = 8$, the curves are almost identical. Figure taken from Ref. [22].

The conditional differential probability for the spin parameter $s_e$ for a peak height $\nu$ is provided by [266]

$$P(s_e|\nu)\mathrm{d}s_e = \frac{N_{\mathrm{pk}}(\nu, s_e)}{N_{\mathrm{pk}}(\nu)}\mathrm{d}s_e, \qquad (2.152)$$

where the comoving differential peak density is expressed as

$$N_{\mathrm{pk}}(\nu)\mathrm{d}\nu = \frac{1}{(2\pi)^2 R_*^3}e^{-\frac{\nu^2}{2}}G(\gamma, x_*)\mathrm{d}\nu, \qquad (2.153)$$

in terms of the function

$$G(\gamma, x_*) = \int_0^\infty \mathrm{d}x f(x)\sqrt{\frac{\Gamma}{2\pi}}e^{-\frac{\Gamma}{2}(x - x_*)^2}, \qquad (2.154)$$

and

$$f(x) = \frac{(x^3 - 3x)}{2}\left[\mathrm{erf}\left(x\sqrt{\frac{5}{2}}\right) + \mathrm{erf}\left(\frac{x}{2}\sqrt{\frac{5}{2}}\right)\right] + \sqrt{\frac{2}{5\pi}}\left[\left(\frac{31x^2}{4} + \frac{8}{5}\right)e^{-\frac{5x^2}{8}} + \left(\frac{x^2}{2} - \frac{8}{5}\right)e^{-\frac{5x^2}{2}}\right]. \qquad (2.155)$$

In Fig. 2.4 we show the comoving peak density of Eq. (2.153) for different values of the power spectrum shape parameter $\gamma$. Finally, using the relation of Eq. (2.129) connecting the parameter $s_e$ with the Kerr parameter $a_s$, one can show in Fig. 2.5 the behaviour of its normalised distribution for different values of the height $\nu$ and $\gamma$. We notice that the conditional probability distribution shows a systematic shift towards smaller values when higher peaks are considered. Furthermore, for a given value of $\nu$, higher values of $\gamma$ provide slightly smaller values of spins, compatible with the overall factor $\sqrt{1 - \gamma^2}$ of Eq. (2.129).

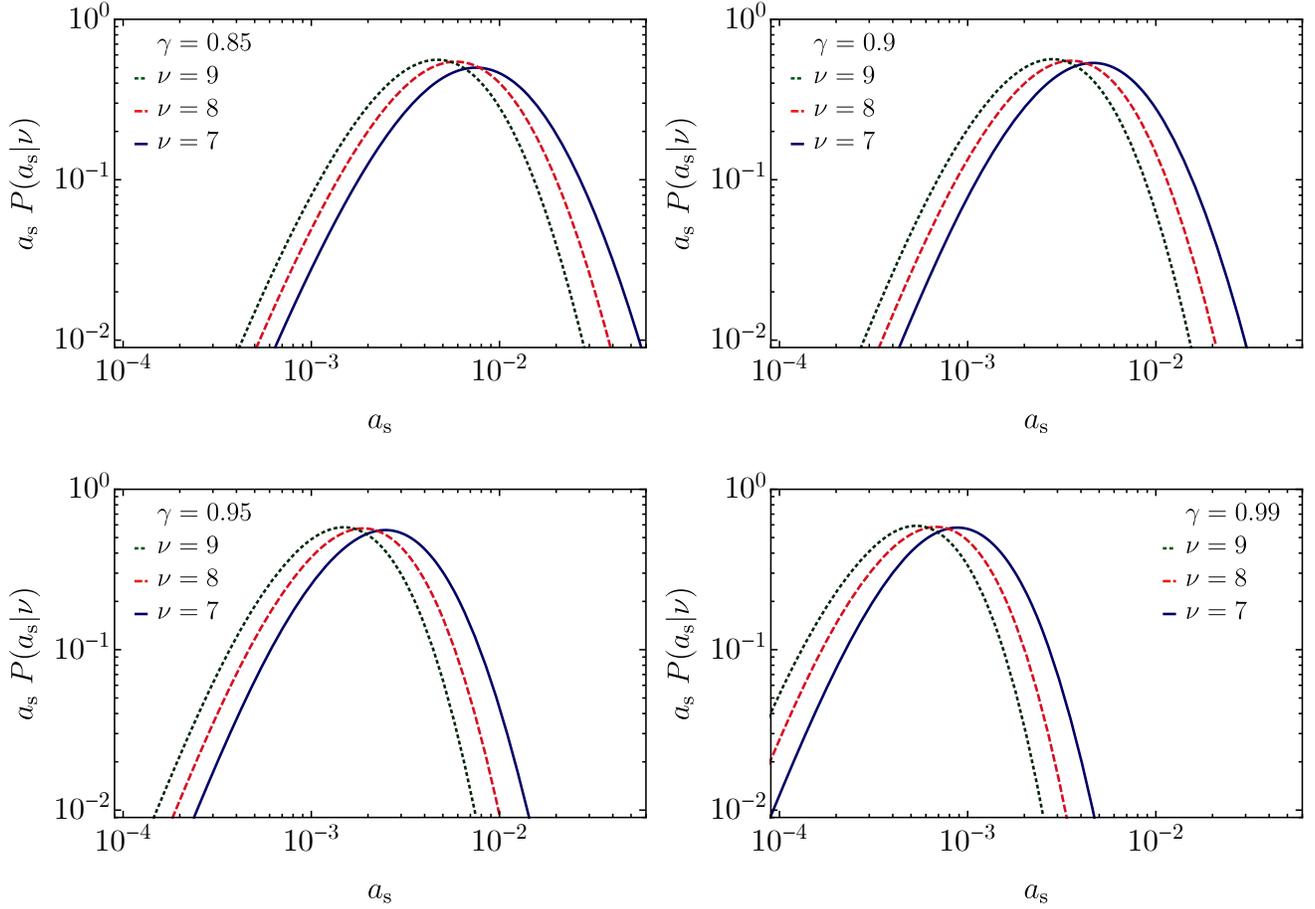

**Figure 2.5**: Normalised distribution function for $a_s$ for different values of the peak height $\nu$ and $\gamma$. Figure taken from Ref. [22].

**The spin distribution in the limit of high peaks**

The probability distribution $P(s_e|\nu)$ can be studied analytically in the limit of large $\nu$, given that PBHs form only for large enough peaks. By defining a normalised dimensionless spin $h$ as

$$s_e \equiv \frac{2^{9/2}\pi}{5\gamma^6\nu}\frac{h}{\Gamma^{1/2}} = \frac{2^{9/2}\pi}{5\gamma^6\nu}\sqrt{1-\gamma^2}h,\tag{2.156}$$

where the factor $\sqrt{1-\gamma^2}$ arises due to the scaling of the velocity shear, then the probability function of the parameter $h$ can be analytically approximated as

$$P(h)\mathrm{d}h = \exp\left(-2.37 - 4.12\ln h - 1.53\ln^2 h - 0.13\ln^3 h\right)\mathrm{d}h.\tag{2.157}$$

As one can appreciate from Fig. 2.6, there is good agreement between the numerical and analytical results in the relevant parameter space related to the physics of PBH formation. The analytical

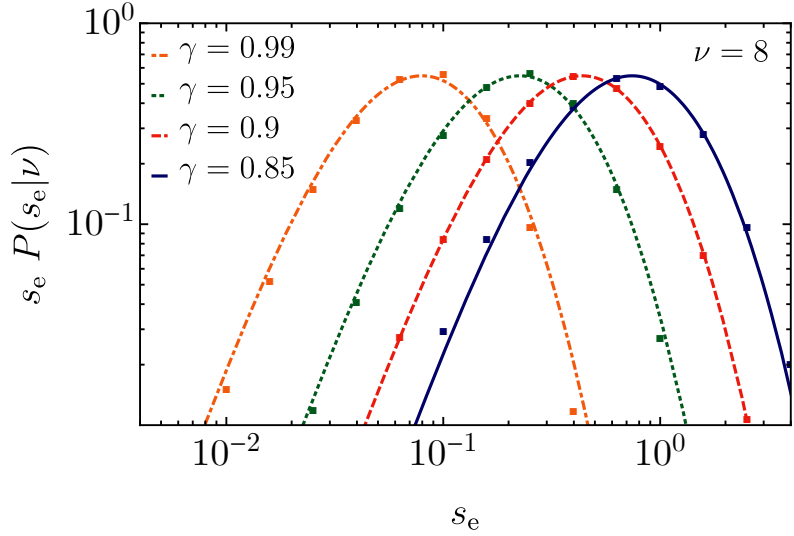

**Figure 2.6**: Comparison between the fit (lines) and the numerical result (dots) for the probability distribution $P(s_e|\nu)$ in the large $\nu$ limit. Figure taken from Ref. [22].

expression for the Kerr parameter distribution is, in the large $\nu$ limit,

$$
\begin{aligned}
P(a_s|\nu)\mathrm{d}a_s = \left(\frac{5\gamma^6\nu}{2^{9/2}\pi}\frac{\Gamma^{1/2}}{A(\eta_H)}\right)\exp\Bigg[&-2.37 - 4.12\ln\left(\frac{5\gamma^6\nu}{2^{9/2}\pi}\frac{\Gamma^{1/2}}{A(\eta_H)}a_s\right)\\
&-1.53\ln^2\left(\frac{5\gamma^6\nu}{2^{9/2}\pi}\frac{\Gamma^{1/2}}{A(\eta_H)}a_s\right) - 0.13\ln^3\left(\frac{5\gamma^6\nu}{2^{9/2}\pi}\frac{\Gamma^{1/2}}{A(\eta_H)}a_s\right)\Bigg]\mathrm{d}a_s,
\end{aligned}
\tag{2.158}
$$

where $A(\eta_H)$ is the parameter entering in Eq. (2.129).

Some remarks follow from these results. In principle, the generation of a first-order non-zero spin during the collapse may have an impact on the threshold at which the collapse takes place. However this effect was shown to be small with negligible impact on the overall PBH abundance [277]. Furthermore, also the impact on the mass distribution is found to be rather negligible [22].

We conclude with possible future prospects. First, it could be interesting to go beyond the Gaussian assumption for the initial distributions of the overdensity and velocity fields, assessing the impact of a primordial non-Gaussianity on the initial PBH spin. Secondly, even though these results show that the spin of PBHs at formation is almost negligible, they do not consider how spins evolve during the cosmic history, for example due to a phase of baryonic mass accretion at late times. This will be discussed in the next chapters of this thesis. Finally, even though the natal spin of PBHs is very small in the formation scenario from the collapse of sizeable density perturbations during a radiation-dominated era, the prediction may be different if one considers alternative PBH formation scenarios like from the collapse in a matter-dominated era [278], from the collapse of Q-balls [279] or for PBHs produced by long-range scalar-mediated Yukawa forces [280, 281].

# Chapter 3

# Gravitational waves from primordial black hole formation

This chapter is dedicated to the investigation of GWs associated to the formation of PBHs. We will mainly focus on the generation of a stochastic gravitational wave background (SGWB) induced at second order in perturbation theory by the same scalar perturbations which are responsible for those overdensities which can collapse into PBHs.

We will discuss the general features of such a SGWB in view of future GW experiments like LISA, by computing the corresponding two-point and three-point correlation functions of the tensor modes, and finally connect this signal to the recent data released by the NANOGrav collaboration.

## 3.1 Scalar-Induced Gravitational Wave Background

In the standard scenario where PBHs are born out of the collapse of sizable overdensities generated during the inflationary era, the same scalar perturbations are unavoidably associated to the emission of induced GWs at second order in perturbation theory due to the nonlinear nature of gravity [214, 282–286]. Several works have been dedicated to investigate this SGWB in contexts related to PBHs [23, 24, 115, 242, 287–299] (for recent reviews see Refs. [300, 301]).

Such a source of gravitational radiation is present at all scales, but the enhancement of the scalar power spectrum around a characteristic scale, responsible for PBH formation, implies that this stochastic background may be dominated at a given frequency. One can relate the peak frequency of such spectrum, close to the characteristic frequency of the corresponding curvature perturbations, to the PBH mass $M_{\rm PBH}$ by using entropy conservation [289]

$$M_{\rm PBH} \simeq 33\,\gamma \left( \frac{10^{-9}\,{\rm Hz}}{f} \right)^2 M_\odot, \tag{3.1}$$

where the constant factor $\gamma$ is introduced to account for the fraction of the horizon mass which ends up in the PBH mass. For example, by choosing the frequency at which the LISA experiment has its maximum sensitivity, i.e. $f_{\rm LISA} \simeq 3.4$ mHz, one finds PBH masses of the order of $M_{\rm PBH} \simeq 10^{-12} M_\odot$ (assuming the analytical estimate $\gamma \simeq 0.2$ from Ref. [92]), while PBH masses around $\sim M_\odot$ correspond to a SGWB with peak frequency around the nHz, that is the relevant frequency range for the PTA and NANOGrav experiments[1].

In this section we are going to focus on the main properties of such a SGWB, especially in the context of the LISA experiment

## 3.1.1 The GW power spectrum

In this section we are going to show the preliminaries to compute the stochastic GW signal sourced by second-order scalar perturbations following the steps shown in Refs. [23, 24].

In a radiation-dominated era, the linearized line element in terms of the Newtonian-gauge scalar metric perturbation $\Psi$ and the transverse-traceless tensor metric perturbation $h_{ij}$ is given by

$$ds^2 = a^2 \left\{ -(1 + 2\Psi)d\eta^2 + \left[ (1 - 2\Psi)\delta_{ij} + \frac{h_{ij}}{2} \right] dx^i dx^j \right\}. \tag{3.2}$$

The tensor modes equation of motion can be determined by expanding the Einstein's equations at second-order in perturbation theory in the linear scalar perturbations as [2]

$$h_{ij}'' + 2\mathcal{H}h_{ij}' - \nabla^2 h_{ij} = -4\mathcal{T}_{ij}{}^{\ell m}\mathcal{S}_{\ell m}, \tag{3.3}$$

where the prime stands for derivatives with respect to conformal time and $\mathcal{H} = a'/a$ is the conformal Hubble parameter in terms of the scale factor $a(\eta)$. The source term $\mathcal{S}_{\ell m}$, intrinsically second-order in the scalar perturbation $\Psi$, takes the form for a radiation-dominated era in the Poisson gauge [283]

$$\mathcal{S}_{ij} = 4\Psi\partial_i\partial_j\Psi + 2\partial_i\Psi\partial_j\Psi - \partial_i\left( \frac{\Psi'}{\mathcal{H}} + \Psi \right)\partial_j\left( \frac{\Psi'}{\mathcal{H}} + \Psi \right), \tag{3.4}$$

---

[1]PBHs with masses around $10^{-12} M_\odot$ could also be investigated with high frequencies experiments searching for GWs from mergers [302].

[2]We neglect the effect of neutrinos free-streaming on the GW signal [303].

with the main contribution to GWs emission coming at the horizon re-entry of the relevant modes. The tensor $\mathcal{T}_{ij}{}^{\ell m}$ acts as a projector selecting the transverse and traceless components, and is defined in Fourier space as

$$\tilde{\mathcal{T}}_{ij}{}^{\ell m}(\vec{k}) = e_{ij}^{\mathrm{L}}(\vec{k}) \otimes e^{\mathrm{L}\ell m}(\vec{k}) + e_{ij}^{\mathrm{R}}(\vec{k}) \otimes e^{\mathrm{R}\ell m}(\vec{k}), \tag{3.5}$$

in terms of the polarisation tensors $e_{ij}^{\lambda}(\vec{k})$ in the chiral basis $(\mathrm{L}, \mathrm{R})$. The scalar perturbation $\Psi(\eta, \vec{k})$ is related to the gauge invariant comoving curvature perturbation $\zeta$ as [71]

$$\Psi(\eta, \vec{k}) = \frac{2}{3} T(\eta, k) \zeta(\vec{k}), \tag{3.6}$$

as a function of the radiation-dominated era transfer function $T(\eta, k)$

$$T(\eta, k) = \mathcal{T}(k\eta), \quad \mathcal{T}(z) = \frac{9}{z^2} \left[ \frac{\sin(z/\sqrt{3})}{z/\sqrt{3}} - \cos(z/\sqrt{3}) \right]. \tag{3.7}$$

Introducing the dimensionless variables $x = p/k$, $y = |\vec{k} - \vec{p}|/k$, and the contracted polarization tensor $e^{\lambda}(\vec{k}, \vec{p}) = e_{ij}^{\lambda}(\vec{k}) p^i p^j$, one can solve the equation of motion (3.3) to get

$$h_{\vec{k}}^{\lambda}(\eta) = \frac{4}{9} \int \frac{\mathrm{d}^3 p}{(2\pi)^3} \frac{1}{k^3 \eta} e^{\lambda}(\vec{k}, \vec{p}) \zeta(\vec{p}) \zeta(\vec{k} - \vec{p}) \Big[ \mathcal{I}_c(x, y) \cos(k\eta) + \mathcal{I}_s(x, y) \sin(k\eta) \Big], \tag{3.8}$$

where [304, 305]

$$\mathcal{I}_c(x, y) = 4 \int_1^{\infty} \mathrm{d}\tau \, \tau(-\sin\tau) \Big[ 2\mathcal{T}(x\tau)\mathcal{T}(y\tau) + \Big( \mathcal{T}(x\tau) + x\tau \, \mathcal{T}'(x\tau) \Big) \Big( \mathcal{T}(y\tau) + y\tau \, \mathcal{T}'(y\tau) \Big) \Big],$$

$$\mathcal{I}_s(x, y) = 4 \int_1^{\infty} \mathrm{d}\tau \, \tau(\cos\tau) \Big\{ 2\mathcal{T}(x\tau)\mathcal{T}(y\tau) + \Big[ \mathcal{T}(x\tau) + x\tau \, \mathcal{T}'(x\tau) \Big] \Big[ \mathcal{T}(y\tau) + y\tau \, \mathcal{T}'(y\tau) \Big] \Big\}. \tag{3.9}$$

The GWs power spectrum is defined in terms of the momentum-space tensor modes as

$$\left\langle h_{\vec{k}_1}^{\lambda_1}(\eta) h_{\vec{k}}^{\lambda_2}(\eta) \right\rangle' \equiv \delta^{\lambda_1 \lambda_2} \frac{2\pi^2}{k_1^3} \mathcal{P}_h(\eta, k_1), \tag{3.10}$$

and evaluating the two-point correlator gives

$$\mathcal{P}_h(\eta, k) = \frac{4}{81} \frac{1}{k^2 \eta^2} \iint_{\mathcal{S}} \mathrm{d}x \, \mathrm{d}y \frac{x^2}{y^2} \left[ 1 - \frac{(1 + x^2 - y^2)^2}{4x^2} \right]^2 \mathcal{P}_{\zeta}(kx) \mathcal{P}_{\zeta}(ky)$$
$$\times \left[ \cos^2(k\eta) \mathcal{I}_c^2 + \sin^2(k\eta) \mathcal{I}_s^2 + \sin(2k\eta) \mathcal{I}_c \mathcal{I}_s \right], \tag{3.11}$$

where $\mathcal{S}$ is the region in the $(x, y)$ plane described by the inequalities

$$d \equiv \frac{1}{\sqrt{3}} |x - y|, \qquad s \equiv \frac{1}{\sqrt{3}} (x + y), \qquad (d, s) \in [0, 1/\sqrt{3}] \times [1/\sqrt{3}, +\infty). \tag{3.12}$$

The GW energy density is then defined as [304]

$$\Omega_{\text{GW}}(\eta, k) = \frac{\rho_{\text{GW}}(\eta, k)}{\rho_c(\eta)} = \frac{1}{24} \left( \frac{k}{\mathcal{H}(\eta)} \right)^2 \overline{\mathcal{P}_h(\eta, k)}, \tag{3.13}$$

in terms of the critical energy density of a spatially flat universe $\rho_c = 3H^2 M_p^2$, where the overline accounts for the average on a timescale much greater than the GW phase oscillations but much smaller than the cosmological time. Its present value can be computed by using the energy density scaling $\propto 1/a^4$ as

$$\Omega_{\text{GW}}(\eta_0, k) = \frac{a_f^4 \rho_{\text{GW}}(\eta_f, k)}{\rho_r(\eta_0)} \Omega_{r,0} = c_g \frac{\Omega_{r,0}}{24} \frac{k^2}{\mathcal{H}(\eta_f)^2} \overline{\mathcal{P}_h(\eta_f, k)}, \tag{3.14}$$

from the epoch $\eta_f$ at which the top quarks start to annihilate, and relating the radiation density $\rho_r(\eta_f) \approx \rho_c(\eta_f)$ to its value today using conservation of entropy. The coefficient

$$c_g \equiv \frac{a_f^4 \rho_r(\eta_f)}{\rho_r(\eta_0)} = \frac{g_*}{g_*^0} \left( \frac{g_{*S}^0}{g_{*S}} \right)^{4/3} \approx 0.4, \tag{3.15}$$

accounts for the change in the effective number of degrees of freedom $g_*$ for the thermal radiation energy density at the top quark annihilation time $\eta_f$, and it is expressed in terms of the effective degrees of freedom for entropy density $g_{*S}$, assuming $g_{*S} \approx g_* \approx 106.75$ from the Standard Model at time $\eta_f$. Finally $\Omega_{r,0}$ indicates the present radiation energy density fraction if neutrinos were massless. One then gets

$$\Omega_{\text{GW}}(\eta_0, k) = c_g \frac{\Omega_{r,0}}{72} \int_{-\frac{1}{\sqrt{3}}}^{\frac{1}{\sqrt{3}}} dd \int_{\frac{1}{\sqrt{3}}}^{\infty} ds \left[ \frac{(d^2 - 1/3)(s^2 - 1/3)}{s^2 - d^2} \right]^2 \mathcal{P}_\zeta \left( \frac{k\sqrt{3}}{2}(s + d) \right) \mathcal{P}_\zeta \left( \frac{k\sqrt{3}}{2}(s - d) \right)$$
$$\times \left[ \mathcal{I}_c^2(x(d, s), y(d, s)) + \mathcal{I}_s^2(x(d, s), y(d, s)) \right], \tag{3.16}$$

where we have introduced the variables as in Eq. (3.12).

An analytical expression for the present GW energy spectrum can be derived making the idealized assumption of a monochromatic curvature perturbation power spectrum with characteristic momentum scale $k_\star$ as

$$\mathcal{P}_\zeta(k) = A_s \, k_\star \delta_D \left( k - k_\star \right), \tag{3.17}$$

such that one gets [288, 289]

$$\Omega_{\text{GW}}(\eta_0, k) = \frac{c_g}{15552} \Omega_{r,0} A_s^2 \frac{k^2}{k_\star^2} \left( \frac{4k_\star^2}{k^2} - 1 \right)^2 \theta \left( 2k_\star - k \right) \left[ \mathcal{I}_c^2 \left( \frac{k_\star}{k}, \frac{k_\star}{k} \right) + \mathcal{I}_s^2 \left( \frac{k_\star}{k}, \frac{k_\star}{k} \right) \right]. \tag{3.18}$$

This spectrum has a resonant effect at $k \sim 2k_\star/\sqrt{3}$, which produces a spike in the GW spectrum [285]. Furthermore, the slow fall-off at low frequencies is an unphysical effect of assuming such a shape of the curvature power spectrum, while for physical spectra one expects a white-noise ($\propto k^3$) behaviour due to causality [306] (see also Refs. [295, 307] for the impact of thermal history on the SGWB spectral shape).

One can also investigate the SGWB shape for the choice of a lognormal curvature perturbation power spectrum

$$\mathcal{P}_\zeta^g(k) = A_\zeta \exp\left(-\frac{\log^2(2k/3k_\star)}{2\sigma_\zeta^2}\right), \tag{3.19}$$

with variance $\sigma_\zeta$.

The fact that the frequency range at which the LISA experiment has its maximum sensitivity corresponds to the generation of PBHs with masses compatible with the scenario in which they comprise the totality of the dark matter represents a serendipity. This implies that LISA will be able to measure the power spectrum of the GWs generated by the production mechanism of PBHs.

In Fig. 3.1 we show a comparison of the LISA sensitivity curve (assuming the proposed design (4y, 2.5 Gm of length, 6 links) and a sensitivity in between the ones dubbed C1 and C2 in Ref. [308]) with the GW abundance generated at second-order assuming both a monochromatic and lognormal shape of the curvature power spectrum. To fix the amplitude of the power spectrum in the two cases we have required PBHs to comprise all the DM of the universe in that mass range. Given that the PBH abundance is exponentially sensitive to the amplitude, even a small decrease of its value may reduce significantly the abundance. This, however, would not affect the expectation of a SGWB, which will be anyway tested by LISA even if $f_{\rm PBH} \ll 1$ [145].

### 3.1.2 The GW primordial bispectrum

From the expression of the source term in the tensor modes equation of motion one can easily see that, being intrinsically at second order, the GWs are non-Gaussian. This implies that their primordial three-point correlator is not vanishing, and can be computed as

$$
\begin{aligned}
\mathcal{B}_{\lambda_i}\left(\vec{k}_i\right) &= \left\langle h_{\lambda_1}(\eta_1, \vec{k}_1) h_{\lambda_2}(\eta_2, \vec{k}_2) h_{\lambda_3}(\eta_3, \vec{k}_3) \right\rangle' \\
&= \left(\frac{8\pi}{9}\right)^3 \int \mathrm{d}^3 p_1 \frac{1}{k_1^3 k_2^3 k_3^3 \eta_1 \eta_2 \eta_3} \cdot e_{\lambda_1}^*(\vec{k}_1, \vec{p}_1) e_{\lambda_2}^*(\vec{k}_2, \vec{p}_2) e_{\lambda_3}^*(\vec{k}_3, \vec{p}_3) \frac{\mathcal{P}_\zeta(p_1)}{p_1^3} \frac{\mathcal{P}_\zeta(p_2)}{p_2^3} \frac{\mathcal{P}_\zeta(p_3)}{p_3^3} \\
&\quad \cdot \left[\left(\cos(k_1\eta_1)\mathcal{I}_c\left(\frac{p_1}{k_1}, \frac{p_2}{k_1}\right) + \sin(k_1\eta_1)\mathcal{I}_s\left(\frac{p_1}{k_1}, \frac{p_2}{k_1}\right)\right) \cdot (1 \rightarrow 2 \,\text{and}\, 2 \rightarrow 3) \cdot (1 \rightarrow 3 \,\text{and}\, 2 \rightarrow 1)\right],
\end{aligned}
\tag{3.20}
$$

where $\vec{p}_2 = \vec{p}_1 - \vec{k}_1$ and $\vec{p}_3 = \vec{p}_1 + \vec{k}_3$.

Some analytical simplifications can be made for a monochromatic shape of the curvature perturbation power spectrum as in Eq. (3.17). Under this assumption the previous equation

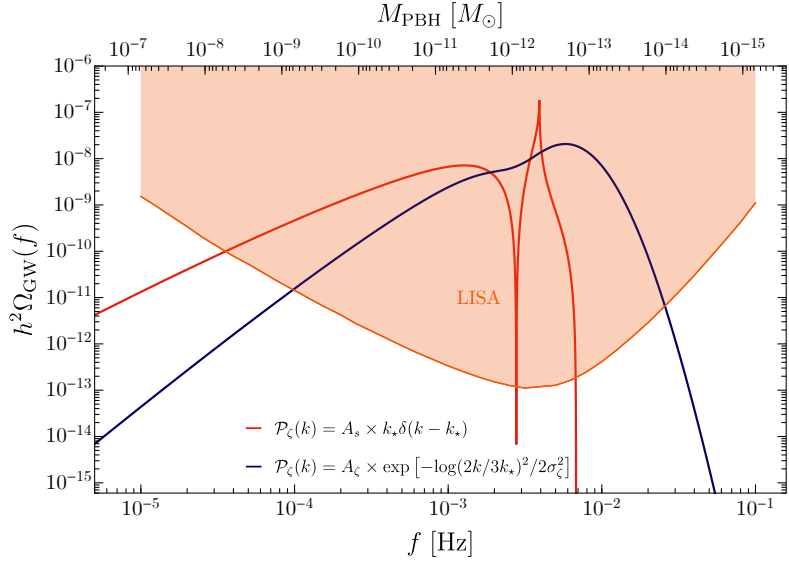

**Figure 3.1**: Comparison of the estimated sensitivity for LISA [203] with the GW abundance generated at second-order for both power spectra in Eqs. (3.17) and (3.19), assuming the following values for the parameters: $A_s = 0.033$, $A_\zeta = 0.044$ and $\sigma_\zeta = 0.5$. Figure taken from Ref. [23].

becomes

$$
\mathcal{B}_{\lambda_i}\left(\vec{k}_i\right) = \left(\frac{8\pi}{9}\right)^3 \frac{A_s^3\, k_\star^3}{k_1^3 k_2^3 k_3^3 \eta_1 \eta_2 \eta_3} \int \mathrm{d}^3 p_1\, e_{\lambda_1}^*\left(\vec{k}_1, \vec{p}_1\right) e_{\lambda_2}^*\left(\vec{k}_2, \vec{p}_1 - \vec{k}_1\right) e_{\lambda_3}^*\left(\vec{k}_3, \vec{p}_1 + \vec{k}_3\right)
$$

$$
\cdot \frac{\delta_D\left(p_1 - k_\star\right)}{k_\star^3} \frac{\delta_D\left(\left|\vec{p}_1 - \vec{k}_1\right| - k_\star\right)}{k_\star^3} \frac{\delta_D\left(\left|\vec{p}_1 + \vec{k}_3\right| - k_\star\right)}{k_\star^3}
$$

$$
\cdot \prod_{i=1}^{3}\left[\cos\left(k_i \eta_i\right) \mathcal{I}_c\left(\frac{k_\star}{k_i}, \frac{k_\star}{k_i}\right) + \sin\left(k_i \eta_i\right) \mathcal{I}_s\left(\frac{k_\star}{k_i}, \frac{k_\star}{k_i}\right)\right]. \tag{3.21}
$$

Following Ref. [309], one can fix the orientation and magnitude of the three vectors $\vec{k}_i$, as

$$
\vec{k}_1 = k_1\, \hat{v}_1, \quad \vec{k}_2 = k_2\, \hat{v}_2, \quad \vec{k}_3 = -\vec{k}_1 - \vec{k}_2, \tag{3.22}
$$

where

$$
\hat{v}_1 = \begin{pmatrix} 1 \\ 0 \\ 0 \end{pmatrix}, \quad \hat{v}_2 = \begin{pmatrix} \frac{k_3^2 - k_1^2 - k_2^2}{2k_1 k_2} \\ \sqrt{1 - \left(\frac{k_3^2 - k_1^2 - k_2^2}{2k_1 k_2}\right)^2} \\ 0 \end{pmatrix}. \tag{3.23}
$$

Going to spherical coordinates for the integration vector and using the orthogonality of the polarization operator $e^\lambda$, one gets the expression

$$
\mathcal{B}_{\lambda_i}\left(\eta_i, \vec{k}_i\right) = \frac{A_s^3 \theta\left(\mathcal{A}\left[r_1,\, r_2,\, r_3\right] - \frac{r_1\, r_2\, r_3}{4}\right)}{k_1^2 k_2^2 k_3^2\, k_\star^3\, \eta_1 \eta_2 \eta_3} \frac{1024\pi^3}{729} \mathcal{D}_{\lambda_i}\left(\hat{k}_i,\, r_i\right)
$$

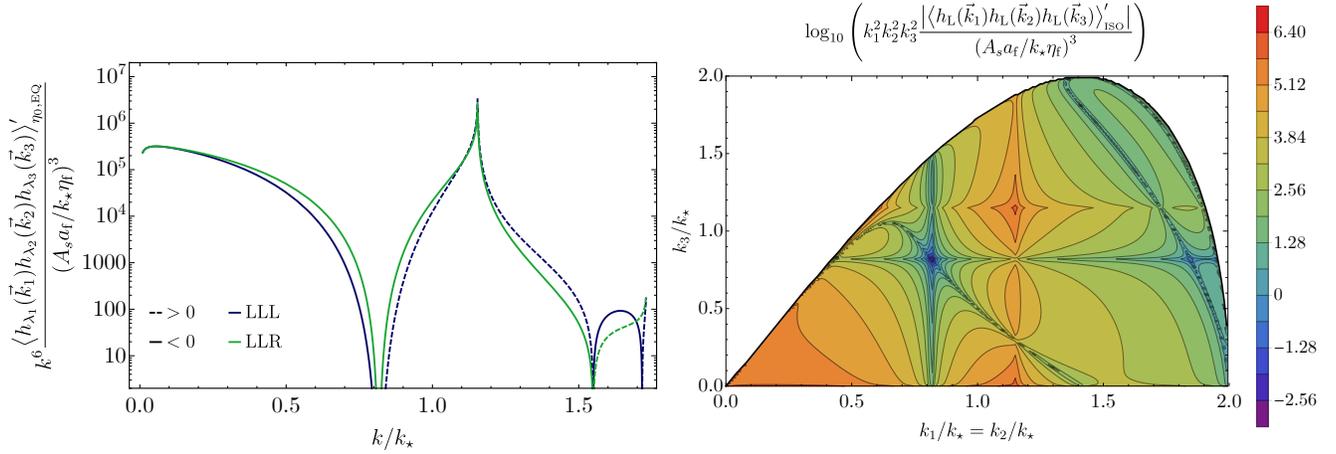

**Figure 3.2**: Contour plot of the rescaled primordial bispectrum in the equilateral (left) and isosceles (right) configuration for the choice of a Dirac delta scalar power spectrum. Figure taken from Ref. [23].

$$\cdot \left( \frac{16\,\mathcal{A}^2\,[r_1,\,r_2,\,r_3]}{r_1^2 r_2^2 r_3^2} - 1 \right)^{-1/2} \frac{r_1^4}{r_2^2 r_3^2} \prod_{i=1}^{3} \left[ \frac{\mathcal{I}_i^*}{2}\,\mathrm{e}^{i\eta_i k_i} + \frac{\mathcal{I}_i}{2}\,\mathrm{e}^{-i\eta_i k_i} \right], \tag{3.24}$$

in terms of the area of the triangle of sides $k_i$

$$\mathcal{A}\,[k_1,\,k_2,\,k_3] \equiv \frac{1}{4}\sqrt{(k_1+k_2+k_3)\,(-k_1+k_2+k_3)\,(k_1-k_2+k_3)\,(k_1+k_2-k_3)} \tag{3.25}$$

with support obtained when

$$\mathcal{A}\,[k_1,\,k_2,\,k_3] > \frac{k_1\,k_2\,k_3}{4\,k_\star}. \tag{3.26}$$

Furthermore, we have introduced the normalised variable $r_i \equiv k_i/k_\star$, the combinations

$$\mathcal{I}_i \equiv \mathcal{I}\left(\frac{1}{r_i}\right) \equiv \mathcal{I}_c\left(\frac{1}{r_i},\,\frac{1}{r_i}\right) + i\,\mathcal{I}_s\left(\frac{1}{r_i},\,\frac{1}{r_i}\right), \tag{3.27}$$

as well as the contractions

$$\mathcal{D}_{\lambda_i}\left(\hat{k}_i,\,r_i\right) \equiv e^*_{ab,\lambda_1}\left(\hat{k}_1\right) e^*_{cd,\lambda_2}\left(\hat{k}_2\right) e^*_{ef,\lambda_3}\left(\hat{k}_3\right) \left\{ \left[ \vec{q}_a\,\vec{q}_b\left(\vec{q}-\hat{k}_1\right)_c\left(\vec{q}-\hat{k}_1\right)_d \vec{q}_e\,\vec{q}_f \right]_I + [\mathrm{same}]_{II} \right\}, \tag{3.28}$$

where we have to sum over the two points

$$(\vec{p}_1)_{I,II} = k_1\left( \frac{1}{2},\,\frac{-k_1^2+k_2^2+k_3^2}{8\,\mathcal{A}\,[k_1,\,k_2,\,k_3]},\,\pm \frac{\sqrt{16\mathcal{A}^2\,[k_1,\,k_2,\,k_3]\,k_\star^2 - k_1^2 k_2^2 k_3^2}}{4\mathcal{A}\,[k_1,\,k_2,\,k_3]\,k_1} \right) \equiv k_1\,\vec{q}_{I,II}. \tag{3.29}$$

The explicit expressions for the isosceles case $r_1 = r_2$ are given by

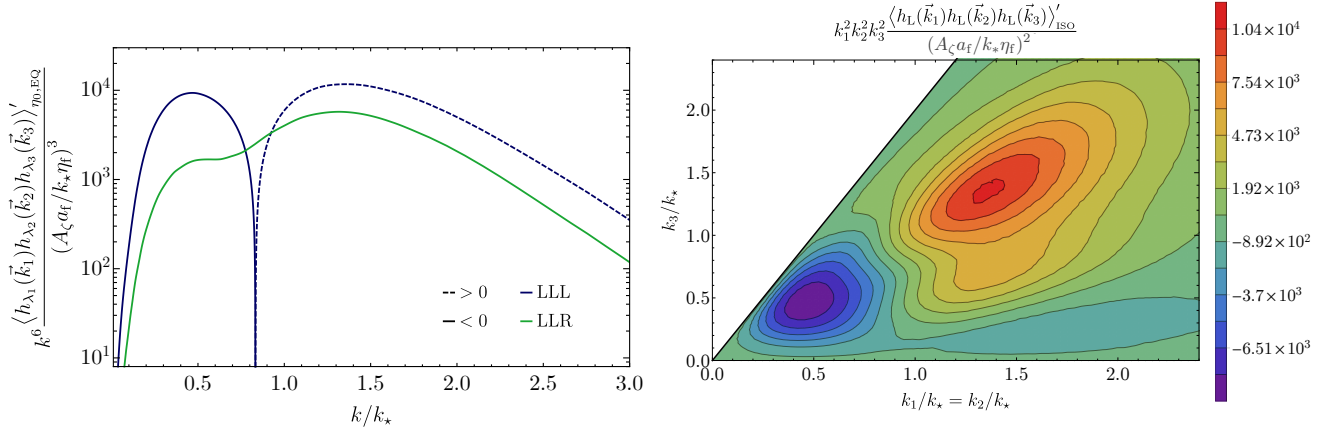

**Figure 3.3**: Contour plot of the rescaled primordial bispectrum in the equilateral (left) and isosceles (right) configuration for the choice of a Gaussian scalar power spectrum. Figure taken from Ref. [23].

$$\mathcal{D}_{\mathrm{RRR}} = \mathcal{D}_{\mathrm{LLL}} = \frac{1}{256}\left[\frac{32 r_1^3}{(2 r_1 + r_3)^3} - \frac{24\left(3 r_1^2 + 8\right)}{(2 r_1 + r_3)^2} + \frac{32\left(r_1^2 + 4\right) r_3}{r_1^5} + \frac{32\left(r_1^2 - 1\right)^2}{r_1^3(2 r_1 - r_3)}\right.$$

$$\left. + \frac{32\left(2\left(r_1^2 + 6\right) r_1^2 + 9\right)}{r_1^3(2 r_1 + r_3)} - \frac{\left(r_1^4 + 24 r_1^2 + 16\right) r_3^2}{r_1^6} - \frac{4\left(33 r_1^4 + 24 r_1^2 + 16\right)}{r_1^6} - 32\right],$$

$$\mathcal{D}_{\mathrm{LRR}} = \mathcal{D}_{\mathrm{RLL}} = \mathcal{D}_{RLR} = \mathcal{D}_{LRL} = \frac{\left(r_1^2 - 4\right)^2 \left(8 r_1^4 - 4 r_1^2\left(r_3^2 + 4\right) + r_3^4 + 4 r_3^2\right)}{256 r_1^6\left(4 r_1^2 - r_3^2\right)},$$

$$\mathcal{D}_{\mathrm{RRL}} = \mathcal{D}_{\mathrm{LLR}} = \frac{1}{256}\left[-\frac{32 r_1^3}{(r_3 - 2 r_1)^3} - \frac{24\left(3 r_1^2 + 8\right)}{(r_3 - 2 r_1)^2} - \frac{32\left(r_1^2 + 4\right) r_3}{r_1^5} - \frac{32\left(2\left(r_1^2 + 6\right) r_1^2 + 9\right)}{r_1^3(r_3 - 2 r_1)}\right.$$

$$\left. + \frac{32\left(r_1^2 - 1\right)^2}{r_1^3(2 r_1 + r_3)} - \frac{\left(r_1^4 + 24 r_1^2 + 16\right) r_3^2}{r_1^6} - \frac{4\left(33 r_1^4 + 24 r_1^2 + 16\right)}{r_1^6} - 32\right], \tag{3.30}$$

where we stress that the contractions are invariant under parity (L $\leftrightarrow$ R), and so it is the bispectrum. In the equilateral case $r_1 = r_2 = r_3$, the equal time bispectrum reads

$$\mathcal{B}_{\lambda_i}^{\mathrm{equil}}\left(\eta, |\vec{k}_i| = k\right) = \frac{A_s^3}{k_\star^3 \eta^3} \cdot \frac{1}{k^6} \frac{1024 \pi^3}{729} \frac{\theta\left(\sqrt{3}\, k_\star - k\right)}{\sqrt{\frac{3 k_\star^2}{k^2} - 1}}\, \left|\frac{1}{\sqrt{2}} \mathcal{I}\left(\frac{k_\star}{k}\right)\right|^3$$

$$\cdot \begin{cases} \frac{365}{6912} - \frac{61}{192}\frac{k_\star^2}{k^2} + \frac{9}{16}\frac{k_\star^4}{k^4} - \frac{1}{4}\frac{k_\star^6}{k^6} & \text{for RRR, LLL,} \\[2ex] \frac{\left[-4 + (k/k_\star)^2\right]^2\left[-12 + 5(k/k_\star)^2\right]}{768(k/k_\star)^6}, & \text{otherwise,} \end{cases} \tag{3.31}$$

where we have averaged over the oscillations of the amplitude and defined $\mathcal{I}^2 = \mathcal{I}_c^2 + \mathcal{I}_s^2$.

In Fig. 3.2 we show the equilateral and isosceles primordial bispectrum for the choice of a monochromatic power spectrum of the curvature perturbation. One can notice that it is peaked in the equilateral configuration at $k = 2k_\star/\sqrt{3}$ and that vanishes for $k > \sqrt{3}k_\star$. In Fig. 3.3 we show its behaviour in the equilateral and isosceles configurations assuming a lognormal power spectrum. Compared to the monochromatic result, the wider shape of the lognormal power spectrum gives rise to a lower peak in the equilateral configuration, making the two peaks with opposite sign in the LLL configuration comparable. Furthermore, the polarization configurations LLR=RRL (and their permutations) are suppressed with respect to the LLL=RRR ones. In the isosceles configuration the profile is more regular and it is not characterised by the cut-off present in the monochromatic case.

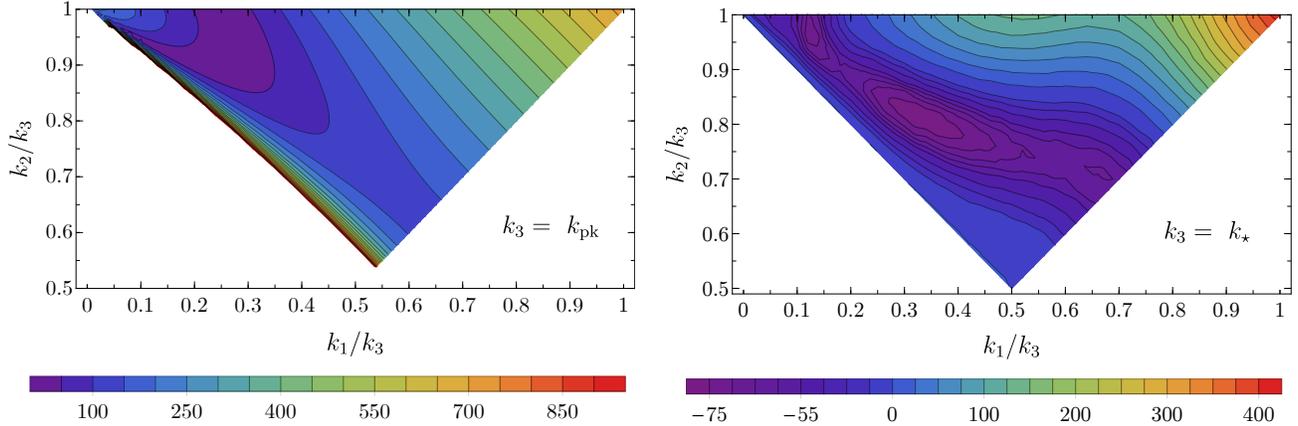

**Figure 3.4**: Contour plot of the shape of the three-point function for the choice of a Dirac delta (left) and Gaussian (right) scalar power spectrum. Figure taken from Ref. [23].

The shape of the primordial bispectrum

$$S_h^{\lambda_1\lambda_2\lambda_3}(\vec{k}_1, \vec{k}_2, \vec{k}_3) = k_1^2 k_2^2 k_3^2 \frac{\langle h_{\lambda_1}(\eta, \vec{k}_1) h_{\lambda_2}(\eta, \vec{k}_2) h_{\lambda_3}(\eta, \vec{k}_3) \rangle'}{\sqrt{\mathcal{P}_h(\eta, k_1)\mathcal{P}_h(\eta, k_2)\mathcal{P}_h(\eta, k_3)}} \tag{3.32}$$

is shown in Fig. 3.4 for the two examples of the curvature perturbation power spectra. One can conclude that its maximum always lies in the equilateral configuration $k_1 \simeq k_2 \simeq k_3$, as expected from the shape of the source term in the tensor modes equation of motion, due to the gradients of the curvature perturbations when the latter re-enter the horizon.

### 3.1.3 GW propagation in a perturbed universe

Even though the GW bispectrum is sizeable at PBH formation, one has still to investigate if it is measurable at the present epoch by GW experiments like LISA. One of the effects which may impact on its detectability is the GWs propagation from the source points towards the detector across the perturbed universe. As we will show in details, this effect largely suppresses the bispectrum and makes it unobservable. The main reason lies in the fact that, in order to detect a non vanishing non-Gaussianity, one needs phase correlations in the GW signal (in turn, random phases correspond to Gaussianity) [310]. Even though the GW phases are correlated at their production during the inflationary stage, their coherence is washed out by the Shapiro time-delays which they experience from travelling along different line of sights, induced by the presence of large-scale gravitational potentials, which destroy the phase correlation. This effect does not impact on the GW power spectrum, but suppresses the primordial bispectrum, and can be interpreted as a consequence of the central limit theorem, for which even though the non-Gaussian signal arriving from a single line of sight is not modified, averaging over many directions makes the signal Gaussian.

In the following we solve the GW propagation in a perturbed universe, and discuss the details of its impact on the GW power spectrum and bispectrum. For simplicity, we are going to assume a monochromatic power spectrum of the curvature perturbation as in Eq. (3.17).

**The GW propagation in the geometrical optic limit**

The GW propagation in a matter-dominated phase can be described assuming the geometrical optic limit, according to which the GW frequency is much larger than the typical momentum associated to the gravitational potential $\Phi$. This implies that one can consider only the leading term in an expansion in gradients of $\Phi$ in the Einstein equations and write [23]

$$h''_{ij} + 2\mathcal{H}h'_{ij} - (1 + 4\Phi) h_{ij,kk} = 0. \tag{3.33}$$

To make the mode function real in coordinate space, one can write it in momentum space as

$$h_{ij}(\vec{k}, \eta) = h^A_{ij}(\vec{k}, \eta) + h^{*A}_{ij}(-\vec{k}, \eta), \tag{3.34}$$

and take the ansatz for the solution

$$h^A_{ij} = A_{ij} e^{ik\eta} e^{i \int^{\eta} d\eta' F_A(\eta')}. \tag{3.35}$$

Neglecting spatial derivatives on $F_A$, the equation of motion becomes

$$\begin{cases} A''_{ij} + 2ikA'_{ij} - k^2 A_{ij} + \frac{4}{\eta} \left[ A'_{ij} + ikA_{ij} \right] + k^2 A_{ij} = 0, \\ 2iF_A A'_{ij} + \left[ iF'_A - 2kF_A - F^2_A + \frac{4}{\eta} iF_A + 4k^2 \Phi \right] A_{ij} \simeq 0, \end{cases} \tag{3.36}$$

which is separated at zeroth order in $F_A$ and the rest. The solution of the first equation is

$$A_{ij} = \frac{C^A_{ij}}{k^2 \eta^2} \left( 1 + \frac{i}{k\eta} \right), \tag{3.37}$$

in terms of an integration constant $C^A$, and we can keep only the first term being interested in the sub-horizon limit. The second equation then becomes

$$iF_A' - 2kF_A - F_A^2 + 4k^2\Phi \simeq 0, \tag{3.38}$$

whose solution at leading order is $F_A = 2k\Phi$. One therefore gets

$$h_{ij}^A = A_{ij} e^{ik\eta + 2ik\int^\eta d\eta'\,\Phi(\eta')}. \tag{3.39}$$

Matching the wave function to Eq. (3.8) in the sub-horizon limit gives

$$h_\lambda\left(\vec{k}\right) = \frac{4}{9}\int \frac{d^3p}{(2\pi)^3}\frac{\eta_{\text{eq}}}{k^3\eta^2}\,e_\lambda^*\left(\vec{k},\,\vec{p}\right)\zeta\left(\vec{p}\right)\zeta\left(\vec{k}-\vec{p}\right)\left[\mathcal{I}_{\text{c}}\left(x,y\right)\cos\left(\Omega_{\eta,\vec{k}}\right) + \mathcal{I}_{\text{s}}\left(x,y\right)\sin\left(\Omega_{\eta,\vec{k}}\right)\right] \tag{3.40}$$

where we have introduced

$$\Omega_{\eta,\vec{k}} = k\,\eta + 2k\int_{\eta_{\text{eq}}}^\eta d\eta'\,\Phi\left(\eta',\,(\eta'-\eta_0)\,\hat{k}\right) = k\,\eta + \frac{6}{5}k\int_{\eta_{\text{eq}}}^\eta d\eta'\,\zeta^L\left((\eta'-\eta_0)\,\hat{k}\right). \tag{3.41}$$

In the last step we have expressed the gravitational potential $\Phi$ in terms of the long-wavelength curvature perturbation $\zeta^L$ as $\Phi = 3\zeta^L/5$ in a matter-dominated universe, and identified with $\hat{k}$ the GW direction of motion. We stress that $\zeta^L$ has a time dependence due to the GW motion, and we assume $\zeta^L$ to be Gaussian distributed with a scalar invariant power spectrum $\mathcal{P}_\zeta^L$. The solution then becomes

$$h_\lambda\left(\vec{k}\right) = \frac{2}{9}\int \frac{d^3p}{(2\pi)^3}\frac{\eta_{\text{eq}}}{k^3\eta^2}\,e_{\lambda,ij}^*\left(\hat{k}\right)\vec{p}_i\vec{p}_j\zeta\left(\vec{p}\right)\zeta\left(\vec{k}-\vec{p}\right)$$

$$\cdot\left[\mathcal{I}^*\left(\frac{p}{k},\frac{|\vec{k}-\vec{p}|}{k}\right)e^{ik\eta + i\frac{6}{5}k\int_{\eta_{\text{eq}}}^\eta d\eta'\,\zeta^L\left((\eta'-\eta_0)\hat{k}\right)} + \mathcal{I}\left(\frac{p}{k},\frac{|\vec{k}-\vec{p}|}{k}\right)e^{-ik\eta - i\frac{6}{5}k\int_{\eta_{\text{eq}}}^\eta d\eta'\,\zeta^L\left(-(\eta'-\eta_0)\hat{k}\right)}\right]. \tag{3.42}$$

**The GW propagation: effect on the power spectrum**

To investigate the effect of the propagation onto the GW power spectrum, one can define the quantity

$$\hat{Z}\left(\eta,\vec{k}\right) \equiv \frac{6}{5}k\int_{\eta_{\text{eq}}}^\eta d\eta'\,\zeta^L\left((\eta'-\eta_0)\,\hat{k}\right). \tag{3.43}$$

The two point correlator of the tensor modes then gives

$$\left\langle h_{\lambda_1}\left(\eta,\vec{k}\right)h_{\lambda_2}\left(\eta,\vec{k}'\right)\right\rangle = \frac{\pi^5}{81}\frac{A_s^2}{k_*^2}\frac{\eta_{\text{eq}}^2}{k^7\eta^4}\,(k-2k_*)^2\,(k+2k_*)^2\,\theta\,(2k_*-k)\,\delta_D^{(3)}\left(\vec{k}+\vec{k}'\right)\delta_{\lambda_1,\lambda_2}$$

$$\cdot \left\langle \left[ \mathcal{I}^* \left( \frac{k_*}{k}, \frac{k_*}{k} \right) \mathrm{e}^{ik\eta + i\hat{Z}\left(\eta, \vec{k}\right)} + \mathcal{I} \left( \frac{k_*}{k}, \frac{k_*}{k} \right) \mathrm{e}^{-ik\eta - i\hat{Z}\left(\eta, -\vec{k}\right)} \right] \right.$$

$$\left. \cdot \left[ \mathcal{I}^* \left( \frac{k_*}{k}, \frac{k_*}{k} \right) \mathrm{e}^{ik\eta + i\hat{Z}\left(\eta, -\vec{k}\right)} + \mathcal{I} \left( \frac{k_*}{k}, \frac{k_*}{k} \right) \mathrm{e}^{-ik\eta - i\hat{Z}\left(\eta, \vec{k}\right)} \right] \right\rangle, \quad (3.44)$$

where we have used that the short and long modes of the curvature perturbation are not correlated. Performing the time average, which leaves only the terms proportional to $|\mathcal{I}|^2$, gives

$$\overline{\mathcal{P}_h\left(\eta, k\right)} = \frac{1}{648} \frac{A_s^2}{k_*^2} \frac{\eta_{\mathrm{eq}}^2}{\eta^4} \left( \frac{4k_*^2}{k^2} - 1 \right)^2 \theta \left( 2 - \frac{k}{k_*} \right) \left[ \mathcal{I}_c^2 \left( \frac{k_*}{k}, \frac{k_*}{k} \right) + \mathcal{I}_s^2 \left( \frac{k_*}{k}, \frac{k_*}{k} \right) \right], \quad (3.45)$$

from which one notices that there are no propagation effects in the power spectrum.

### The GW propagation: effect on the bispectrum

The contraction of three tensor modes in the equilateral configuration $|\vec{k}_1| = |\vec{k}_2| = |\vec{k}_3| = k$ gives the bispectrum

$$B_h^{\lambda_i}\left(\eta_i, \vec{k}_i\right) = \mathcal{N} \left\langle \left[ \mathcal{I}^* \left( \frac{k_*}{k}, \frac{k_*}{k} \right) \mathrm{e}^{ik\eta_1 + i\hat{Z}\left(\eta_1, \vec{k}_1\right)} + \mathcal{I} \left( \frac{k_*}{k}, \frac{k_*}{k} \right) \mathrm{e}^{-ik\eta_1 - i\hat{Z}\left(\eta_1, -\vec{k}_1\right)} \right] \right.$$

$$\left. \cdot \left[ \mathcal{I}^* \left( \frac{k_*}{k}, \frac{k_*}{k} \right) \mathrm{e}^{ik\eta_2 + i\hat{Z}\left(\eta_2, \vec{k}_2\right)} + \mathcal{I} \left( \frac{k_*}{k}, \frac{k_*}{k} \right) \mathrm{e}^{-ik\eta_2 - i\hat{Z}\left(\eta_2, -\vec{k}_2\right)} \right] \right.$$

$$\left. \cdot \left[ \mathcal{I}^* \left( \frac{k_*}{k}, \frac{k_*}{k} \right) \mathrm{e}^{ik\eta_3 + i\hat{Z}\left(\eta_3, \vec{k}_3\right)} + \mathcal{I} \left( \frac{k_*}{k}, \frac{k_*}{k} \right) \mathrm{e}^{-ik\eta_3 - i\hat{Z}\left(\eta_3, -\vec{k}_3\right)} \right] \right\rangle, \quad (3.46)$$

where we have defined the overall factor

$$\mathcal{N} \equiv \frac{\theta\left(\sqrt{3}k_* - k\right)}{k^6} \times \frac{A_s^3 \, \eta_{\mathrm{eq}}^3}{k_*^3 \eta_1^2 \eta_2^2 \eta_3^2} \times \frac{1024\pi^3}{729} \times \mathcal{D}_{\lambda_i} \left( \sqrt{\frac{3k_*^2}{k^2} - 1} \right)^{-1/2} \times \frac{1}{8}. \quad (3.47)$$

Given that the relative variations of the three times are much smaller than the age of the universe, one can study the equal-time bispectrum setting $\eta_1 = \eta_2 = \eta_3 = \eta$. Introducing the definition

$$\mathcal{G}_{c_1, c_2, c_3} \left[ \vec{k}_1, \, \vec{k}_2, \, \vec{k}_3 \right] \equiv \left\langle \mathrm{e}^{ic_1 \hat{Z}\left(\eta, \vec{k}_1\right)} \mathrm{e}^{ic_2 \hat{Z}\left(\eta, \vec{k}_2\right)} \mathrm{e}^{ic_3 \hat{Z}\left(\eta, \vec{k}_3\right)} \right\rangle, \quad (3.48)$$

the three-point function can be recast as

$$B_h^{\lambda_i}\left(\eta, \vec{k}_i\right) = \mathcal{N} \left\{ \mathcal{I}^{*3} \, \mathrm{e}^{3ik\eta} \, \mathcal{G}_{+++} \left[ \vec{k}_1, \, \vec{k}_2, \, \vec{k}_3 \right] \right.$$

$$+ \mathcal{I}^{*2} \, \mathcal{I} \, \mathrm{e}^{ik\eta} \left[ \mathcal{G}_{++-} \left[ \vec{k}_1, \, \vec{k}_2, \, -\vec{k}_3 \right] + \mathcal{G}_{+-+} \left[ \vec{k}_1, \, -\vec{k}_2, \, \vec{k}_3 \right] + \mathcal{G}_{-++} \left[ -\vec{k}_1, \, \vec{k}_2, \, \vec{k}_3 \right] \right]$$

$$+ \mathcal{I}^* \, \mathcal{I}^2 \, \mathrm{e}^{-ik\eta} \left[ \mathcal{G}_{+--} \left[ \vec{k}_1, \, -\vec{k}_2, \, -\vec{k}_3 \right] + \mathcal{G}_{-+-} \left[ -\vec{k}_1, \, \vec{k}_2, \, -\vec{k}_3 \right] + \mathcal{G}_{--+} \left[ -\vec{k}_1, \, -\vec{k}_2, \, \vec{k}_3 \right] \right]$$

$$+ \mathcal{I}^3 \, \mathrm{e}^{-3ik\eta} \, \mathcal{G}_{---} \left[ -\vec{k}_1, \, -\vec{k}_2, \, -\vec{k}_3 \right] \bigg\}. \tag{3.49}$$

The following identity valid for a general Gaussian operator

$$\left\langle \mathrm{e}^{\varphi_1} \mathrm{e}^{\varphi_2} \mathrm{e}^{\varphi_3} \right\rangle = \mathrm{e}^{\frac{\langle \varphi_1^2 \rangle}{2} + \frac{\langle \varphi_2^2 \rangle}{2} + \frac{\langle \varphi_3^2 \rangle}{2} + \langle \varphi_1 \varphi_2 \rangle + \langle \varphi_1 \varphi_3 \rangle + \langle \varphi_2 \varphi_3 \rangle} \tag{3.50}$$

can be used to write

$$\mathcal{G}_{c_1, c_2, c_3} \left[ \vec{k}_1, \, \vec{k}_2, \, \vec{k}_3 \right] = \exp^{-\frac{1}{2} c_1^2 \mathcal{C}(\vec{k}_1, \vec{k}_1) - \frac{1}{2} c_2^2 \mathcal{C}(\vec{k}_2, \vec{k}_2) - \frac{1}{2} c_3^2 \mathcal{C}(\vec{k}_3, \vec{k}_3) - c_1 c_2 \mathcal{C}(\vec{k}_1, \vec{k}_2) - c_1 c_3 \mathcal{C}(\vec{k}_1, \vec{k}_3) - c_2 c_3 \mathcal{C}(\vec{k}_2, \vec{k}_3)}, \tag{3.51}$$

where we have introduced the correlation

$$\mathcal{C} \left( \vec{k}_1, \, \vec{k}_2 \right) \equiv \left\langle \hat{Z} \left( \eta, \, \vec{k}_1 \right) \hat{Z} \left( \eta, \, \vec{k}_2 \right) \right\rangle = \frac{36}{25} k_1 k_2 \int_{\eta_{\mathrm{eq}}}^{\eta} d\eta' \int_{\eta_{\mathrm{eq}}}^{\eta} d\eta'' \left\langle \zeta^L \left( (\eta' - \eta_0) \, \hat{k}_1 \right) \zeta^L \left( (\eta'' - \eta_0) \, \hat{k}_2 \right) \right\rangle. \tag{3.52}$$

By Fourier transforming the long mode, one deduces its two-point correlator as

$$\left\langle \zeta^L \left( (\eta' - \eta_0) \, \hat{k}_1 \right) \zeta^L \left( (\eta'' - \eta_0) \, \hat{k}_2 \right) \right\rangle = \int \frac{d^3 p}{4\pi p^3} \, \mathrm{e}^{i(\eta' - \eta_0) \hat{k}_1 \cdot \vec{p}} \mathrm{e}^{-i(\eta'' - \eta_0) \hat{k}_2 \cdot \vec{p}} \, \mathcal{P}_\zeta^L (p) \,, \tag{3.53}$$

from which

$$\mathcal{C} \left( \vec{k}_1, \, \vec{k}_2 \right) = \frac{36}{25} k_1 k_2 \int_{\eta_{\mathrm{eq}}}^{\eta} d\eta' \int_{\eta_{\mathrm{eq}}}^{\eta} d\eta'' \int \frac{d^3 p}{4\pi p^3} \, \mathrm{e}^{-i(\eta_0 - \eta') \hat{k}_1 \cdot \vec{p}} \mathrm{e}^{i(\eta_0 - \eta'') \hat{k}_2 \cdot \vec{p}} \, \mathcal{P}_\zeta^L (p) \,. \tag{3.54}$$

Using the expansion of the exponential terms in spherical harmonics

$$\mathrm{e}^{i\vec{k} \cdot \vec{r}} = 4\pi \sum_{\ell=0}^{\infty} \sum_{m=-\ell}^{\ell} \, i^\ell \, j_\ell \left( k \, r \right) Y_{\ell m} \left( \hat{r} \right) Y_{\ell m}^* \left( \hat{k} \right), \tag{3.55}$$

after some simplifications one gets

$$\begin{aligned} \mathcal{C} \left( \vec{k}_1, \, \vec{k}_2 \right) = \frac{36}{25} k_1 k_2 \int_{\eta_{\mathrm{eq}}}^{\eta} d\eta' \int_{\eta_{\mathrm{eq}}}^{\eta} d\eta'' \, 4\pi \sum_{\ell=0}^{\infty} \sum_{m=-\ell}^{\ell} Y_{\ell m}^* \left( \hat{k}_1 \right) Y_{\ell m} \left( \hat{k}_2 \right) \\ \times \int \frac{dp}{p} \, \mathcal{P}_\zeta^L (p) \, j_\ell \left( (\eta_0 - \eta') \, p \right) j_\ell \left( (\eta_0 - \eta'') \, p \right). \end{aligned} \tag{3.56}$$

This integral would diverge logarithmically due to the $\ell = 0$ constant zero mode of the gravitational potential. However, being unphysical, it can be removed following the same procedure done in Ref. [311]. Furthermore, using the approximation valid for $\ell \gg 1$ [312]

$$\int_0^\infty dp \, p^2 \, f(p) \, j_\ell (p\eta) \, j_\ell (p\eta') \simeq \frac{\pi}{2\eta^2} f \left( \frac{\ell + 1/2}{\eta} \right) \delta_D \left( \eta - \eta' \right), \tag{3.57}$$

where we set $f(p) = \mathcal{P}_\zeta^L(p)/p^3 = A_L/p^3$, one gets

$$\mathcal{C}\left(\vec{k}_1, \vec{k}_2\right) = \frac{36}{25} k_1 k_2 \times 2\pi^2 \int_{\eta_{\text{eq}}}^{\eta} d\eta' \ (\eta_0 - \eta') \sum_{\ell=1}^{\infty} \sum_{m=-\ell}^{\ell} \frac{A_L}{\left(\ell + \frac{1}{2}\right)^3} Y_{\ell m}^*\left(\hat{k}_1\right) Y_{\ell m}\left(\hat{k}_2\right). \tag{3.58}$$

We assume $\eta = \eta_0$ in the extremum of time integration and therefore disregard the equality time. Orienting the vector $\hat{k}_1$ along the $z-$axis, so to include only the $m = 0$ term in the sum, and labelling $\hat{k}_1 \cdot \hat{k}_2 \equiv \mu$, gives

$$Y_{\ell m}^*\left(\hat{k}_1\right) Y_{\ell m}\left(\hat{k}_2\right) = \delta_{m0} \frac{2\ell+1}{4\pi} P_\ell\left(1\right) P_\ell\left(\mu\right) = \delta_{m0} \frac{2\ell+1}{4\pi} P_\ell\left(\mu\right), \tag{3.59}$$

where in the last step we used $P_\ell(1) = 1$. One therefore obtains

$$\mathcal{C}\left(\vec{k}_1, \vec{k}_2\right) = A_L\, k_1\, k_2\, \eta_0^2 \times \frac{18\,\pi}{25} \sum_{\ell=1}^{\infty} \frac{P_\ell\left(\hat{k}_1 \cdot \hat{k}_2\right)}{\left(\ell + \frac{1}{2}\right)^2} \equiv A_L\, k_1\, k_2\, \eta_0^2 \times \mathcal{S}\left(\hat{k}_1 \cdot \hat{k}_2\right), \tag{3.60}$$

which differs from zero in the following momentum configurations

$$i = j : \quad \mathcal{C}\left(\vec{k}_i, \vec{k}_j\right) = A_L\, k^2\, \eta_0^2 \times \mathcal{S}\left(1\right) \simeq \frac{18\,\pi}{25} \left(\frac{\pi^2}{2} - 4\right) A_L\ (k\,\eta_0)^2 \simeq 2.11\ A_L\ (k\,\eta_0)^2,$$

$$i \neq j : \quad \mathcal{C}\left(\pm\vec{k}_i, \pm\vec{k}_j\right) = A_L\, k^2\, \eta_0^2 \times \mathcal{S}\left(-1/2\right) \simeq -0.50\ A_L\ (k\,\eta_0)^2,$$

$$i \neq j : \quad \mathcal{C}\left(\pm\vec{k}_i, \mp\vec{k}_j\right) = A_L\, k^2\, \eta_0^2 \times \mathcal{S}\left(1/2\right) \simeq 0.37\ A_L\ (k\,\eta_0)^2. \tag{3.61}$$

The GW bispectrum then becomes

$$B_h^{\lambda_i}\left(\eta_0, \vec{k}_i\right) = \mathcal{N}\left\{ \mathcal{I}^{*3}\, \mathrm{e}^{3ik\eta_0}\, \mathrm{e}^{-1.67\,A_L\,(k\,\eta_0)^2} + \mathcal{I}^{*2}\, \mathcal{I}\, \mathrm{e}^{ik\eta_0}\left[\mathrm{e}^{-1.93\,A_L\,(k\,\eta_0)^2} \times 3\right] \right.$$

$$\left. + \mathcal{I}^*\, \mathcal{I}^2\, \mathrm{e}^{-ik\eta_0}\left[\mathrm{e}^{-1.93\,A_L\,(k\,\eta_0)^2} \times 3\right] + \mathcal{I}^3\, \mathrm{e}^{-3ik\eta_0}\, \mathrm{e}^{-1.67\,A_L\,(k\,\eta_0)^2} \right\}$$

$$\simeq \mathcal{N}\, \mathrm{e}^{-1.67\,A_L\,(k\,\eta_0)^2}\left[\mathcal{I}^{*3}\, \mathrm{e}^{3ik\eta_0} + \mathcal{I}^3\, \mathrm{e}^{-3ik\eta_0}\right], \tag{3.62}$$

where we kept only the first and last term, being the least suppressed. Introducing the root-mean-square of the (relative) time delay [311]

$$d_{\text{rms}}^2 = \frac{4}{\eta_0^2}\left\langle \int d\eta' d\eta'' \Phi\left(\eta'\right) \Phi\left(\eta''\right) \right\rangle = \frac{4}{\eta_0^2 k^2} \times \frac{9}{25} k^2 \left\langle \int d\eta' d\eta'' \zeta_L\left(\eta'\right) \zeta_L\left(\eta''\right) \right\rangle$$

$$= \frac{1}{\eta_0^2\, k^2}\, \mathcal{C}\left(\vec{k}, \vec{k}\right) = 2.11\, A_L, \tag{3.63}$$

one can compute the ratio between the unperturbed and perturbed equilateral bispectrum as

$$\frac{B_h^{\lambda_i}\left(\eta_0, \vec{k}_i\right)\Big|_{\text{inhom.}}}{B_h^{\lambda_i}\left(\eta_0, \vec{k}_i\right)\Big|_{\text{no inhom.}}} = \mathrm{e}^{-1.67\,A_L\,(k\,\eta_0)^2\,\frac{d_{\text{rms}}^2}{2.11\,A_L}} = \mathrm{e}^{-0.8\,k^2\eta_0^2\,d_{\text{rms}}^2}, \tag{3.64}$$

which is largely suppressed being $k\eta_0\, d_{\rm rms} \sim 10^9$, if we take $k \sim 10^{-3}$ Hz.

In the squeezed limit $k_1 \sim k_2 \gg k_3$, the bispectrum should reduce to the average of the small-scale power spectrum in a background modulated by the large-scale mode $k_3$. One can show that

$$B_h^{\lambda_i}\left(\eta_0,\, \vec{k}_i\right) \propto |\mathcal{I}_1|^2 \left(\mathcal{I}_3^* {\rm e}^{i\eta_0 k_3} + \mathcal{I}_3 {\rm e}^{-i\eta_0 k_3}\right) {\rm e}^{-\frac{1}{2}k_3^2 \eta_0^2\, d_{\rm rms}^2}. \tag{3.65}$$

A sizeable suppression arises from the average, over many directions, over the long mode. We therefore conclude that propagation effects are present for arbitrary shapes.

### 3.1.4 Anisotropies in the SGWB spectrum

In this subsection we are going to compute the amount of anisotropy which characterise the SGWB inherited at PBH formation. We will see that, to have anisotropies on large scales, one needs long-wavelength perturbations to produce correlations on scales much bigger than those associated to the formation of PBHs. In particular, being the generation of GWs a local event, according to the equivalence principle its physics cannot be affected by modes of wavelength much greater than the PBH horizon. However, the presence of a local primordial non-Gaussianity in the curvature perturbation can lead to small-long scale correlations, such that the amount of GWs produced within each region experience a large-scale modulation.

Following the result of Ref. [20], we first show that, under the assumption of Gaussian curvature perturbations, the amount of anisotropies inherited in the SGWB is almost negligible. However, this conclusion changes in the presence of primordial non-Gaussianity, correlating short and long scales, such that anisotropies are present in the GW spectrum.

**The Gaussian case**

In the following we sketch the main steps to compute the amount of anisotropies in the SGWB assuming a Gaussian curvature perturbation $\zeta$. For simplicity, we consider two relevant scales: the short scale $k_*^{-1}$, of the order of the horizon scale at PBH production, and the long wavelength scale $q^{-1}$, such that we can expand the curvature perturbation in the two components as $\zeta = \zeta_s + \zeta_L$ following the peak-background split picture. The absence of a local non-Gaussianity correlating the two regimes implies that the scalar modes with wavelength of cosmological size do not affect the local physics on small scales according to the Equivalence Principle.

This result can be confirmed from the computation of the two-point correlation function of the GW energy density operator [290, 304]

$$\rho_{\rm GW}\left(\eta,\, \vec{x}\right) = \frac{M_p^2}{81\eta^2 a^2} \int \frac{d^3k_1 d^3k_2 d^3p_1 d^3p_2}{(2\pi)^{12}} \frac{1}{k_1^2 k_2^2} {\rm e}^{i\vec{x}\cdot(\vec{k}_1 + \vec{k}_2)} T[\hat{k}_1,\, \hat{k}_2,\, \vec{p}_1,\, \vec{p}_2]$$

$$\times \zeta(\vec{p_1})\zeta(\vec{k_1}-\vec{p_1})\zeta(\vec{p_2})\zeta(\vec{k_2}-\vec{p_2})\left\langle \prod_{i=1}^{2}\left[\mathcal{I}_s(\vec{k_i},\,\vec{p_i})\cos(k_i\eta)-\mathcal{I}_c(\vec{k_i},\,\vec{p_i})\sin(k_i\eta)\right]\right\rangle_T, \tag{3.66}$$

in terms of the function

$$T\left[\hat{k}_1,\,\hat{k}_2,\,\vec{p_1},\,\vec{p_2}\right]\equiv\sum_{\lambda_1,\lambda_2}e_{ij,\lambda_1}(\hat{k}_1)e^*_{ab,\lambda_1}(\hat{k}_1)e_{ij,\lambda_2}(\hat{k}_2)e^*_{cd,\lambda_2}(\hat{k}_2)\vec{p}_{1a}\vec{p}_{1b}\vec{p}_{2c}\vec{p}_{2d}, \tag{3.67}$$

obtained from the contraction between the internal momenta and the GW polarization operators, and where the angular brackets denote an average over time $T$ [185, 313, 314]. The two-point correlation function $\langle\rho_{\rm GW}(\vec{x})\,\rho_{\rm GW}(\vec{y})\rangle$ depends spatially on the distance $|\vec{x}-\vec{y}|$ as a consequence of statistical isotropy and homogeneity, and it is given by

$$\langle\rho_{\rm GW}(\eta_1,\,\vec{x})\,\rho_{\rm GW}(\eta_2,\,\vec{y})\rangle = \left(\frac{M_p^2}{81a^2(\eta_1)a^2(\eta_2)\eta_1\eta_2}\right)^2\int\frac{d^3k_1d^3k_2d^3p_1d^3p_2}{(2\pi)^{12}}\frac{1}{k_1^2k_2^2}\,{\rm e}^{i\vec{x}\cdot(\vec{k_1}+\vec{k_2})}T\left[\hat{k}_1,\,\hat{k}_2,\,\vec{p_1},\,\vec{p_2}\right]$$

$$\times\int\frac{d^3k_3d^3k_4d^3p_3d^3p_4}{(2\pi)^{12}}\frac{1}{k_3^2k_4^2}\,{\rm e}^{i\vec{y}\cdot(\vec{k_3}+\vec{k_4})}T\left[\hat{k}_3,\,\hat{k}_4,\,\vec{p_3},\,\vec{p_4}\right]\left\langle\zeta_{\vec{p_1}}\zeta_{\vec{k_1}-\vec{p_1}}\zeta_{\vec{p_2}}\zeta_{\vec{k_2}-\vec{p_2}}\zeta_{\vec{p_3}}\zeta_{\vec{k_3}-\vec{p_3}}\zeta_{\vec{p_4}}\zeta_{\vec{k_4}-\vec{p_4}}\right\rangle$$

$$\times\left\langle\left[\mathcal{I}_s(\vec{k_1},\,\vec{p_1})\cos(k_1\eta_1)-\mathcal{I}_c(\vec{k_1},\,\vec{p_1})\sin(k_1\eta_1)\right]\left[\mathcal{I}_s(\vec{k_2},\,\vec{p_2})\cos(k_2\eta_1)-\mathcal{I}_c(\vec{k_2},\,\vec{p_2})\sin(k_2\eta_1)\right]\right\rangle_T$$

$$\times\left\langle\left[\mathcal{I}_s(\vec{k_3},\,\vec{p_3})\cos(k_3\eta_2)-\mathcal{I}_c(\vec{k_3},\,\vec{p_3})\sin(k_3\eta_2)\right]\left[\mathcal{I}_s(\vec{k_4},\,\vec{p_4})\cos(k_4\eta_2)-\mathcal{I}_c(\vec{k_4},\,\vec{p_4})\sin(k_4\eta_2)\right]\right\rangle_T. \tag{3.68}$$

The stochastic average of the 8-point correlator of the curvature perturbation

$$\left\langle\zeta_{\vec{p_1}}\zeta_{\vec{k_1}-\vec{p_1}}\zeta_{\vec{p_2}}\zeta_{\vec{k_2}-\vec{p_2}}\zeta_{\vec{p_3}}\zeta_{\vec{k_3}-\vec{p_3}}\zeta_{\vec{p_4}}\zeta_{\vec{k_4}-\vec{p_4}}\right\rangle$$

$$=4\left\langle\zeta_{\vec{p_1}}\zeta_{\vec{p_2}}\right\rangle\left\langle\zeta_{\vec{k_1}-\vec{p_1}}\zeta_{\vec{k_2}-\vec{p_2}}\right\rangle\left\langle\zeta_{\vec{p_3}}\zeta_{\vec{p_4}}\right\rangle\left\langle\zeta_{\vec{k_3}-\vec{p_3}}\zeta_{\vec{k_4}-\vec{p_4}}\right\rangle$$

$$+8\left\langle\zeta_{\vec{p_1}}\zeta_{\vec{p_3}}\right\rangle\left\langle\zeta_{\vec{k_1}-\vec{p_1}}\zeta_{\vec{k_3}-\vec{p_3}}\right\rangle\left\langle\zeta_{\vec{p_2}}\zeta_{\vec{p_4}}\right\rangle\left\langle\zeta_{\vec{k_2}-\vec{p_2}}\zeta_{\vec{k_4}-\vec{p_4}}\right\rangle$$

$$+32\left\langle\zeta_{\vec{p_1}}\zeta_{\vec{p_2}}\right\rangle\left\langle\zeta_{\vec{p_3}}\zeta_{\vec{p_4}}\right\rangle\left\langle\zeta_{\vec{k_1}-\vec{p_1}}\zeta_{\vec{k_3}-\vec{p_3}}\right\rangle\left\langle\zeta_{\vec{k_2}-\vec{p_2}}\zeta_{\vec{k_4}-\vec{p_4}}\right\rangle$$

$$+16\left\langle\zeta_{\vec{p_1}}\zeta_{\vec{p_3}}\right\rangle\left\langle\zeta_{\vec{k_1}-\vec{p_1}}\zeta_{\vec{k_4}-\vec{p_4}}\right\rangle\left\langle\zeta_{\vec{k_2}-\vec{p_2}}\zeta_{\vec{k_3}-\vec{p_3}}\right\rangle\left\langle\zeta_{\vec{p_2}}\zeta_{\vec{p_4}}\right\rangle \tag{3.69}$$

includes the disconnected (case A) and connected terms (case B, C, D, respectively), which are shown diagrammatically in Fig. 3.5. The disconnected diagram A gives rise to the square of the expectation value of the energy density field as

$$\langle\rho_{\rm GW}(\eta,\,\vec{x})\,\rho_{\rm GW}(\eta,\,\vec{y})\rangle_A = \langle\rho_{\rm GW}(\eta,\,\vec{x})\rangle\,\langle\rho_{\rm GW}(\eta,\,\vec{y})\rangle = \langle\rho_{\rm GW}(\eta)\rangle^2, \tag{3.70}$$

which is homogeneous.

We now evaluate the example case of diagram B to show that also the remaining connected diagrams are suppressed on large scales. It is given by

$$\langle\rho_{\rm GW}(\eta,\,\vec{x})\,\rho_{\rm GW}(\eta,\,\vec{y})\rangle_B = \frac{1}{2^7\pi^4}\left(\frac{M_p^2}{81a^2\eta^2}\right)^2\int d^3k_1d^3p_1d^3k_4d^3p_4\,{\rm e}^{i(\vec{x}-\vec{y})\cdot(\vec{k_1}-\vec{k_4})}$$

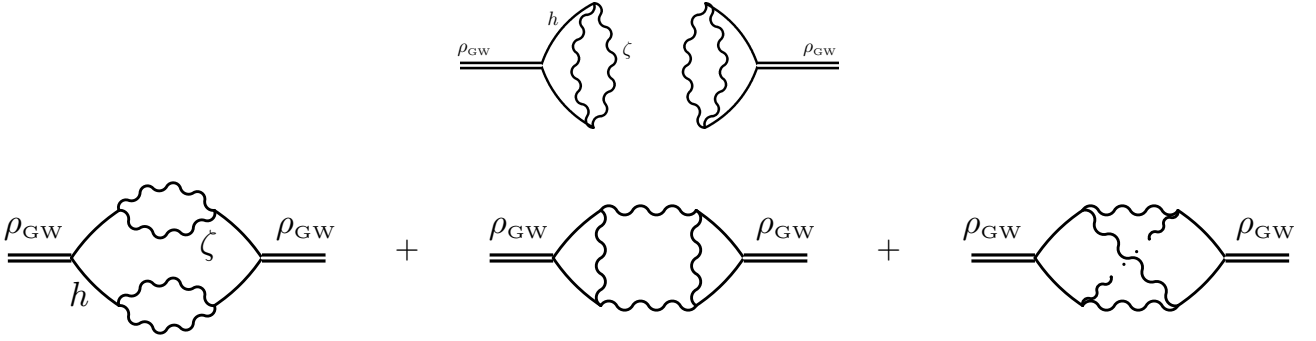

**Figure 3.5**: Feynman diagrams for the disconnected (top) and connected terms (bottom) in the energy density two-point function (case A,B,C,D). The double lines identify the energy density field, the solid lines identify the gravitational waves and the wiggly lines identify the curvature field. Figure taken from Ref. [20].

$$
\begin{aligned}
\times\ & \frac{1}{k_1^4 k_4^4}\frac{1}{p_1^3}\frac{1}{p_4^3}\frac{1}{\left|\vec{k}_1-\vec{p}_1\right|^3}\frac{1}{\left|\vec{k}_4-\vec{p}_4\right|^3}T\left[\hat{k}_1,\,-\hat{k}_4,\,\vec{p}_1,\,-\vec{p}_4\right]T\left[-\hat{k}_1,\,\hat{k}_4,\,-\vec{p}_1,\,\vec{p}_4\right]\\
\times\ & \mathcal{P}_\zeta(p_1)\mathcal{P}_\zeta(p_4)\mathcal{P}_\zeta\left(\left|\vec{k}_1-\vec{p}_1\right|\right)\mathcal{P}_\zeta\left(\left|\vec{k}_4-\vec{p}_4\right|\right)\\
\times\ & (\mathcal{I}_{s1}\mathcal{I}_{s2}+\mathcal{I}_{c1}\mathcal{I}_{c2})\,(\mathcal{I}_{s3}\mathcal{I}_{s4}+\mathcal{I}_{c3}\mathcal{I}_{c4})\frac{\sin\left(\Delta_{34}T\right)}{T\Delta_{34}}\frac{\sin\left(\Delta_{12}T\right)}{T\Delta_{12}},
\end{aligned}
\tag{3.71}
$$

where $\mathcal{I}_{c,s,i}=\mathcal{I}_{c,s}(\vec{k}_1,\vec{p}_1)$ and $\Delta_{ij}=k_i-k_j$. Assuming a Dirac delta shape for the curvature perturbation power spectrum and, given that we are looking at anisotropies on scales $|\vec{x}-\vec{y}|\gg 1/k_*$, after some simplifications one gets

$$
\langle\rho_{\mathrm{GW}}\left(\eta,\,\vec{x}\right)\rho_{\mathrm{GW}}\left(\eta,\,\vec{y}\right)\rangle_B\simeq\frac{1}{4(2\pi)^3}\left(\frac{M_p^2 A_s^2}{81a^2\eta^2}\right)^2\times\pi\left(\frac{1}{k_*\left|\vec{x}-\vec{y}\right|}\right)^3\times\mathcal{S}_B,
\tag{3.72}
$$

where we defined

$$
\begin{aligned}
\mathcal{S}_B=\int d\Omega_s\int d\Omega_{p_1}\int d\Omega_{p_4}\delta\left(\left|\hat{s}-\hat{p}_1+\hat{p}_4\right|-1\right)\frac{1}{\left|\hat{p}_1-\hat{s}\right|^8}\\
\times\ T^2\left[\frac{\hat{p}_1-\hat{s}}{\left|\hat{p}_1-\hat{s}\right|},\,-\frac{\hat{p}_1-\hat{s}}{\left|\hat{p}_1-\hat{s}\right|},\,\hat{p}_1,\,-\hat{p}_4\right]\left[\mathcal{I}^2\left(\frac{1}{\left|\hat{p}_1-\hat{s}\right|},\frac{1}{\left|\hat{p}_1-\hat{s}\right|}\right)\right]^2.
\end{aligned}
\tag{3.73}
$$

Performing the angular integral numerically, one finds that this contribution is highly suppressed by the term $\propto 1/(k_*\left|\vec{x}-\vec{y}\right|)^3$. A similar result holds for the other two connected diagrams, implying no appreciable contribution to the GWs anisotropy.

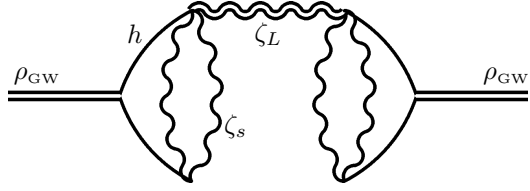

**Figure 3.6**: Feynman diagram for the energy density two-point function connected by a $f_{\mathrm{NL}}$ bridge. The double wiggly line indicates a $\zeta_L$ long mode. Figure taken from Ref. [20].

**The non-Gaussian case**

Even though the effects discussed above lead to a homogeneous and isotropic distribution of the GW energy density, there are additional effects which may give rise to a large anisotropy. One possibility is provided by propagation effects [315, 316], for which GWs coming from different regions experience disconnected large scales density perturbations and arrive anisotropic. The other possibility is given by primordial non-Gaussianity correlating short and long scales, which may modulate the power of scales $k_*^{-1}$. Starting from the latter possibility, we assume the following parametrization for the curvature perturbation

$$\zeta(\vec{k}) = \zeta_g(\vec{k}) + \frac{3}{5} f_{\mathrm{NL}} \int \frac{d^3 p}{(2\pi)^3} \zeta_g(\vec{p}) \, \zeta_g(\vec{k} - \vec{p}), \tag{3.74}$$

in which the non-Gaussian contribution comes into an expansion around the large Gaussian contribution $\zeta_g$ in the so-called local shape, $\zeta = \zeta_g + \frac{3}{5} f_{\mathrm{NL}} \zeta_g^2$, in real space. The value of the local non-linear parameter is constrained by the Planck collaboration to [317]

$$-11.1 \leq f_{\mathrm{NL}} \leq 9.3 \qquad \text{at 95\% C.L.} \tag{3.75}$$

Adding the non-Gaussian term induces the presence of trilinear vertices $\zeta^3$, each proportional to $f_{\mathrm{NL}}$, such that the disconnected diagram A considered before now becomes the one in Fig. 3.6 involving two short-scale and one long-scale mode, that we will denote as "$f_{\mathrm{NL}}$ bridge". The energy density before correlating over the long modes can be written in the effective way

$$\rho_{\mathrm{GW}}(\eta, \vec{x}) = \bar{\rho}_{\mathrm{GW}}(\eta) \left[ 1 + \frac{24}{5} f_{\mathrm{NL}} \int \frac{d^3 q}{(2\pi)^3} \, \mathrm{e}^{i \vec{q} \cdot \vec{x}} \, \zeta_L(\vec{q}) \right], \tag{3.76}$$

in terms of the energy density field at zeroth order in the non-linear parameter, $\bar{\rho}_{\mathrm{GW}}$. This results into a GW abundance

$$\Omega_{\mathrm{GW}}(\eta, \vec{x}, k) = \bar{\Omega}_{\mathrm{GW}}(\eta, k) \left[ 1 + \frac{24}{5} f_{\mathrm{NL}} \int \frac{d^3 q}{(2\pi)^3} \, \mathrm{e}^{i \vec{q} \cdot \vec{x}} \, \zeta_L(\vec{q}) \right]. \tag{3.77}$$

The amount of anisotropy in the GW abundance can be computed introducing the contrast as in [316]

$$\delta_{\mathrm{GW}}(\eta, \vec{x}, \vec{k}) = \frac{\Omega_{\mathrm{GW}}(\eta, \vec{x}, \vec{k}) - \bar{\Omega}_{\mathrm{GW}}(\eta, k)}{\bar{\Omega}_{\mathrm{GW}}(\eta, k)} \equiv \Gamma_I(\eta, \vec{x}, \vec{k}) \left( 4 - \frac{\partial \ln \bar{\Omega}_{\mathrm{GW}}(\eta, k)}{\partial \ln k} \right), \tag{3.78}$$

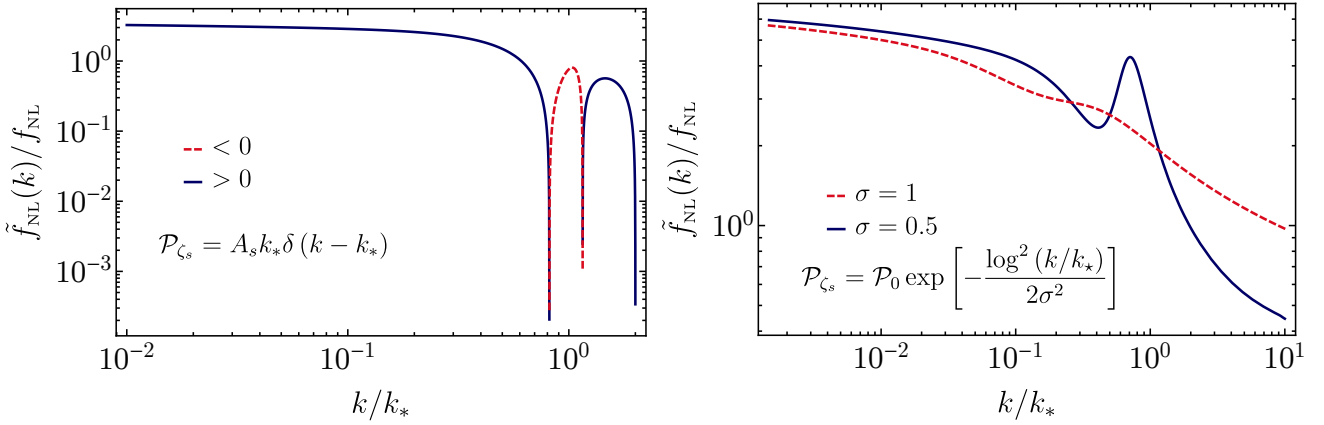

**Figure 3.7**: Plot of the rescaled non-linear parameter $\tilde{f}_{\rm NL}/f_{\rm NL}$ in terms of the momentum $k/k_*$ for a Dirac delta and lognormal power spectrum (Eq. (3.17) and Eq. (3.19), respectively). Figure taken from Ref. [20].

in terms of

$$\Gamma_I(\eta, \vec{x}, \vec{k}) = \frac{3}{5}\tilde{f}_{\rm NL}(k) \int \frac{d^3q}{(2\pi)^3} e^{i\vec{q}\cdot\vec{x}} \zeta_L(\vec{q}), \qquad \tilde{f}_{\rm NL}(k) \equiv \frac{8\,f_{\rm NL}}{4 - \frac{\partial \ln \bar{\Omega}_{\rm GW}(\eta, k)}{\partial \ln k}}. \tag{3.79}$$

This term carries all the information about the amount of anisotropy due to the initial conditions (suffix $I$). The behaviour of the rescaled non-linear parameter $\tilde{f}_{\rm NL}(k)$ for the choice of a Dirac delta and gaussian power spectrum is shown in Fig. 3.7. It can be expanded using spherical harmonics as[3]

$$\Gamma_{\ell m, I}(k) = 4\pi\,(-i)^\ell\,\frac{3}{5}\tilde{f}_{\rm NL}(k) \int \frac{d^3q}{(2\pi)^3}\zeta_L(\vec{q})\,Y^*_{\ell m}(\hat{q})\,j_\ell(q\,(\eta_0 - \eta_{\rm in})), \tag{3.80}$$

where we have evaluated the source term in $\vec{x} = \hat{n}(\eta_{\rm in} - \eta)$ (setting our location at the origin and fixing $\vec{k} = k\hat{n}$), in terms of the emission time $\eta_{\rm in}$ at which the mode $k_*$ re-enters the horizon.

An additional source of anisotropy in the GW background comes from the propagation of the signal across scalar sources ($S$) in the universe and is captured in the term

$$\Gamma_S(\eta_0, \vec{q}) = \mathcal{T}^S(q, \eta_0, \eta_{\rm in})\,\zeta_L(\vec{q}), \tag{3.81}$$

where

$$\mathcal{T}^S(q, \eta_0, \eta_{\rm in}) = \int_{\eta_{\rm in}}^{\eta_0} d\eta' e^{-i\vec{k}\cdot\hat{q}q(\eta_0 - \eta')}\left[T_\Phi(\eta', q)\,\delta_D(\eta' - \eta_{\rm in}) + \frac{\partial\left[T_\Psi(\eta', q) + T_\Phi(\eta', q)\right]}{\partial \eta'}\right]. \tag{3.82}$$

---

[3]We are adopting the normalization for the spherical harmonics $\int d\hat{n}\,Y_{\ell m}Y^*_{\ell' m'} = \delta_{\ell \ell'}\delta_{mm'}$.

The two terms inside the square bracket identify the Sachs-Wolfe and integrated Sachs-Wolfe effects, and are expressed in terms of the gravitational potentials

$$\Phi(\eta, \vec{k}) \equiv T_\Phi(\eta, k)\zeta(\vec{k}), \qquad \Psi(\eta, \vec{k}) \equiv T_\Psi(\eta, k)\zeta(\vec{k}), \tag{3.83}$$

where the transfer functions $T_\Phi(\eta_{\text{in}}, q) = T_\Psi(\eta_{\text{in}}, q) = 3/5$ during matter domination.

Adding the two contributions together one gets

$$\Gamma_{\ell m, I+S}(k) = 4\pi(-i)^\ell \int \frac{d^3 q}{(2\pi)^3} \zeta_L(\vec{q}) Y_{\ell m}^*(\hat{q}) \, \mathcal{T}_\ell^{I+S}(k, q, \eta_0, \eta_{\text{in}}), \tag{3.84}$$

where

$$\begin{aligned}
\mathcal{T}_\ell^{I+S}(k, q, \eta_0, \eta_{\text{in}}) &\equiv \frac{3}{5}\left[1 + \tilde{f}_{\text{NL}}(k)\right] j_\ell(q(\eta_0 - \eta_{\text{in}})) \\
&\quad + \int_{\eta_{\text{in}}}^{\eta_0} d\eta \, \frac{\partial\left[T_\Psi(\eta, q) + T_\Phi(\eta, q)\right]}{\partial\eta} j_\ell(q(\eta_0 - \eta)).
\end{aligned} \tag{3.85}$$

Introducing the variable $\eta' = \eta/\eta_0$ and parametrising the scalar transfer functions as [318]

$$T_\Phi(\eta, q) = T_\Psi(\eta, q) = \frac{3}{5}g(\eta) \qquad \text{where} \qquad \frac{\partial g(\eta')}{\partial\eta'} = -1.25\eta'^5, \tag{3.86}$$

one gets the expression

$$\mathcal{T}_\ell^S(k, q, \eta_0, \eta_{\text{in}}) = \frac{3}{5}\left[\left[1 + \tilde{f}_{\text{NL}}(k)\right] j_\ell(q\eta_0) + 2\int_0^1 d\eta' \frac{\partial g(\eta')}{\partial\eta'} j_\ell(q\eta_0(1 - \eta'))\right]. \tag{3.87}$$

Performing the integral numerically, one finds that the integrated Sachs-Wolfe effect is subdominant, such that the total contribution of the long mode at leading order in $\tilde{f}_{\text{NL}}$ is given by

$$\Gamma_{\ell m, I+S}(k) \simeq 4\pi(-i)^\ell \int \frac{d^3 q}{(2\pi)^3} \zeta_L(\vec{q}) Y_{\ell m}^*(\hat{q}) \frac{3}{5}\left[1 + \tilde{f}_{\text{NL}}(k)\right] j_\ell(q(\eta_0 - \eta_{\text{in}})). \tag{3.88}$$

Its two-point function can be computed as

$$\begin{aligned}
\langle \Gamma_{\ell_1 m_1, I+S}(k) \, \Gamma_{\ell_2 m_2, I+S}^*(k) \rangle &= (4\pi)^2 (-i)^{\ell_1 - \ell_2} \int \frac{d^3 q_1}{(2\pi)^3} \frac{d^3 q_2}{(2\pi)^3} Y_{\ell_1 m_1}^*(\hat{q}_1) Y_{\ell_2 m_2}(\hat{q}_2) \\
&\quad \times \left(\frac{3}{5}\right)^2 \left[1 + \tilde{f}_{\text{NL}}(k)\right]^2 j_{\ell_1}(q_1(\eta_0 - \eta_{\text{in}})) j_{\ell_2}(q_2(\eta_0 - \eta_{\text{in}})) \langle \zeta_L(\vec{q}_1) \zeta_L^*(\vec{q}_2) \rangle.
\end{aligned} \tag{3.89}$$

Assuming a scale invariant power spectrum of the long modes $\mathcal{P}_{\zeta_L}(q) = \mathcal{P}_{\zeta_L}$, one gets

$$\langle \Gamma_{\ell_1 m_1, I+S}(k) \, \Gamma_{\ell_2 m_2, I+S}^*(k) \rangle = \delta_{\ell_1 \ell_2} \delta_{m_1 m_2} 4\pi \left(\frac{3}{5}\right)^2 \left[1 + \tilde{f}_{\text{NL}}(k)\right]^2 \frac{1}{2\ell_1(\ell_1 + 1)} \mathcal{P}_{\zeta_L}. \tag{3.90}$$

The two-point function can then be recast as [316]

$$\left\langle \Gamma_{\ell_1 m_1, I+S}\left(k\right) \Gamma^*_{\ell_2 m_2, I+S}\left(k\right) \right\rangle = \delta_{\ell_1 \ell_2} \delta_{m_1 m_2}\, C_{\ell, I+S}\left(k\right), \tag{3.91}$$

such that one finally gets

$$\sqrt{\frac{\ell\left(\ell+1\right)}{2\pi}\, C_{\ell, I+S}\left(k\right)} \simeq \frac{3}{5} \left| 1 + \tilde{f}_{\text{NL}}\left(k\right)\right| \mathcal{P}^{1/2}_{\zeta_L} \simeq 2.8 \cdot 10^{-4} \left| \frac{1 + \tilde{f}_{\text{NL}}\left(k\right)}{10} \right| \left(\frac{\mathcal{P}_{\zeta_L}}{2.2 \cdot 10^{-9}}\right)^{1/2}, \tag{3.92}$$

which in the last step has been evaluated for values of the non-linear parameter close to its upper bound from the Planck constraint and using the CMB value for the power spectrum of the long modes.

**Constraints from measuring anisotropies in the GW spectrum at LISA**

In this subsection we are going to discuss the detectability of the anisotropies in the SGWB generated at PBH formation at LISA.

As previously discussed, given the large hierarchy of scales between the wavelengths responsible for the GWs formation and the spatial points at which the GW energy density are compared $k_* |\vec{x} - \vec{y}| \gg 1$, being the latter separated by non-negligible fractions of the present horizon, a negligible amount of anisotropies is expected to be detectable at LISA (given its capacity of resolving multipoles up to $\ell \sim 10$ [35, 319]). Indeed, the measurement of the GW energy density at some angular scale implies that we effectively coarse grain it with a resolution related to that scale. This results into the average of a large number of patches of size $k_*^{-1}$, and the resulting energy density becomes homogeneous due to the central limit theorem.

This conclusion can however change when primordial non-Gaussianity in the comoving curvature perturbation is introduced, correlating short and long scales and providing an amount of GW anisotropy possibly detectable at LISA. Even though the presence of a primordial non-Gaussianity has a small impact on the threshold for PBH formation [223], it induces a sizable large-scale variation of the PBH abundance through the modulation of the power on small scales induced by the long modes. This implies the production of isocurvature modes in the fraction of DM fluid composed by PBHs, which are strongly constrained by CMB observations. In particular, bounds on the relative abundance of isocurvature modes are given by the Planck experiments at 95% CL as [317]

$$100\beta_{\text{iso}} < 0.095 \qquad \text{for fully correlated,}$$
$$100\beta_{\text{iso}} < 0.107 \qquad \text{for fully anti-correlated,} \tag{3.93}$$

where fully correlated or fully anti-correlated imply a positive (negative) $f_{\text{NL}}$.

To transfer this bound on the PBH abundance, one can compute the PBH mass fraction in the presence of non-Gaussianity as [241]

$$\bar{\beta} \equiv \frac{\rho_{\text{PBH}}(\eta_{\text{in}})}{\rho_{\text{c}}(\eta_{\text{in}})} = \begin{cases} \sqrt{\dfrac{2}{\pi\sigma_s^2}} \left[ \displaystyle\int_{\zeta_+}^{\infty} d\zeta \exp\left(-\dfrac{\zeta^2}{2\sigma_s^2}\right) + \int_{-\infty}^{\zeta_-} d\zeta \exp\left(-\dfrac{\zeta^2}{2\sigma_s^2}\right) \right] & \text{for} \quad f_{\text{NL}} > 0, \\[3ex] \sqrt{\dfrac{2}{\pi\sigma_s^2}} \left[ \displaystyle\int_{\zeta_+}^{\infty} d\zeta \exp\left(-\dfrac{\zeta^2}{2\sigma_s^2}\right) - \int_{\zeta_-}^{\infty} d\zeta \exp\left(-\dfrac{\zeta^2}{2\sigma_s^2}\right) \right] & \text{for} \quad f_{\text{NL}} < 0, \end{cases} \tag{3.94}$$

in terms of [241]

$$\zeta_\pm = \frac{-5 \pm \sqrt{25 + 60\zeta_c f_{\text{NL}} + 36 f_{\text{NL}}^2 \sigma_s^2}}{6 f_{\text{NL}}}, \tag{3.95}$$

where $\zeta_c$ is the PBH threshold in the presence of non-Gaussianity [320], and $\sigma_s^2$ is the variance of the short modes. With respect to its average $\bar{\beta}$, the perturbation to the mass fraction at leading order in the long modes is

$$\delta_\beta \equiv \frac{\beta - \bar{\beta}}{\bar{\beta}} = \left( \frac{25 + 30\zeta_c f_{\text{NL}} + 36 f_{\text{NL}}^2 \sigma_s^2 - 5\sqrt{25 + 60\zeta_c f_{\text{NL}} + 36 f_{\text{NL}}^2 \sigma_s^2}}{3 f_{\text{NL}} \sigma_s^2 \sqrt{25 + 60\zeta_c f_{\text{NL}} + 36 f_{\text{NL}}^2 \sigma_s^2}} \right) \zeta_L \equiv b\,\zeta_L, \tag{3.96}$$

where we have introduced the bias $b$ induced by the long modes. The relative abundance of isocurvature modes is then given by

$$\beta_{\text{iso}} \equiv \frac{\mathcal{P}_{\text{iso}}}{\mathcal{P}_{\text{iso}} + \mathcal{P}_{\zeta_L}} = \frac{b^2 f_{\text{PBH}}^2}{b^2 f_{\text{PBH}}^2 + 1}, \tag{3.97}$$

where we have used $\mathcal{P}_{\text{iso}} = b^2 f_{\text{PBH}}^2 \mathcal{P}_{\zeta_L}$. The bounds from Planck in Eq. (3.93) can be recast as

$$-0.0327 < b f_{\text{PBH}} < 0.0308. \tag{3.98}$$

The colored region in Fig. 3.8 shows the allowed region by this constraint, according to which a large non-linear parameter implies that only a small fraction of DM can be composed by PBHs (the lower bound $f_{\text{NL}} \geq -1/3$ has been introduced to account for the breaking of the perturbative approach in the PBH abundance computation [228, 321]).

To correlate this information with the amount of anisotropy in the GWs abundance, we report the two-point function $\hat{C}_\ell$ of the GW density contrast $\delta_{\text{GW}}$ rather than of $\Gamma$ as [20]

$$\sqrt{\frac{\ell(\ell+1)}{2\pi} \hat{C}_\ell(k)} \simeq \frac{3}{5} \left| 1 + \tilde{f}_{\text{NL}}(k) \right| \left| 4 - \frac{\partial \ln \bar{\Omega}_{\text{GW}}(\eta, k)}{\partial \ln k} \right| \mathcal{P}_{\zeta_L}^{1/2}, \tag{3.99}$$

which, in the limit of dominant propagation terms ($\tilde{f}_{\text{NL}} \to 0$) [316] or dominant initial condition term ($\tilde{f}_{\text{NL}} \to \infty$), takes the values

$$\sqrt{\frac{\ell(\ell+1)}{2\pi} \hat{C}_\ell(k)} \simeq \begin{cases} \dfrac{3}{5} \left| 4 - \dfrac{\partial \ln \bar{\Omega}_{\text{GW}}(\eta, k)}{\partial \ln k} \right| \mathcal{P}_{\zeta_L}^{1/2} & \text{for} \quad \tilde{f}_{\text{NL}} \to 0, \\[3ex] \dfrac{24}{5} \left| f_{\text{NL}} \right| \mathcal{P}_{\zeta_L}^{1/2} & \text{for} \quad \tilde{f}_{\text{NL}} \to \infty. \end{cases} \tag{3.100}$$

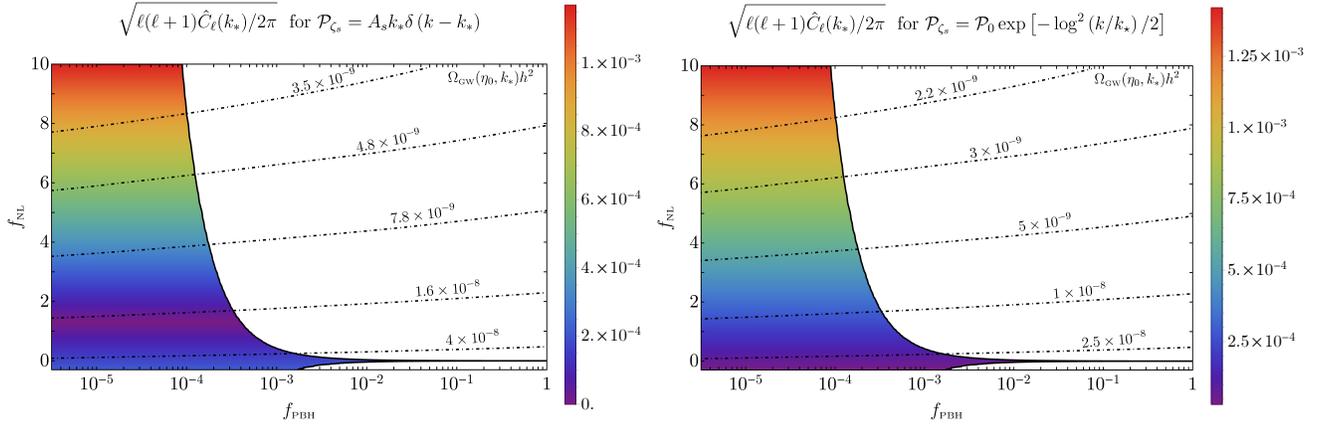

**Figure 3.8**: Contour plot of the GW two-point function anisotropy in the region allowed by the Planck constraints, assuming a Dirac delta (left) and lognormal (right) power spectrum of the short modes, respectively. The characteristic frequency of the power spectrum has been chosen to correspond to PBH masses of $M_{\rm PBH} = 10^{-12} M_\odot$. The dot-dashed lines identify the corresponding GWs abundance. Figure taken from Ref. [20].

In Fig. 3.8 we plot the GW anisotropy assuming a Dirac delta and lognormal power spectrum of the curvature perturbation on small scales, where the peak frequency has been chosen as the one corresponding to PBH masses of $M_{\rm PBH} = 10^{-12} M_\odot$. One can finally conclude that, if PBHs compose a sizable fraction of the dark matter, the SGWB must be highly isotropic and Gaussian, up to propagation effects. By contrast, a large amount of anisotropy and non-Gaussianity would imply a PBH population well below the measured dark matter abundance.

## 3.2 Detectability of the SGWB at NANOGrav

The NANOGrav experiment is one of the Pulsar Timing Arrays experiments which look for delays in the time of arrival of radio pulses coming from pulsars, with a period of the millisecond, due to the passing of a GW with a frequency around the nHz.

In the latest 12.5 yrs data release, the NANOGrav Collaboration has recently reported a strong evidence for a stochastic common-spectrum process across fourty-five pulsars [322], strongly preferred against independent red-noise signals, which may be interpreted as due to a SGWB with strain amplitude $\sim 10^{-15}$ at frequencies $f \sim 3 \cdot 10^{-8}$ Hz with an almost flat GW spectrum, $\Omega_{\rm GW}(f) \sim f^{(-1.5-0.5)}$ at $1\sigma$-level. A similar result has subsequently been found in other PTA experiments like PPTA [323].

Even though this signal seems to be in partial contrast with other bounds on the SGWB, the NANOGrav Collaboration stresses that it may be due to an improved treatment of the intrinsic pulsar red noise. Furthermore, the Collaboration did not claim for a detection since the signal does not possess the quadrupolar, Hellings-Down, angular correlation patterns [324].

Nonetheless, one may ask if the NANOGrav signal, if interpreted as a GW background, can be explained as generated at second-order in perturbation theory during the formation of PBHs, in the standard scenario in which they form from the collapse of sizeable curvature perturbations produced during inflation upon horizon re-entry. The resulting model has been developed in Ref. [13] (with details outlined also in Ref. [19]), whose main results will be outlined in the following sections. We will show how this signal is compatible with a model where PBHs with masses in the range $(10^{-15} - 10^{-11})M_\odot$ comprise the totality of the dark matter in the universe and the corresponding SGWB propagates to frequencies testable by future experiments like LISA [203].

Let us also stress that other interpretations of the NANOGrav signal have been shown in the literature, for example through super-massive BH coalescences [325], through GWs generated by cosmic strings [326–329], phase transitions in a dark sector [330–332], different PBH models [179, 333–336] and other scenarios [337–341].

### 3.2.1 Flat curvature power spectrum

In this subsection we provide the details of the power spectrum of the curvature perturbation we have assumed to explain the signal observed by NANOGrav. We consider a class of models with a broad and flat power spectrum parametrised as

$$\mathcal{P}_\zeta(k) \approx \mathcal{P}_0 \, \theta\left(k_s - k\right) \theta\left(k - k_l\right), \quad k_s \gg k_l, \tag{3.101}$$

where $\mathcal{P}_0$ is the amplitude of the power spectrum, and we denote with $\lambda_s \sim k_s^{-1}$ and $\lambda_l \sim k_l^{-1}$ the short and long wavelengths, respectively, which bound the power spectrum. When these modes re-enter the horizon, they give rise to masses which we denote $M_s$ and $M_l$. Such a shape can be generated for modes which exit the Hubble radius during a non-attractor phase, obtained through an ultra slow-roll regime of the inflaton potential, as a result of a duality transformation which maps the non-attractor phase into a slow-roll phase [238, 342–344].

Our aim is to deduce the PBH mass function which arises from such a power spectrum. The picture describing the formation of PBHs is the following. During the radiation-dominated era, our current horizon contained several Hubble patches which were growing with time. At that epoch, there is a certain probability that PBHs with different masses form within each Hubble horizon at any instant of time. The corresponding mass function may be determined according to the following physical arguments:

1) each perturbation re-entering the horizon at different times is characterised by the same threshold for collapse;

2) PBHs with different masses form with the same probability, which is therefore independent of time;

3) the absorption of PBHs from heavier ones ("cloud-in-cloud" problem) occurs with low probability;

4) PBHs formed earlier dominate the mass fraction due to time evolution.

We will show that the mass function has a peak at the small mass scale $M_s$ and a tail which scales like $M^{-3/2}$ for larger masses. We finally stress that we are neglecting any non-Gaussianity at formation, so that PBHs are not clustered initially, and also the evolution of the mass function due to merging and accretion (which we will investigate in the next chapters).

One of the most important ingredients to characterise the probability of collapse is the critical threshold. It is understood that, in order to formulate a criterion for collapse, the overdensities peaks must be described in real space taking into account contributions from all Fourier modes.

**Same threshold for all PBH masses**

Let us start with the computation of the critical threshold of collapse. In linear approximation, the density contrast can be expressed in terms of the curvature perturbation $\zeta$ as

$$\frac{\delta\rho}{\rho_{\rm b}}(k,\eta) = \frac{4}{9}\left(\frac{k}{\mathcal{H}(\eta)}\right)^2 T(k\eta)\zeta_{\vec{k}}, \qquad (3.102)$$

in terms of the Hubble rate in conformal time $\mathcal{H}(\eta)$ and the radiation linear transfer function $T(k\eta)$, which accounts for the time evolution of the curvature field and effectively smooths out subhorizon modes. Assuming spherical symmetry, the average profile of the overdensities is given by

$$\overline{\frac{\delta\rho}{\rho_{\rm b}}}(r,\eta) = \delta_0 \frac{\xi(r,\eta)}{\sigma^2(\eta)}, \qquad (3.103)$$

where $\xi(r,\eta)$ is the two-point correlator in real space

$$\xi(r,\eta) = \int \frac{\mathrm{d}k}{k}\frac{\sin kr}{kr}\mathcal{P}_{\delta\rho/\rho_{\rm b}}(k,\eta), \qquad \sigma^2(\eta) = \xi(0,\eta) \qquad (3.104)$$

and $\delta_0$ the amplitude of the perturbation. The compaction function can then be estimated in terms of the volume averaged density contrast $\delta(r,\eta)$ as

$$\mathcal{C}(r) = r^2\mathcal{H}^2\delta(r,\eta) \equiv r^2\mathcal{H}^2\left[\frac{3}{r^3}\int_0^r \mathrm{d}r\, r^2\,\overline{\frac{\delta\rho}{\rho_{\rm b}}}(r,\eta)\right], \qquad (3.105)$$

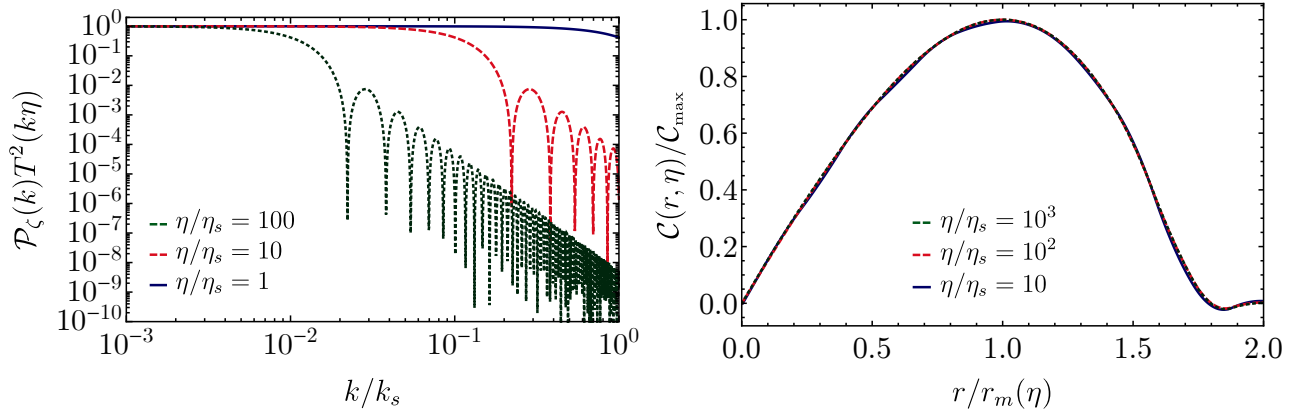

**Figure 3.9**: Plot of the curvature power spectrum (left) and rescaled compaction function (right) at different times $\eta$. Each curve has been normalised to its corresponding $r_m(\eta)$. Figure taken from Ref. [19].

which is time independent on super-Hubble scales. The scale at which the compaction function is maximum, $r_m(\eta)$, denotes the typical size of the perturbation in real space, and it is found to grow in time as $r_m(\eta) \sim \eta$ due to the smoothing of modes done by the transfer function. From it, one can compute the shape parameter $\alpha = -(\mathcal{C}''(r_m)r_m^2/4\mathcal{C}(r_m))$ and then the threshold for collapse $\delta_c(\alpha)$.

The crucial point is now that, at different times, the change in the mean profile is not enough to result in a significant change of the threshold. Indeed, given that a broad curvature power spectrum results into a blue tilted density power spectrum $\mathcal{P}_{\delta\rho/\rho_b}(k, \eta) \sim k^4\theta(\mathcal{H}(\eta) - k)$ at each time $\eta$, see Fig. 3.9, then the rescaled compaction function at different times does not change. In particular, the transfer function effectively smooths out modes $k > 1/\eta = \mathcal{H}$, giving rise to profile with characteristic peaks size of the order of $r_m \sim 1/\mathcal{H}$. This implies that all mean profiles are close to $\overline{\delta\rho}/\rho_b(r, \eta) \sim \mathrm{sinc}(\mathcal{H}(\eta)r)$, which results into a threshold to collapse that, for presentation purposes, we fix to $\delta_c = 0.5$.

**Same formation probability for all PBH masses**

Given that a PBH forms if $\delta(r_m, \eta) \geq \delta_c$, one can compute the PBH abundance using the Press-Schechter formalism [345], under the simplified assumption of Gaussian probability distribution for the curvature perturbation as

$$\beta(r_m(\eta)) = \int_{\delta_c} \frac{\mathrm{d}\delta}{\sqrt{2\pi\sigma^2}} \mathrm{e}^{-\delta^2/2\sigma^2} = \frac{1}{2}\mathrm{erfc}\left[\frac{\nu_c}{\sqrt{2}}\right], \tag{3.106}$$

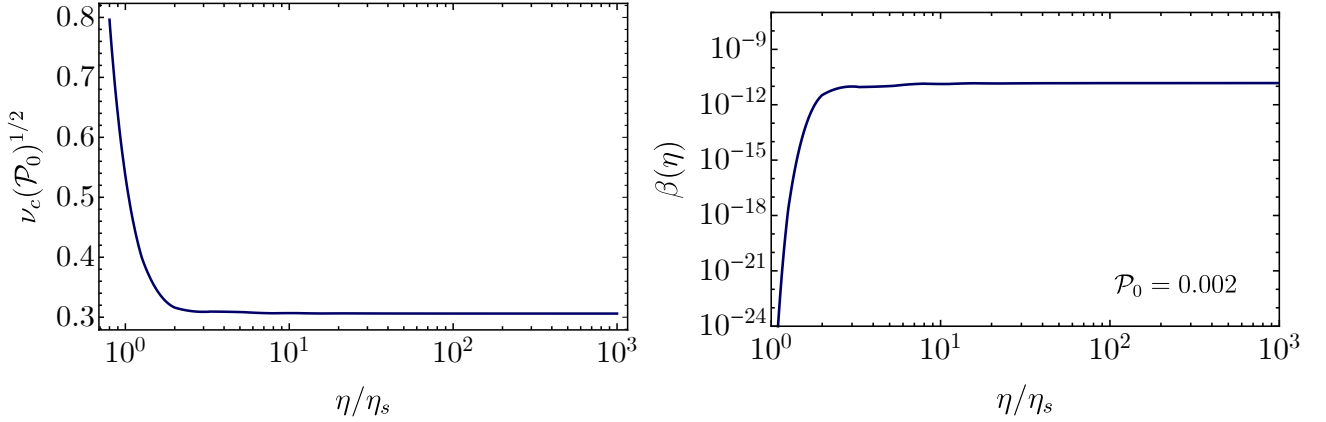

**Figure 3.10**: Time dependence of the rescaled parameter $\nu_c$ (left) and probability of formation $\beta$ (right) as a function of time. Figure taken from Ref. [19].

in terms of $\nu_c \equiv \delta_c / \sigma$ and the variance $\sigma$

$$\sigma^2(\eta) = \frac{16}{81} \int \frac{dk}{k} \frac{k^4}{\mathcal{H}^4} \mathcal{P}_\zeta(k) T^2(k\eta) W^2(k, r_m(\eta)) \tag{3.107}$$

with $r_m(\eta) \simeq 3/\mathcal{H}(\eta)$ and a top-hat window function in real space

$$W(k, r_m(\eta)) = 3 \frac{\sin(k r_m) - (k r_m) \cos(k r_m)}{(k r_m)^3}. \tag{3.108}$$

The time independence of the threshold and the balance between the horizon growth with time and the action of the transfer function smoothing out subhorizon modes in the computation of the variance results into a constant $\nu_c(\mathcal{P}_0)^{1/2} \sim 0.3$, see Fig. 3.10. In particular, at early times this combination scales like $\eta^{-2}$ due to the scaling of the variance as $\sigma \sim 1/\mathcal{H}^2$, which suppresses the formation probability for masses smaller than $M_s \equiv M(r_m(\eta_s))$, being $\eta_s$ the horizon crossing time of the smallest scale $\lambda_s$. This implies that the formation probability of a PBH with mass $M(r_m(\eta))$ is the same for fixed value of the curvature perturbation amplitude, see the right plot in Fig. 3.10.

**Cloud-in-cloud problem**

Given that the formation of a PBH is an extremely rare event, the probability to form two or even more PBHs in the same horizon volume, in which the lighter ones would be subsequently absorbed by bigger ones, is almost negligible. This point can be shown using the excursion set method [225], which consists in the random walk performed by density perturbations in terms of the smoothing scale. In particular, when the overdensity becomes equal to the threshold for PBH formation, at the so-called first-passage time in the presence of a barrier, a PBH may form. The cloud-in-cloud problem is thus taken care of by considering only the trajectories which cross the threshold for the first-time. In the following we will give only the main ideas behind this approach and the final result. The interested reader can find the details of the computation in Refs. [19, 224].

In the Press-Schechter approach to compute the probability of collapse, one can notice that it depends only on the ratio between the threshold and the variance of the smoothed field, $\delta_c/\sigma(R)$. In the excursion set scheme, one usually absorbs the time evolution of the density contrast in the threshold, which becomes a time dependent quantity denoted $\omega(a) \equiv \delta_c/a^2$, such that the barriers at two different times are found by

$$\omega(a_1) = \left(\frac{a_2}{a_1}\right)^2 \omega(a_2). \tag{3.109}$$

This implies that the barrier decreases with the passage of time, and that the evolution with respect to the real time is captured by the time evolution of the barrier. On the other hand, the variance stays constant and depends only on the smoothing scale $S(R)$, such that its scaling is given by

$$S(R_1) = S(R_2) \left(\frac{M_{H_1}}{M_{H_2}}\right)^{-1}, \tag{3.110}$$

where $M_H$ indicates the mass contained within the Hubble volume at the time when the corresponding smoothing scale $R \sim \mathcal{H}^{-1}$.

For the broad and flat power spectrum adopted in this section, one would have the following hierarchy between thresholds and variances as

$$\omega(a_s) \gg \omega(a_l) \quad \text{and} \quad S_s \gg S_l, \tag{3.111}$$

with

$$\sqrt{\frac{S_s}{S_l}} = \frac{\omega(a_s)}{\omega(a_l)}, \tag{3.112}$$

which implies and confirms that light PBHs are generated with the same abundance as heavy ones.

In the "two-barriers" problem [346] one has therefore to solve the Fokker-Planck equation

$$\frac{\partial P}{\partial S} = \frac{1}{2} \frac{\partial^2 P}{\partial \delta^2}, \tag{3.113}$$

to compute the probability to form a PBH, given the conditional probability that a certain trajectory, with an initial up-crossing of the barrier $\omega_1$ and variance $S_1$, will have a second up-crossing of $\omega_2$ between $S_2$ and $S_2 + \mathrm{d}S_2$ with $S_1 \gg S_2$ and $\omega_1 > \omega_2$. This implies that the probability that a heavy PBH absorbs an already formed smaller PBH with $S(< S_s)$ at later times is given by [346]

$$P(S, \omega(a)|S_s, \omega(a_s)) = \frac{1}{2}\mathrm{Erfc}\left(\frac{\omega(a)}{\sqrt{2S}}\right)\left[1 + \exp\left(2\frac{\omega(a)}{\omega(a_s)}\frac{\omega^2(a_s)}{S_s}\right)\right] \simeq \beta(M(r_m)) \ll 1, \tag{3.114}$$

where we used that $\omega(a_s)/\sqrt{S_s} = \mathcal{O}(6-8)$ in the typical PBHs scenario, and that the exponential factor is teamed by the small coefficient $\omega(a)/\omega(a_s) \ll 1$. This result can be understood by realising that the probability that a small PBH is eaten by a larger one scales like $\beta(M(r_m))\beta(M_s) \simeq \beta^2(M_s)$ for a flat broad spectrum [224], and it is therefore totally negligible.

**The impact of time evolution**

Given that the PBH density grows like non-relativistic matter and that $\beta = \rho_{\mathrm{PBH}}/\rho_{\mathrm{tot}}$ at PBH formation time, the abundance of light PBHs with mass $M_s$ at the formation time $\eta_l$ of the heavy ones is given by [347]

$$\frac{\rho_s(\eta_l)}{\rho_l(\eta_l)} = \frac{\rho_s(\eta_s)(a_s/a_l)^3}{\rho_l(\eta_l)} = \frac{\beta(M_s)}{\beta(M_l)}\frac{a_l}{a_s} = \frac{\beta(M_s)}{\beta(M_l)}\frac{k_s}{k_l} \simeq \frac{k_s}{k_l} = \frac{M_l^{1/2}}{M_s^{1/2}} \gg 1. \tag{3.115}$$

This implies that lighter PBHs give a larger contribution to the mass distribution. The latter is usually defined as the fraction of PBHs with $M_{\mathrm{PBH}}$ as [83]

$$f_{\mathrm{PBH}}(M_{\mathrm{PBH}}) = \frac{1}{\Omega_{\mathrm{DM}}}\frac{\mathrm{d}\Omega_{\mathrm{PBH}}}{\mathrm{d}\ln M_{\mathrm{PBH}}}, \tag{3.116}$$

normalised in terms of the total abundance of dark matter $\Omega_{\mathrm{DM}}$, such that the total PBH abundance is given by

$$f_{\mathrm{PBH}} = \int f_{\mathrm{PBH}}(M_{\mathrm{PBH}})\mathrm{d}\ln M_{\mathrm{PBH}}. \tag{3.117}$$

After matter-radiation equality, it can be expressed in terms of the mass fraction $\beta$ as [348]

$$f_{\mathrm{PBH}}(M_{\mathrm{PBH}}) = \frac{1}{\Omega_{\mathrm{DM}}}\left(\frac{M_{\mathrm{eq}}}{M_{\mathrm{PBH}}}\right)^{1/2}\beta(M_{\mathrm{PBH}}), \tag{3.118}$$

where the overall factor, which depends on the horizon mass at matter-radiation equality $M_{\mathrm{eq}} = 2.8 \cdot 10^{17}M_\odot$, accounts for the evolution of the energy density during the remaining radiation-dominated phase after PBH formation. Given that for a flat power spectrum $\beta(M)$ is almost constant, one gets a $M_{\mathrm{PBH}}^{-1/2}$ tail at large masses, see also Refs. [99, 148, 348].

We finally stress that the PBH masses respect a scaling relation near the critical threshold for collapse $\delta_c$ as [257–259]

$$M_{\mathrm{PBH}}(\delta) = \kappa M_H \left(\delta - \delta_c\right)^{\gamma_c}, \tag{3.119}$$

where we assume the reference values $\kappa = 4$ and $\gamma_c = 0.36$ in a radiation dominated universe [209, 349–352]. This results into a mass function which possesses a tail at low masses scaling like $\sim M_{\mathrm{PBH}}^{2.8}$ [258]. Even though an exponential fall-off after the peak located around $M_{\mathrm{PBH}} \sim M_H$ would also be expected from the critical collapse, this scaling does not play any major role in shaping the high-mass range of the final mass function, which for a broad power spectrum will then scale like $\sim M_{\mathrm{PBH}}^{-1/2}$.

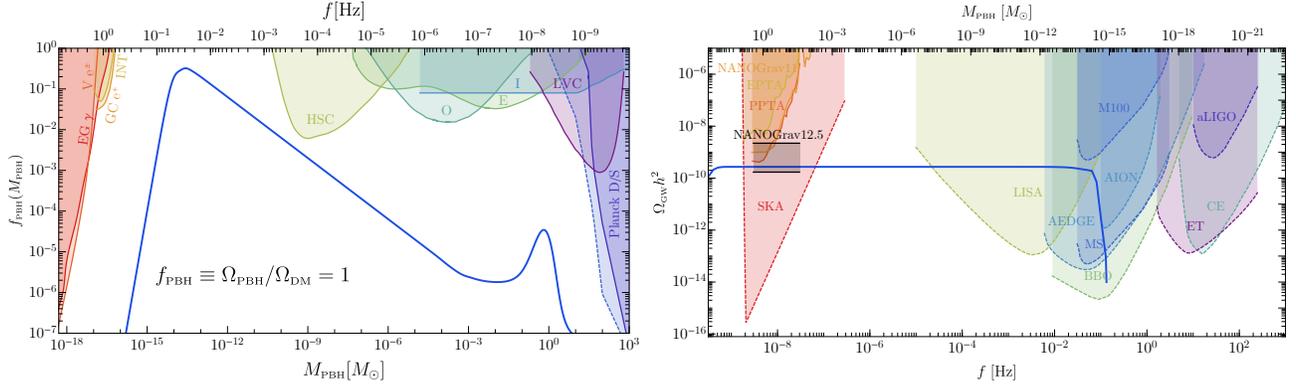

**Figure 3.11**: *Left:* Mass function resulting from a broad and flat curvature power spectrum with a peak at $\simeq 10^{-14} M_\odot$, assuming $\mathcal{P}_0 \simeq 5.8 \cdot 10^{-3}$ and $k_s = 10^9 k_l \simeq 1.6$ Hz. The bump around solar masses is due to the decrease of the threshold by a softening of the QCD equation of state [348, 353]. *Right:* The SGWB abundance generated at second order in perturbation theory according to our PBH model. In black we show the 95% C.I. from the NANOGrav 12.5 yrs experiment. Figure taken from Ref. [13].

## 3.2.2 Comparison with the 12.5 yrs data

Once the mass function from our assumed curvature perturbation power spectrum has been computed, we can compare it with the present observational constraints on the PBH abundance as shown in the left panel of Fig. 3.11. The amplitude of the power spectrum has been fixed by requiring that the formed PBHs contribute to the totality of the dark matter in the universe as

$$f_{\mathrm{PBH}} = \int f_{\mathrm{PBH}}(M_{\mathrm{PBH}}) \mathrm{d} \ln M_{\mathrm{PBH}} = 1. \tag{3.120}$$

As stressed in the previous section, the peak of the mass function corresponds to the shortest scale $\sim 1/k_s$ which re-enter the horizon first and give rise to the lightest PBHs. For smaller masses the dynamics of the critical collapse gives the growing behaviour $\sim M_{\mathrm{PBH}}^{2.8}$, while at larger masses the mass function falls down like $\sim M_{\mathrm{PBH}}^{-1/2}$, as proven in the previous section. A sub-dominant peak around $\sim M_\odot$ is present due to the change of equation of state during the QCD phase transition [348, 353]. The absence of constraints in the mass range of support of the PBH mass function signals that this model is compatible with current constraints and the PBHs generated can comprise the dark matter in the universe.

Given that the amplitude of the power spectrum has been fixed, one can predict the amount of GWs generated at second order in perturbation theory. The resulting SGWB spectrum is shown in the right panel of Fig. 3.11, as a function of the frequency, along with the constraints coming from experiments like EPTA [354], PPTA [355], NANOGrav 11 yrs [356, 357], and the sensitivity curves of future experiments like SKA [358], LISA [203] (power-law integrated sensitivity curve expected to fall in between the designs named C1 and C2 in Ref. [308]), DECIGO/BBO [359], Cosmic Explorer [360], Einstein Telescope [361, 362], Advanced LIGO + Virgo [363], Magis-space and Magis-100 [364], AEDGE [365] and AION [366].

At frequencies around the nHz, corresponding to the re-entering of modes with wavelength $\sim 1/k_l$, the GW abundance falls within the 95% C.I. from the NANOGrav 12.5 yrs observation, showing that this spectrum may explain the signal observed by the collaboration (we stress that the improved priors assumed for the intrinsic pulsar red noise relaxes the NANOGrav 11 yrs bound [322]). The flat shape of the GW spectrum propagates to larger frequencies, reaching the LISA detectable region and decays rapidly at the frequency corresponding to the shortest scale $1/k_s$. Our scenario therefore predicts that the SGWB could also be detected by the forthcoming experiments like LISA, MS and BBO.

We therefore conclude that the signal seen by NANOGrav, if interpreted as a stochastic gravitational wave background produced at second-order in perturbation theory within the PBH model, is in agreement with the possibility that all the dark matter is in the form of extremely light PBHs. We finally stress that the present scenario is consistent with the candidate event found by the HSC collaboration [142], and its possible confirmation may come from future HSC observations, as forecasted in Ref. [367].

# Part II

# Signatures from primordial black holes: mergers

# Chapter 4

# Primordial black holes across cosmic history

At the time of formation in the very early universe PBHs are characterised by some specific features in their mass and spin spectrum, which strongly depend on the considered formation mechanism. However, these cosmic ghosts evolve through the cosmic history, and experience different processes which may modify their initial properties.

We dedicate this chapter to the study of those physical processes which may affect a PBH navigating the history of the universe. In particular, we will focus on PBHs accreting baryonic particles from the intergalactic medium, on the possibility that PBHs assemble in binary systems before the matter-radiation equality and on the generation of PBH clusters if they comprise a large fraction of the dark matter in the universe. We will see how these phenomena modify the PBH properties, such as their masses, spins and merger rate.

## 4.1   On the accretion onto PBHs

During the cosmic history, PBHs may experience a phase of baryonic mass accretion from the intergalactic medium, which may impact onto their masses, spins and merger rate, and thus modify their properties with respect to the ones at formation.

In the following we are going to provide the details of the model of accretion onto PBHs following the results obtained in Refs. [16–18], where the interested reader can find additional details.

### 4.1.1 The accretion model

Once a PBH is formed in the very early universe, its mass may evolve during the cosmological history due to the effect of baryonic mass accretion, which can occur in the redshift range $z \in (10 - 100)$ depending on the PBH masses.

Since the accretion rate may reach super-Eddington values depending on the relative velocity between the accreting system and the surrounding baryonic gas, one can distinguish betweeen accretion onto an isolated PBH or onto a binary system. In the first case the relative velocity is smaller than or comparable to the speed of sound in the gas, while in the second case the orbital velocities are typically much larger than the sound speed. This implies that, being the accretion rate dominated at $z \lesssim 100$ [172, 368], the former case is relevant for PBH binaries formed at smaller redshifts, while the latter case is relevant for PBH binaries formed at $z \gg 100$.

**Accretion onto isolated PBHs**

Isolated PBHs immersed in the intergalactic medium can experience accretion processes which can be modeled through the Bondi-Hoyle mass accretion rate [172, 368, 369]

$$\dot{M}_{\rm B} = 4\pi \lambda m_H n_{\rm gas} v_{\rm eff} r_{\rm B}^2, \tag{4.1}$$

in terms of the PBH mass $M$, the cosmic gas density

$$n_{\rm gas} \simeq 200 \, {\rm cm}^{-3} \left( \frac{1+z}{1000} \right)^3, \tag{4.2}$$

dominated by hydrogen with mass $m_H$, the speed of sound in the gas

$$c_s \simeq 5.7 \left( \frac{1+z}{1000} \right)^{1/2} \left[ \left( \frac{1+z_{\rm dec}}{1+z} \right)^{\beta} + 1 \right]^{-1/2\beta} {\rm km \, s}^{-1}, \tag{4.3}$$

in equilibrium at the temperature of the intergalactic medium, and the redshift of decoupling of baryonic matter from the radiation fluid, $z_{\rm dec} \simeq 130$, with $\beta = 1.72$. The Bondi-Hoyle radius

$$r_{\rm B} \equiv \frac{M}{v_{\rm eff}^2} \simeq 1.3 \times 10^{-4} \left( \frac{M}{M_\odot} \right) \left( \frac{v_{\rm eff}}{5.7 \, {\rm km \, s}^{-1}} \right)^{-2} {\rm pc}, \tag{4.4}$$

denotes the region within which the accretion is efficient, in terms of the effective velocity $v_{\rm eff} = \sqrt{v_{\rm rel}^2 + c_s^2}$ which depends on the PBH relative velocity $v_{\rm rel}$ with respect to the surrounding matter.

The relative velocity strongly affects the dynamics of the accretion process and depends on the amplitude of the inhomogeneities of the dark matter and baryon fluids. Assuming that PBHs behave like dark matter particles, one can identify two main regimes of interest, the linear and non-linear regime. In the linear regime before the decoupling redshift, the Silk damping suppresses the growth of inhomogeneities on small scales, such that the PBH peculiar velocity is of the order of the gas sound speed. At smaller redshift, the gas flow lags behind the dark matter with a relative velocity $v_{\rm rel} = v_{\rm DM} - v_{\rm b}$, such that the PBH peculiar velocity follows a Maxwellian distribution with variance $\sigma = \langle v_{\rm rel} \rangle$ and expectation value [172]

$$\langle v_{\rm eff} \rangle_{\rm A} \sim c_s \left( \frac{16}{\sqrt{2\pi}} \mathcal{M}^3 \right)^{\frac{1}{6}} \theta(\mathcal{M} - 1) + c_s \left( 1 + \mathcal{M}^2 \right)^{\frac{1}{2}} \theta(1 - \mathcal{M}),$$

$$\langle v_{\rm eff} \rangle_{\rm B} \sim c_s \mathcal{M} \left[ \sqrt{\frac{2}{\pi}} \ln \left( \frac{2}{e} \mathcal{M} \right) \right]^{-\frac{1}{3}} \theta(\mathcal{M} - 1) + c_s \left( 1 + \mathcal{M}^2 \right)^{\frac{1}{2}} \theta(1 - \mathcal{M}), \tag{4.5}$$

in terms of the Mach number $\mathcal{M} = \langle v_{\rm rel} \rangle / c_s$, where scenario A refers to low efficient accretion rate, while scenario B refers to an efficient accretion rate. On the other hand, in the non-linear regime, non-linear perturbations like halos may suppress the accretion onto PBHs due to an enhancement of the velocities of PBHs which fall in the structure potential wells. In the next sections we will introduce a cut-off redshift $z_{\rm cut-off}$, below which accretion is negligible, to keep into account the uncertainties in the accretion model, also related to the fate of PBHs in halos.

The accretion eigenvalue $\lambda$ keeps into account the effects of the Hubble expansion, gas viscosity and the coupling of the CMB radiation to the gas through Compton scattering. Its analytical expression is given by [368]

$$\lambda = \exp \left( \frac{9/2}{3 + \hat{\beta}^{0.75}} \right) x_{\rm cr}^2, \tag{4.6}$$

as a function of the sonic radius

$$x_{\rm cr} \equiv \frac{r_{\rm cr}}{r_{\rm B}} = \frac{-1 + (1 + \hat{\beta})^{1/2}}{\hat{\beta}}, \tag{4.7}$$

and the gas viscosity parameter

$$\hat{\beta} = \left( \frac{M}{10^4 M_\odot} \right) \left( \frac{1+z}{1000} \right)^{3/2} \left( \frac{v_{\rm eff}}{5.74 \, {\rm km \, s^{-1}}} \right)^{-3} \left[ 0.257 + 1.45 \left( \frac{x_e}{0.01} \right) \left( \frac{1+z}{1000} \right)^{5/2} \right], \tag{4.8}$$

which depends on the redshift, PBH mass, effective velocity, and ionization fraction of the cosmic gas $x_e$.

Current observational constraints imply that PBHs with masses larger than the solar mass can comprise only a fraction of the dark matter in the universe [84]. This implies that an additional halo made by the second DM component has to be considered in the accretion process. Even though direct accretion of dark matter particles onto the PBH is negligible [368, 370], the halo acts as a catalyst and enhances the gas accretion rate. Assuming a spherical density profile for the DM halo $\rho \propto r^{-\alpha}$, with $\alpha \simeq 2.25$ [371, 372], truncated at a radius $r_h \simeq 0.019\,\mathrm{pc}(M/M_\odot)^{1/3}(1+z/1000)^{-1}$, one has that the PBH+DM system has a total mass of

$$M_h(z) = 3M\left(\frac{1+z}{1000}\right)^{-1},\tag{4.9}$$

which grows with time as long as PBHs are isolated, and eventually stops when all the available DM has been accreted, i.e. approximately when $3f_{\mathrm{PBH}}(1+z/1000)^{-1} = 1$. To keep into account the effect of the DM halo, one can introduce the parameter

$$\kappa \equiv \frac{r_{\mathrm{B}}}{r_h} = 0.22\left(\frac{1+z}{1000}\right)\left(\frac{M_h}{M_\odot}\right)^{2/3}\left(\frac{v_{\mathrm{eff}}}{\mathrm{km\,s^{-1}}}\right)^{-2}.\tag{4.10}$$

In the regime $\kappa \geq 2$, that is when the typical size of the halo is smaller than the Bondi radius, the dark halo behaves the same as a point mass $M_h$, enhancing the accretion rate. In the opposite regime $\kappa < 2$, in which the halo extends much more than the Bondi radius, the effect of a dark clothing is taken into account by the accretion parameter $\lambda$ by correcting the quantities with respect to the naked case as [368]

$$\hat{\beta}^h \equiv \kappa^{\frac{p}{1-p}}\hat{\beta}, \quad \lambda^h \equiv \tilde{\Upsilon}^{\frac{p}{1-p}}\lambda(\hat{\beta}^h), \quad r_{\mathrm{cr}}^h \equiv \left(\frac{\kappa}{2}\right)^{\frac{p}{1-p}}r_{\mathrm{cr}},\tag{4.11}$$

where $p = 2 - \alpha$ and

$$\tilde{\Upsilon} = \left(1 + 10\hat{\beta}^h\right)^{\frac{1}{10}}\exp(2-\kappa)\left(\frac{\kappa}{2}\right)^2.\tag{4.12}$$

It is customary to define a dimensionless accretion rate, normalised to the Eddington one,

$$\dot{m} = \frac{\dot{M}_{\mathrm{B}}}{\dot{M}_{\mathrm{Edd}}} \quad \text{with} \quad \dot{M}_{\mathrm{Edd}} = 1.44 \times 10^{17}\left(\frac{M}{M_\odot}\right)\mathrm{g\,s^{-1}},\tag{4.13}$$

which is shown in the left panel of Fig. 4.1. The relevant accretion time scale is given by the Salpeter time

$$\tau_{\mathrm{acc}} \equiv \frac{\tau_{\mathrm{Salp}}}{\dot{m}} = \frac{\sigma_{\mathrm{T}}}{4\pi m_{\mathrm{p}}}\frac{1}{\dot{m}} = \frac{4.5 \times 10^8\,\mathrm{yr}}{\dot{m}},\tag{4.14}$$

in terms of the Thompson cross section $\sigma_{\mathrm{T}}$ and the proton mass $m_{\mathrm{p}}$. As one can appreciate from the right panel of Fig. 4.1, at redshift smaller than $z \lesssim 100$, $\tau_{\mathrm{acc}}$ is smaller than the typical age of the universe.

**Accretion onto PBHs assembled in binaries**

Baryonic mass accretion onto binary PBHs is characterised by both global (i.e., on the binary as a whole) and local accretion processes (i.e., onto the individual binary components). We will consider a binary system with total mass $M_{\text{tot}} = M_1 + M_2$, reduced mass $\mu = M_1 M_2/(M_1 + M_2)$, mass ratio $q = M_2/M_1 \leq 1$, semi-major axis $a$, and eccentricity $e$. A schematic illustration is shown in Fig. 4.2. Its Bondi radius is given by

$$r_{\text{B}}^{\text{bin}} = \frac{M_{\text{tot}}}{v_{\text{eff}}^2},\tag{4.15}$$

in terms of the effective velocity $v_{\text{eff}} = \sqrt{c_s^2 + v_{\text{rel}}^2}$, which depends on the relative velocity $v_{\text{rel}}$ between the binary center of mass and the surrounding gas.

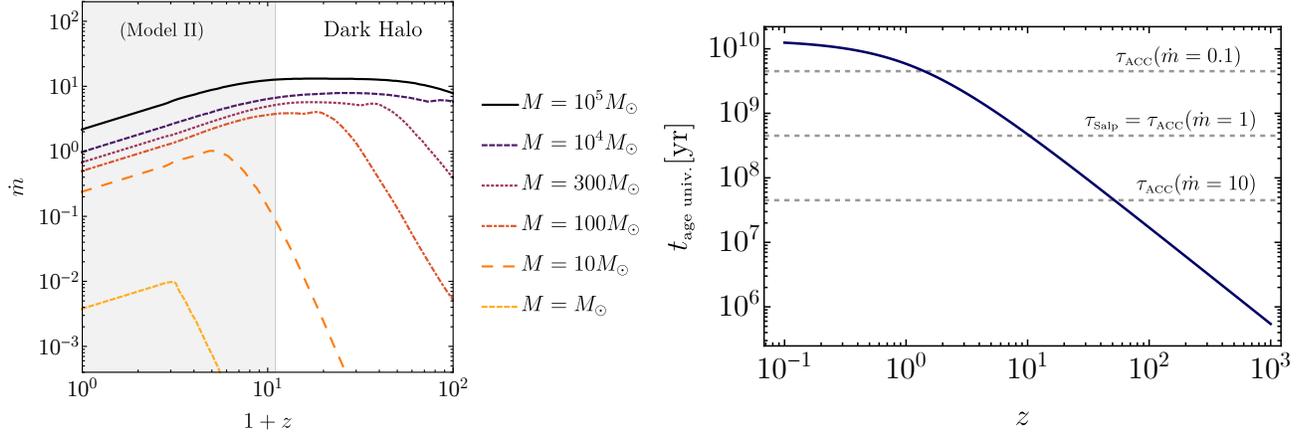

**Figure 4.1**: *Left:* The dimensionless accretion rate as a function of redshift for various isolated PBH masses. The shaded gray region below $z \sim 10$ denotes the redshifts at which uncertainties in the accretion model may reduce its efficiency. Neglecting the impact of these uncertainties on the efficiency leads to an extension of the accretion rate to smaller redshifts (dubbed Model II). In the following we parametrise the accretion efficiency introducing a parameter called $z_{\text{cut-off}}$. *Right:* Accretion time scale $\tau_{\text{ACC}} \equiv \tau_{\text{Salp}}/\dot{m}$ compared to the age of the universe at a given redshift $z$, assuming a $\Lambda$CDM universe. Figure taken from Ref. [18].

If the semi-major axis of the binary is larger than the binary Bondi radius, $a \gg r_{\text{B}}^{\text{bin}}$, then accretion will occur onto the individual PBH components independently, each characterised by its proper velocity, and the accretion model follows the one discussed in the previous section. However, as the binary hardens, the orbital separation becomes smaller than the binary Bondi radius, $a \ll r_{\text{B}}^{\text{bin}}$, and in this case one has to describe the accretion process as occurring onto the whole binary system. We will see that this may happen before the emission of GWs drives the binary inspiral. The accretion rate then reads

$$\dot{M}_{\text{bin}} = 4\pi \lambda m_H n_{\text{gas}} v_{\text{eff}}^{-3} M_{\text{tot}}^2 \,.\tag{4.16}$$

From this expression one can deduce the amount of accretion that each component of the binary

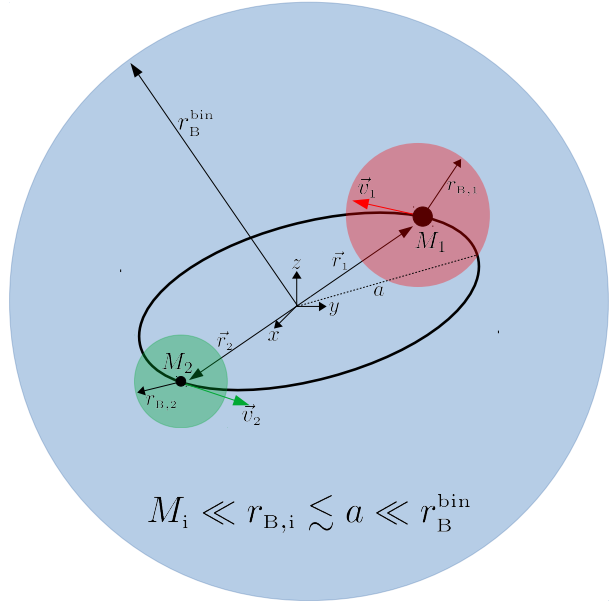

**Figure 4.2**: Illustration of the relevant scales characterising the accretion process onto a PBH binary system, whose Bondi radius is much larger than the orbital separation. Figure taken from Ref. [16].

feels. The PBH positions and velocities with respect to the center of mass are given by [373]

$$r_1 = \frac{q}{1+q}r, \qquad v_1 = \frac{q}{1+q}v; \qquad r_2 = \frac{1}{1+q}r, \qquad v_2 = \frac{1}{1+q}v \qquad (4.17)$$

as a function of their relative distance and velocity

$$r = a(1 - e\cos u), \qquad v = \sqrt{M_{\rm tot}\left(\frac{2}{r} - \frac{1}{a}\right)}, \qquad (4.18)$$

and angle $u$. The angle time evolution is described by $\sqrt{a^3/M_{\rm tot}}(u(t) - e\sin u(t)) = t - T$, in terms of the integration constant $T$ [373]. The effective velocities of the individual PBHs with respect to the cosmic gas are given by

$$v_{\rm eff,1} = \sqrt{v_{\rm eff}^2 + v_1^2}, \qquad v_{\rm eff,2} = \sqrt{v_{\rm eff}^2 + v_2^2}. \qquad (4.19)$$

Given that the binary Bondi radius is much larger than the typical binary semi-axis (as depicted in Fig. 4.2), the total infalling flow of baryons onto the binary system is constant, i.e.

$$4\pi m_H n_{\rm gas}(R) v_{\rm ff}(R) R^2 = {\rm const} = \dot{M}_{\rm bin}, \qquad (4.20)$$

where the gas free fall velocity, $v_{\rm ff}$, is calculated by assuming that at large distances, $R \sim r_{\rm B}^{\rm bin}$, it reduces to the usual effective velocity $v_{\rm eff}$, i.e.

$$v_{\rm ff}(R) = \sqrt{v_{\rm eff}^2 + \frac{2M_{\rm tot}}{R} - \frac{2M_{\rm tot}}{r_{\rm B}^{\rm bin}}}, \qquad (4.21)$$

while $n_{\rm gas}(R)$ identifies the density profile at a distance $R$ from the center of mass of the binary,

$$n_{\rm gas}(R) = \frac{\dot{M}_{\rm bin}}{4\pi m_H v_{\rm ff}(R) R^2}. \qquad (4.22)$$

This implies that, being the infalling flow of baryons constant, their density near the binary system increases with respect to its mean cosmic value at the Bondi radius of the binary.

The accretion rates for the individual binary components can then be computed in terms of their effective Bondi radii[1] $r_{\rm B,i} = M_i/v_{\rm eff,i}^2$ as

$$\dot{M}_1 = 4\pi m_H n_{\rm gas}(r_{\rm B,1}) v_{\rm eff,1}^{-3} M_1^2, \qquad \dot{M}_2 = 4\pi m_H n_{\rm gas}(r_{\rm B,2}) v_{\rm eff,2}^{-3} M_2^2. \qquad (4.23)$$

In this description we have considered the individual PBHs as naked, i.e. without a DM halo surrounding them, for which the accretion parameter $\lambda \approx 1$ at low redshift. On the other hand, we have described the binary as clothed by a dark halo, in which $\lambda$ takes into account the ratio of the Bondi radius of the binary with respect to the dark halo radius as described in the previous section [172].

The accretion rates (4.23) can then be recast in terms of the orbital parameters as

$$\dot{M}_1 = \dot{M}_{\rm bin} \sqrt{\frac{1 + \zeta + (1-\zeta)\gamma^2}{2(1+\zeta)(1+q) + (1-\zeta)(1+2q)\gamma^2}}, \qquad (4.24)$$

$$\dot{M}_2 = \dot{M}_{\rm bin} \sqrt{\frac{(1+\zeta)q + (1-\zeta)q^3\gamma^2}{2(1+\zeta)(1+q) + (1-\zeta)q^2(2+q)\gamma^2}}, \qquad (4.25)$$

in terms of $\zeta = e\cos u$ and $\gamma^2 = av_{\rm eff}^2/\mu q$. In the limit $q \to 0$, in which there is no dependence on the binary parameters $e$ nor $u$, one gets

$$\dot{M}_1 = \dot{M}_{\rm bin} + \mathcal{O}(q), \qquad \dot{M}_2 = \sqrt{\frac{q}{2}} \dot{M}_{\rm bin} + \mathcal{O}(q^{3/2}). \qquad (4.26)$$

In the general case, since the orbital period is much smaller than the accretion time scale $\tau_{\rm ACC}$, one can average over the angle $u$ and eliminate the explicit time dependence in Eqs. (4.24) and (4.25).

---

[1] We stress that the Bondi radii $r_{\rm B,i}$ of the individual PBH components are much smaller than the Bondi radius $r_{\rm B}^{\rm bin}$ of the binary.

After this procedure, even the dependence on the eccentricity is negligible. Therefore, one gets

$$\dot{M}_1 = \dot{M}_{\mathrm{bin}} \sqrt{\frac{M_1 q^2 + a(1+q)v_{\mathrm{eff}}^2}{(1+q)[2M_1 q^2 + a(1+2q)v_{\mathrm{eff}}^2]}},$$

$$\dot{M}_2 = \dot{M}_{\mathrm{bin}} \sqrt{\frac{q[M_1 + a(1+q)v_{\mathrm{eff}}^2]}{(1+q)[2M_1 + a(2+q)v_{\mathrm{eff}}^2]}}, \tag{4.27}$$

which can be further simplified given that

$$M_1 q^2 \gg a v_{\mathrm{eff}}^2, \tag{4.28}$$

for the interesting regime $M_1 \sim \mathcal{O}(M_\odot)$ and $a \sim \mathcal{O}(10^6) M_1$, for any $q > 10^{-2}$. This implies that the individual accretion rates reduce to

$$\dot{M}_1 = \dot{M}_{\mathrm{bin}} \frac{1}{\sqrt{2(1+q)}},$$

$$\dot{M}_2 = \dot{M}_{\mathrm{bin}} \frac{\sqrt{q}}{\sqrt{2(1+q)}}, \tag{4.29}$$

that recover the behaviour $\dot{M}_1 = \dot{M}_2 = \dot{M}_{\mathrm{bin}}/2$ in the limit $q \to 1$ and do not depend on the orbital parameters of the binary system.

One can finally introduce the normalised accretion rate in terms of the Eddington rate, $\dot{m}_i = \tau_{\mathrm{Salp}} \dot{M}_i / M_i$ as

$$\dot{m}_1 = \dot{m}_{\mathrm{bin}} \sqrt{\frac{1+q}{2}}, \qquad \dot{m}_2 = \dot{m}_{\mathrm{bin}} \sqrt{\frac{1+q}{2q}}. \tag{4.30}$$

The evolution equation for the mass ratio is given by

$$\dot{q} = q \left( \frac{\dot{M}_2}{M_2} - \frac{\dot{M}_1}{M_1} \right) = \frac{q}{\tau_{\mathrm{Salp}}} (\dot{m}_2 - \dot{m}_1). \tag{4.31}$$

From the last equation one can appreciate that, being $\dot{m}_2 > \dot{m}_1$, the growth rate of the mass ratio is always positive, i.e. the mass ratio grows until it reaches a fixed point with $q = 1$ and $\dot{m}_1 = \dot{m}_2$.

In Fig. 4.3 we show the evolution of the PBH individual masses $M_1$, $M_2$ and their mass ratio in a binary system, for different choices of the cut-off redshift $z_{\mathrm{cut\text{-}off}}$, below which accretion is negligible. As shown in the figure, the accretion onto a PBH binary system is such that the PBH masses can increase by one or two orders of magnitude and that the secondary body always experiences a stronger accretion rate compared to the primary body. Furthermore, for strongly-accreting binaries the mass ratios are expected to be close to unity. Finally, we stress that, due to a strong accretion phase in the redshift range $z \approx (10 - 30)$, PBHs formed with initial mass $M^{\mathrm{i}} \gtrsim (20 - 100) M_\odot$ can become much heavier and can represent natural candidates for intermediate-mass BHs.

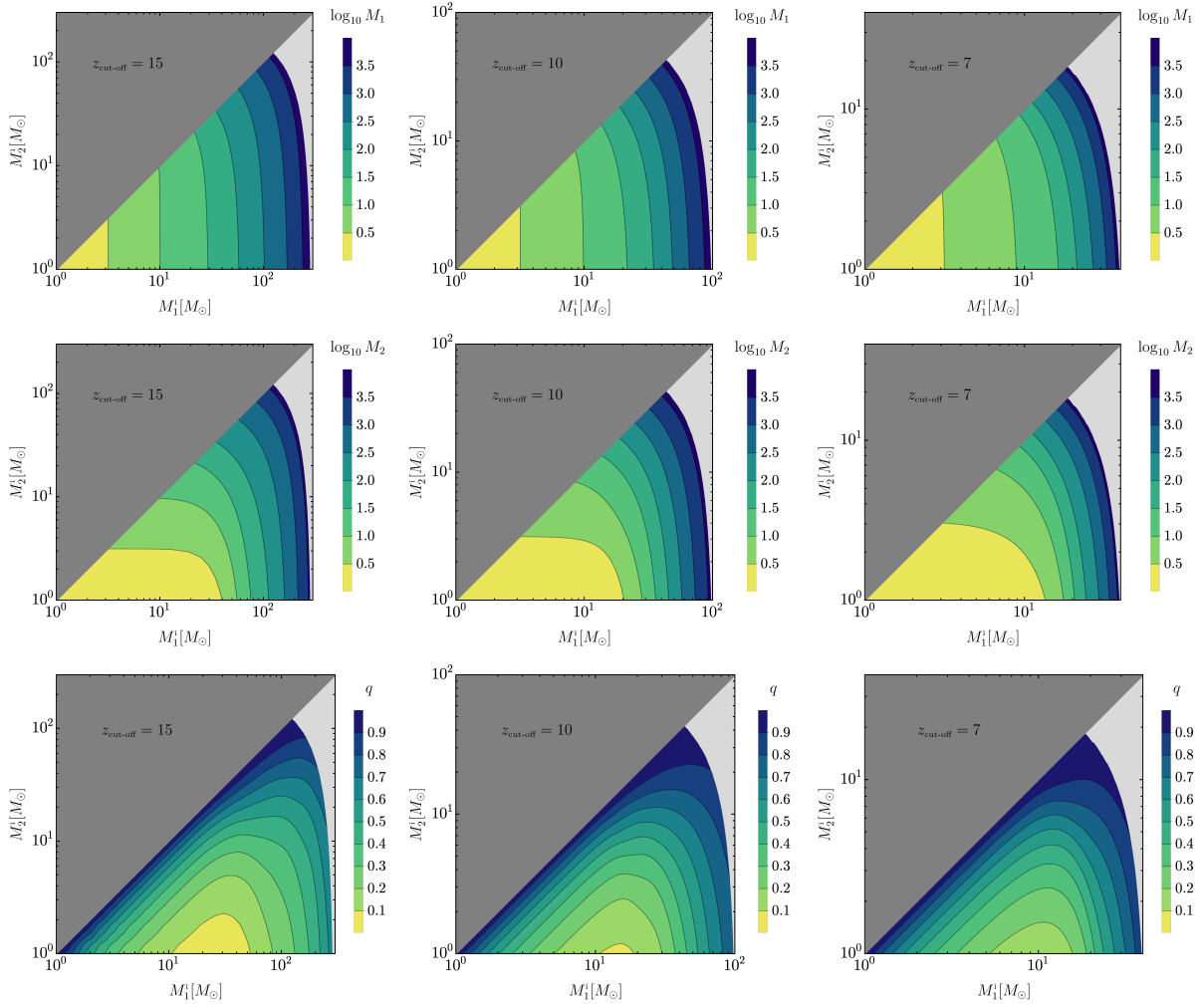

**Figure 4.3**: From top to bottom: final masses $M_1$, $M_2$ and final mass ratio $q$ of a binary system with initial masses $(M_1^i, M_2^i)$, assuming three accretion scenarios parametrised by $z_{\text{cut-off}} = 15$ (left panels), $z_{\text{cut-off}} = 10$ (middle panels), and $z_{\text{cut-off}} = 7$ (right panels). Figure taken from Ref. [16].

### 4.1.2 Impact of accretion on the PBH population

The evolution of the PBH masses up to the present epoch due to accretion implies that the mass distribution of PBHs at low redshifts might be significantly different from that at high redshifts. Furthermore, changes in the PBH masses would consequently modify the abundance of PBHs in the universe. In this subsection we are going to investigate these effects.

**Effect on the PBH mass function**

The mass function of PBHs at high redshifts critically depends on the details of the considered formation mechanism, and different shapes may therefore arise.

Let us define the initial mass function $\psi(M, z_i)$ as the fraction of PBHs with mass in the interval $(M, M + \mathrm{d}M)$ at the formation redshift $z_i$. Motivated by the collapse of scale invariant perturbations, an interesting possibility is provided by a power-law mass function [19, 92, 148, 348]

$$\psi(M, z_i) = \frac{1}{2}\left(M_{\min}^{-\frac{1}{2}} - M_{\max}^{-\frac{1}{2}}\right)^{-1} M^{-\frac{3}{2}}, \tag{4.32}$$

in terms of the free parameters $M_{\min}$ and $M_{\max}$. An alternative mass function is the lognormal one

$$\psi(M, z_i) = \frac{1}{\sqrt{2\pi}\sigma M}\exp\left(-\frac{\log^2(M/M_c)}{2\sigma^2}\right), \tag{4.33}$$

as a function of a width $\sigma$ and peak reference mass $M_c$ [99]. This mass function may arise in scenarios in which PBHs form from the collapse of perturbations characterised by a symmetric peak in the primordial power spectrum [374].

From the initial shape $\psi(M_i, z_i)$, its evolution at redshift $z$ is given by [17, 18]

$$\psi(M(M_i, z), z)\mathrm{d}M = \psi(M_i, z_i)\mathrm{d}M_i, \tag{4.34}$$

where $M(M_i, z)$ is the final mass at redshift $z$ for a PBH with mass $M_i$ at redshift $z_i$. In accounting for the impact of accretion onto the PBH mass function we have considered the contribution coming from isolated PBHs. As we will see, this is the dominant contribution when we want to consider how existing constraints on the PBH abundance (mainly due to PBHs) are modified in the presence of accretion.

The main effect of accretion on the mass function is to make the latter broader at high masses, with a high-mass tail that can be orders of magnitude above its corresponding value at formation [17]. In Fig. 4.4 we compare the initial and evolved mass functions for the choices of power-law and lognormal initial shape.

**Effect on the PBH abundance and constraints**

The evolution of the PBH masses, and consequently their energy densities, would imply that also the value of $f_{\mathrm{PBH}}$ is affected by accretion. Making the simplifying assumption that a non-relativistic dominant DM component (whose energy density scales as the inverse of the volume) is present, one can show that the PBH abundance evolves as [17]

$$f_{\mathrm{PBH}}(z) = \frac{\rho_{\mathrm{PBH}}}{(\rho_{\mathrm{DM}} - \rho_{\mathrm{PBH}}) + \rho_{\mathrm{PBH}}} = \frac{\langle M(z)\rangle}{\langle M(z_i)\rangle(f_{\mathrm{PBH}}^{-1}(z_i) - 1) + \langle M(z)\rangle}, \tag{4.35}$$

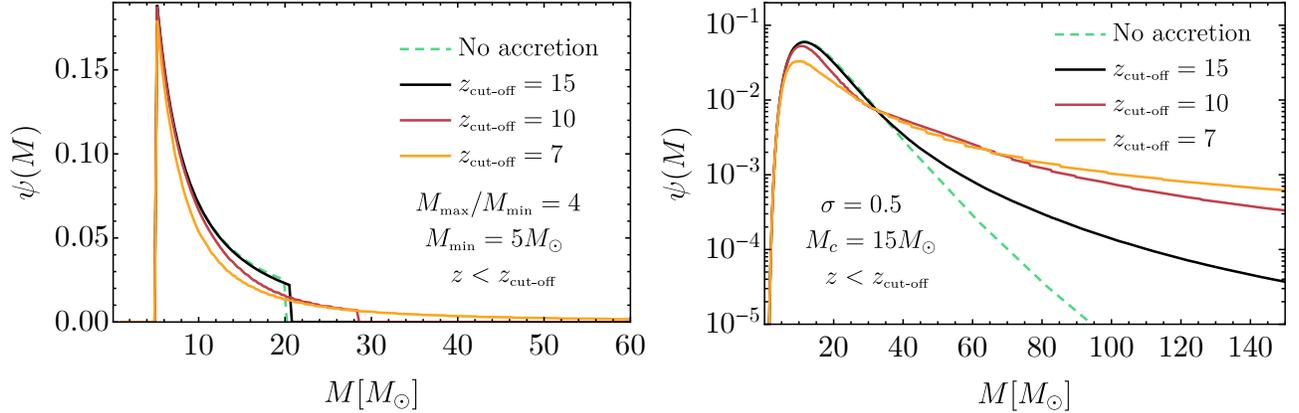

**Figure 4.4**: Evolution of the power-law (left) and lognormal (right) mass functions due to the presence of accretion, for different choices of the cut-off redshift. Figure taken from Ref. [16].

in terms of the average mass

$$\langle M(z) \rangle = \int \mathrm{d}M M \psi(M, z). \tag{4.36}$$

Due to the presence of accretion, $f_{\mathrm{PBH}}(z)$ can be significantly larger than $f_{\mathrm{PBH}}(z_i)$. An example is shown in Fig. 4.5 for the same choices of initial mass function of the previous section.

The evolution of the PBH abundance due to accretion is important also for the interpretation of observational constraints on its value. As we have seen in the introduction, usually these constraints are expressed in terms of the present maximum fraction $f_{\mathrm{PBH}}$ of PBHs allowed in a given mass range. However, given that the PBH mass distribution and abundance may be different at high and low redshifts, these limits have to be properly revisited.

One example is provided by the constraints coming from CMB. In particular, CMB temperature and polarization fluctuations depend on the energy injection up to redshift $z \sim 300$ [171, 172]. If PBHs experience a phase of accretion at lower redshifts, the present PBH masses and distribution do not coincide with those at redshifts $z \sim (300 - 600)$ when the energy deposition has the most crucial effect on the CMB.

To include this effect in the constraints, we can follow the prescription introduced in Ref. [374] to estimate the bound on the abundance $f_{\mathrm{PBH}}(z_e)$ of PBHs at the redshift $z_e$ of a given experiment as [17]

$$f_{\mathrm{PBH}}(z_e) \lesssim \left( \int_{M_{\min}(z_e)}^{M_{\max}(z_e)} \mathrm{d}M \frac{\psi(M, z_e)}{f_{\max}(M, z_e)} \right)^{-1}, \tag{4.37}$$

in terms of the mass range affected by the constraint with limits $M_{\min}(z_e)$ and $M_{\max}(z_e)$, the maximum allowed abundance $f_{\max}(M, z_e)$ for a monochromatic mass function at the redshift of the experiment obtained neglecting accretion [84, 374] and taking into account the evolution of the mass function from formation to $z_e$ through Eq. (4.34). Once $f_{PBH}(z_e)$ has been computed, one can map it into the present fraction of PBHs as DM $f_{PBH}(z = 0)$ using Eq. (4.35) for a given $\langle M(z = 0) \rangle$.

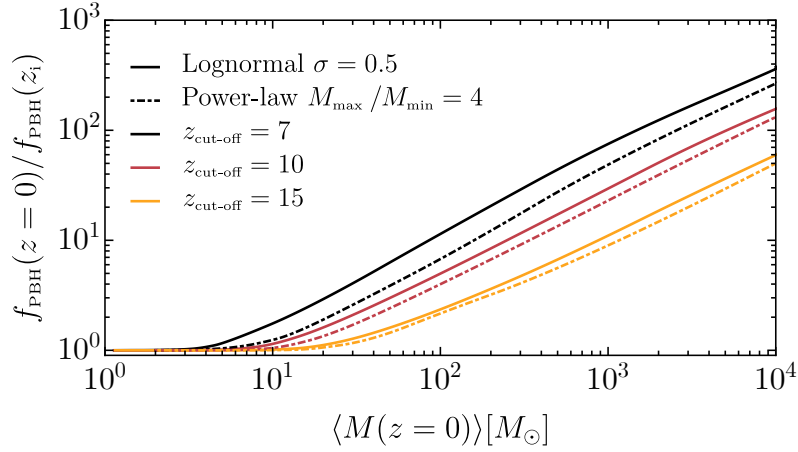

**Figure 4.5**: Evolution of the PBH abundance assuming either a power-law or a lognormal mass function at formation, for different choices of the cut-off redshift. Figure taken from Ref. [16].

In Fig. 4.6 we show an envelope resulting from the most stringent constraints on the PBH abundance for the choices of a monochromatic and lognormal mass function, for different values of $\langle M(z = 0) \rangle$ (obtained by varying the initial mass function reference scale $M_c$). The envelope is shown both without the effect of accretion and considering three different cut-off redshifts $z_{cut-off} = 15$, 10 and 7. Fig. 4.6 shows that the observational constraints are drastically weakened at the present epoch with respect to the case without considering the effect of accretion. In particular, for the case of a monochromatic mass function, the bounds relaxation depends on the shift of the distribution's peak and the evolution of $f_{PBH}(z)$. In this case, since the bounds refer to the present value of the average PBH mass, only the constraints obtained with observations at high redshifts are affected (e.g. Planck D). We stress that, even though the constraints obtained with observations at small redshifts are unaffected for a monochromatic distribution, they still refer to a different mass $\langle M(z = 0) \rangle$ relative to the case in which accretion is absent, for which $\langle M(z = 0) \rangle = M_c$. For extended distributions, the relaxation of the constraints depends also on the mass function evolution, which affects both bounds at high and low redshifts, since the latter are inferred assuming a broader distribution.

### 4.1.3 Spin evolution

Given that the accretion rate and the geometry of the accretion flow are highly intertwined, they both crucially determine the evolution of the accreting PBHs. An additional ingredient which affects their evolution is the amount of angular momentum carried by the infalling accreting gas onto the PBH, which strongly impacts onto the geometry of the accretion flow and may be responsible for the evolution of the PBH spin. In this section we are going to provide the details related to the evolution of the PBH spin due to accretion following the results of Ref. [16, 18].[2]

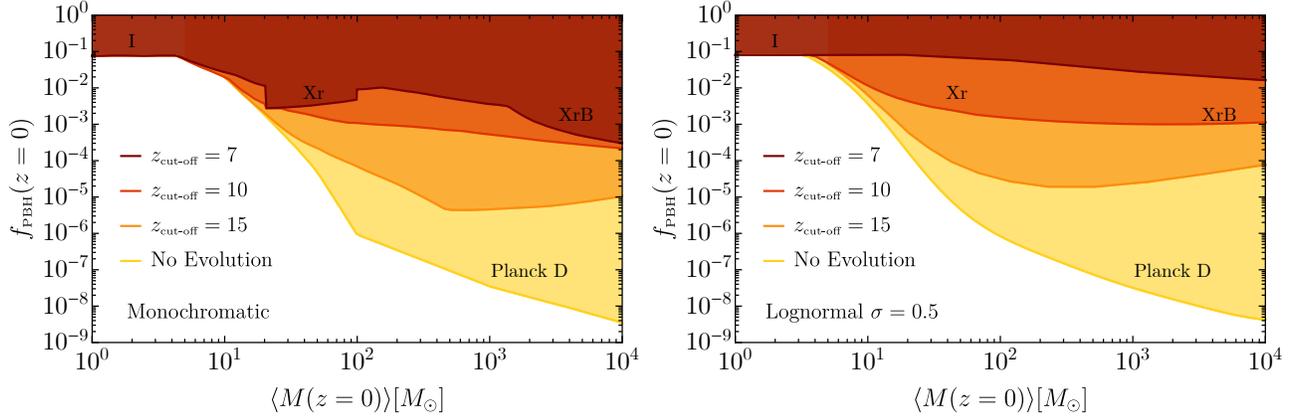

**Figure 4.6**: Envelope of the most stringent constraints on the present PBH abundance in terms of $\langle M(z=0)\rangle$ for different accretion cut-off redshifts $z_{\text{cut-off}} = 15$, 10 and 7 compared to the case in which accretion is neglected ("No Evolution"). The labels identify the corresponding experiment dominating each portion of the graphs. *Left*: Monochromatic case. *Right*: Lognormal mass function with width $\sigma = 0.5$ at formation. Figure taken from Ref. [17].

We first investigate under which conditions there may be the formation of an accreting disk around the PBH, and then determine how the PBH spins evolve.

**Formation of a disk**

Angular momentum may be carried by the accreting gas infalling onto the PBH. This can crucially impact the geometry of the accreting flow and induce the formation of an accreting disk. The latter may in turn spin up the PBHs.

The formation of an accreting disk can be understood by studying the relative velocity between the accreting gas and the PBH. In particular, the expression for the baryon velocity

---

[2]We stress that the PBH spin may be modified also by plasma-driven superradiant instabilities [375–377], which strongly depends upon the geometry of the plasma surrounding the BH. Given that for realistic systems [378] it is negligible, we will neglect it in the following. Furthermore, we also neglect the presence of ultralight bosons for which superradiance could be effective [376].

variance is given by [172]

$$\sigma_{\rm b} \simeq \sigma_{\rm b,0}\,\xi^{-1.7}(z) \left(\frac{1+z}{1000}\right)^{-1} \left(\frac{M_h}{M_\odot}\right)^{0.85},$$ (4.38)

in terms of $\sigma_{\rm b,0} = 3.8 \times 10^{-7}{\rm km\,s^{-1}}$. The Mach number $\xi(z) = {\rm Max}[1, \langle v_{\rm eff}\rangle/c_s]$ keeps into account the effect of a (relatively small) PBH proper motion in reducing the Bondi radius. The formation of an accretion disk, and the consequent transfer of angular momentum from the baryonic particles to the PBH, is effective only if the typical gas velocity is larger than the Keplerian velocity close to the PBH

$$\sigma_{\rm b} \gtrsim 2D\xi^2(z)c_s^2,$$ (4.39)

where the constant $D \sim \mathcal{O}(1-10)$ takes into account relativistic corrections. In the opposite regime, the accretion geometry is quasi-spherical. Using this condition and the expression of the dark halo mass in terms of the PBH one, Eq. (4.9), one can determine the minimum PBH mass for which a disk geometry appears

$$M \gtrsim 6 \times 10^2 M_\odot\, D^{1.17} \xi^{4.33}(z)\, \frac{(1+z/1000)^{3.35}}{\left[1 + 0.031\,(1+z/1000)^{-1.72}\right]^{0.68}}.$$ (4.40)

While this occurs for baryonic particles, for the case of accreting DM the situation is slightly different. In particular, its angular momentum is typically much smaller than the one of the gas and thus does not lead to the formation of a DM disk. However, due to the presence of the dark halo which envelops the PBH, it affects the baryonic disk formation.

The above condition, even though necessary for the formation of a disk, is not sufficient for an effective transfer of angular momentum from baryons to the PBH, which also depends on the accretion rate. In particular, if $\dot{m} < 1$ and accretion is non-spherical, an advection-dominated accretion flow (ADAF) may form [379], while if $\dot{m} \gg 1$ the accretion luminosity might be strong enough that the disk "puffs up" and becomes thicker. The interesting regime in which a geometrically thin accretion disk forms is when [172, 380]

$$\dot{m} \gtrsim 1.$$ (4.41)

One can numerically check that the latter condition is always more stringent than condition (4.40), implying that $\dot{m} \gtrsim 1$ can be regarded as the sufficient condition for the formation of a thin disk around an isolated PBH.

If accretion occurs onto a PBH binary system, the transfer of angular momentum on each PBH is much more efficient [18]. In particular, in this case each component of the binary is characterised by velocities comparable to the orbital velocity, and even though the Bondi radii of the single PBHs are much smaller than the binary Bondi radius, they are still parametrically larger than the radius of the innermost stable circular orbit (ISCO, see Eq. (4.44) below). This implies that accretion onto a binary system is never spherical and a disk can form.

We can therefore summarize and state that, for both isolated and binary PBHs, a thin accretion disk forms whenever $\dot{m} \gtrsim 1$ during their cosmic evolution.

**Spin dynamics**

Once a thin accretion disk is formed, transfer of angular momentum from the baryonic particles to the PBH is efficient and mass accretion is accompanied by an increase of the PBH spin.

Given that the accretion disk is located along the equatorial plane [380, 381], with the PBH spin perpendicularly aligned with respect to the disk plane, one can apply the geodesic model to describe the angular-momentum accretion [382]. Assuming a circular disk motion, the evolution rate of the PBH angular momentum is given in terms of the mass accretion rate by [382–385]

$$\dot{J} = \frac{L(M,J)}{E(M,J)}\dot{M},\qquad(4.42)$$

where

$$E(M,J) = \sqrt{1 - 2\frac{M}{3r_{\text{ISCO}}}}\qquad\text{and}\qquad L(M,J) = \frac{2M}{3\sqrt{3}}\left(1 + 2\sqrt{3\frac{r_{\text{ISCO}}}{M} - 2}\right),\qquad(4.43)$$

and the ISCO radius is given by

$$r_{\text{ISCO}}(M,J) = M\left[3 + Z_2 - \sqrt{(3 - Z_1)\left(3 + Z_1 + 2Z_2\right)}\right],\qquad(4.44)$$

with

$$Z_1 = 1 + \left(1 - \chi^2\right)^{1/3}\left[(1 + \chi)^{1/3} + (1 - \chi)^{1/3}\right]\qquad\text{and}\qquad Z_2 = \sqrt{3\chi^2 + Z_1^2}.\qquad(4.45)$$

By introducing the Kerr parameter for the PBH, $\chi \equiv |\vec{J}|/M^2$, its time evolution is described by

$$\dot{\chi} = (\mathcal{F}(\chi) - 2\chi)\frac{\dot{M}}{M},\qquad(4.46)$$

in terms of the combination (which is only $\chi$-dependent)

$$\mathcal{F}(\chi) \equiv \frac{L(M,J)}{ME(M,J)}.\qquad(4.47)$$

This equation implies that the PBH spin grows over a typical accretion time scale until it reaches extremality, $\chi = 1$. However, radiation effects constrain the maximum value to be $\chi_{\max} = 0.998$ [383]. Furthermore, with the help of magnetohydrodynamic simulations of accretion disks around Kerr BHs, it has been suggested that the maximum spin might be even smaller, $\chi_{\max} \simeq 0.9$ [386], even though this limit may not apply to geometrically thin disks. The geometrically thin-disk approximation is expected to be valid for each PBH when $\dot{m}_i \gtrsim 1$. For larger accretion rates, the disk might be geometrically thicker and the transfer of angular momentum might be less efficient. However, it is expected that the spin evolution time scale does not change significantly in more realistic accretion models [386].

To conclude, we have showed that an important feature of the PBH scenario is the expectation of high spins for sufficiently massive PBHs, and that the conditions for efficient angular momentum transfer are more easily satisfied by binary PBHs than by the isolated ones, since in the former case accretion is enhanced by the larger total mass of the binary.

In Fig. 4.7 we show the final spins of the components of a PBH binary in terms of their final masses for different choices of the cut-off redshift $z_{\text{cut-off}}$. We have assumed, following the results of the previous chapter, that the initial PBH spin is of the order of the percent level or smaller. One can appreciate that the final PBH mass and spin are correlated: PBHs with low-mass are slowly spinning or non-spinning, whereas PBHs with large masses are rapidly spinning. The transition mass scale depends on the cut-off redshift and it is always around $(10 - 40)\, M_\odot$. In particular, smaller values of $z_{\text{cut-off}}$ favour a transition at low masses, given that in this case the accreting phase lasts longer during the cosmic evolution.

In Fig. 4.8 we show the distribution of the effective spin parameter $\chi_{\text{eff}}$ in terms of the final PBH mass $M_1$, for different values of the binary mass ratio and cut-off redshifts. To obtain the probability distribution functions (PDFs) one has to perform a statistical ensemble over the masses of the binary components and the relevant angles of the spin vectors with respect to the angular momentum. For the latter, we have assumed a uniform distribution for the spin vectors orientations on a unit two-sphere [387], given that the initial PBHs spin orientations are uncorrelated at the formation epoch. When accretion is efficient, a transition from initially vanishing values of $\chi_{\text{eff}}$ to large ones occurs, with the transition point dependent on the value of $z_{\text{cut-off}}$. As a comparison, we have superimposed some of the data from GWTC-1 which are compatible (within their reported errors) with the corresponding chosen value of $q$.

## 4.1.4 Uncertainties in the accretion model

In this section we are going to describe some of the main uncertainties in the accretion model, which may affect its efficiency during the cosmic history, related for example to its dependence on the velocity or to the geometry of the accretion flow [170, 172, 368].

a) *Local feedback*: in our analysis we did not consider the effect of local feedback on the accretion flow. For the range of masses of ground-based GW interferometers, this effect may be safely neglected [170, 368].

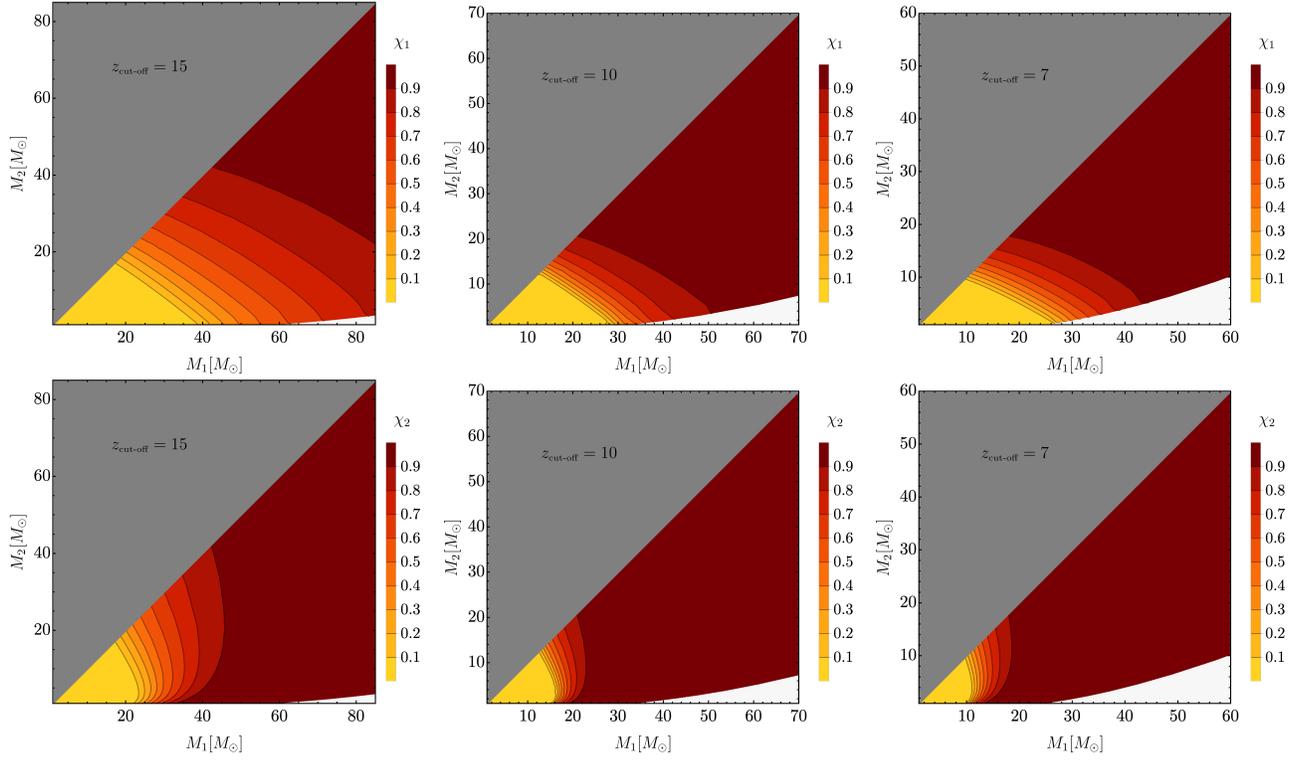

**Figure 4.7**: Contour plot of the PBH final spins $\chi_1$ (top panels) and $\chi_2$ (bottom panels) in terms of the final PBH masses ($M_1, M_2$), for different cut-off redshifts. Figure taken from Ref. [16].

b) *Global feedback & X-ray pre-heating*: after the seminal paper of Ref. [368], in which X-ray heating of the gas and its impact on PBH accretion were investigated (but neglecting other possible sources of X-ray heating [393]), the more recent analysis of Ref. [170] has shown that global feedbacks are much less important for the mass range of LVKC. Indeed, Ref. [170] found an estimate for the accretion rate which is compatible with the one of Ref. [368] without the effect of global heating. A proper model of the temperature of the intergalactic medium at redshift $10 \lesssim z \lesssim 30$ is particularly relevant, since higher temperatures imply an increase in the sound speed and, in turn, a reduction of the accretion rate.

c) *DM halo*: the inclusion of a DM halo is inevitable in the model of accretion if PBHs form only a fraction of the DM in the universe, at least if the latter is made by particles.

d) *Structure formation*: after redshift $z \simeq 10$, a fraction of the PBH population starts falling in the gravitational potential well of large-scale structures. This implies that the relative velocity may increase up to one order of magnitude [394], which results in a suppression of the accretion rate [172, 177, 395]. Furthermore, the fraction of PBHs that stop accreting efficiently enough due to this effect may be hardly estimated; in particular, dynamical friction effects may induce captured PBHs to settle at the center of the halo within a Hubble

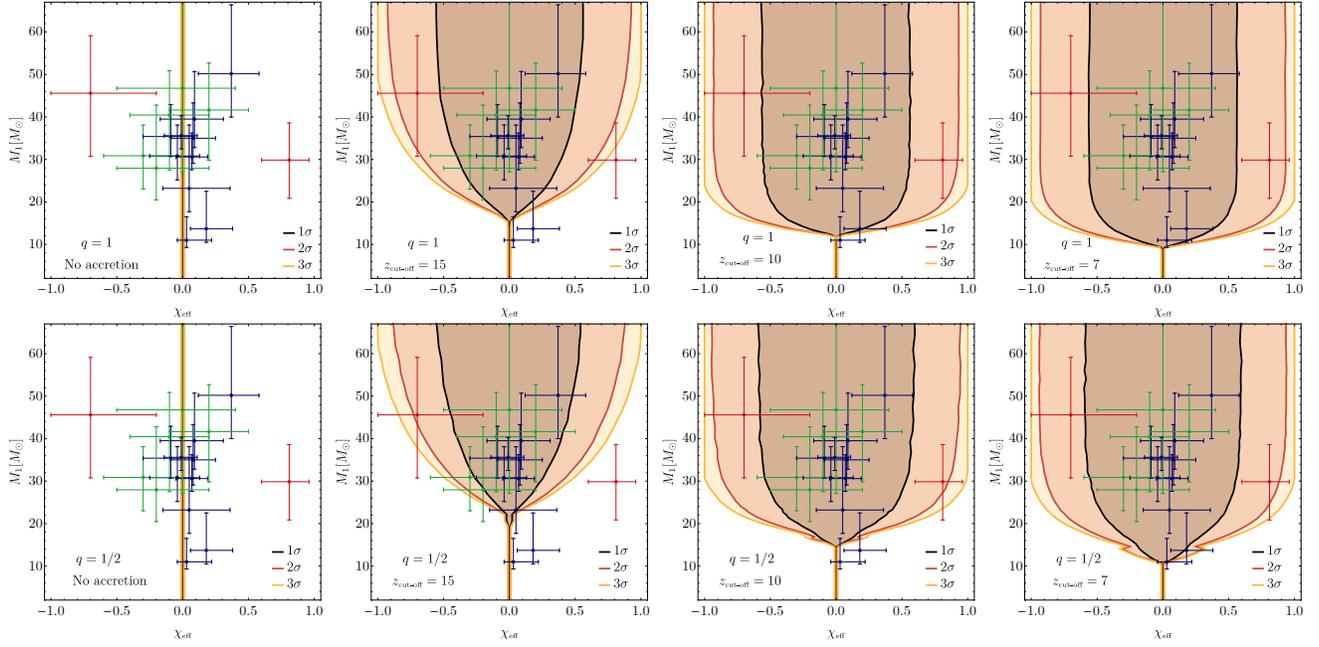

**Figure 4.8**: Contour plot of the probability distribution of the effective spin parameter $\chi_{\rm eff}$ in terms of the final PBH mass $M_1$ for different values of the mass ratio $q$. Blue data points refer to the events listed in Ref. [388], while green and red data points refer to the events found in Refs. [389, 390]; the red data points refer to GW151216 and GW170403, for which the measured value of $\chi_{\rm eff}$ strongly depends on the prior on the spin angles [391], while the cyan data point refers to GW190412 reported in Ref. [392]. Figure taken from Ref. [16].

time and keep accreting efficiently.

e) *Spherical accretion & disk geometry*: we have assumed a quasi-spherical description for the accretion flow onto compact objects, which however might break down in the presence of outflows [396]. Their efficiency in reducing the accretion rate depends on the geometry of the accretion flow and on the relative direction of the outflow.

f) *Angular momentum transfer*: when the Eddington normalised accretion rate $\dot{m} \sim 1$, a geometrically thin disk may form, which can be described with a geodesic model. When $\dot{m} \ll 1$ or $\dot{m} \gg 1$, the disk geometry is different, with a consequent impact on the accretion luminosity and feedback. In the super-Eddington regime, the disk is expected to "puff up" and become geometrically thicker. In this case the transfer of angular momentum is more complex, although numerical simulations indicate that the time scale for spin evolution does not change significantly [386].

All the above uncertainties would require a complex and model-dependent simulation to be kept into account. In our model of accretion, we have incorporated their effects by introducing a cut-off redshift $z_{\text{cut-off}}$, whose value is left unconstrained, and parametrises in an agnostic way these uncertainties.

## 4.2 On the PBH clustering

The other important property of PBHs across the cosmic history is the evolution of the PBH spatial distribution at any redshift $z$.

This feature is extremely relevant to the calculation of the GW signals emitted from PBH mergers, given that the merger rate of PBHs in clusters may be strongly different from the one in isolated environments, due to the occurring of close encounters with other surrounding PBHs. It is therefore important to address the question if PBHs cluster or not, and if yes, to what extent they do so. Furthermore, a proper understanding of the PBH clustering is relevant also for its possible impact on some of the constraints coming from microlensing searches, which bound the PBH abundance to be smaller than unity in a given range of masses [397] (see also Refs. [398, 399] for recent works along this direction).

In this section we are going to investigate the PBH clustering in the standard scenario in which PBHs form from the collapse of large overdensities produced during inflation. Assuming that the scalar curvature perturbation follows a Gaussian probability distribution, several works have shown that PBHs are not clustered at formation [224, 400–402]. However, after the matter-radiation equality, they can start forming clusters if the PBH abundance $f_{\text{PBH}}$ is large enough [178, 403]. Following Ref. [14], we are going to shown some analytical insights on the PBH clustering which are compatible with the results from the most recent N-body cosmological simulation of Ref. [403] and to some numerical results shown in Ref. [178].

### 4.2.1 Preliminaries and initial conditions

To characterise the spatial distribution of PBHs one has to compute the PBH two-point correlation function. Given that PBHs are discrete tracers, one can evaluate this quantity, in terms of the comoving separation $x = |\vec{x}|$, by introducing the PBH overdensity at position $\vec{x}_i$ with respect to the DM mean energy density $\bar{\rho}_{\text{DM}}$ as

$$\frac{\delta\rho_{\text{PBH}}(\vec{x}, z)}{f_{\text{PBH}}\bar{\rho}_{\text{DM}}} = \frac{1}{\bar{n}_{\text{PBH}}}\sum_i \delta_D(\vec{x} - \vec{x}_i(z)) - 1, \tag{4.48}$$

in terms of the average PBH number density per comoving volume

$$\bar{n}_{\text{PBH}} \simeq 3.2\, f_{\text{PBH}}\left(\frac{20\, M_\odot/h}{M_{\text{PBH}}}\right)(h/\text{kpc})^3 \tag{4.49}$$

and the index $i$ running over the PBH positions. The corresponding two-point correlation function takes the form

$$\left\langle \frac{\delta \rho_{\text{PBH}}(\vec{x}, z)}{\overline{\rho}_{\text{DM}}} \frac{\delta \rho_{\text{PBH}}(0, z)}{\overline{\rho}_{\text{DM}}} \right\rangle = \frac{f_{\text{PBH}}^2}{\overline{n}_{\text{PBH}}} \delta_D(\vec{x}) + \xi(x, z), \tag{4.50}$$

where the term proportional to the Dirac delta identifies the Poisson noise, while $\xi(x, z)$ is the reduced PBH correlation function, which will carry the information on the evolution of PBH clustering with time. The corresponding PBH power spectrum then reads

$$\Delta^2(k, z) = \frac{k^3}{2\pi^2} \int \mathrm{d}^3 x \, e^{i\vec{k}\cdot\vec{x}} \left\langle \frac{\delta \rho_{\text{PBH}}(\vec{x}, z)}{\overline{\rho}_{\text{DM}}} \frac{\delta \rho_{\text{PBH}}(0, z)}{\overline{\rho}_{\text{DM}}} \right\rangle, \tag{4.51}$$

with respect to the total cold DM average density.

In the following we will make the standard assumption of hierarchical clustering, according to which, above a characteristic clustering length, the Poisson fluctuations dominate the two-point correlation function, while on smaller scales fluctuations in the PBH number are dominated by the reduced PBH correlation function $\xi(x, z)$.

In the standard scenario of PBH formation from the collapse of density perturbations generated during inflation, the clustering length at formation would in principle depend on the shape of the primordial curvature perturbation power spectrum. However, assuming that the scalar perturbations follow a Gaussian statistics, this characteristic length scale is found to be significantly smaller than the mean comoving PBH separation, implying that PBH clustering is not relevant at the time of formation [224, 400–402]. This result may drastically change in the presence of primordial non-Gaussianity (for example in the form of a local, scale invariant, term in the curvature perturbation field), which is responsible for a correlation between the long- and short-wavelength fluctuations.

Under the Gaussian assumption, and considering the simplifying case of a monochromatic PBH mass function, one gets that the initial PBH power spectrum at the formation redshift $z_i$ takes the form [403]

$$\Delta_i^2(k) = \frac{k^3}{2\pi^2} \int \mathrm{d}^3 x \, e^{i\vec{k}\cdot\vec{x}} \left\langle \frac{\delta \rho_{\text{PBH}}(\vec{x}, z_i)}{\overline{\rho}_{\text{DM}}} \frac{\delta \rho_{\text{PBH}}(0, z_i)}{\overline{\rho}_{\text{DM}}} \right\rangle \approx f_{\text{PBH}}^2 \left( \frac{k}{k_*} \right)^3, \tag{4.52}$$

in terms of the characteristic wavenumber

$$k_* = (2\pi^2 \bar{n}_{\text{PBH}})^{1/3} \simeq 4 \, f_{\text{PBH}}^{1/3} \left( \frac{20 \, M_\odot/h}{M_{\text{PBH}}} \right)^{1/3} h/\text{kpc}, \tag{4.53}$$

that is inversely proportional to the initial PBH mean spatial separation.

We will now follow the time evolution of PBH clustering across the linear, quasi-linear and non-linear regimes.

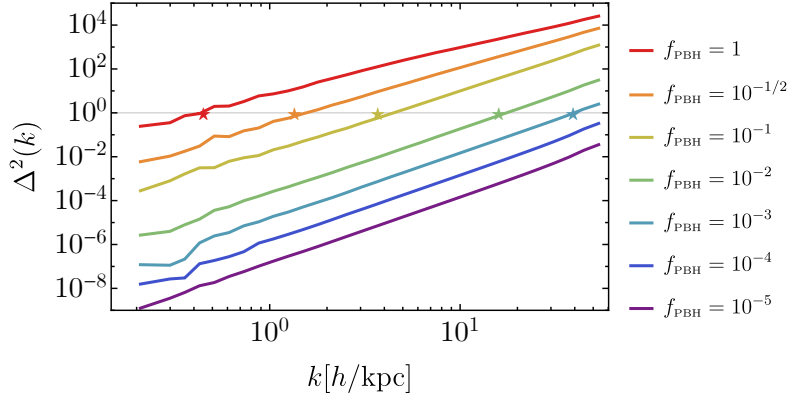

**Figure 4.9**: The PBH power spectra at $z = 99$ for different values of the PBH abundance $f_{\mathrm{PBH}}$ superimposed with the data taken from Ref. [403]. The stars identify the transition from the linear to the quasi-linear regime as predicted by Eq. (4.56). Figure taken from Ref. [14].

### 4.2.2 The linear regime

Before the matter-radiation equivalence, the PBH density contrast is essentially frozen in time; then it starts growing linearly according to [403]

$$\Delta_{\mathrm{L}}^2(k, z) \simeq \left(1 + \frac{3}{2} f_{\mathrm{PBH}} \frac{1 + z_{\mathrm{eq}}}{1 + z}\right)^2 \Delta_{\mathrm{i}}^2(k), \tag{4.54}$$

where we have adopted the matter-dominated epoch behaviour $(1 + z)^{-1}$. These linear (L) perturbations in the PBH number density grow until they enter the quasi-linear (QL) regime, when the density contrast reaches values of order unity. Such a transition may occur at different redshifts, depending on the fluctuation wavenumber $k$, and is achieved approximately when

$$\Delta_{\mathrm{L}}^2(k = k_{\mathrm{L\text{-}QL}}(z), z) \simeq 1, \tag{4.55}$$

which implies

$$k_{\mathrm{L\text{-}QL}}(z) \simeq \frac{4}{f_{\mathrm{PBH}}^{1/3}} \left(\frac{20 \, M_\odot / h}{M_{\mathrm{PBH}}}\right)^{1/3} \left[1 + 26 f_{\mathrm{PBH}} \left(\frac{100}{1 + z}\right)\right]^{-2/3} h/\mathrm{kpc}. \tag{4.56}$$

In Fig. 4.9 we show the PBH power spectra extracted from the N-body numerical simulations of Ref. [403] at redshift $z = 99$ for different values of the PBH abundance. As one can appreciate, the numerical results are fitted very well from the corresponding values of $k_{\mathrm{L\text{-}QL}}$, which are indicated by stars.

### 4.2.3 The quasi-linear regime

Once the PBH overdensities have become of order unity and have entered the quasi-linear regime, they will decouple from the Hubble flow, collapse under the effect of gravity and virialize in structures like halos with a virial density about 200 times larger the background density at the virialization time. As it happens in standard cold dark matter cosmology, the dynamics of halo formation is hierarchical, meaning that small PBH halos form at earlier times and act as progenitors of more massive ones which form at later epochs. During the quasi-linear regime, the PBH density power spectrum $\Delta^2$ grows from unity up to values around 200 (while the linearly evolved PBH averaged two-point correlator, to be defined below, grows from unity until 5.85 [404]), at which the evolution enters the fully non-linear regime.

The collapse in virialised structures occurs when the (integrated) overdensity within a given comoving radius $R$ reaches the critical value $\delta_c \simeq 1.68$. We therefore introduce the volume averaged correlation function

$$\overline{\xi}(R, z) = \frac{3}{4\pi R^3} \int_0^R \mathrm{d}s\, 4\pi s^2 \xi(s, z),$$

$$\xi(x, z) \simeq \int \frac{\mathrm{d}k}{k}\, e^{i\vec{k}\cdot\vec{x}} \Delta^2(k, z),$$ 

(4.57)

where in the second equation we have used the fact that the large-scale Poisson noise in the correlation function is subdominant in the quasi-linear regime with respect to the reduced correlation function. The volume averaged correlation function can also be thought of as measuring the power at some effective wavenumber

$$\Delta^2(k, z) \simeq \overline{\xi}(1/k, z).$$ 

(4.58)

To connect the quasi-linear correlation function to the linear theory, one can make use of the conservation of particle pairs. By introducing the mean number of neighbours [46]

$$N(x, z) = \bar{n}_{\mathrm{PBH}} \int_0^x \mathrm{d}s\, 4\pi s^2 \left[1 + \xi(s, z)\right],$$ 

(4.59)

and using its conservation, one gets the equation for the reduced correlation function [46]

$$\frac{\partial \xi}{\partial t} + \frac{1}{ax^2} \frac{\partial}{\partial x} \left[x^2(1 + \xi)v\right] = 0,$$ 

(4.60)

in terms of the scale factor $a$ and the mean relative velocity of pairs at separation $x$ and time $t$, $v(x, t)$. This equation has been obtained neglecting the two-body relaxation and evaporation from the PBH cluster, on which we will come back in the following paragraphs. The conservation of pairs gives the mass conservation relation [46]

$$x^3(1 + \overline{\xi}) = R^3,$$ 

(4.61)

as a function of the initial shell radius $R$. In the linear regime, $\overline{\xi} \ll 1$ and $R \sim x$, while as clustering becomes relevant in the quasi-linear regime, $\overline{\xi}$ increases and the scale $x$ becomes smaller than $R$. This implies that the reduced correlation function in the quasi-linear regime $\overline{\xi}_{\mathrm{QL}}(x)$ can be expressed as a function of the one in the linear regime, $\overline{\xi}_{\mathrm{L}}(R)$. Indeed, by considering a region surrounding a density peak in the linear regime, around which clustering takes place [404], one can show that the density profile is proportional to the underlying correlation function [213] (ignoring the contribution from gradients). This implies that the linear volume averaged squared density contrast scales with the initial shell radius $R$ as $\overline{\xi}_{\mathrm{L}}(R)$ as long as we are in the linear regime. Using the spherical collapse model, this perturbation expands to a maximum radius $x_{\max}$ proportional to $R/\overline{\xi}_{\mathrm{L}}(R)$ [46]. By considering a halo of mass $M$ and such radius, one has

$$\overline{\xi}_{\mathrm{QL}}(x) \sim \frac{M}{x^3} \sim \frac{R^3}{(R/\overline{\xi}_{\mathrm{L}}(R))^3} \sim \overline{\xi}_{\mathrm{L}}^3(R),$$
$$R^3 \sim x^3 \overline{\xi}_{\mathrm{L}}^3 \sim x^3 \overline{\xi}_{\mathrm{QL}}. \tag{4.62}$$

Given that $\overline{\xi}_{\mathrm{L}}(x) \sim x^{-3}$ for PBHs, one gets the scaling result $\overline{\xi}_{\mathrm{QL}}(x) \sim x^{-9/4}$ or

$$\Delta_{\mathrm{QL}}^2(k) \simeq \left( \frac{k}{k_{\mathrm{L\text{-}QL}}(z)} \right)^{9/4} \simeq 0.04 \, f_{\mathrm{PBH}}^{3/4} \left( \frac{20 \, M_\odot/h}{M_{\mathrm{PBH}}} \right)^{-3/4} \left[ 1 + 26 f_{\mathrm{PBH}} \left( \frac{100}{1+z} \right) \right]^{3/2} \left( \frac{k}{h/\mathrm{kpc}} \right)^{9/4}. \tag{4.63}$$

From Fig. 4.10 one can see that this analytical prediction matches the simulation results found in Ref. [403].

The time dependence of the density power spectrum in the quasi-linear regime may be found using arguments of self-similarity, in the case in which the DM represents the total energy density of the universe (which is a good approximation until dark energy dominates). Indeed, if the initial power spectrum has a power law behaviour, the evolution has to be self-similar [46], with a solution to the Boltzmann equation for the self-gravitating PBHs of the form $\xi(x,t) = f(x/t^\alpha)$ [405]. Such a solution is consistent with the linear regime only for $\alpha = 4/9$ and, given that $\overline{\xi}_{\mathrm{QL}}(x,z)$ can only depend upon the combination $x(1+z)^{2/3}$, $\Delta_{\mathrm{QL}}^2(k)$ must have a scaling behaviour of $(1+z)^{-3/2}$, which is weaker than that in the linear regime. As one can see from Eq. (4.63), this scaling is manifest if one goes in the regime $26 f_{\mathrm{PBH}}(10^2/(1+z)) \gtrsim 1$.

## 4.2.4 The non-linear regime

Once the power spectrum of the PBH density perturbations increases to values larger than $\sim 200$, the system evolves in the non-linear regime. To describe the evolution during this stage we use the stable clustering hypothesis, according to which, even though the separation between clusters is modified by the expansion of the universe, their internal structure does not change with time and they do not expand.

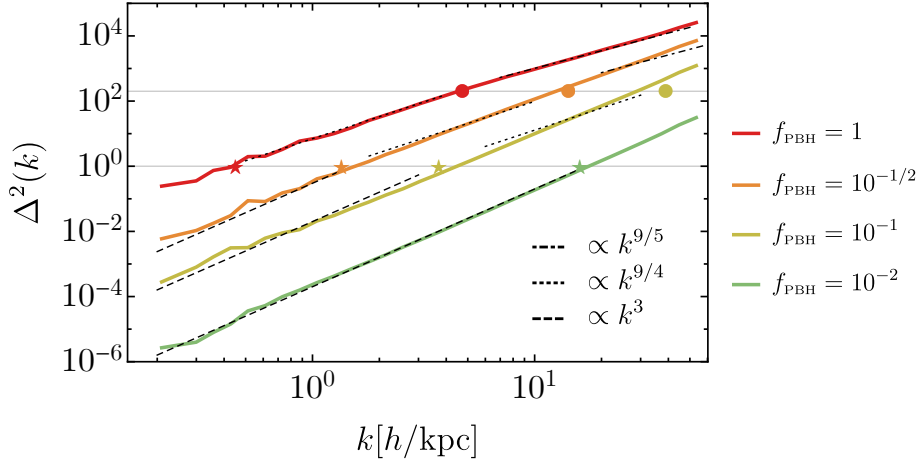

**Figure 4.10**: The PBH power spectra at redshift $z = 99$ for different values of the PBH abundance $f_{\mathrm{PBH}}$ from [403]. The dots indicate the predictions of equation (4.66) and the power laws are indicated by straight lines. Figure taken from Ref. [14].

Under this assumption, the pair conservation equation in the non-linear regime $\xi \gg 1$ can be written in the form [46]

$$\frac{\partial}{\partial t}(1+\xi) = \frac{H}{x^2}\frac{\partial}{\partial t}\left[a^3(1+\xi)\right] \tag{4.64}$$

and admits a solution with a power-law behaviour like

$$\xi_{\mathrm{NL}}(x,z) \sim \frac{x^{-m}}{(1+z)^{3-m}}. \tag{4.65}$$

The power-law index $m$ can be determined using arguments from self-similarity (with the assumption that the total DM density contributes to all the energy budget of the universe). Indeed, the latter implies that the correlation function has the functional form $\xi(x,t) = f(x/t^\alpha)$ with $\alpha = 4/9$ to reproduce the results in the linear regime or, equivalently, $\xi(x,z) \sim f(x(1+z)^{2/3})$. This implies that in the non-linear regime we have $\xi_{\mathrm{NL}}(x,z) \sim \left(x(1+z)^{2/3}\right)^{-m}$, from which one deduces $m = 9/5$.

The transition from the quasi-linear to non-linear regime is then obtained when $\bar{\xi} \sim 200$, or $(k_{\mathrm{L\text{-}QL}}/k_{\mathrm{QL\text{-}NL}})^{-9/4} \sim 200$, from which one deduces that

$$k_{\mathrm{QL\text{-}NL}}(z) \simeq 42 f_{\mathrm{PBH}}^{-1/3}\left(\frac{M_{\mathrm{PBH}}}{20 M_\odot/h}\right)^{-1/3}\left[1 + 26 f_{\mathrm{PBH}}\left(\frac{100}{1+z}\right)\right]^{-2/3} h/\mathrm{kpc}. \tag{4.66}$$

The corresponding density power spectrum then reads

$$\Delta_{\mathrm{NL}}^2(k) \simeq 200\left(\frac{k}{k_{\mathrm{QL\text{-}NL}}(z)}\right)^{9/5} \simeq 0.2\, f_{\mathrm{PBH}}^{3/5}\left(\frac{M_{\mathrm{PBH}}}{20 M_\odot/h}\right)^{3/5}\left[1 + 26 f_{\mathrm{PBH}}\left(\frac{100}{1+z}\right)\right]^{6/5}\left(\frac{k}{h/\mathrm{kpc}}\right)^{9/5}, \tag{4.67}$$

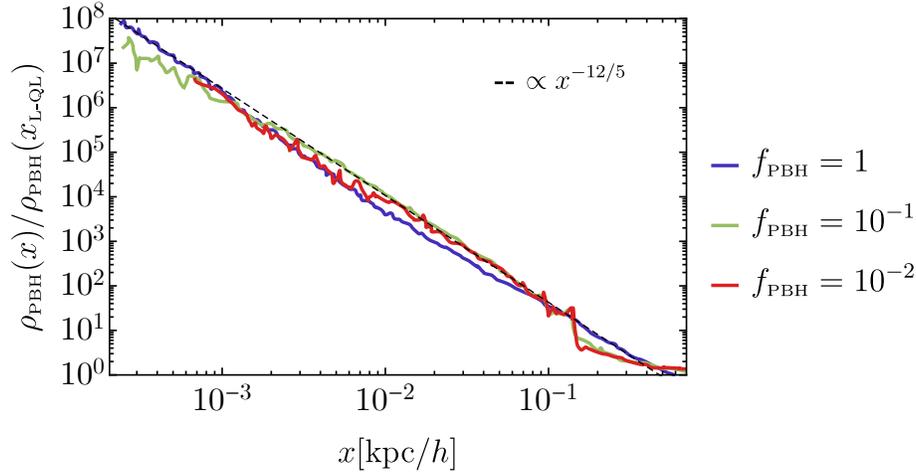

**Figure 4.11**: The PBH profile at redshift $z \simeq 1100$ for different values of the PBH abundance $f_{\text{PBH}}$ obtained from Fig. 7 of Ref. [178], compared with our prediction in Eq. (4.68) shown in dashed lines. Figure taken from Ref. [14].

whose behaviour with redshift is given by $(1+z)^{-6/5}$ for $26 f_{\text{PBH}}(10^2/(1+z)) \gtrsim 1$.

In Fig. 4.10 we compare the quasi-linear and non-linear predictions of our model with the numerical results of Ref. [403] at redshift $z = 99$, where the dots indicate the position of $k_{\text{QL-NL}}$. As one can appreciate, the model predictions are in excellent agreement with the numerical N-body data for a large PBH abundance, while some differences start to appear at smaller values of $f_{\text{PBH}}$. This occurs since we did not take into account the backreaction of the other DM component when $f_{\text{PBH}} \ll 1$.

We can now investigate the validity of our model predictions by studying the PBH density profile. Indeed, at small scales, both members of a PBH pair are probably drawn from the same PBH halo. If the PBH density profile is $\rho_{\text{PBH}}(x) \sim x^{-\epsilon_{\text{PBH}}}$, then the two-point correlation function behaves like $\sim x^{-2\epsilon_{\text{PBH}}+3}$ [406, 407] given that it is proportional to the square density profile. We stress that the density profile carries the information of the correlation function with the additional constraint that one member of each pair lies at the center of a halo, while the correlation function is obtained when the location of each pair member is unconstrained. Imposing $(-2\epsilon_{\text{PBH}} + 3) = -9/5$, one gets that the PBH density profile satisfies the behaviour

$$\rho_{\text{PBH}}(x) \sim x^{-12/5}. \tag{4.68}$$

In Fig. 4.11, we compare our model predictions with the numerical data shown in Fig. 7 of Ref. [178] on the spatial profile of PBHs surrounding a central binary at $z \simeq 1100$, finding that there is an excellent agreement with our findings and the simulation results.

To understand what are the typical halos which give the largest contribution to the PBH

correlation function, one can apply the Press-Schechter theory [345] to an initial Poisson power spectrum, such that the number density of PBH halos with mass between $M$ and $(M + \mathrm{d}M)$ reads

$$\frac{\mathrm{d}n(M,z)}{dM} = \frac{\overline{\rho}_{\mathrm{PBH}}}{\sqrt{\pi}} \left(\frac{M}{M_*(z)}\right)^{1/2} \frac{e^{-M/M_*(z)}}{M^2}, \tag{4.69}$$

in terms of the average PBH energy density $\overline{\rho}_{\mathrm{PBH}} = M_{\mathrm{PBH}} n_{\mathrm{PBH}}$ and the typical mass of halos collapsing at redshift $z$ [408]

$$M_*(z) = N_*(z) \cdot M_{\mathrm{PBH}} \simeq f_{\mathrm{PBH}}^2 \left(\frac{2600}{1+z}\right)^2 M_{\mathrm{PBH}}. \tag{4.70}$$

In the framework of the halo model [409], the non-linear correlation function can be written as [407, 410]

$$\xi(x,z) = \frac{1}{\overline{\rho}_{\mathrm{DM}}^2} \int \mathrm{d}M \frac{\mathrm{d}n(M,z)}{dM} M^2 \lambda_M(x,z), \tag{4.71}$$

where

$$\lambda_M(x,z) = \int \mathrm{d}^3 s \, \rho_{\mathrm{PBH}}(s,M,z) \rho_{\mathrm{PBH}}(|\vec{s}+\vec{x}|,M,z) \simeq \frac{1.22}{4\pi R_{\mathrm{vir}}^3} \left(\frac{x}{R_{\mathrm{vir}}}\right)^{-9/5}, \tag{4.72}$$

assuming our behaviour for the average density profile of a halo of mass $M$ [410]

$$\rho_{\mathrm{PBH}}(x,M,z) = \left(\frac{3}{5 \cdot 4\pi R_{\mathrm{vir}}^3}\right) \left(\frac{x}{R_{\mathrm{vir}}}\right)^{-12/5}. \tag{4.73}$$

The virial radius $R_{\mathrm{vir}}$ is defined as

$$R_{\mathrm{vir}}^3 = \left(\frac{3M}{4\pi \cdot 200 \, \overline{\rho}_{\mathrm{PBH}}}\right), \tag{4.74}$$

assuming an average overdensity $\sim 200$ within each virialized halo. One can easily deduce that the dominant contribution to the halo mass function comes from masses around $(11/10)M_*(z)$ at a given redshift $z$.

Our results can be improved in several aspects. First of all, one should consider that the stable clustering hypothesis breaks down when a cosmological constant- or a dark energy-dominated period begins. Secondly, one should take into account binary segregation, i.e. binaries sinking towards the halo centers since they are heavier than single PBHs, and eventual binary-PBHs interactions which can heat up the core and modify its shape. Finally, additional phenomena may occur on very small scales, like encounters of light PBHs with heavier ones, which may eject them from the halos and form a shallower profile [411].

## 4.2.5 Evaporation and the halo survival time

In the analysis shown in the previous paragraphs we have neglected the contribution coming from the evaporation of PBHs from the edges of the virialised halos. In the following, focusing on the interesting case $f_{\mathrm{PBH}} = 1$, we show that this effect is not relevant due to the competing effect of accretion of smaller halos into bigger ones.

The redshift of formation of a cluster of $N = M/M_{\mathrm{PBH}}$ PBHs can be deduced from Eq. (4.70) to be

$$1 + z_{\mathrm{form}} = \frac{2600}{\sqrt{N}}. \tag{4.75}$$

If a PBH experiences random encounters, it may receive enough energy to escape from the host halo and evaporate. The characteristic evaporation time scale of a cluster subject to gravity is given by [412]

$$t_{\mathrm{ev}} \simeq 14 \frac{N}{\log N} \frac{R}{v}, \tag{4.76}$$

where $v \simeq \sqrt{GNM_{\mathrm{PBH}}/R}$ and $R$ identifies the size of the clustered region. Therefore, for the typical cluster virialization radius $R_{\mathrm{vir}}$, one gets

$$t_{\mathrm{ev}} \simeq \frac{8 \cdot 10^{20}\mathrm{s}}{\log N} \left( \frac{N}{100} \right)^{1/2} \left( \frac{M_{\mathrm{PBH}}}{20 M_{\odot}/h} \right)^{-1/2} \left( \frac{R_{\mathrm{vir}}}{\mathrm{kpc}/h} \right)^{3/2}. \tag{4.77}$$

However, given that clustering is hierarchical, before evaporating a given halo may be included in a bigger one. One should therefore investigate if the survival time and the probability to be absorbed by a bigger halo formed at a later redshift are efficient enough for evaporation not to occur.

The survival time of a given halo of mass $M$ can be computed using the Press-Schechter formalism. In this formalism, one can introduce a time dependent threshold for collapse $\omega(z) \equiv \delta_c/a = \delta_c(1+z)$, assuming a matter-dominated period, and a time independent variance as [346]

$$S(R) = (1+z)^2 \int \frac{\mathrm{d}k}{k} \Delta_{\mathrm{L}}^2(k,z) W^2(kR), \tag{4.78}$$

in terms of the Fourier transform of a top-hat window function $W(kR)$ and the linear density power spectrum $\Delta_{\mathrm{L}}$. Using the relation

$$\frac{\omega}{\sqrt{S}} = \left( \frac{M}{M_*} \right)^{1/2} = 2 \cdot 10^{-4} \left( \frac{M}{M_{\mathrm{PBH}}} \right)^{1/2} \cdot 1.68 \,(1+z)\,, \tag{4.79}$$

one deduces that

$$S(M) = 2.5 \cdot 10^7 \left( \frac{M}{M_{\mathrm{PBH}}} \right)^{-1} = 2.5 \cdot 10^7 N^{-1}. \tag{4.80}$$

The probability that a halo of mass $M_1$ formed at redshift $z_{\mathrm{form}}(M_1)$ (corresponding to a variance

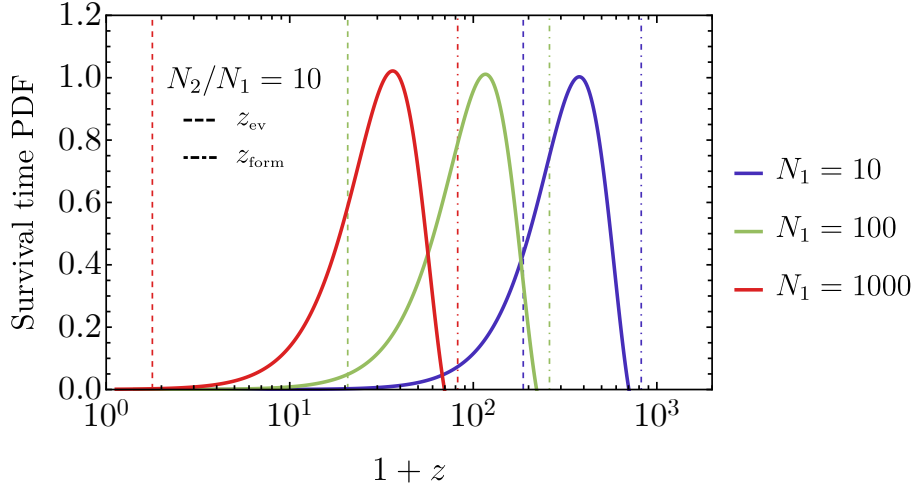

**Figure 4.12**: Probability distribution for the survival time of a halo with $N_1$ PBHs at redshift $z = z_{\text{form}}$ until it is absorbed into a heavier halo with $N_2 = 10N_1$ PBHs. Dashed (dot-dashed) vertical lines indicate the characteristic evaporation (formation) redshift of the cluster of $N_1$ objects. Figure taken from Ref. [14].

$S_1$ and time $\omega_1$) is incorporated in a bigger halo of mass $M_2$ at a subsequent redshift (corresponding to $S_2$ and $\omega_2$) is given by [346]

$$g(S_2, \omega_2 | S_1, \omega_1) \mathrm{d}\omega_2 = \sqrt{\frac{2}{\pi}} \frac{1}{\omega_1} \sqrt{\frac{S_1}{S_2(S_1 - S_2)}} \exp\left[\frac{2\omega_2(\omega_1 - \omega_2)}{S_1}\right] \left\{ \frac{-S_2(\omega_1 - 2\omega_2) - S_1(\omega_1 - \omega_2)}{S_1 e^{X^2}} \right.$$
$$\left. + \sqrt{\frac{\pi}{2}} \sqrt{\frac{S_2(S_1 - S_2)}{S_1}} \left[1 - \frac{(\omega_1 - 2\omega_2)^2}{S_1}\right] \left[1 - \mathrm{erf}(-X)\right] \right\} \mathrm{d}\omega_2, \tag{4.81}$$

where $\omega_2 < \omega_1$, $S_2 < S_1$ and

$$X = \frac{S_2(\omega_2 - 2\omega_1) + S_1\omega_2}{\sqrt{2S_1 S_2(S_1 - S_2)}}. \tag{4.82}$$

In Fig. 4.12 we show the probability that a halo containing $N_1$ PBHs at redshift $z_{\text{form}}$ is absorbed into a larger halo containing $N_2 = 10N_1$ PBHs at redshift $z < z_{\text{form}}$. One can see that, in all cases, the peak of the distribution occurs before the evaporation redshift $z_{\text{ev}}$, which is represented as vertical dashed lines. This implies that the evaporation time of PBH halos is typically larger than their survival time, which implies that PBH halos are stable against evaporation. From the figure one can also notice that the characteristic survival redshift of the progenitor halo of mass $N_1 M_{\text{PBH}}$ is broadly compatible with the formation redshift of the descendant halo of mass $N_2 M_{\text{PBH}}$ (for $N_1 = 10$ and 100), which confirms the consistency of the excursion set approach.

We therefore conclude that the correlation function is not altered by PBH evaporation if PBHs comprise the full dark matter in the universe, $f_{\text{PBH}} = 1$. For smaller abundances, the dynamics would be more complicated due to the presence of an additional component of DM, even though clustering is expected to be less relevant.

### 4.2.6 The correlation function for initially clustered PBHs

We devolve this section to the analysis of a different scenario in which PBHs are clustered at formation, so that the reduced PBH correlation function $\xi_{\mathrm{PBH}}(x)$ is important also at initial times and for the formation of binaries. In the following we report the results obtained in Ref. [8].

We consider the scenario in which, in order to introduce a sizeable PBH clustering on large-scales, the curvature perturbation $\zeta(\vec{x})$ is non Gaussian [336, 413]

$$\zeta(\vec{x}) = \zeta_{\mathrm{g}}(\vec{x}) + f_{\mathrm{NL}}\zeta_{\mathrm{g}}^2(\vec{x}), \tag{4.83}$$

in terms of the Gaussian component $\zeta_{\mathrm{g}}(\vec{x})$ and the local-type non-Gaussian parameter $f_{\mathrm{NL}}$. We assume that the Gaussian curvature perturbation has three components, one at small-scales $\sim k_s^{-1}$ responsible for PBH production, one at large scales $\sim k_l^{-1}$ at which the PBH clustering is sourced, and the almost scale-invariant contribution responsible for CMB anisotropies

$$\mathcal{P}_{\mathrm{g}}(k) = k_s A_s \delta_{\mathrm{D}}(k - k_s) + k_l A_l \delta_{\mathrm{D}}(k - k_l) + \mathcal{P}_{\mathrm{CMB}}(k). \tag{4.84}$$

In the following we consider a monochromatic (Dirac delta) shape for the curvature perturbation power spectrum on both small (large)-scales, with corresponding amplitudes $A_s(A_l)$. In this case, the PBH density power spectrum on large-scales $\sim k_l^{-1}$ reads [240]

$$\mathcal{P}_{\delta_{\mathrm{PBH}}}(k) \simeq 4\nu^4 f_{\mathrm{NL}}^2 A_l k_l \delta(k - k_l), \tag{4.85}$$

in terms of the bias factor $\nu = \delta_c/\sigma$, given that PBHs are born from maxima of the underlying radiation energy density perturbation, and $\delta_c \simeq 0.59$ is the threshold for PBH formation [211, 222]. The corresponding variance of the density field is given by

$$\sigma^2 = \frac{16}{81} \int_0^\infty \mathrm{d}\ln k\, T^2(k, r_m) W^2(k, r_m)(k r_m)^4 \mathcal{P}_{\mathrm{g}}(k), \tag{4.86}$$

where $W(k, r_m)$ denotes the Fourier transform of the real space top-hat window function, $T(k r_m)$ the radiation transfer function and $r_m = 2.74/k_s$ the relevant scale for PBH formation [211]. The corresponding initial PBH correlation function is [213]

$$\xi_{\mathrm{PBH}}(x) = \int_0^\infty \frac{\mathrm{d}k}{k} \mathcal{P}_{\delta_{\mathrm{PBH}}}(k)\, j_0(kx) \simeq 4\nu^4 f_{\mathrm{NL}}^2 A_l j_0(k_l x), \tag{4.87}$$

as a function of the zeroth spherical Bessel function $j_0$. The correlation function is therefore spatially flat on a scale $\sim k_l^{-1}$. This result is valid for $\xi_{\mathrm{PBH}} \lesssim 1$, which will be consistent with the results found in the following.

Given that for relative separations $x$ larger than the horizon scale no physical process may have an effect, the PBH correlation function does not change when $k_l^{-1}$ is outside the horizon. Upon the corresponding horizon re-entry, the PBH density contrast is essentially frozen until matter-radiation equality, and then grows linearly according to the relation [14, 403]

$$\xi_{\text{PBH}}(x, z) \simeq \left(1 + \frac{3}{2} f_{\text{PBH}} \frac{1 + z_{\text{eq}}}{1 + z}\right)^2 \xi_{\text{PBH}}(x). \tag{4.88}$$

Given that PBH binaries form before the matter-radiation equality, the corresponding correlation function does not change significantly between the PBH and binary formation epochs.

Even though the PBH spatial correlation function does not change much during this epoch, the corresponding radiation correlation function, whose peaks may give rise to PBHs, grows as $(1 + z)^{-4}$ until the scale $k_l^{-1}$ enters the horizon, and afterwards remains almost constant in time. This growth would be responsible for the injection of a large amount of energy in the thermal bath due to the dissipation of acoustic waves through Silk damping as they re-enter the horizon and start oscillating, giving rise to the known CMB $\mu$-distortions [414, 415]. This phenomenon occurs on length scales in the range $(10^{-4} - 2 \cdot 10^{-2})$ Mpc (which are not accessible from CMB anisotropies observations or from $y$-distortions, that involve larger comoving scales).

There is therefore a strong overlap with the initial comoving distances relevant for the formation of PBH binaries with masses around $\sim 30 \, M_\odot$, $(4 \cdot 10^{-5} - 10^{-3})$ Mpc [83], given that only PBHs separated by distances smaller than $\sim 10^{-3}$ Mpc can form a binary, while a minimum separation $\sim 4 \cdot 10^{-5}$ Mpc is needed to have merger time scales compatible with the reach of present GW experiments. One can therefore conclude that exists the possibility of enhancing the PBH initial clustering through primordial non-Gaussianity (and consequently the PBH merger rate), which could be tested by present and future CMB $\mu$-measurements.

We can now go into the details by computing the amount of energy injection into the CMB as [416, 417]

$$\mu = 1.4 \int_{z_1}^{z_2} \mathrm{d}z \frac{\mathrm{d}Q/\mathrm{d}z}{\overline{\rho}_r} e^{-(z/z_{\text{DC}})^{5/2}}, \tag{4.89}$$

in terms of the mean radiation energy density $\overline{\rho}_r$ and the redshift for double Compton scattering $z_{\text{DC}} \simeq 2.6 \cdot 10^6$. The energy release per unit redshift is given by

$$\frac{\mathrm{d}Q/\mathrm{d}z}{\overline{\rho}_r} = -\int \frac{\mathrm{d}k}{k} \mathcal{P}_r(k, z) \frac{\mathrm{d}\Delta_Q^2}{\mathrm{d}z}, \tag{4.90}$$

with

$$\Delta_Q^2(k) = \frac{9c_s^2}{2} e^{-2k^2/k_{\text{D}}^2}, \tag{4.91}$$

as a function of the sound speed $c_s$, the diffusion scale

$$k_{\text{D}} = A_{\text{D}}^{-1/2}(1 + z)^{3/2}, \quad A_{\text{D}} \simeq 6 \cdot 10^{10} \, \text{Mpc}^2, \tag{4.92}$$

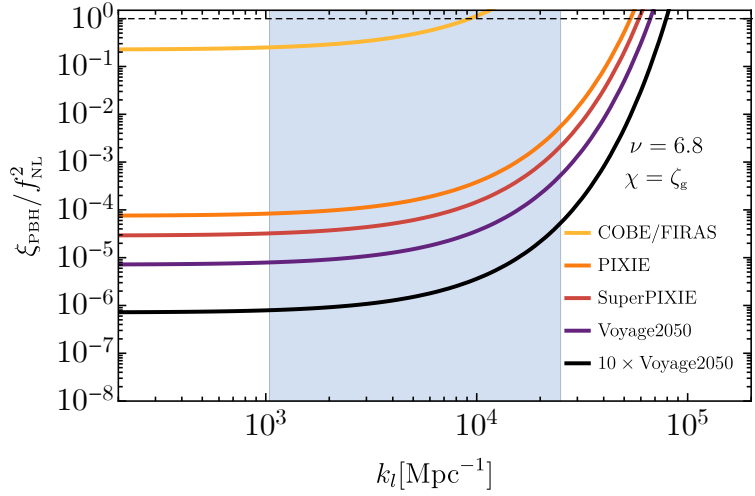

**Figure 4.13**: Forecasted limits on the PBH correlation function from CMB $\mu$-distortions. The blue band indicates the range of scales relevant for the formation of PBH binaries. For a representative monochromatic mass function with central scale $M_{\rm PBH} = 30 M_\odot$, we have fixed the value of $\nu \simeq 6.8$ to have a PBH abundance of $f_{\rm PBH} = 10^{-3}$, with small-scales momentum $k_s \simeq 2.4 \cdot 10^5 {\rm Mpc}^{-1}$ and amplitude $A_s \simeq 0.0063$. Figure taken from Ref. [8].

and the radiation power spectrum

$$\mathcal{P}_r(k, z) \simeq \left(\frac{4}{9}\right)^2 \left(\frac{k}{aH}\right)^4 T^2(k, a) \mathcal{P}_\zeta(k). \tag{4.93}$$

We stress that, on the relevant length scales, the dominant contribution from the curvature perturbation power spectrum comes from the peaked piece proportional to the large-scale amplitude $A_l$ in Eq. (4.84).

The redshift interval is determined by the double Compton scattering decoupling $z_1 = 2 \cdot 10^6$, given that at earlier times the universe content is modelled as a photon-baryon fluid in thermal equilibrium, thanks to elastic and double Compton scatterings, and the thermalization decoupling by Compton scattering $z_2 = 5 \cdot 10^4$, after which double Compton scattering is no longer efficient whereas single Compton scattering still provides equilibrium. The $\mu$-distortion is finally found to be

$$\mu \simeq \frac{16}{81} A_l \mathcal{I}(k_l) \simeq \frac{4}{81} \frac{\xi_{\rm PBH}}{\nu^4 f_{\rm NL}^2} \mathcal{I}(k_l), \tag{4.94}$$

where

$$\mathcal{I}(k_l) = \frac{189}{5} A_{\rm D} k_l^2 c_s^2 \int_{z_2}^{z_1} \frac{{\rm d}z}{(1+z)^4} e^{-(z/z_{\rm DC})^{5/2}} e^{-2k_l^2/k_{\rm D}^2}, \tag{4.95}$$

where the PBH correlation function has been assumed constant for $x \lesssim k_l^{-1}$.

In Fig. 4.13 we show the forecasted limits on the PBH correlation function, at the range of scales relevant for PBH binary assembling, coming from CMB $\mu$-distortions. We consider the constraints from the COBE/FIRAS ($\mu < 9 \cdot 10^{-5}$) [146] and the forecasted ones from PIXIE ($\mu < 3 \cdot 10^{-8}$) [418], SuperPIXIE ($\mu < 7 \cdot 10^{-9}$) [419], Voyage2050 ($\mu < 1.9 \cdot 10^{-9}$) and $10 \times$ Voyage2050 ($\mu < 1.9 \cdot 10^{-10}$) [420].

From Eq. (4.94) one can see that the $\mu$-distortion depends on the amplitude $A_l$ at large-scales and that only large value of $f_{NL}$ could give a PBH correlation $\xi_{PBH} \gtrsim 1$. This implies that if, for example, PIXIE does not find any CMB $\mu$-distortion, corresponding to $\xi_{PBH}/f_{NL}^2 \lesssim 10^{-2}$, then to have a significant amount of clustering at formation $\xi_{PBH} \gtrsim 1$ would require $|f_{NL}| \gtrsim 10$. Currently, COBE/FIRAS limits $A_l \lesssim 10^{-4}$, corresponding to a necessary value of $|f_{NL}| \gtrsim 1$.

As long as $\xi_{PBH} \lesssim 1$, the PBH abundance is not modified by the presence of non-Gaussianities, given that the short-scale variance is significantly shifted only for $f_{NL} \gtrsim A_l^{-1/2}$. This justifies the use of the Gaussian approximation to compute the PBH abundance (we have fixed the parameter $\nu \sim 6.8$ to have $f_{PBH} = 10^{-3}$). Since the PBH abundance $f_{PBH}$ is exponentially sensitive to $\nu$ as $f_{PBH} \sim \exp(-\nu^2/2)$, a small change in $\nu$ would largely modify it, without however affecting our results. Let us also stress that the unavoidable source of non-Gaussianity due to the non-linear relation between the density contrast and the curvature perturbation would change the short-scale amplitude $A_s$ of a factor of order unity in order to keep the same PBH abundance [21, 219], without affecting our results.

There is however another important effect one has to consider. The introduction of a local-type non-Gaussianity results into a coupling between small and large scales, which in turn introduces an isocurvature dark matter anisotropy from the PBHs in the CMB anisotropies, which is severely constrained by Planck data. This implies that one cannot consider arbitrarily large values of $|f_{NL}|$ to have a strong PBH clustering. In particular, for the current lower bound $|f_{NL}| \gtrsim 1$ from COBE/FIRAS, the isocurvature bound imposes $f_{PBH} \lesssim 5 \cdot 10^{-4}$ [20, 240, 241], implying that PBHs cannot comprise a large fraction of the dark matter in the universe. On the other hand, for large PBH abundances $f_{PBH} = 1$, the isocurvature bound imposes $|f_{NL}| \lesssim 4 \cdot 10^{-4}$.

Our results therefore indicate that future experiments looking for CMB $\mu$-distortions would be able to constrain the hypothesis of initial PBH clustering induced by local non-Gaussianity, with consequences in the interpretation of the merger events currently detected at GW experiments.

## 4.3 On the PBH merger rate

In the most standard scenario, PBHs are generated from the gravitational collapse of sizable curvature fluctuations in the very early universe. At formation time, they are dynamically coupled to cosmic expansion, with negligible peculiar velocities. However, due to the presence of large

Poisson fluctuations at small scales, they may decouple and become gravitationally bound to each other before the matter-radiation equality. Such a two-body system will experience head-on collision unless the gravitational field of surrounding PBHs and other matter inhomogeneities prevent it. As a result, a population of PBH binaries is formed, which will later evolve due to GWs emission and possibly merge. On the other hand, PBH binaries may also form in the late universe by gravitational capture, which will rapidly merge and generate GWs.

We dedicate this section to the discussion of PBHs assembling in binary systems in the early universe, followed by their evolution during the cosmic history including the effect of baryonic mass accretion, and finally compute the PBH merger rate.

## 4.3.1 Assemble in binaries

There are two main mechanisms for the formation of PBH binaries. The first takes place in the early universe before the matter-radiation equality, while the other takes place in the late time universe in present-day halos. In the next subsections we are going to provide details on these two binary formation scenarios following the results of Ref. [18]. For simplicity, we will assume a monochromatic PBH mass function.

**Early Universe**

A couple of neighboring PBHs with masses $M_{\rm PBH}$, separated by a physical distance $x(z)$ at redshift $z$, can decouple from the Hubble flow if their gravitational interaction is strong enough

$$M_{\rm PBH} x^{-3}(z) > \rho(z) \tag{4.96}$$

in terms of the background cosmic energy density $\rho$. By rescaling the quantities in terms of those at matter-radiation equality $z_{\rm eq}$, one finds that the decoupling occurs at redshift $z_{\rm dec}$ if

$$\frac{1 + z_{\rm dec}}{1 + z_{\rm eq}} = f_{\rm PBH} \left( \frac{\bar{x}}{x} \right)^3 - 1 > 0, \tag{4.97}$$

in terms of the physical mean separation $\bar{x}$ between PBHs

$$\bar{x}(z_e) = \left( \frac{M_{\rm PBH}}{\rho_{\rm PBH}(z_{\rm eq})} \right)^{1/3} = \frac{1}{(1 + z_{\rm eq}) f_{\rm PBH}^{1/3}} \left( \frac{8 \pi G}{3 H_0^2} \frac{M_{\rm PBH}}{\Omega_{\rm DM}} \right)^{1/3}. \tag{4.98}$$

One can immediately appreciate how the characteristic decoupling redshift is of the order $z_{\rm dec} > 10^4$ for the relevant range of PBH masses and abundance considered [177].

The motion of the infalling pair components may be affected by the presence of surrounding DM overdensities and especially by close PBHs, which can exert tidal forces and give angular momentum to the pair, forming therefore a binary system. The corresponding binary semi-major axis $a$ and eccentricity $e$ are given by

$$a = \frac{\alpha}{f_{\text{PBH}}} \frac{x^4}{\bar{x}^3}, \qquad e = \sqrt{1 - \beta^2 \left(\frac{x}{y}\right)^6}, \tag{4.99}$$

in terms of the physical distance $y$ to the third PBH at the matter-radiation equivalence $z_{\text{eq}}$. We will assume the parameter values $\alpha = \beta = 1$ [97, 421] (see Ref. [422] for more precise estimates of their value). The geometrical condition $x < y < \bar{x}$ translates into an upper bound on the eccentricity as

$$e_{\text{max}} = \sqrt{1 - f_{\text{PBH}}^{3/2} \left(\frac{a}{\bar{x}}\right)^{3/2}}. \tag{4.100}$$

Assuming a uniform probability distribution for the physical distances $x$ and $y$ in three dimensional space, one can deduce the corresponding probability function for the binary parameters as

$$\mathrm{d}P = \frac{3}{4} f_{\text{PBH}}^{3/2} \bar{x}^{-3/2} a^{1/2} e (1 - e^2)^{-3/2} \mathrm{d}a \mathrm{d}e. \tag{4.101}$$

As one can easily notice, the probability distribution peaks at large values of the eccentricity, meaning that PBHs are expected to form eccentric binaries in the early universe.

**Late Universe**

The formation of PBH binaries can also occur in the late universe, in particular in present-day halos [95, 423], and relies on the GW capture mechanism. If a PBH with velocity $v$ passes close to another PBH, the energy loss due to the sudden GW emission can make the former object to loose its kinetic energy and become bound to the latter. The amount of energy loss during the encounter can be estimated as

$$\Delta E = \frac{85\pi \sqrt{GM_{\text{PBH}}} G^3 M_{\text{PBH}}^4}{12 r_{\text{p}}^{7/2}}, \tag{4.102}$$

in terms of the periastron $r_{\text{p}}$. Adopting the Newtonian approximation, the impact parameter of the collision is given by

$$b(r_{\text{p}}) = \sqrt{r_{\text{p}}^2 + 2GM r_{\text{p}}/v^2}, \tag{4.103}$$

such that the corresponding cross section for binary formation reads

$$\sigma_{\text{bin}} \simeq \left(\frac{85\pi}{3}\right)^{2/7} \frac{\pi \left(2GM_{\text{PBH}}\right)^2}{v^{18/7}}. \tag{4.104}$$

As one can see, smaller relative velocities will induce larger cross section and therefore enhance the production of PBH binaries.

## 4.3.2 Evolution of the binary system

After their formation in the early universe, binaries will evolve in time due to the energy loss through gravitational radiation, possibly merging. One has however to keep into account the role of accretion during the binary evolution, as discussed in the previous sections of this chapter. In the following we report the results obtained in Ref. [16], studying first the evolution under GW emission, and then investigating the role of accretion.

**GW-dominated evolution**

In the absence of GW radiation-reaction, the eccentricity $e$ and semi-major axis $a$ of the binary system are constants of motion and can be expressed in terms of the angular momentum $L$ and energy $E$ as

$$e^2 = 1 + \frac{2EL^2}{M_{\rm tot}^2\mu^3}, \qquad a = \frac{GM_{\rm tot}\mu}{2|E|}, \tag{4.105}$$

where $M_{\rm tot} = M_1 + M_2$ is the total mass of the binary and $\mu = M_1M_2/(M_1 + M_2)^2$ is its reduced mass.

To the leading order in the weak-field/slow-motion approximation, one can adopt the quadrupole formula to compute the energy and angular momentum losses through the emission of GWs as [424, 425]

$$\begin{aligned}
\frac{{\rm d}E}{{\rm d}t} &= -\frac{32}{5}\frac{\mu^2 M_{\rm tot}^3}{a^5}\frac{1}{(1-e^2)^{7/2}}\left(1 + \frac{73}{24}e^2 + \frac{37}{96}e^4\right), \\
\frac{{\rm d}L}{{\rm d}t} &= -\frac{32}{5}\frac{\mu^2 M_{\rm tot}^{5/2}}{a^{7/2}}\frac{1}{(1-e^2)^2}\left(1 + \frac{7}{8}e^2\right),
\end{aligned} \tag{4.106}$$

which can be recast in terms of the binary parameters $e$ and $a$ to get

$$\begin{aligned}
\frac{{\rm d}a}{{\rm d}t} &= -\frac{64}{5}\frac{\mu M_{\rm tot}^2}{a^3}\frac{1}{(1-e^2)^{7/2}}\left(1 + \frac{73}{24}e^2 + \frac{37}{96}e^4\right), \\
\frac{{\rm d}e}{{\rm d}t} &= -\frac{304}{15}\frac{\mu M_{\rm tot}^2}{a^4}\frac{e}{(1-e^2)^{5/2}}\left(1 + \frac{121}{304}e^2\right).
\end{aligned} \tag{4.107}$$

This system of equations can be solved to estimate the merging time $t_c$, defined to be the time at which $a(t_c) = 0$. For an initial orbit with $e(t_{\rm i}) = e_{\rm i}$ and $a(t_{\rm i}) = a_{\rm i}$, one finds

$$t_c(a_{\rm i}, e_{\rm i}) = t_c(a_{\rm i})\frac{48}{19}\frac{1}{g^4(e_{\rm i})}\int_0^{e_{\rm i}}{\rm d}e\,\frac{g^4(e)(1-e^2)^{5/2}}{e(1 + 121e^2/304)}, \tag{4.108}$$

in terms of the function

$$g(e) = \frac{e^{12/19}}{1-e^2}\left(1 + \frac{121}{304}e^2\right)^{870/2299}. \tag{4.109}$$

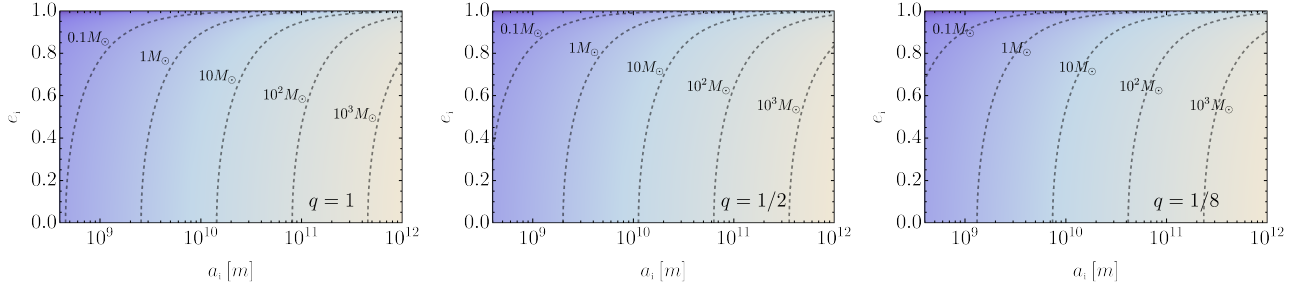

**Figure 4.14**: Contour lines for the binary parameters $e_i$ and $a_i$ to have a coalescence time scale equal to the age of the universe, $t_c(a_i, e_i) = t_0$, for several values of the mass ratio $q$ and primary mass $M_1$. Figure taken from Ref. [16].

For a circular orbit $e_i = 0$, the coalescence time scale reduces to

$$t_c(a_i, e_i = 0) \equiv t_c(a_i) = \frac{5}{256} \frac{a_i^4}{M_{tot}^2 \mu}, \qquad (4.110)$$

while in the limit of highly eccentric binaries, $e_i \to 1$, one finds that

$$t_c(a_i, e_i \to 1) \simeq t_c(a_i) \frac{768}{429} (1 - e_i^2)^{7/2}. \qquad (4.111)$$

One can immediately notice that the initial value of the semi-major axis $a_i$ would diverge as $e_i$ tends towards unity, since the coalescence time scale shrinks rapidly in that limit.

In Fig. 4.14 we show the parameter space for the binary components requiring a coalescence time scale equal to the age of the universe, $t_0 = 13.7 \, \text{Gyr}$.

**Accretion-dominated evolution**

If the binary components experience a phase of accretion during their evolution, the orbital parameters will further change in addition to GW radiation-reaction [426–428]. Given the characteristic time scales involved in the evolution of PBH binaries, one can study the accretion-driven phase and the GW-driven phase separately. Indeed, since at redshifts smaller than the cut-off redshift, $z < z_{cut\text{-}off}$, accretion is drastically suppressed due to the several feedbacks effects discussed in the previous section, the evolution of the binary system from $z_{cut\text{-}off}$ to the redshift of detection is dictated by GW emission only. In the following we will consider detection redshifts close to $z \approx 0$, which is the proper range for GW experiments like LVKC.

To have an idea of the time scales involved in the process, one can have a look at some explicit examples. From Eq. (4.108), a merger event occurring today corresponds to a binary with orbital separation $a = \mathcal{O}(10^{11} \, \text{m})$ at $z = 10$. In this case the variation time scale of the binary semi-major axis due to GW emission reads

$$T_{\mathrm{GW}} \sim \frac{a}{\dot{a}}\Big|_{\mathrm{GW}} = 4 \times 10^{17} \left(\frac{a}{1.4 \times 10^{11}\mathrm{m}}\right)^4 \left(\frac{M}{30M_\odot}\right)^{-3} \left(\frac{1-e^2}{0.1}\right)^{7/2} \mathrm{s}, \qquad (4.112)$$

where we have chosen the value $a = 1.4 \times 10^{11}$m to have a merger event in a time equal to the age of the universe, for a binary with components of equal mass $M = 30M_\odot$ and eccentricity $e = 0.95$. One can therefore realise that at redshifts $z > z_{\mathrm{cut\text{-}off}}$, due to the dependence $T_{\mathrm{GW}} \propto a^4$, the typical accretion time scale (4.14) (given by the Salpeter time) is much smaller than the one governing GW radiation-reaction. This implies that one can parametrise the inspiral as being driven by accretion when $z > z_{\mathrm{cut\text{-}off}}$ and by GW radiation-reaction when $z < z_{\mathrm{cut\text{-}off}}$. If the detected events correspond to redshifts larger than $z > z_{\mathrm{cut\text{-}off}}$, which could be relevant for third-generation ground-based detectors like Einstein Telescope [429] and for the space mission LISA [203], a different analysis would be needed.

For the relevant range of parameters, the mass accretion time scale in Eq. (4.14) is always much longer than the orbital time scale

$$T_{\mathrm{orbital}} \sim \left(\frac{M_{\mathrm{tot}}}{a^3}\right)^{-1/2} \sim 8 \times 10^5 \left(\frac{M}{30M_\odot}\right)^{-1/2} \left(\frac{a}{1.4 \cdot 10^{11}\,\mathrm{m}}\right)^{3/2} \mathrm{s}, \qquad (4.113)$$

which implies that mass accretion can be treated as an adiabatic process, in which the adiabatic invariants of the elliptical motion are kept constant. For a Keplerian two-body problem, the action variables $I_k = \int p\,dk/(2\pi)$, where $k = r, \phi$ are the polar coordinates, are adiabatic invariants and are given by [430]

$$I_\phi = \frac{1}{2\pi}\int_0^{2\pi} p_\phi \mathrm{d}\phi = L_z, \qquad (4.114)$$

$$I_r = \frac{1}{2\pi}\int_{r_{\min}}^{r_{\max}} p_r \mathrm{d}r = -L_z + \sqrt{M_{\mathrm{tot}}\mu^2 a}, \qquad (4.115)$$

in terms of the binary energy and angular momentum

$$E = -\frac{\mu M_{\mathrm{tot}}}{2a}, \qquad L_z = \sqrt{\frac{M_{\mathrm{tot}}^2\mu^3(e^2-1)}{2E}} = \sqrt{1-e^2}\sqrt{M_{\mathrm{tot}}\mu^2 a}. \qquad (4.116)$$

Their invariance implies that

$$\frac{\mathrm{d}I_\phi}{\mathrm{d}t} = \frac{\partial L_z}{\partial e}\frac{\partial e}{\partial t} + \frac{\partial L_z}{\partial a}\frac{\partial a}{\partial t} + \frac{\partial L_z}{\partial \mu}\frac{\partial \mu}{\partial t} + \frac{\partial L_z}{\partial M_{\mathrm{tot}}}\frac{\partial M_{\mathrm{tot}}}{\partial t} = 0,$$

$$\frac{\mathrm{d}I_r}{\mathrm{d}t} = \frac{\partial I_r}{\partial e}\frac{\partial e}{\partial t} + \frac{\partial I_r}{\partial a}\frac{\partial a}{\partial t} + \frac{\partial I_r}{\partial \mu}\frac{\partial \mu}{\partial t} + \frac{\partial I_r}{\partial M_{\mathrm{tot}}}\frac{\partial M_{\mathrm{tot}}}{\partial t} = 0, \qquad (4.117)$$

which can be recast as

$$\frac{\partial a}{\partial t} = -\frac{2a}{\mu}\frac{\partial \mu}{\partial t} - \frac{a}{M_{\mathrm{tot}}}\frac{\partial M_{\mathrm{tot}}}{\partial t},$$

$$\frac{\partial e}{\partial t} = 0 \,. \tag{4.118}$$

This implies that the eccentricity is a constant of motion for an accretion-driven inspiral, as expected given that it can be expressed in terms of the adiabatic invariants as [430]

$$e = \sqrt{1 - \left(\frac{I_\phi}{I_\phi + I_r}\right)^2} \,. \tag{4.119}$$

The other dynamical equation gives

$$\frac{\dot{a}}{a} + 2\frac{\dot{\mu}}{\mu} + \frac{\dot{M}_{\text{tot}}}{M_{\text{tot}}} = 0, \tag{4.120}$$

which recovers the result of Ref. [428] obtained in the absence of GW emission and for circular binaries. This equation can be expressed in terms of the binary component masses as

$$\frac{\dot{a}}{a} + \frac{M_2(M_1 + 2M_2)\dot{M}_1 + M_1(2M_1 + M_2)\dot{M}_2}{M_1 M_2(M_1 + M_2)} = 0 \,, \tag{4.121}$$

and it is coupled with the masses evolution equations in Eqs. (4.27). In the equal mass limit $q \to 1$, this equation finally becomes

$$\frac{\dot{a}}{a} + 3\frac{\dot{M}_1}{M_1} = 0 \,. \tag{4.122}$$

It implies that, as the binary components accrete baryons, the semi-major axis of the binary shrinks, facilitating the inspiral and enhancing the merger rate, as we will see in the following paragraph.

### 4.3.3 Computation of the merger rate

After the discussion on the assemble of PBHs in binary systems, both at early and late times, and their evolution due to the emission of GWs and the impact of baryonic mass accretion, we can move to the computation of the PBH merger rate. We first analyse the merger rate of early binaries, with further investigation of the impact of accretion and clustering. Then we study the merger rate of late binaries, and see how clustering may enhance it. Finally, we move to a different scenario and compute the merger rate assuming that PBHs are initially spatially clustered. In the following, we report the results of Refs. [14, 16].

**Merger rate of early-binaries**

Let us focus on the computation of the merger rate for binaries formed in the very early universe, before the matter-radiation equality epoch. We first focus on the standard scenario, and then investigate the impact of accretion and evolution of PBH clustering.

*Standard scenario:* Once the probability distribution of the orbital parameters is known, as shown in Eq. (4.101), and keeping into account the orbital evolution of the binary eccentricity and semi-major axis, the merger rate can be computed by performing the integral [178]

$$dR(t) = \int da\, de \frac{d^2 P}{da\, de} \delta\left(t - t_c(\mu, M_{tot}, a, e)\right).\qquad(4.123)$$

Following the results of Ref. [178], one can write the differential merger rate of PBH binaries at the time of coalescence $t$ in the form

$$dR = \frac{1.6 \times 10^6}{\mathrm{Gpc}^3\,\mathrm{yr}} f_{\mathrm{PBH}}^{\frac{53}{37}}(z_i)\left(\frac{t}{t_0}\right)^{-\frac{34}{37}} \eta_i^{-\frac{34}{37}}\left(\frac{M_{tot}^i}{M_\odot}\right)^{-\frac{32}{37}} S\left(M_{tot}^i, f_{\mathrm{PBH}}(z_i)\right)\psi(M_1^i, z_i)\psi(M_2^i, z_i)dM_1^i dM_2^i,$$
$$(4.124)$$

where $f_{\mathrm{PBH}}(z_i)$ denotes the PBH abundance at the formation time $z_i$, $\eta_i = \mu^i/M_{tot}^i$ is the symmetric mass ratio, $\mu^i = M_1^i M_2^i/M_{tot}^i$ the reduced mass and $M_{tot}^i = M_1^i + M_2^i$ the total mass of the binary components at the formation time. The merger rate depends on the initial PBH mass function $\psi(M^i, z_i)$, normalised to unity. We have made explicit use of the formation redshift $z_i$ to highlight the difference when accretion will be included.

The suppression factor $S$ is introduced to keep into account the effect of the matter density perturbations, the presence of additional surrounding PBHs and possible modifications due to the size of the empty region around the binary. Its expression is given by [178]

$$S\left(M_{tot}^i, f_{\mathrm{PBH}}(z_i)\right) = \frac{e^{-\bar{N}(y)}}{\Gamma(21/37)}\int dv\, v^{-\frac{16}{37}}\exp\left[-\bar{N}(y)\langle m\rangle\int\frac{dm}{m}\psi(m, z_i)F\left(\frac{m}{\langle m\rangle}\frac{v}{\bar{N}(y)}\right) - \frac{3\sigma_M^2 v^2}{10 f_{\mathrm{PBH}}^2(z_i)}\right],$$
$$(4.125)$$

as a function of the generalised hypergeometric function

$$F(z) = {}_1F_2\left(-\frac{1}{2}; \frac{3}{4}, \frac{5}{4}; -\frac{9z^2}{16}\right) - 1,\qquad(4.126)$$

the matter density perturbations rescaled variance $\sigma_M^2 = (\Omega_M/\Omega_{\mathrm{DM}})^2\langle\delta_M^2\rangle \simeq 3.6\cdot 10^{-5}$ at the binary formation time, and the mean mass

$$\langle m\rangle = \left(\int\frac{1}{m}\psi(m, z_i)dm\right)^{-1}.\qquad(4.127)$$

The characteristic number $\bar{N}(y)$ of PBHs in a spherical volume of radius $y$, corresponding to the distance to a third PBH, is determined in order for the binaries to not get destroyed by other PBHs, i.e.

$$\bar{N}(y) = \frac{M_{tot}^i}{\langle m\rangle}\frac{f_{\mathrm{PBH}}(z_i)}{f_{\mathrm{PBH}}(z_i) + \sigma_M}.\qquad(4.128)$$

The mass integral in the expression of the suppression factor gives an estimate of the typical mass of the perturber PBHs which prevent the two PBHs to directly collide, exerting a torque on the system, and is responsible for the formation of the binary.

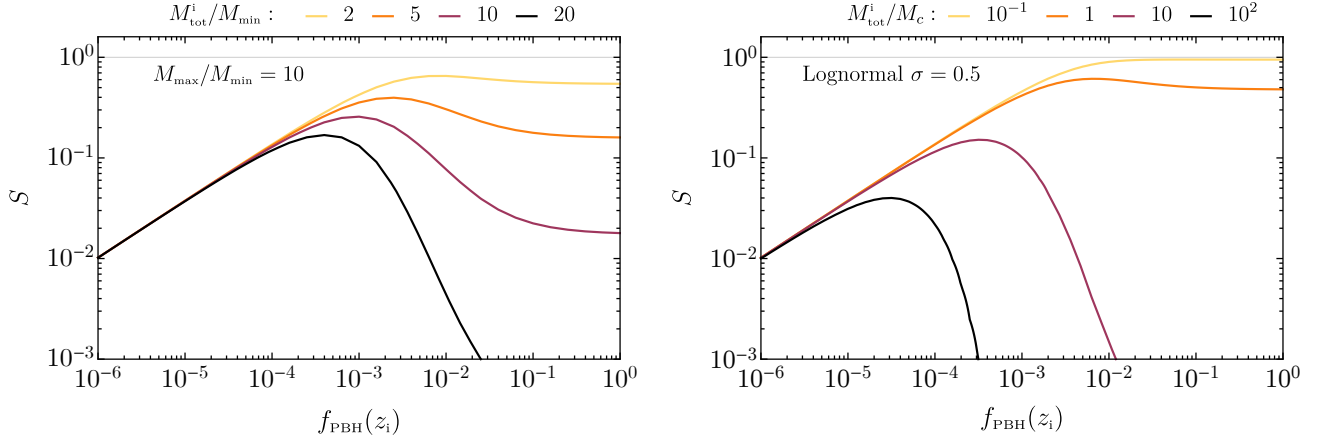

**Figure 4.15**: Suppression factor for a power-law mass distribution with $M_{max} = 10M_{min}$ (left panel) and a lognormal mass distribution with $\sigma = 0.5$ (right panel) for various values of $M_{tot}^i$. Figure taken from Ref. [16].

In Fig. 4.15 we plot the behaviour of the suppression factor assuming either a power-law or a lognormal mass function.

*Impact of accretion:* We can now describe the main impact of accretion on the PBH merger rate. Following the discussion of the previous section, one can study the impact of accretion relying on the different time scales involved in the problem. Indeed, one finds that accretion is the main driver of the binary evolution up to the cut-off redshift $z_{cut-off}$, while at smaller redshifts GW radiation-reaction is dominant.

Even though the ellipticity is an adiabatic invariance, the time evolution of the PBH masses induces a change in the binary semi-major axis $a$, as one can see from Eq. (4.120), such that its coalescence time is modified. Given that the accretion phase occurs earlier and independently from the emission of GWs, the merging time becomes

$$t_c^{acc} = \frac{3}{85} \frac{\mathcal{N}^4 a_i^4 (1 - e_i^2)^{7/2}}{\eta(z_{cut-off}) M_{tot}^3(z_{cut-off})} \equiv \frac{\mathcal{N}^4}{\mathcal{S}} t_c(M_j^i), \tag{4.129}$$

in terms of the factor

$$\mathcal{S} = \frac{\eta(z_{cut-off}) M_{tot}^3(z_{cut-off})}{\eta(z_i) M_{tot}^3(z_i)}, \tag{4.130}$$

which keeps into account the evolution of the masses from the initial redshift $z_i$ to the cut-off redshift $z_{cut-off}$, and the orbit shrinking factor

$$\mathcal{N} \equiv \frac{a(z_{cut-off})}{a_i} = \exp\left[-\int_{t_i}^{t_{cut-off}} \mathrm{d}t \left(\frac{\dot{M}_{tot}}{M_{tot}} + 2\frac{\dot{\mu}}{\mu}\right)\right], \tag{4.131}$$

which properly considers the evolution of the semi-major axis. After the cut-off redshift, the binary proceeds in the standard GW radiation-reaction scenario, but with different masses with respect to the case without accretion.

Given that the suppression factor does not depend on time, and using Eq. (4.34) for the evolution of the PBH mass function, one can compute the final differential merger rate including the effect of accretion as

$$
\begin{aligned}
\mathrm{d}R_{\mathrm{acc}}(t, M_j, f_{\mathrm{PBH}}(z_i)) &= \frac{\mathcal{S}}{\mathcal{N}^4} \mathrm{d}R\left(t\mathcal{S}/\mathcal{N}^4, M_j^{\mathrm{i}}, f_{\mathrm{PBH}}(z_i)\right) \\
&= \mathcal{N}^{-12/37} \mathcal{S}^{3/37} \mathrm{d}R\left(t, M_j^{\mathrm{i}}, f_{\mathrm{PBH}}(z_i)\right),
\end{aligned} \tag{4.132}
$$

where in the last line the merger rate is expressed in terms of the initial quantities. The set of masses $M_j^{\mathrm{i}}$ and $M_j$ identify the couple $(M_1, M_2)$ at the formation and final time, respectively. As one can appreciate from the explicit expressions for the factors $\mathcal{N}$ and $\mathcal{S}$, the merger rate is enhanced in the presence of accretion, as expected due to the growth of the PBH masses and the shrinking of the binary semi-major axis.

*Impact of clustering:* In the standard scenario PBH binaries form before the matter-radiation equality, when fluctuations in the PBH number counts are still Poissonian. Assuming Gaussian perturbations, PBHs are not clustered at their formation epoch [224, 400–402], which implies that the dynamics of binary assembling is not altered by clustering.

After the matter-radiation equality, however, for large PBH abundance, early-universe PBH binaries may end up in highly clustered PBH regions, inside which there is a high probability that they get perturbed by close encounters with surrounding PBHs. The effect of these three-body interactions is to modify the binary semi-major axis and eccentricity, moving the latter away from its initial high values [431]. Given that the merging time in Eq. (4.111) depends on large powers of $(e - 1)$, clustering significantly increases the coalescence time, thus affecting the PBH merger rate.

Even though one may expect that when clustering is strong, for $f_{\mathrm{PBH}} \simeq 1$, the merger rates are highly suppressed, one has to consider that there are other effects which may reduce the frequency of binary-PBH encounters [432] and change this prediction. Examples are the disruption of binaries residing in dark matter halos, which decreases as smaller halos merge into bigger ones [177] and their expansion due to the heating provided by binary-PBH collisions.

We can adopt a conservative approach and focus on the effects that may increase the disruption of PBH binaries and decrease the merger rate. The physical picture is the following: given that PBH binaries are heavier than single PBHs, they sink towards the center of the host structure (halo), while individual objects remain in its outer part. Then, PBH evaporation from the cluster triggers a gravothermal instability (due to the negative heat capacity of self-gravitating systems [433]), for which the halo core may collapse. During its contraction, the halo central density increases, leading to more frequent binary-PBH encounters, which may eventually halt the collapse. In this conservative approach we assume that all PBH binaries are perturbed on a timescale smaller than the age of the Universe at a given redshift, as done in Ref. [432].

To estimate the suppression to the merger rate, we compute the gravothermal instability timescale as [434]

$$t_{\mathrm{GI}} = \frac{v^3(r)}{G^2 M_{\mathrm{PBH}} \rho_{\mathrm{PBH}}(r) \log(M(<r)/M_{\mathrm{PBH}})}, \tag{4.133}$$

assuming a monochromatic population of PBHs with mass $M_{\mathrm{PBH}}$, and a density profile for the PBH halo of mass $M_h$ as

$$\rho_{\mathrm{PBH}}(r) \simeq \frac{3 M_h}{20\pi} R_{\mathrm{vir}}^{-3/5} r^{-12/5}, \tag{4.134}$$

where the normalisation constant has been deduced by assuming that the halo is composed only of PBHs (hereafter we assume $f_{\mathrm{PBH}} \simeq 1$), and the virial radius by imposing that $\rho_{\mathrm{PBH}}(< R_{\mathrm{vir}}) = 200 M_{\mathrm{PBH}} \bar{n}_{\mathrm{PBH}}$. The relative velocity at a radius $r$ is given by

$$v(r) = \sqrt{\frac{GM(<r)}{r}} \simeq \sqrt{GM_h R_{\mathrm{vir}}^{-3/5}} r^{-1/5}. \tag{4.135}$$

Assuming the characteristic value for the halo mass $M_h$ to be $M_*(z)$ in Eq. (4.70), one can compute the critical radius below which the gravothermal instability occurs in a timescale smaller than the Hubble time, $t_{\mathrm{GI}} \lesssim H^{-1}$. For example, at $z = 0$, the critical radius is computed to be $\sim 3 \cdot 10^{-3}$ kpc$/h$, corresponding to a critical number of PBHs within that distance of $N_{\mathrm{c}} \sim 4.6 \cdot 10^4$.

Once the critical number of PBHs beyond which the instability occurs is known, one can compute the fraction of initial PBH binaries contained in gravothermally unstable cores. Assuming an initial Poisson distribution, the probability of finding a PBH within a halo made up of $N$ PBHs at redshift $z$ is [435] (see also [436])

$$p_N(z) \propto N^{-1/2} e^{-N/N_*(z)}. \tag{4.136}$$

Given that the probability of finding a binary in a halo of $N$ PBHs is proportional to $p_N$, the probability of having a PBH in a subhalo of $N$ PBHs embedded in a parent halo of $N' > N$ PBHs is proportional to $p_N \cdot p_{N'}$ [432]. This implies that the present fraction of unperturbed binaries is larger than

$$P_{\mathrm{np}} \gtrsim 1 - \sum_{N=3}^{N_{\mathrm{c}}} \overline{p}_N(z_{\mathrm{form}}^{\mathrm{c}}) - \sum_{N' > N_{\mathrm{c}}} \left[ \sum_{N=3}^{N_{\mathrm{c}}} \widetilde{p}_N(z_{\mathrm{form}}^{\mathrm{c}}) \right] \overline{p}_{N'}(z_{\mathrm{form}}^{\mathrm{c}}) \simeq 10^{-2}, \tag{4.137}$$

in terms of the formation redshift of the halo with $N_{\mathrm{c}}$ PBHs, $z_{\mathrm{form}}^{\mathrm{c}}$, and

$$\sum_{N \geq 2} \overline{p}_N = 1 \ \text{ and } \ \sum_{N=2}^{N'} \widetilde{p}_N = 1. \tag{4.138}$$

This probability therefore provides the information of the impact of clustering on the early-time merger rate.

In Ref. [437], this procedure has been extended to different values of the PBH abundance and time, getting an effective suppression factor to be multiplied to Eq. (4.124) as

$$S_2(x) \approx \min\left[1, 9.6 \cdot 10^{-3} x^{-0.65} \exp\left(0.03 \ln^2 x\right)\right] \qquad \text{with} \qquad x \equiv (t(z)/t_0)^{0.44} f_{\text{PBH}}, \qquad (4.139)$$

which tends to unity for PBH abundances smaller than $f_{\text{PBH}} \lesssim 0.003$. This is compatible to the results obtained in the cosmological N-body simulation of Ref. [403], in which PBHs are essentially isolated for small enough abundances.

Finally, one should remember that there are additional contributions to the PBH merger rate. First of all, even though some PBH binaries are perturbed by binary interactions, it may happen that their orbital parameters still allow for a coalescence time comparable to the current age of the universe. Secondly, not all the binaries end up inside halos. In particular assuming $f_{\text{PBH}} \simeq 1$, one can estimate using the probability $p_N$ that about $\sim 10^{-3}$ PBHs are not in clusters (defining halos to have at least ten PBHs as in Ref. [436]). A deeper investigation of the early-time merger rate is therefore needed to account for these effects.

**Merger rate of late-binaries**

Let us now focus on the case of binaries formed in the late universe halos due to GW capture.

*Standard scenario:* Once formed, these binaries can merge in less than the age of the universe with a rate given by [95]

$$\mathcal{R}_h(M_h) = 2\pi \int_0^{R_{\text{vir}}} \mathrm{d}r \, r^2 \left(\frac{\rho_{\text{PBH}}(r)}{M_{\text{PBH}}}\right)^2 \langle \sigma_{\text{bin}} v \rangle, \qquad (4.140)$$

in terms of the halo mass $M_h$ and virial radius $R_{\text{vir}}$. The brackets stand for the thermally averaged cross-section, i.e. the mean of the combination $\sigma_{\text{bin}} v$ with velocities drawn from a Maxwell-Boltzmann distribution and the cross section given in Eq. (4.104).

The total merger rate is obtained by convolving the merger rate per halo $\mathcal{R}_h$ with the halo mass function $\mathrm{d}n/\mathrm{d}M_h$ derived in Eq. (4.69) using the Press-Schechter formalism [345]

$$\mathcal{V}_{\text{LU}} = \int_{M_{\text{min}}} \mathrm{d}M_h \frac{\mathrm{d}n}{\mathrm{d}M_h} \mathcal{R}_h(M_h), \qquad (4.141)$$

where $M_{\text{min}}$ denotes the minimum halo mass [83].

To give some values, let us assume that the PBH local density profile $\rho_{\text{PBH}}(r)$ follows a Navarro-Frenk-White behaviour, and consider halos with minimum mass and virial velocities of Ref. [95]. In this case one gets

$$\mathcal{V}_{\text{LU}} \simeq 2 \, f_{\text{PBH}}^{53/21} \, \text{Gpc}^{-3} \, \text{yr}^{-1}, \qquad (4.142)$$

which is much smaller than the one obtained from early-universe binaries.

*Impact of clustering:* Assuming $f_{\mathrm{PBH}} \simeq 1$, one can investigate the role of clustering on the late-time PBH binary merger rate. Assuming a constant PBH density profile within the core of size $r_s$ (to be determined below), which scales like $\rho_{\mathrm{PBH}}(r) \sim r^{-12/5}$ for $r > r_s$ as found in the previous section on the PBH clustering, the present-day halo merger rate reads [14]

$$\mathcal{R}_h(M_h) \simeq 22 \left( G M_h^{4/5} M_{\mathrm{PBH}}^{1/5} \bar{n}_{\mathrm{PBH}}^{1/5} \right)^{17/14} r_s^{-52/35},$$  (4.143)

such that the total one becomes

$$\mathcal{V}_{\mathrm{LU}} \simeq 1.5 \cdot 10^3 \, G^{17/14} (M_{\mathrm{PBH}} \bar{n}_{\mathrm{PBH}})^{73/42} M_*^{-11/21} \mathcal{R}_{\mathrm{cl}}^{52/35}.$$  (4.144)

In the previous expression we have introduced the dimensionless "cluster factor"

$$\mathcal{R}_{\mathrm{cl}} = \frac{R_*}{r_s},$$  (4.145)

in terms of the characteristic scale $R_* \simeq 9 \left( h M_{\mathrm{PBH}}/20 M_\odot \right)^{1/3}$ kpc$/h$, corresponding to the virial radius of an halo of mass $M_*$. At present time, the mean mass of halos is $M_* = 6.8 \cdot 10^6 M_{\mathrm{PBH}}$, such that one finally gets

$$\mathcal{V}_{\mathrm{LU}} \simeq 10^{-4} \left( \frac{M_{\mathrm{PBH}}}{20 M_\odot / h} \right)^{-11/21} \mathcal{R}_{\mathrm{cl}}^{52/35} \mathrm{Gpc}^{-3} \mathrm{yr}^{-1}.$$  (4.146)

In getting this result we have made the assumption that the power law behaviour of the PBH correlation function (and its corresponding profile) remain valid until late times and down to small scales.

To compute the cluster factor, one has to estimate first the value of $r_s$, which depends on the details of the dynamical processes responsible for core collapse. In our setting, the evaporation of fast moving PBHs would imply the contraction of the core to conserve energy, which, as the collapse proceeds, becomes hotter. When the core becomes small enough, binaries can form and harden, which may halt and possibly reverse the collapse and set a minimum radius $r_s$. By requiring that the gravothermal instability occurs within the Hubble time, one can estimate the minimum radius to be of order $r_s \simeq 3 \cdot 10^{-3}$kpc$/h$ for the present mean halo mass $M_*$. This gives a cluster factor of $\mathcal{R}_{\mathrm{cl}} \simeq 3 \cdot 10^3$.[3]

One conclude that, for large PBH abundances $f_{\mathrm{PBH}} \simeq 1$, clustering may help in increasing the late time merger rate, even though a deeper understanding of the minimum radius $r_s$ and dedicated N-body simulations down to low redshifts are needed before drawing any firm conclusion.

---

[3]A similar estimate would be obtained in the setup in which a halo core of radius $r_s$ is formed by the balance between the gravitational PBH interactions and the kinetic energies [438], and imposing the core relaxation time to be smaller than the age of the universe, one gets $\mathcal{R}_{\mathrm{cl}} \simeq 10^2$.

**Hierarchical PBH mergers**

In this section we are going to investigate the possibility that PBHs experience second-generation mergers, that are coalescences in which a PBH binary might merge giving rise to a new PBH, which later undergoes a further merger event with another PBH, forming a binary which is eventually detected at GW experiments. Following the approach adopted in Refs. [439, 440] and the results in Ref. [18], we show the main formalism to compute hierarchical merger rates. We stress that we do not consider the impact of suppression factors in the merger rate expressions, due to early- and late-time binary disruptions, which would make the conclusions even stronger.

To neglect the impact of linear density perturbations on the merger rate, we assume that the PBH abundance is larger than a critical value found to be

$$f_c = 1.63 \times 10^{-4} \left(\frac{M_c}{M_\odot}\right)^{\frac{5}{21}} \left(\frac{t}{t_0}\right)^{\frac{1}{7}}, \tag{4.147}$$

in terms of the corresponding reference mass $M_c$ associated to horizon re-entry. The fraction of the average PBH number density with mass $M$ with respect to the total average at present time is given by

$$F(M) = \frac{\psi(M)}{M} \left[\int \mathrm{dln} M' \, \psi(M')\right]^{-1}, \tag{4.148}$$

in terms of the present PBH mass distribution $\psi(M)$. This implies that the fraction of PBHs that have undergone a coalescence before the time $t$ is given by [439]

$$P_{\mathrm{PBH}}^{(1)}(t) = 1.34 \times 10^{-2} \left(\frac{M_c}{M_\odot}\right)^{\frac{5}{37}} \left(\frac{t}{t_0}\right)^{\frac{3}{37}} f_{\mathrm{PBH}}^{\frac{16}{37}} \Upsilon_1, \tag{4.149}$$

where the dependence on the shape of the mass function is captured in the dimensionless factor

$$\Upsilon_1 = \left(\int \mathrm{dln} x \, \tilde{\psi}(x)\right)^{\frac{16}{37}} \int \mathrm{d}x_i \mathrm{d}x_j \mathrm{d}x_l \tilde{F}(x_i) \tilde{F}(x_j) \tilde{F}(x_l)(x_i + x_j)^{\frac{36}{37}} x_i^{\frac{3}{37}} x_j^{\frac{3}{37}} x_l^{-\frac{21}{37}}, \tag{4.150}$$

in terms of $\tilde{F}(x = M/M_c) = M_c F(M, M_c)$ and $\tilde{\psi}(x = M/M_c) = M_c \psi(M, M_c)$.

The fraction of PBHs that have merged in a second-generation coalescence process at time $t$ is then given by [439]

$$P_{\mathrm{PBH}}^{(2)}(t) = 1.21 \times 10^{-4} \left(\frac{M_c}{M_\odot}\right)^{\frac{10}{37}} \left(\frac{t}{t_0}\right)^{\frac{6}{37}} f_{\mathrm{PBH}}^{\frac{32}{37}} \Upsilon_2, \tag{4.151}$$

in terms of the shape-dependent factor

$$\Upsilon_2 = \left(\int \mathrm{dln} x \, \tilde{\psi}(x)\right)^{\frac{32}{37}} \int \mathrm{d}x_i \mathrm{d}x_j \mathrm{d}x_k \mathrm{d}x_l \tilde{F}(x_i) \tilde{F}(x_j) \tilde{F}(x_k) \tilde{F}(x_l)(x_i + x_j)^{\frac{6}{37}} x_k^{\frac{6}{37}} x_l^{-\frac{42}{37}} (x_i + x_j + x_k)^{\frac{72}{37}}. \tag{4.152}$$

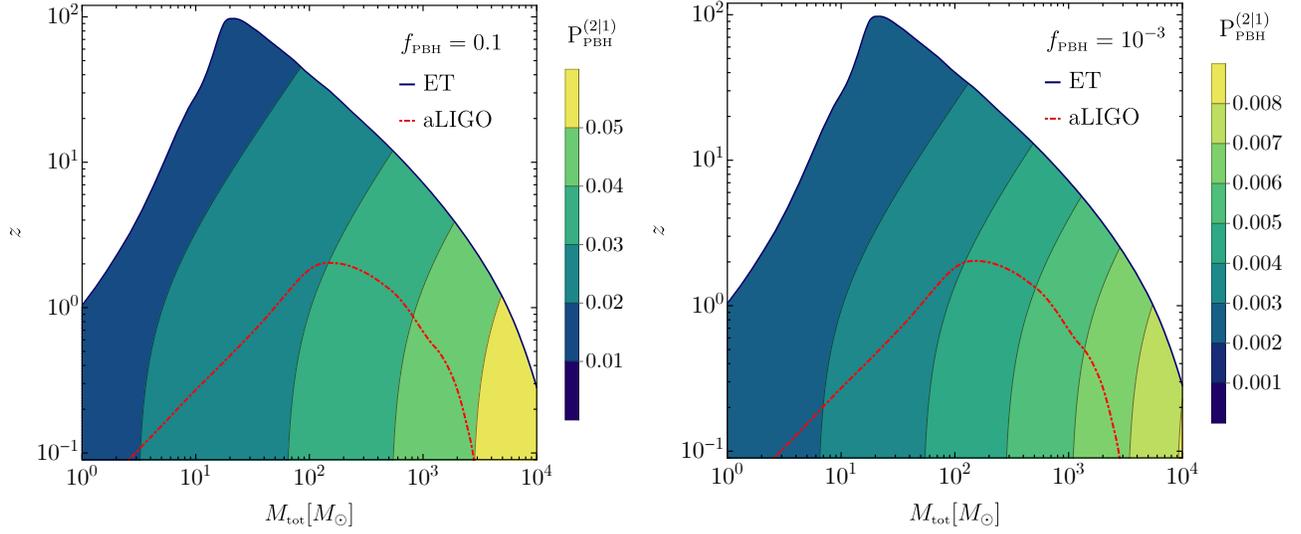

**Figure 4.16**: Total fraction of PBH binaries formed out of second-generation mergers in terms of the binary total mass $M_{tot}$ and redshift $z$, with fixed $f_{PBH}$, superimposed with the horizon redshifts of aLIGO (red line) and ET (blue line). Figure taken from Ref. [18].

This implies that the conditional probability that merging PBHs come from a second-merger process at time $t$ is given by

$$P_{PBH}^{(2|1)}(t) = 9 \times 10^{-3} \left(\frac{M_c}{M_\odot}\right)^{\frac{5}{37}} \left(\frac{t}{t_0}\right)^{\frac{3}{37}} f_{PBH}^{\frac{16}{37}} \Upsilon \qquad \text{with} \qquad \Upsilon = \frac{\Upsilon_2}{\Upsilon_1}. \tag{4.153}$$

The value of the shape parameter $\Upsilon$ will strongly depend on the shape of the PBH mass function. For example, for a monochromatic mass function its value is given by $\Upsilon = 4.8$, while for a lognormal one [374] its value lies in the range $\Upsilon \sim (5 - 15)$ depending on the value of the width $\sigma$ of the mass distribution. For a power-law mass function [19], obtained in models with a broad curvature perturbation power spectrum, its value is $\Upsilon = 4.75$. Finally, for the mass function obtained by the dynamics of the critical collapse [231, 258]

$$\psi(M) = \frac{3.2}{M} \left(\frac{M}{M_c}\right)^{3.85} e^{-\left(\frac{M}{M_c}\right)^{2.85}}, \tag{4.154}$$

one has $\Upsilon = 6$. In general, therefore, the values of the shape parameter $\Upsilon$ are always $\sim \mathcal{O}(10)$.

We show in Fig. 4.16 the conditional probability $P_{PBH}^{(2|1)}(z)$ in terms of the binary total mass $M_{tot}$ and redshift $z$, for fixed values of the PBH abundance and assuming a critical scaling mass function, superimposed with the horizon reaches of aLIGO and ET [429, 441]. The results for other mass functions would give rise to $\mathcal{O}(1)$ corrections. One concludes that only a small fraction of detectable PBHs at aLIGO or ET may have experienced a previous merger. Furthermore, we stress that our estimate does not consider the effect of kicks received during the first coalescence onto the remnant PBHs due to GW emission, which will reduce the rate of hierarchical mergers [442–444].

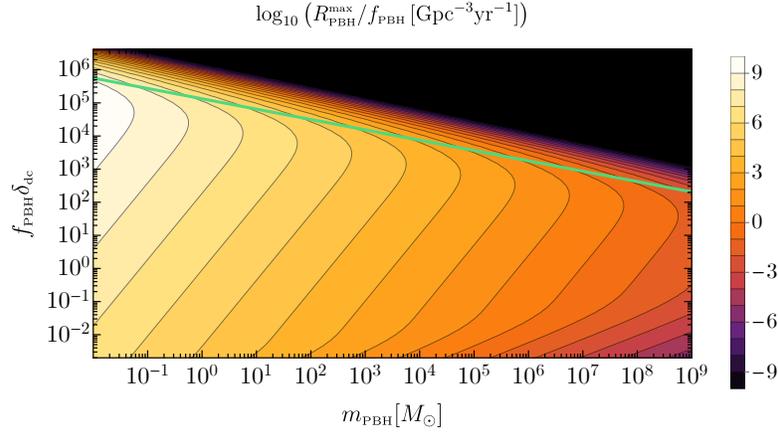

**Figure 4.17**: Contour plot of the maximum PBH merger rate at present time in terms of the PBH mass and local abundance $f_{\mathrm{PBH}}\delta_{\mathrm{dc}}$. The green line identifies the region above which the exponential suppression takes over. Figure taken from Ref. [5].

**Maximum merger rate for initially clustered PBHs**

To close this chapter, let us now investigate how the predictions for the PBH merger rate change if PBHs are spatially clustered at formation. Given that a proper investigation of the dynamical suppression of the merger rate due to binary disruption in sub-structures is still missing if PBHs are initially clustered, we can maximise the impact of clustering and neglect these suppression factors. This implies that we estimate the maximum possible merger rate of clustered PBHs. We report the results of Ref. [5].

In the standard scenario in which PBHs are generated from Gaussian curvature perturbations, they end up Poisson distributed at formation. However, the presence of primordial non-Gaussianity correlating long and short modes is responsible for a different spatial distribution, such that PBHs are initially clustered. Even though this phenomenon catalyses the formation of binaries (and in principle enhance the merger rate), for a large PBH abundance and strong clustering the typical binary semi-major axis may become so small that the binary mergers happen much before the present epoch, eventually suppressing the corresponding merger rate.

To be more general, we follow the results of Ref. [395] and assume a PBH correlation function $\xi_{\mathrm{PBH}}$ that is constant up to the binary scale $\bar{x}$ at the epoch of decoupling as

$$1 + \xi_{\mathrm{PBH}}(x) \approx \delta_{\mathrm{dc}}, \quad \text{if} \quad x < \bar{x}. \tag{4.155}$$

This situation may be easily obtained in non-Gaussian models like those of Refs. [8, 240, 413, 445].

The maximum PBH merger rate, that is the one without accounting for the dynamical suppression due to the disruption of binaries in sub-structures, takes the form [395]

$$
R_{\text{PBH}}^{\text{max}} = \frac{6.2 \cdot 10^4}{\text{Gpc}^3 \text{yr}} \delta_{\text{dc}}^{16/37} f_{\text{PBH}}^{53/37} \left(\frac{t}{t_0}\right)^{-34/37} \left(\frac{m_{\text{PBH}}}{30 M_\odot}\right)^{-32/37}
$$
$$
\times \left\{ \Gamma \left[ \frac{58}{37}, 9.5 \cdot 10^{-5} \delta_{\text{dc}} f_{\text{PBH}} \left(\frac{m_{\text{PBH}}}{30 M_\odot}\right)^{5/16} \left(\frac{t}{t_0}\right)^{3/16} \right] \right.
$$
$$
\left. - \Gamma \left[ \frac{58}{37}, 850 \delta_{\text{dc}} f_{\text{PBH}} \left(\frac{m_{\text{PBH}}}{30 M_\odot}\right)^{-5/21} \left(\frac{t}{t_0}\right)^{-1/7} \right] \right\},
\tag{4.156}
$$

assuming a monochromatic mass distribution of PBHs with reference value $m_{\text{PBH}}$, in terms of the incomplete Gamma functions $\Gamma$. For large values of the local abundance $\delta_{\text{dc}} f_{\text{PBH}} \gg 1$, the maximum merger rate simplifies to

$$
R_{\text{PBH}}^{\text{max}} \simeq \frac{1.9 \times 10^6}{\text{Gpc}^3 \text{yr}} f_{\text{PBH}} \left(\frac{t}{t_0}\right)^{-1} \left(\frac{m_{\text{PBH}}}{30 M_\odot}\right)^{-1} \left(1 + 1.7 \cdot 10^{-4} \Delta_{\text{dc}}\right) \exp \left[-9.5 \cdot 10^{-5} \Delta_{\text{dc}}\right],
\tag{4.157}
$$

in terms of

$$
\Delta_{\text{dc}} = \delta_{\text{dc}} f_{\text{PBH}} \left(\frac{t}{t_0}\right)^{3/16} \left(\frac{m_{\text{PBH}}}{30 M_\odot}\right)^{5/16}.
\tag{4.158}
$$

This limit shows that, for large values of $\delta_{\text{dc}}$, there is an enhancement in the merger rate until the exponential suppression becomes dominant. For fixed values of the PBH abundance and mass scale, the exponential suppression is more efficient at low redshift (i.e. $\Delta_{\text{dc}} \propto (t/t_0)^{3/16}$), given that most of the binaries are produced with small separation and merger times. The maximum merger rate can therefore be obtained, before the exponential suppression takes over, by maximising Eq. (4.157) and approaching the value $\Delta_{\text{dc}} \approx 4.6 \cdot 10^3$. Let us also stress that, for strongly clustered PBHs, the overall merger rate evolution with redshift follows the behaviour $R_{\text{PBH}}^{\text{max}} \approx (t/t_0)^{-1}$.

In Fig. 4.17 we show the maximum merger rate of clustered PBHs in terms of their mass and local abundance. The merger rate increases for small masses and large values of the PBH correlation function up to the green line, identified by the critical value where the exponential suppression takes over and substantially reduces the merger rate.

# Chapter 5

# Gravitational waves from primordial black hole mergers

Gravitational wave astronomy is reshaping our understanding of the universe. Individual detections of BH binary coalescence events by the LVKC [388, 392, 446, 447] have firmly established that BHs may merge within a Hubble time, that at least some of them have nonzero spin and that they can exist in mass ranges that challenge the current stellar-formation paradigm [392, 446, 447]. These observations have therefore renewed interest in understanding the nature of the observed BH population, from an astrophysical, cosmological, and theoretical standpoint [448].

However, the large uncertainties in both the astrophysical and primordial channels make it hard to draw definite conclusions at a population level and to confidently claim for the detection of a BH of primordial origin. It is therefore important to investigate all the properties of PBHs as GW sources to disentangle this channel from the astrophysical foreground, and thus discriminating between them. The advent of next-generation GW detectors, such as the third-generation ground-based interferometers Cosmic Explorer and Einstein Telescope, and the future space mission LISA, will provide much larger detection rates and much more accurate measurements with high signal-to-noise ratio (SNR), that could provide golden events for PBHs.

The main purpose of this chapter is to compare the predictions we developed in the previous chapter for the PBH merger rate with the current GW data detected by the LVKC experiments, and then investigate the role of PBHs at future GW experiments.

Table 5.1:: Binary event parameters $\boldsymbol{\theta}$ and hyperparameters $\boldsymbol{\lambda}$ of the PBH model.

| Event parameters $\boldsymbol{\theta}$ | |
| --- | --- |
| $m_1$ | Source-frame primary mass |
| $m_2$ | Source-frame secondary mass |
| $\chi_{\text{eff}}$ | Effective spin |
| $z$ | Merger redshift |
| Hyperparameters $\boldsymbol{\lambda}$ | |
| $M_c$ | Peak reference mass of the log-normal distribution |
| $\sigma$ | Variance of the log-normal mass distribution |
| $f_{\text{PBH}}$ | Fraction of PBHs in DM at formation |
| $z_{\text{cut-off}}$ | Accretion cut-off redshift |

## 5.1 PBHs and GWTC-2

In the second gravitational-wave transient catalog (GWTC-2) [197, 449] of compact binary mergers released by the LVKC [450, 451], the total number of binary black holes grew to 47. Recently, this figure increased further with the latest GWTC-3 catalog [452].

In their population analysis, the LVKC has built up phenomenological models to capture the crucial features of the mass, spin, and redshift distributions of the detected binary BHs (for example a power-law mass distribution), but not the physical mechanisms which are responsible for these characteristics [449, 453]. The model that best-fit the GWTC-2 data is described by a distribution of the primary BH in the binary as the sum of a power-law and a Gaussian distribution, called "POWER LAW + PEAK" in Ref. [449]. Such a model is characterised by many free parameters and it is preferred to a simpler power-law function, which might point towards the possibility that multiple formation channels are at play.

In the following we will discuss the comparison between the GWTC-2 data and the PBH model we have outlined in the previous chapter.

### 5.1.1 The scenario with only PBHs

In this section we want to focus on the possibility that all BH binaries detected by the LVKC in the GWTC-2 catalog are of primordial origin, meaning that their formation took place in the early stages of the universe. Our main purpose is to set constraints on the PBH population model we developed in the previous section using the dataset reported in the GWTC-2 catalog. To do so, we will apply hierarchical Bayesian inference, whose formalism is summarised in Appendix A, based on deep learning techniques, to determine the best-fit parameters of the PBH model. We report

Table 5.2:: PBH model hyperparameters inferred using the GWTC-2 data.

| | |
|---|---|
| $M_c[M_\odot]$ | $15.86^{+2.36}_{-2.35}$ |
| $\sigma$ | $0.56^{+0.09}_{-0.11}$ |
| $\log f_{\mathrm{PBH}}$ | $-2.53^{+0.09}_{-0.11}$ |
| $z_{\mathrm{cut\text{-}off}}$ | $20.76^{+3.09}_{-2.68}$ |

the results of Ref. [12], where the reader can find additional details.

The hyperparameters of the PBH model we consider are shown in Table 5.1. The reference mass $M_c$ describes the central mass of the PBH mass function, which we assume to have a log-normal shape as in Eq. (4.33), while $\sigma$ denotes its width. $f_{\mathrm{PBH}}$ indicates the PBH abundance in the universe, while $z_{\mathrm{cut\text{-}off}}$ is the cut-off redshift, after which baryonic mass accretion is negligible. We let this parameter to vary to capture the uncertainties in the accretion model we have described in the previous chapter.

Within the GWTC-2 catalog, we have decided to use the same subset of binary events adopted in the population analysis by Ref. [449]. In particular, we exclude events with large false-alarm rate (GW190426, GW190719, GW190909) and events where the secondary component of the binary has mass smaller than $m_2 < 3M_\odot$ (GW170817, GW190425, GW190814). For these light mass events, the presence of an electromagnetic counterpart implies that the binary component is most likely a neutron star (see for example GW170817 [454]). However, in the absence of such a counterpart, it is much more uncertain to assess if the light components are indeed neutron stars or have a different origin. We therefore exclude these events, which the LVKC identifies as neutron-star or mixed binaries, and consider only the remaining 44 events. However, let us stress that the inclusion of only two additional and not particularly loud events like GW190425 and GW190814 would not change our conclusions.

We have used the OVERALL_POSTERIOR provided in [455] for the events in GWTC-1, and the PUBLICATIONSAMPLES provided in [456] for those in the GWTC-2 catalog. To stress the role of the spins in the analysis, we have applied both models with and without the effective spin to analyze the data. We sample Eq. (A.10) using the MCMC package `emcee` [457].

**Bayesian inference from the GWTC-2 catalog**

In this section we describe the results of our Bayesian inference on the GWTC-2 events. The best-fit hyperparameters obtained are summarized in Table 5.2.

In Fig. 5.1 we show the posterior distribution of the PBH hyperparameters obtained by applying hierarchical Bayesian inference to the GWTC-2 dataset. We compare the results of the analysis with and without the use of spin information to highlight the information content coming from effective spin measurements. In particular, when the spin is not included, higher values of the cut-off redshift $z_{\text{cut-off}}$ are favored, making accretion less relevant. On the other hand, when the spin is included, the best-fit value of $M_c$ decreases slightly and the posterior of $z_{\text{cut-off}}$ gets narrower and peaks at smaller values. This happens since many events in the third observational run have effective spin not compatible with zero, which implies that accretion is necessary to spin up PBHs. This also affects the posterior of $\sigma$, which shows a similar trend, because accretion broadens the mass function. The PBH abundance is instead found to be relatively stable with respect to changes of the other hyperparameters, with best fit value $f_{\text{PBH}} \simeq 3 \cdot 10^{-3}$, indicating that this population of PBHs can comprise at most a subpercent fraction of the dark matter.

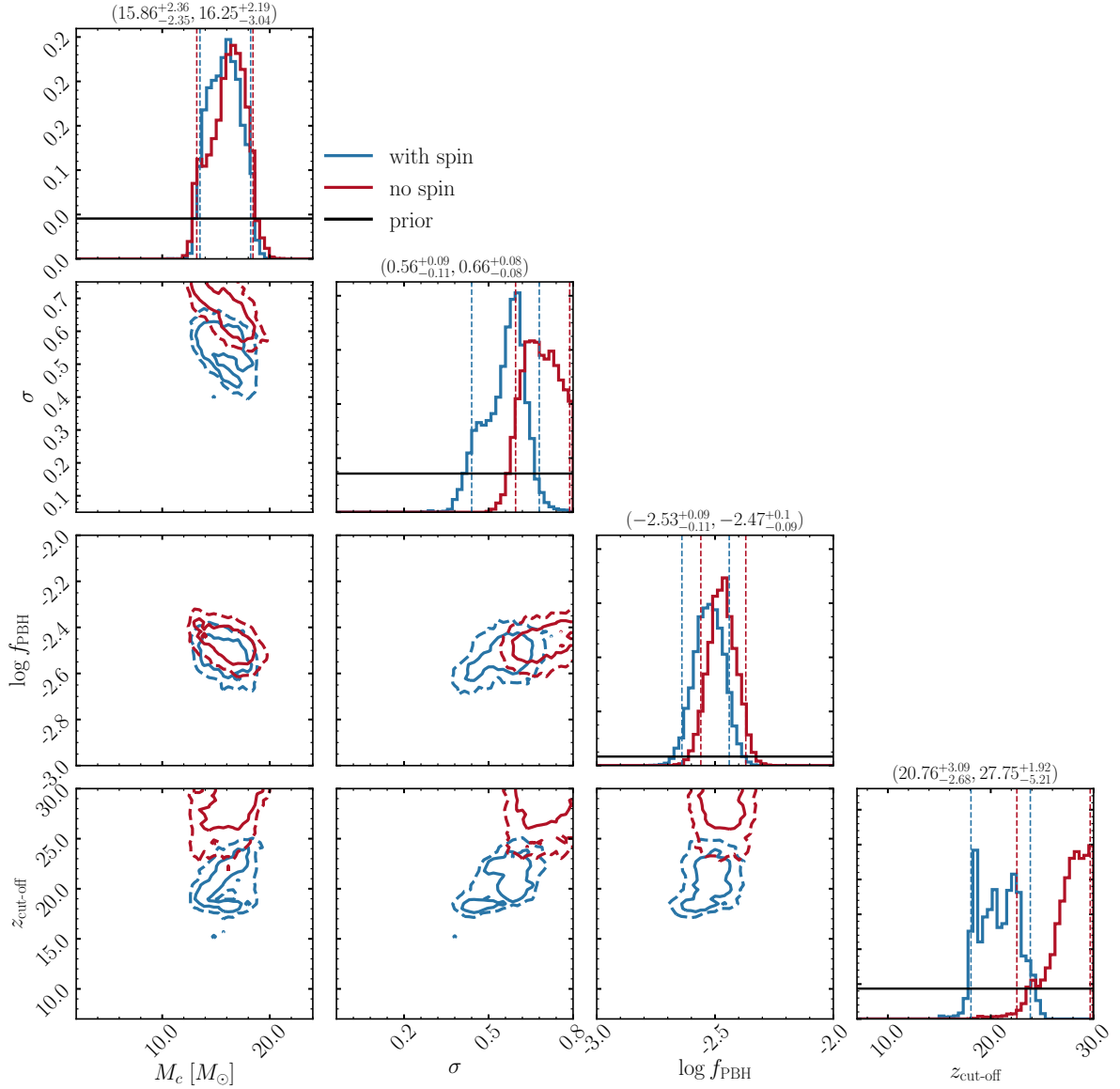

**Figure 5.1**: Population posterior of the PBH hyperparameters using the 44 GWTC-2 BH binary events. Blue lines are obtained using $(m_1, m_2, \chi_{\text{eff}}, z)$, while red lines do not consider the effective spin in the inference. Solid (dashed) contours represent the 68% (95%) confidence intervals. The solid black lines indicate the priors assumed for the population hyperparameters. The first (second) number in parentheses indicates the hyperparameter range inferred by including (omitting) the effective spin from the inference. Figure taken from Ref. [12].

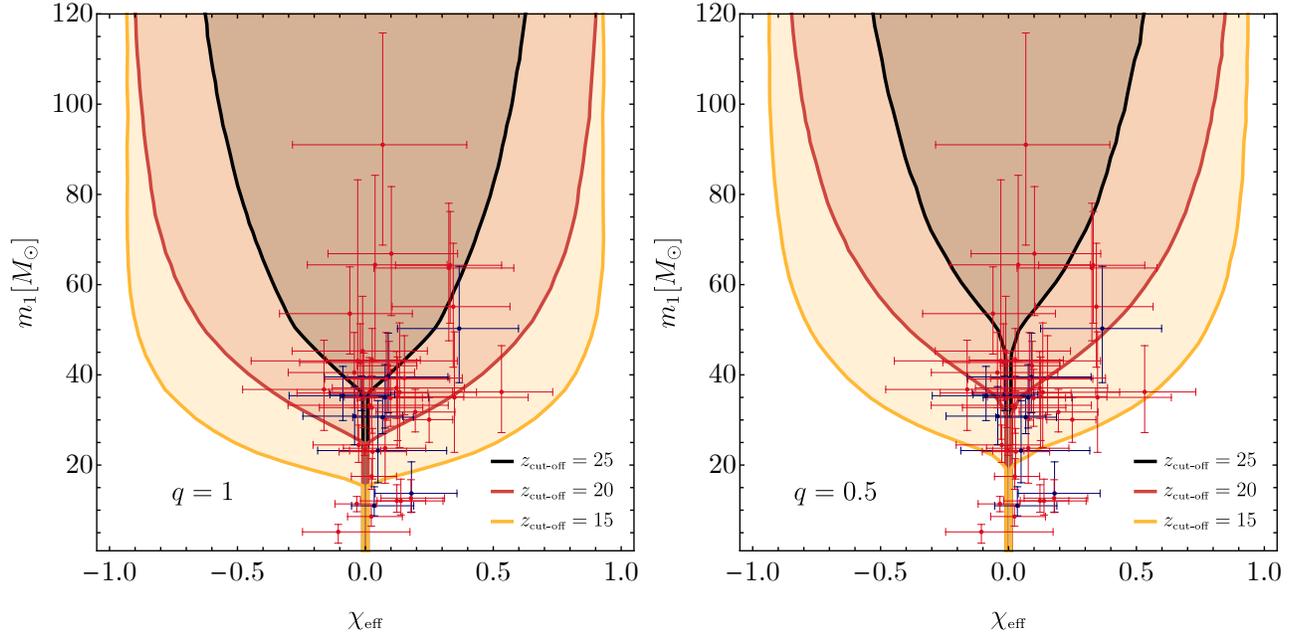

**Figure 5.2**: Prediction for the effective spin distribution in terms of the primary BH mass $m_1$ and cut-off redshift, at $2\sigma$ confidence level, for the best PBH scenario inferred from the GWTC-2 dataset. In blue we show the events from the GWTC-1 and in red those reported after the O3a observing run. Figure taken from Ref. [12].

In Fig. 5.2 we show the $2\sigma$ confidence intervals of the predicted distribution of the effective spin parameter in terms of the mass of the primary component of the binary, for different values of the mass ratio. The values of the cut-off redshift have been chosen around the $2\sigma$ range obtained from the inference analysis. Following Refs. [16, 18], for each value of $m_1$, we have performed an average over the individual spin directions with respect to the total angular momentum assuming isotropic and independent distributions. Given that accretion is more relevant for larger total masses, one finds that only above a certain threshold the PBH spins evolve from their values at formation. Such a transition is pushed towards smaller masses as the cut-off is reduced, corresponding to stronger accretion. This implies that the presence of several spinning events in GWTC-2 leads to a preference towards smaller values of $z_{\rm cut\text{-}off}$. We have superimposed to the effective spin distribution the data from GWTC-1, as blue points with error bars, and from the O3a run, as red points, adopting agnostic priors.

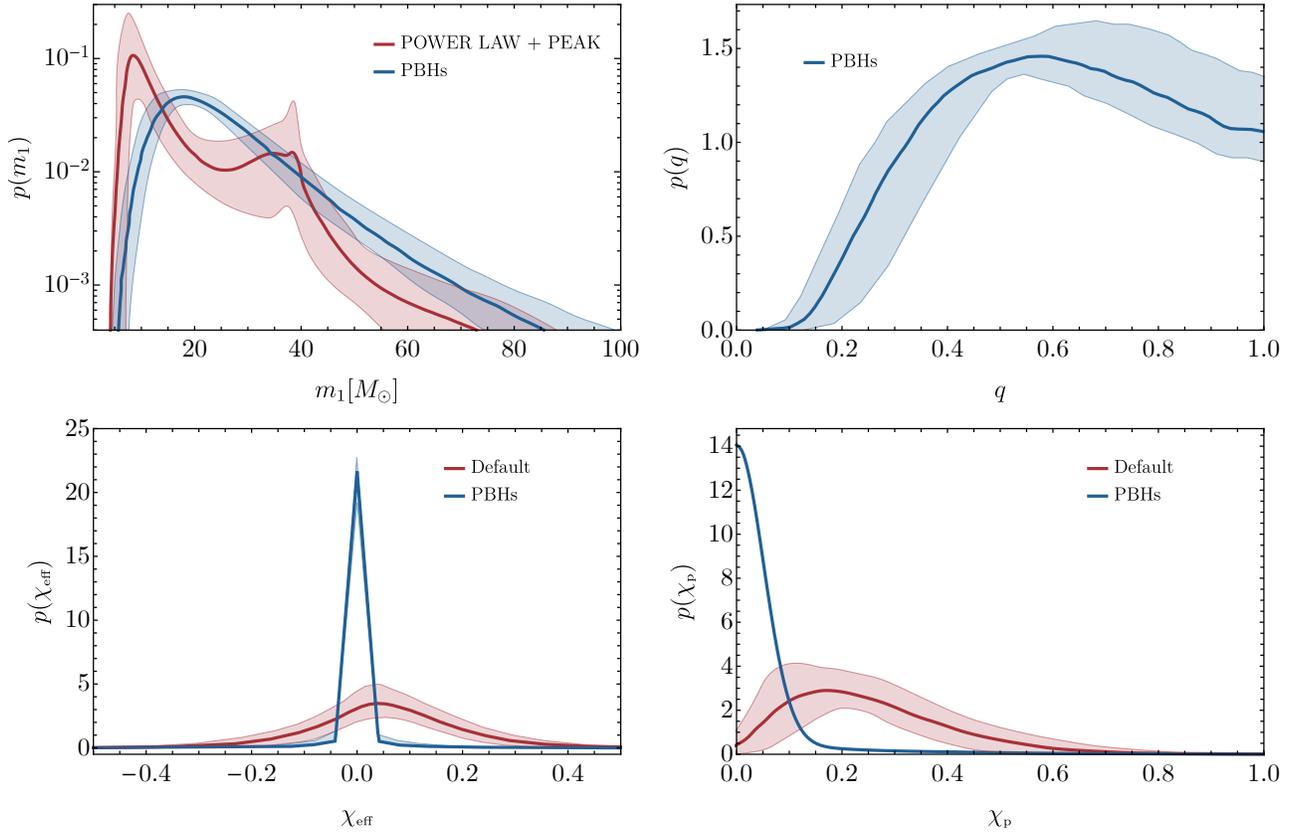

**Figure 5.3**: Probability distribution of the primary mass $m_1$ (top left), mass ratio $q$ (top right), effective spin $\chi_{\rm eff}$ (bottom left) and precession spin $\chi_{\rm p}$ (bottom right) from the best-fit model when the effective spin information is included in the analysis. For comparison, we also show the 90% CL distributions found by the LVKC in Ref. [449]. Figure taken from Ref. [12].

In Fig. 5.3 we plot the distribution of the primary mass $m_1$, mass ratio $q$, effective spin $\chi_{\rm eff}$, and precession spin $\chi_{\rm p}$ parameters inferred from our best-fit model. On the top left, we show the marginalised posterior probability for the primary mass, also adding the corresponding preferred result found in Ref. [449] assuming a "POWER-LAW + PEAK" mass function. On the top right, we show the marginalised distribution for the mass ratio which, due to the preferred relatively high value of the cut-off redshift and the significant width of the mass function, is peaked at $q \sim 0.5$ (for smaller values of $z_{\rm cut\text{-}off}$ the distribution would have peaked at higher values, as predicted by our model of PBH accretion [16, 18]). On the bottom, we show the marginalised distributions for the effective and precession spin parameters, both for the PBH scenario and the so-called "Default" model (see Appendix D.1 of Ref. [449]). In both cases, the probability distributions show a narrow peak around zero since the best-fit PBH mass function is dominated by low masses, that are correlated with small spins. This result is not in contrast with the preference we found for an accreting PBH model due to the presence of several (moderately) spinning binaries in the catalog. Indeed, we stress that Fig. 5.3 shows the intrinsic population distribution, that does not take into account selection effects (for which larger masses are more easily observed by current detectors, see Appendix B for details). This explains the difference with the "Default" model, where masses and spins are not correlated, giving rise to a peak at nonvanishing spins and broader distributions.

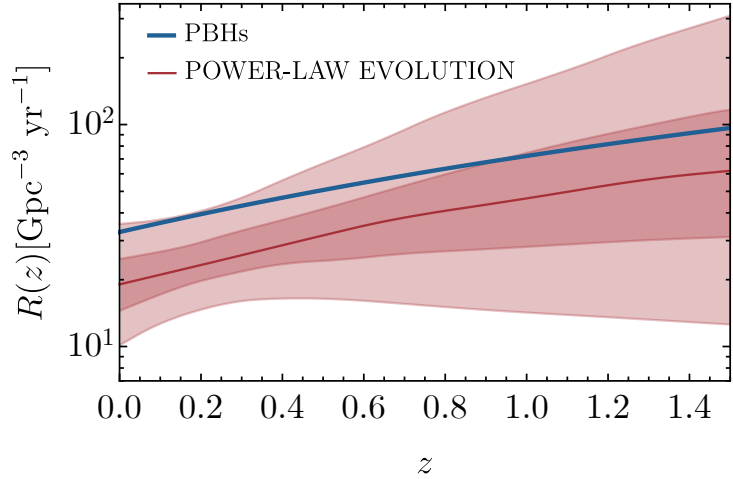

**Figure 5.4**: Redshift evolution of the PBH merger rate for the best-fit PBH population (blue line). For comparison, we also show in red the 50% (90%) confidence level for the merger rate found by the LVKC [449] adopting a power-law evolution model for astrophysical sources. Figure taken from Ref. [12].

In Fig. 5.4 we finally compare the best-fit prediction for the evolution of the merger rate density $R(z)$ in terms of redshift, given by Eq. (4.132), with the power-law evolution model for astrophysical sources found in Ref. [449]. The latter is less steep than the behavior $R(z) \propto (1+z)^{2.7}$ predicted from the star formation rate [458], although present observational errors are still too large to draw any firm conclusion. The measured merger rate evolution is compatible with the PBH scenario, that predicts less mergers compared to the stellar-origin scenario in the high redshift side of the LVKC horizon. Furthermore, the PBH scenario predicts a monotonically increasing merger rate, even at redshifts beyond the LVKC horizon, while the astrophysical one is expected to decrease after the peak $z \sim 2$ for the star formation rate (unless Population III binaries give a strong contribution and induce a second peak at large redshift). This difference is particularly important for third-generation GW detectors, that could detect merger events up to redshift $z \simeq \mathcal{O}(10^2)$ [429, 459].

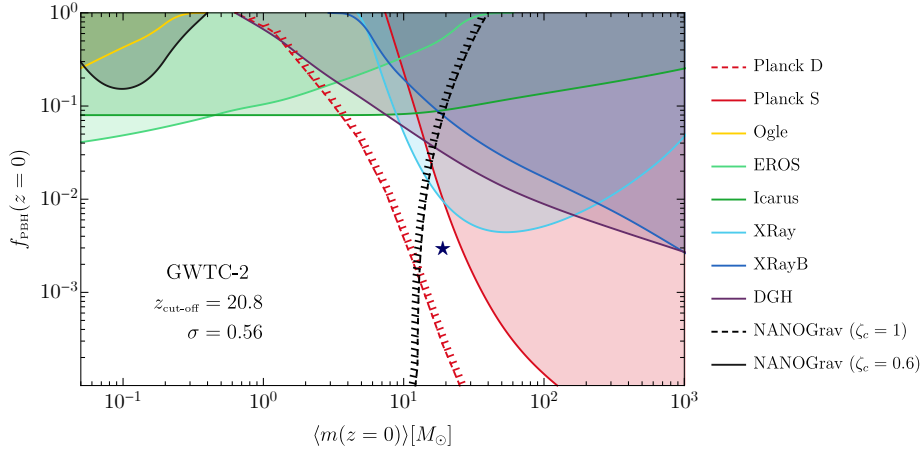

**Figure 5.5**: Constraints on the present PBH abundance $f_{\rm PBH}(z=0)$ in terms of the mean present PBH mass $\langle m(z=0) \rangle$. The blue star indicates the median values for the population parameters $\{M_c, \sigma, f_{\rm PBH}, z_{\rm cut-off}\}$ inferred from the analysis with the GWTC-2 dataset. The red and black dashed lines denote the uncertain bounds from Planck D and NANOGrav, respectively. The perpendicular dashes point towards the would-be excluded regions. Figure taken from Ref. [12].

## Confront with the PBH constraints

Finally, we compare the PBH abundance determined by the inference to explain the observed events to the other existing constraints described in the introduction. The result is shown in Fig. 5.5.

Given that we are considering an extended mass function, we have adapted the bounds derived for a monochromatic PBH population using the techniques described in [374, 460]. We stress that we consider the effect of accretion in alleviating the early universe constraints by shifting them to higher late-time range of masses and making them weaker due to the growth of the PBH abundance $f_{\rm PBH}$, following the discussion described in detail in Ref. [17]. Furthermore, given the large systematic uncertainties in the NANOGrav 11-yr dataset from Ref. [150], in particular for the choice of the PBH threshold $\zeta_c = 1$, we have decided to show this constraint also assuming the threshold value $\zeta_c = 0.6$, that is motivated by state-of-art numerical simulations [320].

By looking at the plot, it may seem that the best-fit result from the GWTC-2 analysis is in tension with Planck D. However, one should remember that the assumption of a thin disk in such a constraint is less reliable at high redshift [172] with respect to spherical accretion, with which GWTC-2 is compatible. In conclusion, the hypothesis that all the GWTC-2 events are originated from PBHs is not in contrast with current observational constraints. We finally stress that this hypothesis provides the maximum overlap with the observational constraints, since it maximizes the PBH merger rate.

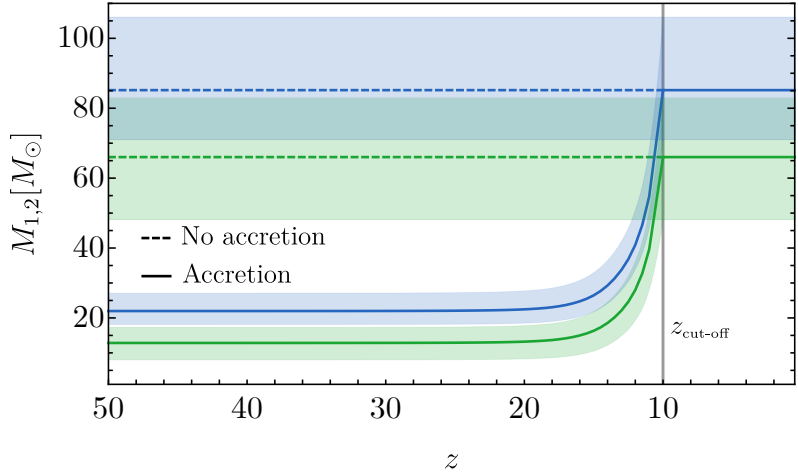

**Figure 5.6**: Redshift evolution of the PBH masses for GW190521 without and with the effect of accretion (assuming a cut-off redshift $z_{\text{cut-off}} = 10$). The green (blue) bands indicate the primary (secondary) PBH mass within 90% CL. Figure taken from Ref. [15].

## 5.1.2   Mass gap events

One of the most interesting event released in the GWTC-2 catalog is the mass gap event GW190521 [461], describing the coalescence of two BHs with masses $M_1 = 85^{+21}_{-14}M_\odot$ and $M_2 = 66^{+17}_{-18}M_\odot$, which fall within the astrophysical mass gap between about $65M_\odot$ and $135M_\odot$ in the spectrum of population I/II stellar-origin BHs. Such a gap arises for BHs formed out of the gravitational collapse of unstable stars with a helium core mass in the range $\sim (32-64)M_\odot$ due to the pulsational pair instability mechanism, which describes the production of electron-positron pairs at high temperature in the stars and the consequent contraction of the unstable stellar cores. On the other hand, stars with masses above $\sim 200M_\odot$ and very low metallicity (typical of population III) can avoid disruption and give rise to a population of intermediate mass BHs above $\sim 135M_\odot$ [462–469].

Even though the presence of astrophysical BHs in the mass gap is not excluded due to possible hierarchical mergers of smaller BHs [470–476], or by mass accretion in primordial dense clusters [477], or by beyond Standard Model physics that modifies the pair instability [478], the discovery of this event has assigned to PBHs even more relevance.

A possible primordial origin for GW190521 can be assessed if the corresponding merger rate is in agreement with observations, with the corresponding value of the PBH abundance $f_{\text{PBH}}$ allowed by the present observational constraints [84]. Following the results of Ref. [15], we will investigate this possibility either by assuming that the PBH mass function is determined only by this event, or assuming that all the detections in the GWTC-2 catalog are of primordial origin, similarly to the analysis in Ref. [12].

**GW190521 as a single PBH event**

In the first case, we assume that the PBH mass function is determined solely by the mass gap event. Let us start neglecting the role of accretion. Then, for the representative choice of a lognormal mass function as in Eq. (4.33), we have fixed the central mass scale to $M_c = 75 M_\odot$ and its width to $\sigma = 0.2$ (a slightly different choice would modify the merger rate of a $\mathcal{O}(1)$ factor). Given that the PBH merger rate scales like $R \propto f_{\rm PBH}^{53/37}$, the corresponding abundance is quite stable with respect to reasonable modifications of the mass distribution.

One can compute the PBH merger rate as shown in Eq. (4.124) and then estimate the corresponding abundance required to match the observed value, which we assume to be $\approx 1/{\rm yr}$ given the current uncertainties [461]. This gives $f_{\rm PBH} \approx 1.2 \cdot 10^{-4}$ which, compared to current observational constraints in Fig. 5.7, is in tension with the limits coming from CMB distortions.

The next step is therefore to consider the role of accretion in the evolution of the PBH masses and its impact on the constraints, as discussed in the previous chapter. Assuming the benchmark value of the cut-off redshift $z_{\rm cut-off} = 10$, one can see in Fig. 5.6 the evolution of the masses starting from the initial values of $M_1^i \simeq 22 M_\odot$ and $M_2^i \simeq 13 M_\odot$, respectively. We stress that our conclusions would not change qualitatively by choosing a different cut-off.

The corresponding initial mass function is assumed to have $M_c = 18.5 M_\odot$ and $\sigma = 0.2$, such that the corresponding value of the abundance to match the observed merger rate is $f_{\rm PBH}(z_i) \approx 2.5 \cdot 10^{-5}$ ($f_{\rm PBH}(z=0) \approx 3.7 \cdot 10^{-5}$), which is smaller than the one found in the non-accreting scenario since accretion enhances the merger rate. As shown in Fig. 5.7, this value is allowed by the present constraints, which are strongly relaxed in the presence of accretion [17].

**GW190521 as an event within the full PBH population**

In the second case, let us consider the scenario when all the observed events in the GWTC-2 catalog are ascribed to PBHs. Following the Bayesian analysis done in Ref. [12] and described in the previous section, one can determine the parameters of the PBH mass function which best-fit the data, also including the effect of accretion. The resulting plot of the constraints on the PBH abundance for the best-fit population is shown in Fig. 5.5.

If accretion does not play any role, one finds that the observable rate of events with masses in the mass gap or at least as extreme as GW190521 is lower than the measured value, implying that GW190521 is an outlier of the population and not generally expected in the full population scenario without accretion. On the other hand, for the scenario with accretion and considering the best-fit value for the cut-off redshift $z_{\rm cut-off} = 20.8$, the corresponding enhancement in the merger rate results in smaller values of the PBH abundance with respect to the non-accreting case. The

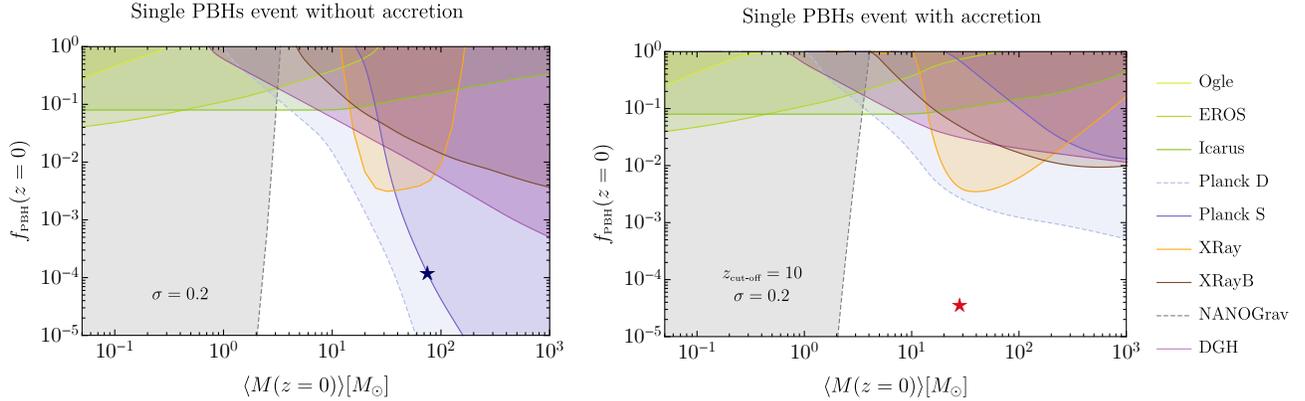

**Figure 5.7**: Constraints on the PBH abundance $f_{\rm PBH}$ in the single event scenario. *Left:* The star shows the value of the PBH abundance needed to explain the event GW190521 in the scenario without accretion. *Right:* Same as in the left panel but in the scenario with accretion. We highlight the change in the early-universe constraints NANOGrav and Planck (D/S) due to the effect of accretion following the discussion in Ref. [17]. We stress that here the NANOGrav constraint has not been reanalysed using a more motivated value of the threshold to collapse, differently from Fig. 5.5. Figure taken from Ref. [15].

corresponding merger rate of mass gap events (or as massive as GW190521) is

$$
\begin{aligned}
R_{\rm PBH}(M_{1,2} > 65 M_\odot) &\simeq 1.1/{\rm yr}, \\
R_{\rm PBH}(M_1 > 85 M_\odot, M_2 > 65 M_\odot) &\simeq 0.8/{\rm yr},
\end{aligned}
\tag{5.1}
$$

which are compatible with the observed rate. This implies that, in the full population scenario with accretion, events like GW190521 are allowed and expected.

The presence of accretion also implies that the binary mass ratio is pushed towards unity and the individual PBH spins should be non vanishing, with the secondary component spin always larger than the one of the primary [16]. Due to the small SNR of the event, no firm conclusion can be drawn, even though it seems that this pattern is in agreement with the parameters measured for GW190521. In particular, there is evidence that the individual spins of the event are large and likely lie on the orbital plane.

Finally, let us stress that the conclusions of our analysis are also conservative since, in more realistic scenarios where some of the detected GW events have an astrophysical origin, their removing from the analysis would make our results on the viability of GW190521 as a primordial binary even more robust.

### 5.1.3 Mixed population analyses

So far we have been focused on the assumption that all binary BH events detected so far have a primordial origin and consequently deduced the properties of the primordial population, meaning its mass function and abundance. Even though this possibility is still allowed by present constraints [12] as we have discussed in the section above, it is clear that the developed PBH model cannot account for all the features in the GW catalogue and it is plausible that all the events may not be explained through a single population, either astrophysical [479] or primordial [437]. It is therefore interesting to extend the analysis to include astrophysical black hole (ABH) models in a multi-channel inference. To disentangle the contribution of different populations in the data one has to rely on their features that could be a smoking gun for either PBHs or ABHs.

From the astrophysical side, BHs may form at the end of the stellar evolution in the late universe and they can assemble in binaries either in isolation through a common envelope phase or dynamically through three-body encounters in dense stellar clusters, see Refs. [183, 184] for some recent reviews.

In this section we summarise the main findings on the comparison between the PBH model with phenomenological or true astrophysical models, trying to infer the corresponding mixing fraction of the considered populations, see Refs [437, 479–484] for works along this direction.

**Analysis with phenomenological astrophysical model**

We focus on the comparison between the PBH model and the LVC "truncated" phenomenological model [449, 485] by performing a hierarchical Bayesian analysis on the GWTC-2 catalog to determine the fraction of the considered events that can be ascribed to PBHs. We report only the main findings of Ref. [10], where all the details of the analysis and a comprehensive discussion of the results are shown.

The main result consists in the mixing fraction, which shows that roughly 80% of the BH mergers in the GWTC-2 catalog are of astrophysical origin. The analysis also stresses that the best-fit mixed ABH+PBH model has a decisive statistical evidence relative to the single-population phenomenological ABH model, supporting the existence of extra features which are not fully explained by the simple truncated ABH model. This points towards the existence of at least two different populations of BH mergers in LVC data and is compatible with a subdominant PBH population. The fraction of putative PBH events could also explain the events with binary components in the mass gap, that are at odds with the standard astrophysical scenarios.

A simple extension could be the inclusion of a PBH population to the LVC "POWER LAW + PEAK" model used for the GWTC-2 catalogue [449]. However, since the latter does not carry any astrophysical inputs, it would be hard to distinguish its phenomenological features, like the Gaussian peak, from physical effects coming from the putative subdominant PBH population. Let us stress, though, that the Bayes factor between the mixed population of our analysis and the single ABH model are comparable to the one of the "POWER LAW + PEAK" model with respect to the "truncated" model used in Ref. [449]. This suggests that PBHs add similar features that those needed to match the phenomenological astrophysical model preferred by the LVC analysis.

**Analysis with motivated astrophysical models**

The next and more comprehensive analysis is to mix the PBH model with several state-of-the-art astrophysical models, that can reproduce many features of the observed BH population, and therefore set statistical constraints on a putative subpopulation of PBHs given our present knowledge of binary BH formation scenarios. We report the main conclusions of Ref. [7], where the interested reader can find all the details of the analysis.

The astrophysical models we considered are taken from Ref. [479], where three field formation channels and two dynamical formation channels were considered. The three field formation scenarios are based on a late-phase common envelope (CE), which is usually invoked to harden the binary and catalyze mergers, the stable mass transfer between the star and the already formed BH (SMT), and chemically homogeneous evolution (CHE). We stress that the dominant channels correspond to the CE and SMT scenarios [479]. The two dynamical models consist on the formation of binaries in globular clusters (GC) and in nuclear star clusters (NSC) through three-body encounters.

The main result of the analysis is that the relative role of PBHs in the GWTC-2 catalogue depends on the considered astrophysical channels. In particular, even though a PBH population is statistically preferred against models like GC and NSC, a dominant contribution from the SMT model along with CE and GC drastically reduces its role, except for the mass-gap events like GW190521, whose observation rates are compatible with the ones predicted by the inferred PBH population. The exclusion of this event in the catalog implies that the Bayesian evidence for CE+GC+SMT becomes comparable to CE+GC+SMT+PBH, showing that the constraining power of the current data set is not sufficient enough to draw firm conclusions. Including the recently reported events in the GWTC-3 catalog [452] would not modify qualitatively our conclusions, since its bulk mass and spin distribution are consistent with GWTC-2 [453].

We therefore conclude that, to set stringent constraints on a primordial population in the GW data, a more complete understanding of astrophysical populations is required. In particular, each of the four state-of-the-art astrophysical models is affected by large uncertainties, and there might exist others which are competitive against the primordial subpopulation. Furthermore, we have adopted a lognormal shape for the PBH mass distribution at formation, but it would be important to test the consequences of other assumptions and of different priors on the hyperparameters of the PBH model motivated by specific formation mechanisms (see e.g. [145, 486]).

The confident claim of the primordial nature of some individual binary BHs would probably require single-event analyses for events with large signal-to-noise ratio, especially through a cross-correlation between the merger rates and masses, spins, and redshift measurements to identify key features of the PBH scenario (see e.g. [28, 487]). Furthermore, to break the degeneracy between the PBH and astrophysical channels, one can perform population studies focusing on spin distributions [488–490], and accounting for the $q - \chi_{\mathrm{eff}}$ correlation induced by accretion effects in PBH models [16, 18].

## 5.2  Future GW experiments

In this section we want to investigate the role of future GW experiments in constraining, and eventually discovering, a putative population of PBHs. We will consider both LISA and third-generation (3G) detectors like the Einstein Telescope [198] and Cosmic Explorer [199] which, due to their larger horizon redshifts and better sensitivity, could be able to probe together high redshifts and span a range in the BH masses from supermassive to subsolar ones.

In Fig. 5.8 we show the horizon redshift (left panel) and the sensitivity curves for the SGWB (right panel) for several experiments. As one can appreciate, the improved features of these experiments could shed light on several properties of a primordial population, if any.

### 5.2.1  Merger events at high redshifts

One promising feature of the PBH model, which can be possibly used to disentangle the primordial population from any astrophysical formation scenario, is the peculiar time evolution of the merger rate, which is monotonically increasing with redshift as [16, 178]

$$R_{\mathrm{PBH}}(z) \approx t^{-34/37}(z), \tag{5.2}$$

up to redshifts around $z \sim \mathcal{O}(10^3)$. Such a behavior is dictated by the dynamics of formation of primordial binaries before the matter-radiation equivalence, that is how pairs of PBHs decouple from the Hubble flow, and is therefore a robust prediction within the PBH model.

This time evolution is different from what one expects for an astrophysical population, whose merger rate should peak around redshift of a few, with a possible second peak coming from Pop III BHs at redshift $z \sim 10$ [491–494]. Indeed, astrophysical models predicts that black holes originated from Pop III stars [495–498] should form at redshift $z \sim 25$ [499], with the corresponding merger time dependent on the binary formation mechanism. In particular, it ranges from $\mathcal{O}(\text{Gyr})$ (corresponding to merger redshift $z \lesssim 6$) to $\mathcal{O}(10\,\text{Myr})$ if they form dynamically in Pop III clusters, in which case they would merge almost at the redshift of BH formation [499].

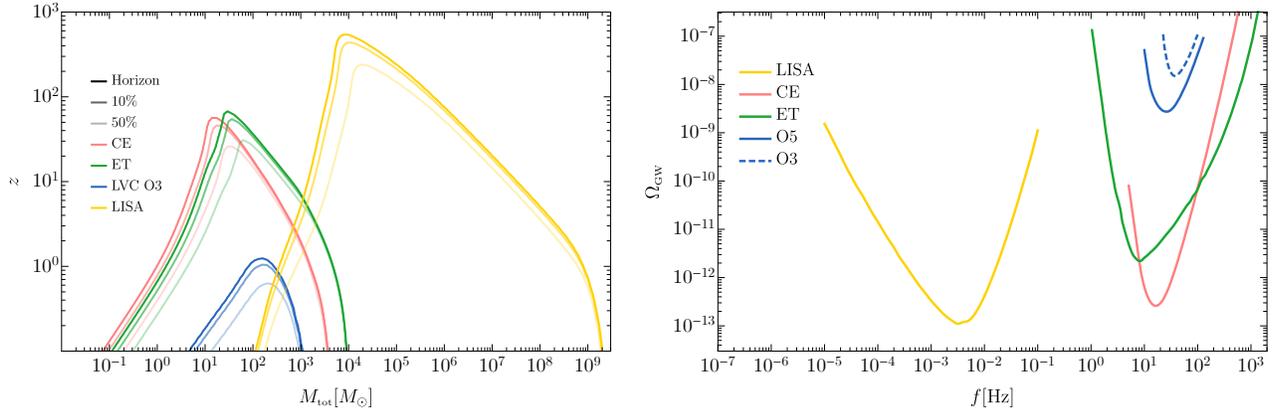

**Figure 5.8**: *Left:* Horizon redshift for some GW experiments in terms of the total mass of a symmetric binary. We refer to the "Horizon" as the maximum distance at which a binary can be detected if optimally oriented with respect to the detectors, while 10%, 50% as the redshift at which those fraction of binaries can be observed (corresponding to values of SNR = {8, 10, 19}). *Right:* Power-law integrated sensitivity curves for the SGWB for several present and future GW experiments. Figure taken from Ref. [10].

Even though present GW experiments are still limited to "local" GW signals generated by sources at redshift up to $z \lesssim 1$, 3G detectors like ET [198] and CE [199], characterised by larger horizons, will allow to use the evolution of the merger rate to distinguish between different binary BH populations [491, 500–502].

Assuming the most conservative scenario in which astrophysical black holes formed from Pop III clusters, the detection of a binary BH at redshift higher than $z \sim 30$ would be a smoking-gun signal in favour of PBHs, given that no ABHs are expected at those or higher redshifts in a standard cosmology [503]. Adopting the PBH mixing fraction and mass function allowed by the mixed inference using the GWTC-2 dataset of Ref. [10], one can make a forecast for the expected number of observable events at high redshifts, as well as the corresponding redshift dependence of the merger rate, at 3G detectors.

The expected detectable number of PBH merger events is defined as (see also the Appendices A and B)

$$N_{\text{det}} = \int \text{d}z \text{d}m_1 \text{d}m_2 \frac{1}{1+z} \frac{\text{d}V_c(z)}{\text{d}z} \frac{\text{d}^2 R_{\text{PBH}}}{\text{d}m_1 \text{d}m_2} p_{\text{det}}(m_1, m_2, z), \qquad (5.3)$$

in terms of the PBH merger rate, the redshift factor $1/(1+z)$ to account for the difference in the

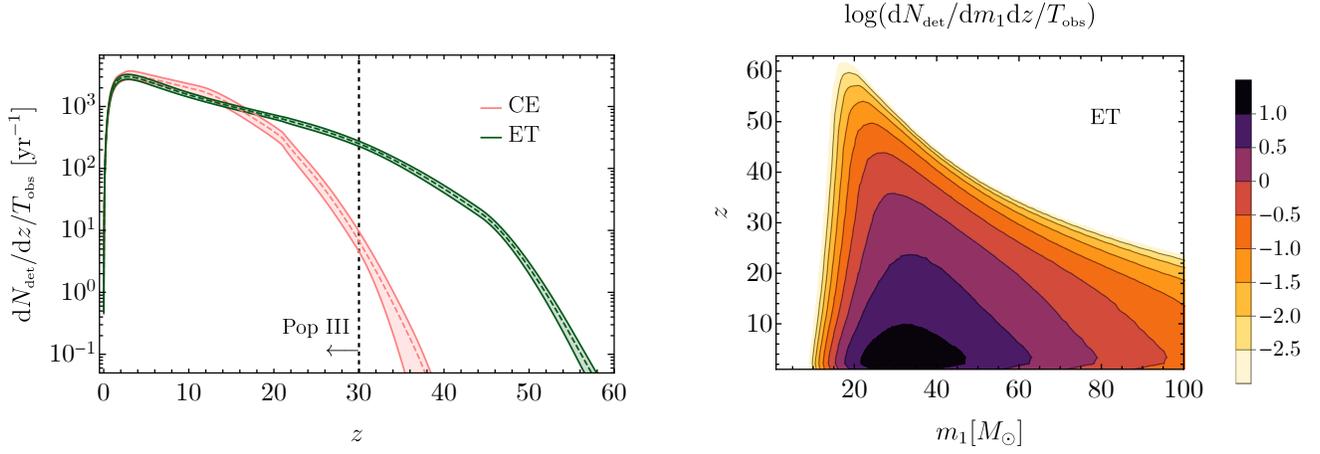

**Figure 5.9**: Distribution of observable events per year coming from the PBH subpopulation in terms of redshift at ET and CE (left panel), and as a function of both primary mass and redshift at ET (right panel). Figure taken from Ref. [10].

clock rates at the time of merger and detection, and the comoving volume per unit redshift [504]

$$\frac{\mathrm{d}V_c(z)}{\mathrm{d}z} = \frac{4\pi}{H_0} \frac{D_c^2(z)}{E(z)} = \frac{4\pi}{H_0^2} \frac{1}{E(z)} \left( \int_0^z \frac{\mathrm{d}z'}{E(z')} \right)^2, \tag{5.4}$$

in terms of the comoving distance

$$D_c(z) = \frac{1}{H_0} \int_0^z \frac{\mathrm{d}z'}{E(z')}, \tag{5.5}$$

with

$$E(z') = \sqrt{\Omega_r(1+z')^4 + \Omega_m(1+z')^3 + \Omega_K(1+z')^2 + \Omega_\Lambda}, \tag{5.6}$$

and $\Omega_K = 0.0007$, $\Omega_r = 5.38 \times 10^{-5}$, $\Omega_\Lambda = 0.685$, $\Omega_m = 0.315$, $h = 0.674$, $H_0 = 1.0227 \times 10^{-10} h \,\mathrm{yr}^{-1}$. The probability of binary detection $p_{\mathrm{det}}(m_1, m_2, z)$ is computed following the details of Appendix B in terms of the signal-to-noise ratio.

In Fig. 5.9 we show the predicted distribution of observable events per year in terms of redshift at ET and CE. The former, at design configuration, is able to observe at least one PBH event per year up to redshift $z \sim 50$. In the right panel of the same figure we also show the distribution as a function of the primary mass at ET. Due to the characteristic shape of the observable horizon, as shown in Fig. 5.8, most of the distant binaries will have their primary mass around the central scale $M_c$ of the PBH mass function.

One can also determine the total number of observable events per year at redshifts larger than $z > 30$ for both ET and CE, finding

$$N_{\mathrm{det}}^{\mathrm{ET}}(z > 30) = 1315^{+305}_{-168} \,/\mathrm{yr}, \qquad N_{\mathrm{det}}^{\mathrm{CE}}(z > 30) = 12^{+22}_{-11} \,/\mathrm{yr}. \tag{5.7}$$

Given that at low-frequencies ET has a larger sensitivity (ET-D) compared to the CE phase-1 design (see Fig. 1.3), being the relevant range for heavy and/or high-redshift mergers, a larger event rate is expected for ET. One can therefore conclude that 3G detectors can unequivocally confirm the presence of a PBH subpopulation in the current dataset and to unveil a novel (primordial) family of binary BHs. We stress, however, that this conclusion depends also on the accuracy of the redshift measurements, which can be low for distant events even in the 3G era [491].

## 5.2.2 Minimum testable abundance

Current GW data measured by the LVKC experiments allow for the possible presence of merger events ascribed to PBHs only if their abundance in the universe in the corresponding mass range is below $10^{-3}$ [7, 12]. Given that smaller values may still be relevant for the cosmological evolution and development of large scale structures, it is important to estimate the minimum value of $f_{\rm PBH}$ which can be detected by future GW experiments like ET, CE and LISA.

In this section we estimate such a value following the results of Ref. [5]. To do so, we consider two different PBH scenarios. In the first case, we consider the standard PBH binary formation and merger rate computation assuming an initial Poisson spatial distribution, as described in Chapter 4 [14, 224, 400–403]. In the second one, we will assume a significant spatial clustering at PBH formation, with the corresponding impact on the PBH merger rate, as shown in Section 4.3.3. In the latter, we however neglect the evolution of clustering and the potential suppression effects coming from binary interactions in local clusters, which are currently poorly understood. Therefore, the clustered scenario we consider should be regarded as corresponding to the maximum possible merger rate, see Eq. (4.157). In particular, we determine the optimal value of $\delta_{\rm dc}$ that leads to the maximum number of events at GW detectors, in order to asses the minimum testable PBH abundance.

Finally, assuming that the value of $f_{\rm PBH}$ is such that the corresponding PBH population is able to produce merger events observable by future GW detectors, one can investigate the possibility to confidently identify the primordial nature of some GW events. This can be achieved if at least one of the components of the binary have a subsolar mass, given that astrophysical compact objects are expected to have larger masses, or if the coalescences occur at large enough redshift, where astrophysical sources do not contribute. One can therefore make a forecast for the minimal value of $f_{\rm PBH}$ required to confidently assign a fraction of the observed signals to the merger of PBHs.

**Event rate of resolvable mergers**

One can start the investigation to determine the minimum PBH abundance by considering the event rate of merger events which can be resolved by GW experiments. Such a value can be determined by requiring the detection of at least one event per year at those experiments, $N_{\text{det}} > 1/\text{yr}$, for both cases of Poisson distribution of Eq. (4.124) and clustered PBHs of Eq. (4.157), where in the latter case we have assumed the theoretical maximum merger rate before the exponential suppression takes over.

The result is shown in the left panel of Fig. 5.10 for a monochromatic PBH population, where the minimum testable PBH abundance has been superimposed with several observational constraints coming from independent searches (obtained by assuming a monochromatic PBH mass function and for an initial Poisson spatial distribution). The enhancement in the merger rate induced by clustering results in a smaller value of the testable PBH abundance with respect to the Poisson case.

One can see that a portion of the ET region is allowed by present constraints, while the LISA region is almost completely ruled out. We stress, however, that for clustered PBHs such a comparison is more delicate, given that some of the constraints might be modified by PBH clustering. Nonetheless, the most stringent ones coming from CMB and X-ray observations are not weakened if PBHs are clustered [14, 174, 408]. We also notice that constraints applying on early universe quantities, like CMB anisotropies sensitive to the redshift range $300 \lesssim z \lesssim 600$, may be potentially evaded if accretion is efficient enough to shift them to larger final masses [17]. This effect may be particularly relevant for masses $m_{\text{PBH}} \gtrsim \mathcal{O}(10)M_\odot$ and for LISA.

Even though a PBH population with abundance larger than the determined minimum would contribute to the observed events, one would not necessarily be able to differentiate it from other astrophysical contributions, unless one focuses on the mergers of subsolar objects (see Refs. [505–510] for constraints with current data). This implies that the considered bounds are necessary, but not sufficient, to test the existence of PBHs. Restricting to high enough redshifts, where no astrophysical contamination is present, would instead represent a tighter condition to confidently ascribe a primordial origin to a merger event.

**Stochastic gravitational wave background**

To determine a minimum PBH abundance one can also investigate the merger of PBHs which are not individually resolved by the experiments and give rise to a SGWB. The abundance of the corresponding spectrum at frequency $\nu$ is given by [178]

$$\Omega_{\text{GW}}(\nu) = \frac{\nu}{\rho_0} \int_0^{\frac{\nu_3}{\nu}-1} \mathrm{d}z \mathrm{d}m_1 \mathrm{d}m_2 \frac{1}{(1+z)H(z)} \frac{\mathrm{d}R_{\text{PBH}}}{\mathrm{d}m_1 \mathrm{d}m_2} \frac{\mathrm{d}E_{\text{GW}}(\nu_s)}{\mathrm{d}\nu_s}, \tag{5.8}$$

as a function of the present energy density $\rho_0 = 3H_0^2/8\pi G$ in terms of the Hubble constant $H_0$ and the redshifted source frequency $\nu_s = \nu(1+z)$. The phenomenological expression of the GW energy

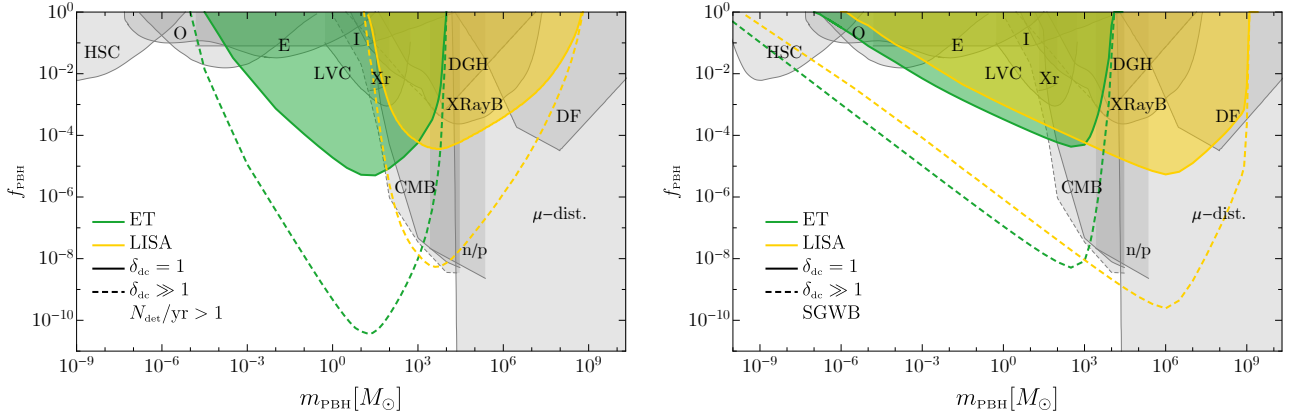

**Figure 5.10**: Forecast on the PBH abundance required to have at least one event per year (left panel) and to detect a SGWB (right panel) at the ET (green) and LISA (yellow) experiments, for both cases of Poisson distributed (solid) and strongly clustered (dashed) PBHs. The minimum abundance is compared with observational constraints (gray) coming from microlensing searches by Subaru HSC [164, 165], MACHO/EROS (E) [166, 167], Ogle (O) [169] and Icarus (I) [168], X-rays (Xr) [174] and X-Ray binaries (XRayB) [175], CMB anisotropies [171], Dwarf Galaxy heating (DGH) [176, 511], dynamical friction (DF) [512], the neutron-to-proton ratio (n/p) [513] and CMB $\mu$-distortions [514]. Figure taken from Ref. [10].

spectrum in the non-spinning limit is given by [515]

$$\frac{dE_{\text{GW}}(\nu)}{d\nu} = \frac{\pi^{2/3}}{3} M_{\text{tot}}^{5/3} \eta \times \begin{cases} \nu^{-1/3} \left[1 + \alpha_2 (\pi M_{\text{tot}} \nu)^{2/3}\right]^2 & \text{for} \quad \nu < \nu_1, \\ w_1 \nu^{2/3} \left[1 + \epsilon_1 (\pi M_{\text{tot}} \nu)^{1/3} + \epsilon_2 (\pi M_{\text{tot}} \nu)^{2/3}\right]^2 & \text{for} \quad \nu_1 \leq \nu < \nu_2, \\ w_2 \nu^2 \frac{\sigma^4}{(4(\nu - \nu_2)^2 + \sigma^2)^2} & \text{for} \quad \nu_2 \leq \nu < \nu_3, \end{cases}$$
$$(5.9)$$

where $\eta = m_1 m_2 / M_{\text{tot}}^2$, $\alpha_2 = -323/224 + \eta \, 451/168$, $\epsilon_1 = -1.8897$, $\epsilon_2 = 1.6557$,

$$w_1 = \nu_1^{-1} \frac{[1 + \alpha_2 (\pi M_{\text{tot}} \nu_1)^{2/3}]^2}{[1 + \epsilon_1 (\pi M_{\text{tot}} \nu_1)^{1/3} + \epsilon_2 (\pi M_{\text{tot}} \nu_1)^{2/3}]^2},$$
$$w_2 = w_1 \nu_2^{-4/3} [1 + \epsilon_1 (\pi M_{\text{tot}} \nu_2)^{1/3} + \epsilon_2 (\pi M_{\text{tot}} \nu_2)^{2/3}]^2,$$
$$(5.10)$$

and

$$\pi M_{\text{tot}} \nu_1 = (1 - 4.455 + 3.521) + 0.6437\eta - 0.05822\eta^2 - 7.092\eta^3,$$
$$\pi M_{\text{tot}} \nu_2 = (1 - 0.63)/2 + 0.1469\eta - 0.0249\eta^2 + 2.325\eta^3,$$
$$\pi M_{\text{tot}} \sigma = (1 - 0.63)/4 - 0.4098\eta + 1.829\eta^2 - 2.87\eta^3,$$
$$\pi M_{\text{tot}} \nu_3 = 0.3236 - 0.1331\eta - 0.2714\eta^2 + 4.922\eta^3.$$
$$(5.11)$$

One can compare the SGWB strength with the sensitivity curves of various GW experiments, shown in the right panel of Fig. 5.8, to make a forecast for the minimum PBH abundance, see for example Refs. [16, 178, 293, 395, 516, 517]. The result is shown in the right panel of Fig. 5.10 for the ET and LISA experiments, for both cases of Poisson distributed and clustered PBHs.

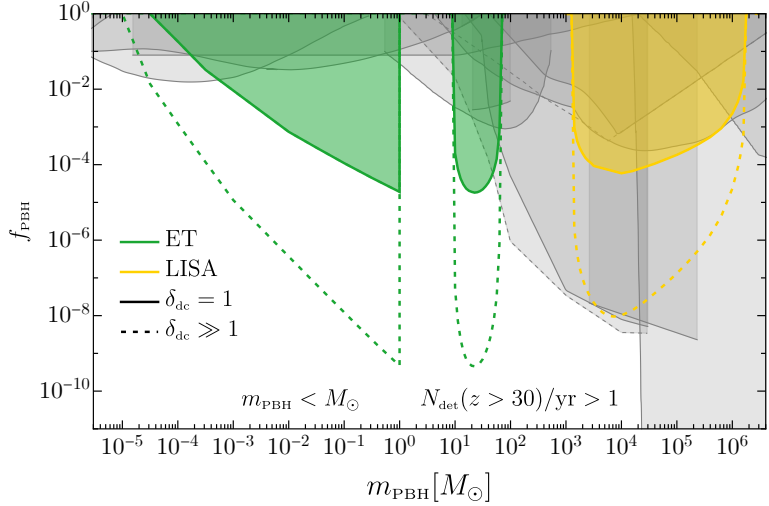

**Figure 5.11**: Minimum abundance for PBH evidence, obtained by requiring at least an observable event per year at redshift larger than 30 and including the contribution from subsolar objects. The gaps between $(1-10)M_\odot$ and $(10^2-10^3)M_\odot$ are due to the inefficiency of the ET to detect high redshift mergers with $z \gtrsim 30$. Figure taken from Ref. [10].

In the presence of strong clustering enhancing the merger rate, the forecast becomes much more stringent with respect to the Poisson distribution case. Furthermore, by comparing this result with the one obtained from the event rate, one deduces that the forecast deduced from the SGWB is less stringent, even though it applies to a broader range of PBH masses. For the LISA experiment, the bound reaches similar values of the PBH abundance, extending the regime of applicability to lighter PBHs, due to their contribution to the frequency tail $\sim f^{2/3}$ in the SGWB which may cross the LISA detectability band.

We finally notice that the SGWB depends on the evolution of the merger rate at high redshift. This implies that, given the monotonic behaviour of the PBH merger rate with redshift, contrarily to the astrophysical one which peaks around $z \simeq 2$, the two populations would give different SGWB strengths that could be used to distinguish between them when cross-correlated with the rate of resolved events [517–519].

**Minimum abundance for PBH evidence**

Finally, let us move to the computation of the minimum PBH abundance required to claim for a PBH evidence. As we have discussed above, the behaviour of the merger rate with redshift is found to follow the power law slopes $R_{\text{PBH}} \sim (t/t_0)^{-34/37}$ for the Poisson case and $R_{\text{PBH}} \sim (t/t_0)^{-1}$ for the clustered one. On the other hand, the astrophysical merger rate is characterised by a peak around redshift of a few, with a possible second peak due to the coalescence of BHs generated from Pop III clusters at redshift $z \lesssim 25$ [499].

In the conservative scenario in which these coalescences occur up to redshift $z = 30$, the detection of a binary system at larger redshifts at future GW experiments would be a smoking-gun in favour of PBHs. Similarly, PBH evidence would come from the detection of PBH masses below the Chandrasekhar's limit, where the results of Fig. 5.10 are sufficient. Even though this portion of masses overlaps with the one for a population of $\mathcal{O}(1) M_\odot$ BHs coming from transmuted neutron stars [520, 521], its mass distribution would be correlated with the neutron star one, possibly helping in distinguish the two populations.

One can therefore make a forecast for the minimum PBH abundance required to detect at least one event per year at ET and LISA considering redshifts $z \gtrsim 30$, in order to get an evidence of PBH mergers[1]. The result is shown in Fig. 5.11 for both cases of Poisson and clustered spatial distributions. Even though the overall minimum value of $f_{\text{PBH}}$ does not strongly differ with the one obtained by integrating over all redshifts, the constraint applies to a smaller range of masses. This occurs because of the shapes of the horizon redshift for the two experiments shown in Fig. 5.8, that become narrower at larger redshifts and limit the experiments capability to probe a large mass range. At the same time, the contribution coming from large masses is strongly suppressed in both cases.

## 5.2.3 Extreme mass-ratio inspirals

A promising avenue to distinguish PBHs from BHs born from the gravitational collapse of stars would be the detection of a subsolar-mass compact object, since astrophysical BHs are expected to have an initial mass larger than the Chandrasekhar one [5, 510, 523, 524].

Given the characteristic horizon redshifts of future GW experiments like LISA and 3G detectors, one can investigate the possible existence of subsolar mass BHs in special systems like extreme mass-ratio inspirals (EMRIs) around supermassive or intermediate mass BHs with unparalleled statistical confidence level (see for example Refs. [525, 526] for related studies). These systems would provide a novel source of GWs at ground-based detectors that are currently undetectable at present GW experiments (see Refs. [507–509, 527–529] for constraints on subsolar objects and Ref. [529] for bounds on the merger rate of binary systems with large mass ratios and a subsolar mass component). In the following we report the results of Ref. [3], where additional

---

[1]See also Ref. [522] for a recent analysis based on the merger rate evolution at high redshift of the primordial and Pop III populations.

details can be found.

### Setup and detectability

Given the large hierarchy between the two masses $q = \mu/M \ll 1$, one can model the evolution of an EMRI assuming BH perturbation theory [530, 531], in which a point-particle with mass lighter than a solar mass, $\mu < M_\odot$ (the secondary), performs a quasi-adiabatic orbital motion around a much heavier BH with mass $M$ (the primary). We focus on the leading-order adiabatic evolution, in which the smaller object moves in a quasicircular, equatorial orbit around a Kerr BH. Given that measurements of the waveform parameters are not strongly modified by the introduction of the spin of the subsolar component [532–536], which enters at first post-adiabatic order, and since the spin of subsolar-mass PBHs is expected to be negligible [18, 22, 261], one can neglect it in the analysis.

Using the BH Perturbation Toolkit [537], one can solve the Teukolsky equation with arbitrary precision to determine the total energy flux $\dot{E}$ emitted by the binary. The emitted flux is then responsible for the adiabatic evolution of the inspiral according to the equations for the binary radius and phase

$$\frac{dr}{dt} = -\dot{E}\frac{dr}{dE_{\rm orb}} \quad , \quad \frac{d\Phi}{dt} = \frac{M^{1/2}}{r^{3/2} + \chi M^{3/2}} \ , \tag{5.12}$$

in terms of the binary orbital energy $E_{\rm orb}$ of a particle orbiting around a Kerr BH with mass $M$ and dimensionless angular momentum $\chi$ [538], focusing on prograde orbits. The equations are integrated with initial conditions $(r_0, \Phi_0)$ in order for the secondary object to reach an orbit within a distance, $r_{\rm plunge}$, of $0.1M$ from the ISCO in a given observation time $T$.

Once the binary evolution equations have been solved, one can compute the corresponding GW signal using the quadrupole approximation [532, 533] including the effect of the detector pattern functions, which can be expressed in terms of the source orientation $(\theta_{\rm s}, \phi_{\rm s})$ and spin direction $(\theta_{\rm l}, \phi_{\rm l})$ in a solar barycentric frame, see Ref. [539] for ET and Refs. [540, 541] for LISA. One should also consider the phase modulation induced by the orbital motion [542], and an effective description of the LISA and ET triangle configuration as a network of two L-shaped detectors, with the second interferometer rotated by 45° with respect to the first one.

The corresponding GW signal in the time domain is therefore determined by the parameters $\vec{\theta} = (\ln M, \ln \mu, \chi, \ln D_L, \theta_{\rm s}, \phi_{\rm s}, \theta_{\rm l}, \phi_{\rm l}, r_0, \Phi_0)$, in terms of the luminosity distance of the source $D_L$. Assuming that an EMRI is observed for an observation time $T$, which we take to be $T = 1\,{\rm yr}$ at LISA and $T = 1\,{\rm hr}$ at ET, one can compute the corresponding SNR, where the minimum frequency is set by requiring the binary to span the frequency band up to a maximum frequency, set when the secondary object reaches $r_{\rm plunge}$.

One can then perform a Fisher analysis to estimate the precision with which one would measure the mass of the subsolar component, following the procedure outlined in Appendix C. Given that the waveforms and their derivatives are generated numerically in the time domain, we have computed the exact GW fluxes and the Fisher matrix with high-precision numerics to guarantee a stable evaluation of the covariance matrix, as done in Ref. [543].

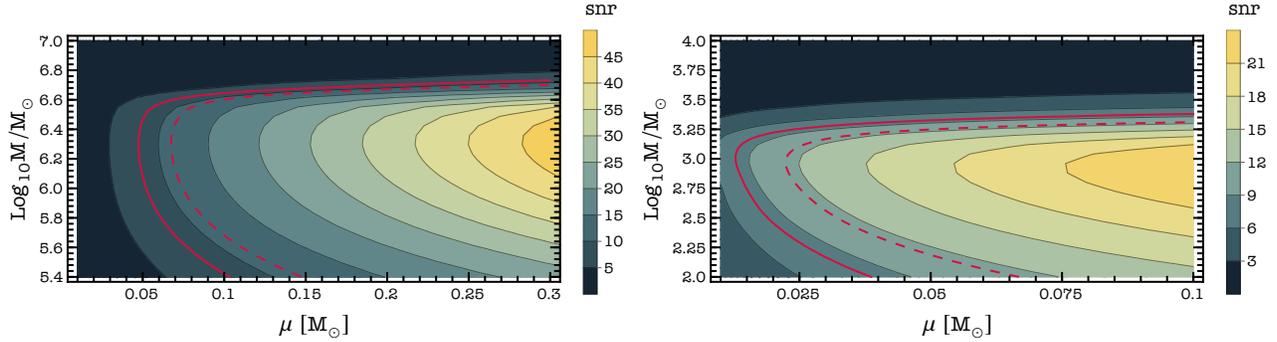

**Figure 5.12**: Signal-to-noise ratio for EMRIs with primary mass $M$ and subsolar secondary mass $\mu$ potentially observable by LISA (left) and ET (right). For both experiments we have fixed the primary spin to $\chi = 0.9$, assuming a distance $D_L = 100$ Mpc ($D_L = 1$ Gpc) and duration of the signal $T = 1$ yr ($T = 1$ hr) for LISA (ET). The red solid (dashed) line identifies systems with SNR = 8 (SNR = 11.3). Figure taken from Ref. [3].

In the left panel of Fig. 5.12 we show contour lines of fixed SNR for binary systems observed by LISA at a luminosity distance $D_L = 100$ Mpc, in terms of their component masses, assuming the spin of the primary component to be $\chi = 0.9^2$. For a given primary mass $M$ the SNR decreases for smaller values of the secondary mass $\mu$, since SNR $\sim q^3$ in the EMRI limit. Assuming a detection threshold SNR = 8 one finds that binaries as distant as $\sim 500$ Mpc can be observed by LISA. For $d = 100$ Mpc only secondary BHs heavier than $\sim 0.05 M_\odot$ can be potentially detected. However, EMRIs with $\mu \sim 0.1 M_\odot$ are characterised by a SNR > 8 for a large range of primary masses and can reach SNR as high as SNR $\approx 40$ for $M \approx 10^6 M_\odot$ and $\mu \approx 0.3 M_\odot$. Assuming a larger threshold SNR = 11.3, corresponding to the strength of a signal observed by two aligned detectors with SNR = 8, one could potentially detect masses heavier than $\sim 0.1 M_\odot$ and reach distances as large as $\sim 400$ Mpc.

The corresponding result for ET is shown in the right panel of Fig. 5.12. Given that ground-based interferometers cover a different frequency range, we focus on intermediate-mass primaries with $M \in [10^2, 10^4] M_\odot$. Due to the larger horizon redshift of ET for these sources with respect to the one of LISA for their supermassive counterparts, one can study more distant binaries (in particular, we set $d = 1\,\mathrm{Gpc}$ in the figure). EMRIs in this mass range would therefore represent a new class of astrophysical sources for 3G interferometers, that can be detected with SNRs larger than those obtained for LISA binaries. At $d \sim 100\,\mathrm{Mpc}$ ET would be able to detect EMRIs with a secondary object as light as $\mu \sim 10^{-2} M_\odot$ with SNR $\gtrsim 20$ for $M \lesssim 2.5 \times 10^3 M_\odot$. One can also notice that ET can observe BHs with a subsolar mass at cosmological distances: a given system with $(M, \mu) = (10^3, 10^{-1}) M_\odot$ would be detected at the SNR threshold up to few gigaparsec. This results from the superior sensitivity, especially at low frequency, of 3G detectors with respect to the one of present GW experiments like LIGO/Virgo. In particular, for the same sources shown in Fig. 5.12, the SNR in LIGO/Virgo is smaller by a factor 10 to 100 depending on the primary mass and spin. This implies that all sources are well below the threshold of detectability.

---

[2]We fix the angles of the source to $\theta_s = \phi_s = \pi/2$ and $\theta_l = \phi_l = \pi/4$, although a random sample of their values does not significantly affect our numerical results.

Due to their long orbital evolution, EMRIs are powerful sources to extract measurements of the binary parameters with high accuracy. One can therefore investigate the detectability of a subsolar object being the secondary component of the binary system, see for example Refs. [532, 534, 536, 542–545] for results at LISA for sources with $\mu \gtrsim M_\odot$. The corresponding constraints are shown in Fig. 5.13 for LISA and ET, for different configurations of the binary system, assuming a conservative threshold value SNR = 8 (larger thresholds result in smaller errors). We find that both ground-based and space-based interferometers can measure the mass of a subsolar secondary component with a precision of subpercent in a wide region of the observable parameter space.

In particular, LISA (left panel) will be able to identify a subsolar component as massive as $\mu \sim 0.1 M_\odot$ for systems where the primary component has a mass $M \lesssim 10^6 M_\odot$, with relative uncertainties well below 0.1%. One can therefore exclude $\mu \gtrsim M_\odot$ for these systems at more than 5-$\sigma$ confidence level. The result does not change for ET (right panel), which is able to constrain values of the secondary component mass down to $10^{-2} M_\odot$ for primary masses in the range $M \in (10^2, 10^4) M_\odot$, with relative errors $\sigma_\mu / \mu$ below 10% for $\mu \gtrsim 0.025 M_\odot$.

The behaviour of the errors with respect to the masses strongly depends on the initial frequency. In particular, for heavier primary components, the initial frequency approaches the detector reach, while for smaller $M$ it grows as $\mu$ decreases for a given observing time $T$.

Let us finally comments on some technicalities of the analysis. First, assuming a primary component with larger spin, e.g $\chi = 0.99$, would result in larger SNRs and in an improvement of the detectability horizon by a factor of a few; on the other hand, assuming $\chi = 0.8$ would decrease the SNR by a factor of a few with respect to $\chi = 0.9$, reducing the observable parameter space. The errors on $\mu$ depend on $\chi$ less significantly. Second, even though we have focused on circular equatorial orbits, it was shown that the estimated errors on the parameters do not differ dramatically when one considers eccentric and inclined orbits [532, 533, 544]. Finally, considering a network of ET plus one/two CE would improve the overall SNR and the measurement errors.

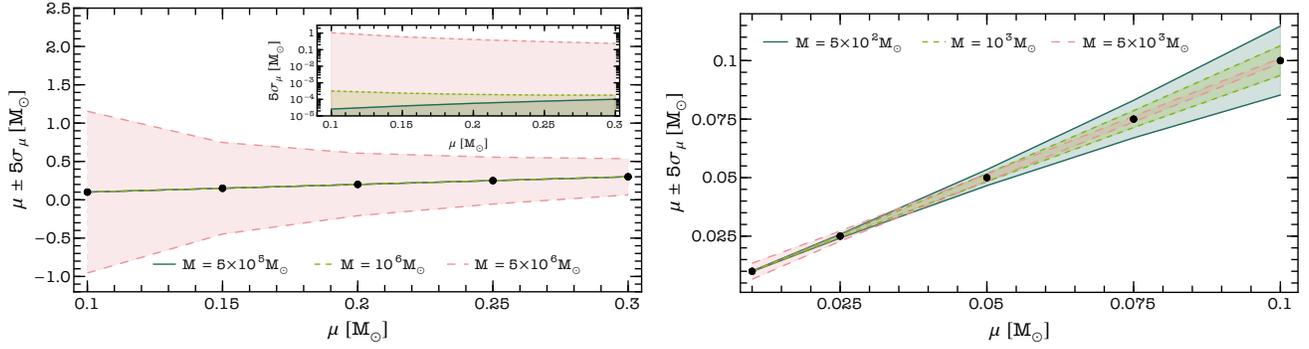

**Figure 5.13**: Injected values of the secondary component mass (black dots) and their corresponding 5-$\sigma$ interval inferred for EMRIs detected by LISA (left panel) and ET (right panel), assuming a spinning primary component with $\chi = 0.9$ and rescaling the distance such that SNR = 8. Different values of the primary mass are identified by color bands (see the inset for the 5-$\sigma$ (half) interval in logarithmic scale when they are too narrow to be resolved). Figure taken from Ref. [3].

**Interpretation of the results**

To connect this result with the detection of a PBH, one has to study the eventual role of other subsolar compact objects. A possibility is provided by white dwarfs and neutron stars, which are formed with masses respectively above $\approx 0.2 M_\odot$ [546] and $\approx M_\odot$ in standard astrophysical scenarios. Given that these objects suffer tidal deformations due to the presence of the primary component, one has to compute their Roche radius and see if they would be tidally disrupted before reaching the ISCO. For the former objects, the Roche radius becomes larger than the ISCO radius when

$$\mu \gtrsim 0.002 \left( \frac{M}{10^3 M_\odot} \right) M_\odot, \tag{5.13}$$

which implies that white dwarfs and less compact stars with masses $\mu \approx 0.2 M_\odot$ would be tidally disrupted before their plunge into a primary BH component with $M \lesssim 10^5 M_\odot$. This conclusion holds also for brown dwarfs with characteristic masses $\approx 10^{-2} M_\odot$ [547]. This implies that a confident measurement of $\mu$ well below the solar-mass scale would necessarily imply the presence of exotic new physics.

Subsolar PBHs would then provide a natural explanation, with their abundance as the dark matter in the universe currently constrained by microlensing limits [548] to be of the order of few percent [84]. In particle-dark-matter scenarios BHs born out of neutron star transmutation [520, 549] can only be super-solar, even though adding dark sector interactions may give a chance of creating such objects [550]. Other subsolar exotic compact objects may be boson stars [551], even though they should be compact enough not to be tidally disrupted. In any case, detecting an EMRI with secondary mass $\mu \ll M_\odot$ would imply new physics and should be considered in the science case for fundamental physics with LISA [36] and 3G GW detectors [429].

An important question concerns the rates of detection for subsolar BHs in EMRIs/IMRIs [544, 552]. By assuming that PBHs follow the dark-matter spatial density distribution and by rescaling standard EMRI rates to subsolar masses, Ref. [525] showed that experiments like LISA could observe sources of this kind if they comprise a few percent of the dark matter, even though further analyses are needed to make stronger conclusions. For IMRIs the situation is even worse, given that the population of intermediate-mass BHs is essentially unknown. However, the recent mass gap event GW190521 [461] shows that BHs with masses around $\mathcal{O}(100)M_\odot$ may form from hierarchical mergers. This implies that a subsolar component with mass $\mu \sim \text{few} \times 10^{-2}M_\odot$ orbiting around a primary with $M \sim \text{few} \times 10^2 M_\odot$ could be observed and identified with high confidence level by ET up to a few gigaparsec.

Finally, the inspiral of subsolar objects in binaries would generate a SGWB which could be observed by LISA [553] and 3G detectors [554]. Resolving a primordial source would break the degeneracy between the PBH abundance and mass in the slope of the stochastic spectrum [553] and possibly unveil the nature of the unresolved sources.

# Part III

# Black hole tests of fundamental physics

# Chapter 6

# Standard Model and beyond in the early universe

In the standard cosmological evolution, the inflationary era ends when reheating occurs, that is when the inflaton field decays in the Standard Model particles, which we observe in the present universe and comprise the baryonic fraction of the total energy density budget, and eventually into some physics beyond the Standard Model. The subsequent era is dominated by radiation and, as described in the previous parts of this thesis, may be characterised by a population of PBHs.

In this chapter we are going to investigate some aspects of the interplay between black holes born in the radiation-dominated era and particle physics within and beyond the Standard Model, in the context of early universe phenomena like phase transitions.

## 6.1   Phase transitions in the early universe

First-order phase transitions may have occurred in the early universe [555] and proceed with the nucleation of bubbles of the true vacuum, within the sea of false vacuum, which collide until all the available volume is occupied and trigger the end of the transition. They can have a crucial role in the evolution of the universe and leave detectable footprints like GWs [186], the baryon asymmetry [556] and PBHs [132], which all critically depend on the typical average distance between the bubbles of the broken phase before the collisions occur. It is therefore crucial to understand how these critical bubbles are spatially correlated to draw firm conclusions on the signatures coming from phase transitions.

With dedicated numerical simulations and assuming $(1 + 1)$-dimensional real-time models at zero temperature [557], it has been shown that critical bubbles are spatially clustered [558], with bubble sites distributed as maxima of a random Gaussian field [213].

In the following we report the main results of Ref. [2], where we have provided theoretical insights on the clustered spatial distribution of bubbles, which is found to be biased with respect to the underlying scalar field spatial distribution, adopting the stochastic approach to tunneling and threshold statistics. We also show that this correlation is relevant only for first-order phase transitions happening either at very high temperatures or energies, and may have implications for the generation of PBHs.

### 6.1.1 Nucleation of critical bubbles through the stochastic approach to tunneling

Let us start by considering the one-loop effective potential for a scalar field at finite temperature $T$ as [559]

$$V(\phi, T) = \frac{m^2(T)}{2}\phi^2 - ET\phi^3 + \frac{\lambda}{4}\phi^4, \tag{6.1}$$

where the thermal mass $m(T)$ is parametrised as

$$m^2(T) = 2D\left(T^2 - T_0^2\right), \tag{6.2}$$

in terms of the reference temperature $T_0$, above which the scalar field does not posses a metastable minimum at the origin, and a model dependent constant $D$. If the condition $\lambda D - E^2 > 0$ is satisfied, then a second, energetically favoured, minimum appears in the potential, see Fig. 6.1 for an example. This occurs when the temperature drops below a critical value, which corresponds to the temperature where the second minimum is degenerate with the false vacuum, i.e. [560]

$$T_c^2 = \frac{T_0^2}{1 - E^2/\lambda D}, \tag{6.3}$$

such that the corresponding thermal mass is given by

$$\frac{m^2(T_c)}{T_c^2} = \frac{2E^2}{\lambda}. \tag{6.4}$$

To understand the main properties of the spatial distribution of critical bubbles, we adopt the stochastic approach to tunnelling [69, 560–562], which we review in the following steps. First of all, critical bubbles are defined as those which are able to expand. Assuming small bubble velocities and O(3) spherically symmetric solutions, the evolution of the scalar field $\phi$ at finite temperature is described by the equation of motion

$$\ddot{\phi} = \mathrm{d}^2\phi/\mathrm{d}r^2 + (2/r)\mathrm{d}\phi/\mathrm{d}r - V'(\phi), \tag{6.5}$$

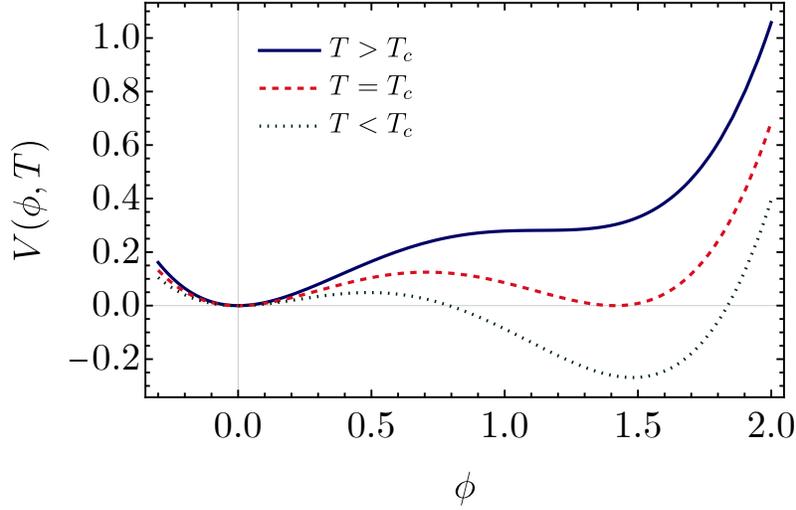

**Figure 6.1**: Effective potential for the scalar field $\phi$ at temperatures close to the phase transition $T_c$.

where the dot denotes a time derivative and the prime a derivative with respect to the field. To have processes which are energetically favourable, the field $\phi$ inside a bubble should be larger than a critical value $\phi_*$, defined as $V(\phi_*, T) = V(0, T)$. When this occurs, the effective potential has a negative derivative $V'(\phi) < 0$. Neglecting the quartic coupling, one finds that $\phi_* = m^2/2ET$ [560] (we will show later that this approximation is consistent with our results). To be conservative, one can therefore require field values beyond $\phi_{\text{th}} \sim 2\phi_* = m^2/ET$. Then the condition of expanding bubble $\ddot{\phi} > 0$ translates into

$$|\mathrm{d}^2\phi/\mathrm{d}r^2 + (2/r)\mathrm{d}\phi/\mathrm{d}r| < -V'(\phi), \tag{6.6}$$

which implies that the bubbles size has to be sufficiently large not to have the gradient terms larger than the potential-induced drift $|V'(\phi)|$.

One can make a rough estimate and get [560]

$$\frac{1}{2}r^{-2} \lesssim \phi^{-1}|V'(\phi)| \sim 2m^2(T), \tag{6.7}$$

which in momentum space gives rise to the requirement for expansion $k \lesssim k_{\text{max}} \simeq 2m(T)$. From the correlation of the scalar field $\phi$,

$$\langle \phi(\vec{k}_1)\phi(\vec{k}_2) \rangle = \frac{1}{\omega_{k_1}} \left( \frac{1}{2} + n_{\text{B}}(k_1) \right) (2\pi)^3 \delta_D(\vec{k}_1 + \vec{k}_2), \tag{6.8}$$

in terms of $\omega_{k_1} = \sqrt{k_1^2 + m^2}$ and the Bose-Einstein distribution $n_{\text{B}}$, one can compute the dispersion of its thermal fluctuations with the additional requirement of the expansion condition (i.e.

$k \lesssim k_{\max}$), to get

$$\sigma^2 \equiv \langle \phi^2 \rangle_{k < k_{\max}} \simeq \frac{1}{2\pi^2} \int_0^{k_{\max}} \frac{k^2 \mathrm{d}k}{\sqrt{k^2 + m^2} \left[ \exp\left( \frac{\sqrt{k^2 + m^2}}{T} \right) - 1 \right]} \simeq \frac{C^2 T m}{\pi^2}. \tag{6.9}$$

In the expression we have neglected the mass term since the integrand is dominated close to $k \sim k_{\max}$, and we have introduced a $\mathcal{O}(1)$ coefficient $C$ to track the uncertainties in the computation of $k_{\max}$ and the integral [560].

Assuming that the field thermal fluctuations follow a Gaussian statistics, then the field value distribution goes like

$$P(\phi) \sim \exp\left( -\frac{\phi^2}{2\sigma^2} \right). \tag{6.10}$$

Therefore, the onset of the phase transition occurs when the scalar field fluctuates, due to thermal effect, to values larger than the critical one $\phi_{\mathrm{th}}$, such that critical bubbles are generated. The corresponding nucleation probability is found by integrating the distribution of field values above the threshold, that is

$$P_1(\phi > \phi_{\mathrm{th}}) = \int_{\phi_{\mathrm{th}}} \mathrm{d}\phi \, P(\phi) \sim \exp\left( -\frac{\phi_{\mathrm{th}}^2}{2\sigma^2} \right) \sim \exp\left( -\frac{m^3 \pi^2}{2C^2 E^2 T^3} \right), \tag{6.11}$$

where we have neglected the overall factor and assumed the large threshold limit. One can also show that this result matches the tunnelling probability obtained by evaluating the Euclidean bounce solution [563]

$$P_1 \simeq \exp\left( -\frac{S_3}{T} \right) \sim \exp\left( -\frac{4.85 m^3}{E^2 T^3} \right), \tag{6.12}$$

in terms of the three-dimensional Euclidean action $S_3$, by setting $C^2 = 1.02$, value which we assume from now on.

In the context of phase transitions, it is useful to introduce some parameters describing its strength, time duration and energy. The strength is usually parametrised as a function of the order parameter $\phi(T_c)/T_c = 2E/\lambda$, which implies that strong phase transitions are obtained for small values of $\lambda$. The inverse time duration $\beta$ and the ratio of the vacuum to the radiation energy density $\alpha$ are instead defined as [564]

$$\frac{\beta}{H(T_c)} \equiv T \frac{\mathrm{d}}{\mathrm{d}T} \left( \frac{S_3}{T} \right) \bigg|_{T=T_c} = 3\sqrt{2}\pi^2 \frac{(\lambda D - E^2)}{E \lambda^{3/2}},$$

$$\alpha \equiv \frac{\rho_{\mathrm{vac}}}{\rho_{\mathrm{rad}}} \bigg|_{T=T_c} = \frac{60}{\pi^2 g_*} \frac{E^2 (\lambda D - E^2)}{\lambda}. \tag{6.13}$$

Here $H(T_c)$ denotes the Hubble parameter at the transition temperature $T_c$, $\rho_{\mathrm{vac}}(T_c) = (\Delta V - T \mathrm{d}V/\mathrm{d}T)|_{T=T_c}$ the corresponding vacuum energy density, $\rho_{\mathrm{rad}}(T_c) = \pi^2 g_* T_c^4 / 30$ the radiation

energy density and $g_*$ the effective number of degrees of freedom. At the critical temperature $T_c$, the first term in $\rho_{\text{vac}}(T_c)$ vanishes since the stable and metastable minima become degenerate. Small values of the quartic coupling $\lambda$ (consistently with the previous assumption) implies large ratio between the vacuum and radiation energy density and a shorter duration of the phase transition.

One can consider as well phase transitions of a scalar field at zero temperature, whose benchmark potential is given by

$$V(\phi) = \frac{M^2}{2}\phi^2 - \frac{\delta}{3}\phi^3 + \frac{\lambda}{4}\phi^4, \tag{6.14}$$

where the stochastic approach to tunneling now considers fluctuations of quantum nature. Neglecting again the quartic coupling, the critical value of the scalar field reads $\phi_{\text{th}} \simeq 3M^2/\delta$ and therefore

$$\sigma^2 = \frac{1}{2\pi^2}\int_0^{k_{\max}}\frac{k^2 \mathrm{d}k}{\sqrt{k^2 + M^2}} = \frac{M^2}{4\pi^2}\left[C\sqrt{1 + C^2} - \text{arc}\sinh C\right], \tag{6.15}$$

where we have parametrised $k_{\max} = CM$. The corresponding tunneling rate determined using threshold statistics matches the result obtained in terms of the four-dimensional Euclidean action $S_4$,

$$P_1 \simeq \exp\left(-S_4\right) \sim \exp\left(-\frac{205M^2}{\delta^2}\right), \tag{6.16}$$

for $C \simeq 1.2$ [69]. In this case the duration of the phase transition $1/\beta$ can be estimated by imposing that, at the end of the phase transition, the fraction of volume in the true vacuum is of order unity, that is $\Gamma/\beta^4 \sim 1$ [565, 566]. Given that the tunneling rate density is $\Gamma \sim M^4 \cdot P_1$, one finds ($M_{\text{pl}}$ being the Planck mass)

$$\frac{\beta}{H} \sim \left(\frac{M}{H}\right)\exp\left(-S_4/4\right) \sim \left(\frac{M_{\text{pl}}}{M}\right)\exp\left(-S_4/4\right). \tag{6.17}$$

## 6.1.2 The bubble correlation function

In this section we move to the estimate of the bubble two-point spatial correlation function $\xi_{\text{b}}(r)$ in terms of the distance $r$, by using the threshold statistics formalism frequently used in large-scale structure. We stress that, in the limit of large thresholds, similar results would have been obtained adopting the formalism of peak theory, as it occurs in cosmology for the galaxy bias.

The spatial density of discrete bubble nucleation centers at position $\vec{r}_i$ ($i$ indicating the various initial positions of bubbles) is given by

$$\delta_{\text{b}}(\vec{r}) = \frac{1}{\bar{n}_{\text{b}}}\sum_i \delta_D(\vec{r} - \vec{r}_i) - 1, \tag{6.18}$$

in terms of the average bubble number density $\bar{n}_{\rm b}$. The corresponding two-point correlation function takes the form [400]

$$\left\langle \delta_{\rm b}(\vec{r})\delta_{\rm b}(\vec{0}) \right\rangle = \frac{1}{\bar{n}_{\rm b}}\delta_D(\vec{r}) + \xi_{\rm b}(r), \tag{6.19}$$

where the first term accounts for the Poisson shot noise, due to the discrete nature of the bubbles, while $\xi_{\rm b}(r)$ is the (reduced) bubble correlation function. At very short distances, which are approximately identified with the critical bubble size $r_{\rm cr}$, the correlation function must satisfy the exclusion requirement, since distinct bubbles cannot form arbitrarily close to each other. This implies that the conditional probability to find a bubble at a distance $r$ from another one must vanish for $r \lesssim r_{\rm cr}$ which, being proportional to $1 + \xi_{\rm b}(r)$, results into

$$\xi_{\rm b}(r) \approx -1 \quad \text{for} \quad r \lesssim r_{\rm cr}, \tag{6.20}$$

i.e. bubbles are anti-correlated on small scales.

Let us now compute the spatial correlation function of critical bubbles. Adopting the stochastic approach to tunnelling and threshold statistics, one can estimate the correlation function of the scalar field, with values above the critical threshold $\phi_{\rm th}$, and then determine the correlation function of nucleation sites. This procedure is very similar to the one for the study of galaxy correlators: indeed, dark matter halos where galaxies end up are discrete objects formed when the dark matter overdensity is larger than a given threshold value and the galaxy correlators are biased with respect to the dark matter ones [567]. In the same way, critical bubbles are formed when the scalar field has a value larger than a given threshold, and critical bubbles will be biased compared to the underlying scalar field spatial distribution.

The threshold correlation function is defined to be

$$1 + \xi_{\rm b}(r) = \frac{P_2}{P_1^2}, \tag{6.21}$$

in terms of the probability $P_1$ of one region being above threshold, estimated in Eq. (6.11), and the probability $P_2$ that two regions separated by $r$ are both above threshold, given by [568]

$$P_2 = \int_\nu^\infty \frac{{\rm d}x_1 {\rm d}x_2}{2\pi} \frac{1}{\sqrt{1-w^2}} \exp\left[-\frac{x_1^2 + x_2^2 - 2wx_1x_2}{2(1-w^2)}\right], \tag{6.22}$$

as a function of the dimensionless threshold $\nu \equiv \phi_{\rm th}/\sigma$, field values $x_i \equiv \phi(r_i)/\sigma$ and rescaled scalar field correlation function $w(r) = \xi_\phi(r)/\sigma^2$. The threshold correlation function takes the form

$$1 + \xi_{\rm b}(r) \approx (1+w)\frac{{\rm erfc}\left(\sqrt{\frac{1-w}{1+w}}\,\nu/\sqrt{2}\right)}{{\rm erfc}(\nu/\sqrt{2})} \quad \text{for} \quad \nu \gg 1, \tag{6.23}$$

which reduces to the well-known result [568]

$$\xi_{\rm b}(r) \equiv b_1^2 \xi_\phi(r) \simeq \left( \frac{\phi_{\rm th}}{\sigma^2} \right)^2 \xi_\phi(r), \tag{6.24}$$

in the limit of $w \ll 1/\nu^2 \ll 1$. One can immediately notice that the bias factor is proportional to the order parameter of the phase transition, which implies that strong first-order phase transitions, with large order parameters, would lead to strong bubble correlations and clustered bubbles. At finite temperature $T$, the scale independent bias factor takes the value

$$b_1^2 = \frac{\pi^4 m^2}{E^2 T^4}, \tag{6.25}$$

such that at the critical temperature $T_c$ becomes

$$b_1^2(T_c) \sim \frac{2\pi^4}{\lambda T_c^2} \gg \frac{1}{T_c^2}. \tag{6.26}$$

To determine the threshold correlation function $\xi_{\rm b}$, one has to compute the scalar field correlation function, that is given by

$$\xi_\phi(r) = \frac{1}{2\pi^2} \int_0^{k_{\rm max}} {\rm d}k k^2 \frac{j_0(kr)}{\sqrt{k^2+m^2} \left[ \exp\left( \frac{\sqrt{k^2+m^2}}{T} \right) - 1 \right]}. \tag{6.27}$$

By using the property of the spherical Bessel function to be constant up to momenta $kr < 1$, $j_0(kr) \sim 1$, and to decrease rapidly otherwise, in the relevant limit $r \gtrsim 1/k_{\rm max} \simeq 1/2m$ one can neglect the momentum scale with respect to the scalar field thermal mass $m$ and get

$$\xi_\phi(r) \simeq \frac{1}{2\pi^2} \int_0^{k_{\rm max}} \frac{k^2 {\rm d}k}{m} \left[ \exp\left( \frac{m}{T} \right) - 1 \right]^{-1} j_0(kr) \simeq \frac{2Tm}{\pi^2} \frac{j_1(2mr)}{mr}. \tag{6.28}$$

We stress that the opposite regime $r < 1/k_{\rm max}$ is not relevant for our discussion, since it corresponds to distances smaller than the minimum size of an expanding bubble.

The same quantities can be computed for the potential at zero temperature given in Eq. (6.15) to find

$$b_1^2 \simeq \frac{192\pi^4}{\delta^2}, \qquad \xi_\phi(r) \simeq \frac{M^2}{2\pi^2} \frac{j_1(Mr)}{Mr}. \tag{6.29}$$

Notice that small values of $\delta$ lead to a strong phase transitions and large bias.

**The bubble clustering scale**

Once the expression for the bubble correlation function is determined, one can estimate the corresponding clustering length by imposing the condition $\xi_{\rm b}(r) \sim 1$. The latter requirement can be understood by considering the counts of neighbours. In particular, the average count of bubble nucleation sites $\langle N \rangle$ in a cell of volume $V$ centered on a bubble is given by the expression

$$\langle N \rangle = \bar{n}_{\rm b} V + \bar{n}_{\rm b} \int_V {\rm d}^3 r \, \xi_{\rm b}(r). \tag{6.30}$$

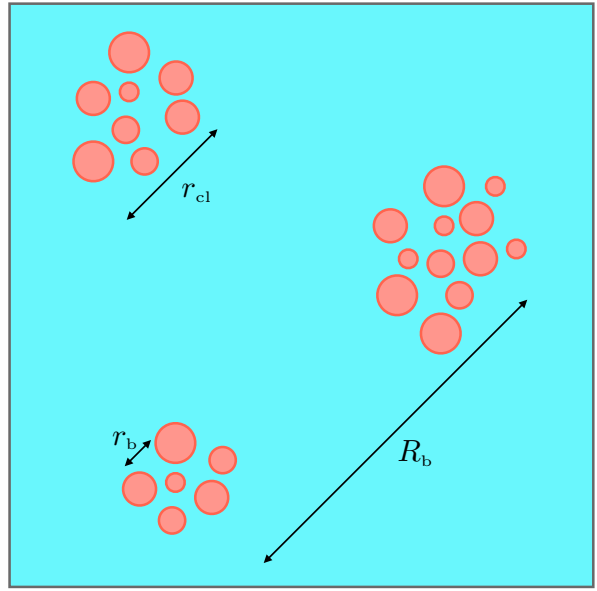

**Figure 6.2**: Pictorial representation of the clustering of bubbles showing the various length scales of the problem. Figure taken from Ref. [2].

It can significantly deviate from the Poisson term if the contribution from the reduced correlation function is larger than the discreteness noise $\bar{n}_{\rm b} V$. This can occur on scales $r \lesssim r_{\rm cl}$, defined to be the relevant clustering length through the relation $\xi_{\rm b}(r_{\rm cl}) = 1$.

Adopting the large scale limit for the peak correlation function $\xi_\phi$, and neglecting the periodic oscillations induced by the Bessel function, one finds

$$\left( \frac{\phi_{\rm th}^2}{\sigma^2} \right) (r_{\rm cl} k_{\rm max})^{-2} \sim 1, \tag{6.31}$$

which results into the values

$$r_{\rm cl} = \begin{cases} \frac{\pi}{2^{1/4}} \left( \frac{E^{1/2}}{\lambda^{3/4}} \right) \frac{1}{m} & \text{at finite temperature,} \\ \frac{6\sqrt{2}\pi}{3^{1/4}\delta} & \text{at zero temperature.} \end{cases} \tag{6.32}$$

### 6.1.3   On the clustering of bubbles

The computation of the clustering length does not automatically imply that the spatial correlation is relevant and has a physical impact on the system. In particular, by looking at Eq. (6.30), one can appreciate that the discrete Poisson term depends on the volume centered around a given bubble and increases with distances like $\sim r^3$, while the term which depends on the reduced correlation function scales with distances like $\sim r$, since the correlation function goes like $\sim 1/r^2$, and the two contributions become comparable at $\sim r_{\rm cl}$. However, one should also compute the average number of bubbles in a volume of size the correlation length and see if it is larger than unity. The latter can be computed as

$$\langle N \rangle \sim \bar{n}_{\rm b} r_{\rm cl}^3 \sim \frac{1}{v^3} \left( \beta r_{\rm cl} \right)^3 = \frac{1}{v^3} \left( \frac{\beta}{H} \right)^3 \left( \frac{r_{\rm cl}}{H^{-1}} \right)^3 \tag{6.33}$$

in terms of the bubble velocity $v$, such that at finite temperatures it has the value

$$\langle N \rangle \sim \frac{1}{v^3} \frac{(\lambda D - E^2)^3}{E^{9/2} \lambda^{21/4}} \left( \frac{T_c}{M_{\rm pl}} \right)^3, \tag{6.34}$$

while at zero temperature it becomes

$$\langle N \rangle \sim \frac{1}{v^3} \left( \frac{M}{\delta} \right)^6 \left( \frac{H}{\delta} \right)^3. \tag{6.35}$$

From these expressions one can immediately understand that only for high-energy first-order phase transitions the mean number of bubbles at the cluster scale is larger than unity. In particular, in the finite temperature case for $v \sim 0.1$, $D \sim E \sim \lambda \sim 10^{-2}$, corresponding to $(\beta/H) \sim 10^2$, one gets

$$T_c \gtrsim 10^{-5} \left( \frac{10^2}{\beta/H} \right) M_{\rm pl} \simeq \left( \frac{10^2}{\beta/H} \right) 10^{13} \, \text{GeV}, \tag{6.36}$$

while in the zero temperature case one finds, for $v \sim 1$ and $\delta \sim \epsilon M$ (with $\epsilon$ a coefficient smaller than unity), $H \sim M^2/M_{\rm pl} \gtrsim \epsilon^3 M$, or

$$M \gtrsim \epsilon^3 M_{\rm pl}. \tag{6.37}$$

These two conditions for the clustering of bubbles are quite tight, since the mean number density of bubbles tends to be small in a volume with size the cluster length, while the typical distance between bubbles is comparable to the Hubble radius and therefore much larger than the clustering length. This picture is similar to the one for the clustering of PBHs at formation. Indeed, since the formation of PBHs is a rare event, characterised by a small average number density, their mean number in a volume of size the clustering length is smaller than unity. PBHs are therefore not clustered at formation [401], and a similar conclusion may occur for critical expanding bubbles if the phase transition does not occur at high enough energies. We finally stress that the effect of bubble clustering is more relevant for strong phase-transitions whose duration is short [558].

## Possible consequences of the clustering of bubbles

The clustering of bubbles may have some direct consequences on phenomena which could occur during first-order phase transitions.

One possibility is provided by the generation of GWs [186, 558], which could be produced by several mechanisms like bubble collisions, turbulence and sound waves close to the bubble walls. Indeed, all of them depend on the average distance between bubbles, both in the amplitude and in the peak frequency of the GWs. If bubbles are spatially correlated at the time of nucleation, they will collide with a typical average distance $r_b$ which is smaller than the Poisson case $R_b$. From Eq. (6.30) one finds

$$r_b \sim \frac{1}{(\bar{n}_b \xi_b)^{1/3}} \ll \frac{1}{\bar{n}_b^{1/3}} \sim R_b. \tag{6.38}$$

Only after these first collisions, the few resulting bubbles will collide at a larger distance $R_b \sim \bar{n}_b^{-1/3}$, see Fig. 6.2 for a pictorial representation. One expects that the GW spectrum has two peaks at frequencies $f_{GW} \sim 1/r_b$ and $\sim 1/R_b$, separated by an amount $\sim \xi_b^{1/3}$, with the GW amplitude at $f_{GW} \sim 1/r_b$ smaller than the one at $f_{GW} \sim 1/R_b$, since it depends on powers of the bubble average distance for all mechanisms. Plugging numbers in, one finds that the clustering of bubbles could be important in generating a peak at frequencies around the GHz range, where detecting GWs is rather challenging [569][1].

A second possibility is provided by the formation of PBHs from the collisions of bubbles [132, 570, 571] in supercooled phase transitions. In particular, PBHs may be generated if a large amount of energy is deposited in bubble walls and it is concentrated within its corresponding Schwarzschild radius, satisfying the hoop conjecture. This condition requires the collision of many bubbles, for which clustering may therefore help. Following the results of Ref. [132], one can write down the condition to form a PBH in a volume of size the clustering length as

$$\langle N \rangle \gtrsim \frac{4}{r_{cl} H}, \tag{6.39}$$

which implies, for our model parameters,

$$M \lesssim \frac{H}{\epsilon^{5/2}} \quad \text{or} \quad M \gtrsim \epsilon^{5/2} M_{pl}. \tag{6.40}$$

As expected, this condition is parametrically more stringent than the one shown in Eq. (6.37). The corresponding PBHs would have a typical mass of the order of $\sim M_{pl}/\epsilon^6 \langle N \rangle$ and therefore quickly evaporate due to Hawking radiation.

## 6.2 Sphaleron transitions in the presence of black holes

One of the features of the Standard Model (SM) of electroweak interactions is that the baryon ($B$) and lepton ($L$) symmetries are accidental, whose corresponding charges cannot be violated at any order in perturbation theory.

---

[1]The typical GW frequency is proportional to $(\beta/H)T_c$ [186] and therefore is not modified by changing $(\beta/H)$ when the condition in Eq. (6.36) is imposed.

There are however non-perturbative effects which could let to the violation of the baryon and lepton numbers. In particular, the non-abelian nature of the gauge group $SU(2)_L$ implies that the ground state of the theory is a sum over an infinite number of vacua, which are degenerate at a classical level and have different baryon (and lepton) numbers. There exist static configurations called sphalerons [572], which are unstable solutions of the equations of motion and saddle points of the energy functional, which interpolate between two nearby vacua and give rise to a non-zero, even though exponentially suppressed, probability of violating baryon number in the vacuum as [573]

$$\Gamma_B \sim e^{-4\pi/\alpha_W} \sim e^{-150}, \tag{6.41}$$

in terms of the $SU(2)_L$ gauge coupling constant $\alpha_W = g_2^2/4\pi$. Such an exponential suppression can be interpreted as the probability of performing a transition from one classical vacuum to the closest one by quantum tunneling, through a barrier of energy about $E_{sph} \sim 10$ TeV, thanks to the formation of a sphaleron.

Even though this picture is valid in the vacuum, the presence of a thermal bath with high temperatures may trigger classical transitions, which could be responsible for baryon and lepton number violation processes. These can occur in the primordial universe and may play a role in the generation of the baryon asymmetry [556].

It has been argued that gravity is responsible for the violation of all global symmetries, with global charges swallowed by BHs according to the no-hair theorems [574]. In particular, quanta with global charge may scatter with a BH, resulting into a BH with a slightly larger mass, but indeterminate global charge. Furthermore, BHs are gravitational impurities and, as such, they can catalyse tunnelling processes. Indeed, in the standard interpretation of Coleman-De Luccia, the nucleation of a critical bubble is the result of a balance between the energy used to form a wall between the false and true vacuum, and the energy gain from having a lower energy in the true vacuum. When a BH is present, the bubble forms around it, and the distortion of space lowers the cost to form bubbles: the instanton has a smaller action and the decay process is significantly enhanced.

This picture has been used to show that BHs can trigger electroweak SM vacuum instability in their vicinity, both at zero temperature [575–578] and in the early universe [579–584], and baryon number violations through interactions with skyrmions [585, 586].

One can therefore investigate baryon number violation processes induced by SM sphaleron transitions in the presence of BHs. Their typical Schwarzschild radius to have an impact on the rate of baryon number violation is of the order of

$$r_s = 2GM_{BH} \sim \frac{1}{M_W} \sim \frac{1}{g_2 v}, \tag{6.42}$$

in terms of the Vacuum Expectation Value (VEV) of the Higgs field $v = 246$ GeV. This leads to

BH masses in the ballpark of

$$M_{\text{BH}} = \mathcal{O}(1) \cdot 10^{-22} M_{\odot} \sim 10^{17} M_{\text{Pl}}, \tag{6.43}$$

which correspond to BHs that evaporate with a typical lifetime of $\mathcal{O}(1)$ yr and might have been present during the evolution of the Universe. In the following we show that these processes in the presence of such BHs can be faster than in the pure vacuum, reporting the results of Ref. [9].

## 6.2.1 Baryon number violation seeded by BHs: formalism

We consider the action of the Higgs doublet field $\phi$ along with a $SU(2)_L$ gauge field $W_\mu^a$ in a curved spacetime

$$S = \int_{\mathcal{M}} \mathrm{d}^4 x \sqrt{-g} \left[ \frac{\mathcal{R}}{16\pi G} - g^{\mu\nu}(D_\mu \phi)^\dagger D_\nu \phi - V(\phi)\frac{1}{4}g^{\mu\rho}g^{\nu\sigma}F^a_{\rho\sigma}F^a_{\mu\nu} \right] + \frac{1}{8\pi G} \int_{\partial\mathcal{M}} \mathcal{K}\sqrt{\gamma}\mathrm{d}^3 y, \tag{6.44}$$

in terms of the Higgs potential $V(\phi)$, the Ricci scalar $\mathcal{R}$ and the Gibbons-Hawking-York boundary term, dependent on the extrinsic curvature $\mathcal{K}$, since we are considering a spacetime manifold $\mathcal{M}$ with a BH horizon. We stress that the inclusion of the abelian hypercharge group $U(1)_Y$ does not affect our results. The geometry of the spacetime around the BH is assumed to be static and spherically symmetric, such that its metric takes a Schwarzschild-like form

$$\mathrm{d}s^2 = -e^{2\delta(r)}A(r)\mathrm{d}t^2 + A^{-1}(r)\mathrm{d}r^2 + r^2\mathrm{d}\Omega^2, \qquad A(r) = 1 - \frac{2GM(r)}{r}, \tag{6.45}$$

where the function $A(r)$ vanishes at the horizon

$$r_{\text{s}} \equiv 2GM(r_{\text{s}}) = 2GM_{\text{BH}}. \tag{6.46}$$

Following Ref. [572], we take the ansatz for the gauge and Higgs field

$$W_i^a \sigma^a \mathrm{d}x^i = -\frac{2i}{g_2}f(g_2 vr)\mathrm{d}U^\infty (U^\infty)^{-1},$$

$$\phi = \frac{v}{\sqrt{2}}h(g_2 vr)U^\infty \begin{pmatrix} 0 \\ 1 \end{pmatrix}, \tag{6.47}$$

in terms of the matrix

$$U^\infty = \frac{1}{r}\begin{pmatrix} z & x+iy \\ -x+iy & z \end{pmatrix}. \tag{6.48}$$

To compute the energy functional, one can analytically continue the action to the Euclidean metric with $t = i\tau$, with $\tau$ periodic with period $1/T$ (which has to be identified with the relevant temperature of the system).

By introducing the dimensionless parameter $\xi = g_2 vr$, and expanding the mass around its value at the horizon

$$M(\xi) = M_{\text{BH}} + \delta M(\xi),\tag{6.49}$$

one gets the equations of motion

$$\delta M' = \frac{4\pi v}{g_2}\left[\frac{\xi^2}{2}A(h')^2 + \epsilon\frac{\xi^2}{4}(h^2-1)^2 + 4A(f')^2 + \frac{8}{\xi^2}f^2(1-f)^2 + \xi^2\mathcal{F}(\xi)\tilde{\epsilon}h^2 + h^2(1-f)^2\right],$$

$$\delta' = 2\alpha^2\left[\frac{8}{\xi}(f')^2 + \xi(h')^2\right],$$

$$(Ae^\delta f')' = \frac{2}{\xi^2}e^\delta f(1-f)(1-2f) - \frac{1}{4}e^\delta h^2(1-f),$$

$$(\xi^2 Ae^\delta h')' = \epsilon\xi^2 e^\delta h(h^2-1) + 2e^\delta h(1-f)^2 + e^\delta\xi^2\mathcal{F}(\xi)\tilde{\epsilon}h,\tag{6.50}$$

in terms of the constants $\alpha = \sqrt{4\pi G}(v/\sqrt{2})$, $\epsilon = \lambda/g_2^2$ and $\tilde{\epsilon} = \tilde{\lambda}/768\pi^2$. We have introduced the Higgs quartic coupling $\lambda$ and wrote its potential as

$$V(h,\xi) \simeq \frac{\lambda}{4}v^4(h^2-1)^2 + \frac{\tilde{\lambda}g_2^2}{768\pi^2}\mathcal{F}(\xi)v^4 h^2.\tag{6.51}$$

The second term in the potential comes from vacuum polarization effects, due to the Hawking radiation, originating at one-loop because of the interactions of the Higgs with the other SM particles close to the BH horizon [587]. It is very similar to the corrections at finite temperature to the Higgs mass squared $\sim T^2 h^2$ in a plasma with temperature $T$. The crucial difference is the dependence of the effective temperature on the distance from the horizon [588, 589] (introducing the dimensionless BH horizon $\xi_{\text{s}} \equiv g_2 vr_{\text{s}}$)

$$\mathcal{F}(\xi) = \begin{cases} \frac{3}{4\xi_{\text{s}}^2} & \xi \simeq \xi_{\text{s}}, \\ \frac{1}{\xi^2} & \xi \gg \xi_{\text{s}}, \end{cases}\tag{6.52}$$

such that, close to the horizon, the correction has the form $T_{\text{H}}^2 h^2$, in terms of the Hawking temperature [590]

$$T_{\text{H}} = \frac{1}{8\pi GM_{\text{BH}}} \simeq 10\left(\frac{M_{\text{BH}}}{5\cdot 10^{-22}\,M_\odot}\right)^{-1}\text{GeV}.\tag{6.53}$$

Since in our physical situation the universe temperature may be different from the Hawking one, the Hartle-Hawking vacuum [591] is not the proper vacuum to consider, since it assumes full and static thermal equilibrium with the surrounding plasma. We will therefore assume the Unruh vacuum [592] as the most appropriate vacuum. The effective coupling $\tilde{\lambda}$ then reads

$$\tilde{\lambda} = 24\left(\frac{3}{16}g_2^2 + \frac{1}{16}g_1^2 + \frac{1}{4}y_t^2 + \frac{\lambda}{2}\right) \sim 9.56,\tag{6.54}$$

as a function of the gauge coupling $g_2$ of $SU(2)_L$, $g_1$ of the $U(1)_Y$ group, and top Yukawa coupling $y_t$, all evaluated at the electroweak scale [593].

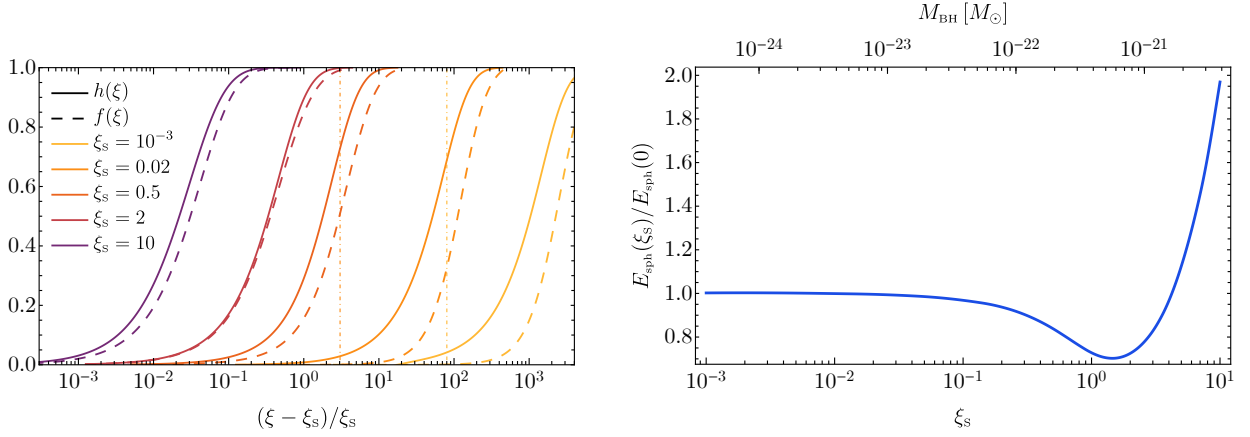

**Figure 6.3**: *Left*: Higgs and $SU(2)_L$ fields in terms of the radial coordinate for different values of the rescaled horizon $\xi_s$. The vertical lines indicate the radial distance at which the Hawking temperature can restore the symmetry. *Right*: Rescaled sphaleron energy for different BH masses. Figure taken from Ref. [9].

Given that $\alpha \ll 1$, one can approximate $\delta' \simeq 0$, whose leading order solution gives $\delta \simeq 0$, since the metric has to approach the Minkowski spacetime at infinity. The equations of motion for the gauge and Higgs fields then read

$$f'' + \frac{\xi_s}{\xi(\xi - \xi_s)} f' = \frac{2}{\xi(\xi - \xi_s)} f(1-f)(1-2f) - \frac{1}{4} \frac{\xi}{\xi - \xi_s} h^2(1-f),$$

$$h'' + \left( \frac{2}{\xi} + \frac{\xi_s}{\xi(\xi - \xi_s)} \right) h' = \frac{\xi}{\xi - \xi_s} \epsilon h(h^2 - 1) + \frac{2\xi \mathcal{F}(\xi)}{(\xi - \xi_h)} \tilde{\epsilon} h + \frac{2}{\xi(\xi - \xi_s)} h(1-f)^2. \tag{6.55}$$

Proper boundary conditions have to be imposed to correctly solve the equations of motion. At spatial infinity the metric has to approach the Minkowski spacetime and the fields have to be in their true vacuum,

$$f(\xi) \to 1, \quad h(\xi) \to 1, \quad \text{when} \quad \xi \to \infty, \tag{6.56}$$

while at the BH horizon $\xi \to \xi_s$, one can set the fields in the false vacuum

$$f(\xi) \to 0, \quad h(\xi) \to 0, \quad \text{when} \quad \xi \to \xi_s. \tag{6.57}$$

The numerical solutions of the equations of motion are shown in the left panel of Fig. 6.3 for different rescaled BH horizons. For light enough BHs, there exists a critical radius below which vacuum polarization effects induced by the Hawking radiation can restore the symmetry close to the BH horizon, nevertheless allowing for a sphaleron solution which interpolates between the false and true vacua.

The mass contribution at spatial infinity is given by

$$\delta M_\infty = \frac{4\pi v}{g_2} \int_{\xi_S}^\infty d\xi \left[ \frac{1}{2}\xi(\xi - \xi_S)(h')^2 + \epsilon\frac{\xi^2}{4}(h^2 - 1)^2 + \frac{4}{\xi}(\xi - \xi_S)(f')^2 + \frac{8}{\xi^2}f^2(1 - f)^2 \right.$$
$$\left. + \xi^2 \mathcal{F}(\xi)\tilde{\epsilon}h^2 + h^2(1 - f)^2 \right], \tag{6.58}$$

which is interpreted as the sphaleron energy in the presence of a BH,

$$\delta M_\infty = E_{\text{sph}}(M_{\text{BH}}). \tag{6.59}$$

In particular, in the absence of BHs and in the limit of flat spacetime, $(r_S \ll 1/g_2 v)$, one gets

$$E_{\text{sph}}(0) \simeq 1.92 \frac{4\pi v}{g_2}, \tag{6.60}$$

that agrees with the standard result for the current value of the Higgs mass (that is, for $\epsilon \simeq 0.3$, see right panel of Fig. 6.3. We stress that vacuum polarization effects in the Higgs potential have a minor role for light BH masses, since the radius of the sphaleron configuration is located away from the Schwarzschild radius.

For light BHs, the sphaleron radius is larger than the Schwarzschild radius, and its energy is only slightly perturbed with respect to the vacuum solution. As one considers heavier BHs, the sphaleron radius approaches the horizon and the BH helps catalysing the sphaleron transitions. For even larger masses, generating sphaleron solutions is even more costly, since the corresponding size has to be larger than the BH horizon and therefore larger than $\sim 1/g_2 v$. One can also appreciate that the minimum BH mass is consistent with the estimate shown in Eq. (6.43).

## 6.2.2 Baryon number violation seeded by BHs: rate

The baryon number violation occurs with a decay rate given by [594]

$$\Gamma_{\text{sph}}(M_{\text{BH}}) \sim \sqrt{\frac{B}{2\pi}}\frac{1}{\ell_{\text{sph}}}e^{-B}, \tag{6.61}$$

in terms of the size of the sphaleron configuration $\ell_{\text{sph}}$. For heavy BHs, it is comparable to the Schwarzschild radius $r_S$, while for light BHs it is of the order of $1/g_2 v$. The dimensionless term $\sqrt{B/2\pi}$ comes from the normalization of the zero mode associated with the symmetry of time translation, as a function of the bounce exponent $B$, given by the difference between the Euclidean action of the bounce solution and the one before the transition. For a static solution, the bounce is equal to the difference between the areas of the BH at the horizon and at infinity. Given that the bulk part of the energy functional vanishes due to the Hamiltonian constraint, the boundary terms give the Bekenstein-Hawking entropy at the BH horizon [594]

$$B = -\frac{\mathcal{A}_S}{4G} + \frac{\mathcal{A}_\infty}{4G} = 4\pi G\left[ M_\infty^2 - M_{\text{BH}}^2 \right]. \tag{6.62}$$

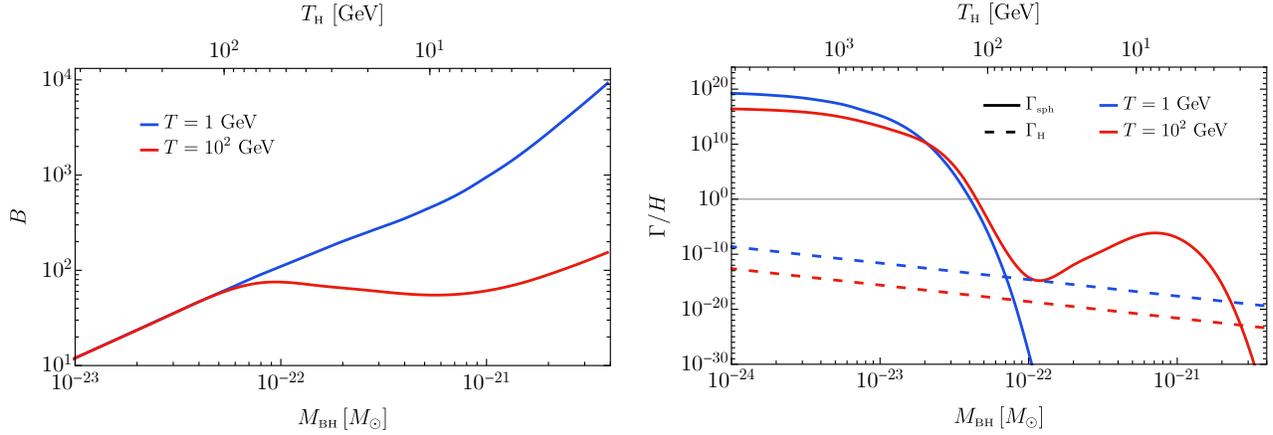

**Figure 6.4**: *Left*: Bounce exponent as a function of the BH mass (and corresponding Hawking temperature) for different plasma temperatures $T$. *Right*: Ratio between the sphaleron rate in the presence of a BH of mass $M_{BH}$ and the Hubble rate $H$, compared to the evaporation rate $\Gamma_H$. For light BH masses, the sphaleron rate decreases as $\sqrt{B}$, while for heavier BHs it has a bump due to the plasma temperature effect. Figure taken from Ref. [9].

For the BH mass range we are interested in, one can expand at leading order $M_\infty = M_{BH} + \delta M_\infty$, such that the previous expression becomes

$$B \simeq 8\pi G M_{BH} \delta M_\infty \simeq \frac{E_{sph}(M_{BH})}{T_H}, \tag{6.63}$$

where in the last step we have introduced the BH temperature $T_H$. This result shows that the exponential bounce factor $\exp(-B)$ for the sphaleron transition may be interpreted as the standard Boltzmann suppression factor in a thermal environment with temperature given by the Hawking temperature. This interpretation allows for a smooth interpolation between the zero and the finite temperature limits. For a BH immersed into a plasma at finite temperature $T$, that is the standard setting for PBHs, the sphaleron baryon number violating rate goes as $\exp(-E_{sph}(M_{BH})/T)$ for $T \gtrsim T_H$, fitting the exponential factor in Eq. (6.61) at temperature $T \sim T_H$. In this limit one must adopt the Higgs field VEV at finite temperature $v(T)$, see Refs. [559, 595].

In Fig. 6.4 we show on the left panel the bounce exponent $B$ in terms of the BH mass and for two choices of the plasma temperature. One can appreciate that, for light enough BHs, the bounce $B$ can be much smaller than the vacuum bounce $4\pi/\alpha_w \sim 150$, reaching values of order unity for masses $M_{BH} \sim 10^{-24} M_\odot$, below which the validity of the computation breaks down. For heavier BHs, the thermal bath temperature dominates over the Hawking one, leading to a suppression of the bounce exponent.

### 6.2.3 Baryon number violation seeded by BHs: some considerations

PBHs with masses around $10^{-22} M_\odot$ could have been generated during the early universe with an abundance $f_{\mathrm{PBH}} \lesssim 10^{-4}$ to avoid constraints from Big Bang nucleosynthesis [84, 596].

If these PBHs are generated from the collapse of sizable overdensities within a horizon volume, their formation temperature is [83]

$$T_{\mathrm{f}} \simeq 10^{10} \left( \frac{M_{\mathrm{BH}}}{5 \cdot 10^{-22} M_\odot} \right)^{-1/2} \mathrm{GeV}. \tag{6.64}$$

PBHs with small masses are therefore born with a Hawking temperature which is smaller than the corresponding temperature of the plasma. The evaporation rate for the considered range of masses is given by [90]

$$\Gamma_{\mathrm{H}} \sim 4 \cdot 10^{-33} \left( \frac{10^{-22} M_\odot}{M_{\mathrm{BH}}} \right)^3 \mathrm{GeV}. \tag{6.65}$$

This implies that, in first approximation, the PBH masses may be considered as constant in time for our considerations. In the right panel of Fig. 6.4 we show a comparison between the baryon number violation rate and the evaporation rate, both in terms of the Hubble rate. One can notice that the evaporation rate becomes important only for BH lighter than $10^{-28} M_\odot$, for which evaporation is effective at temperatures around $100\,\mathrm{GeV}$.

Let us now outline the main picture of sphaleron transitions during the cosmic history. At very high temperatures, thermal fluctuations induce unsuppressed baryon number violation processes through sphaleron transitions before the electroweak phase transition [556] which, in the SM, occurs at $T_{\mathrm{EW}} \simeq 163$ GeV for the current value of the Higgs mass. At smaller temperatures and far from the PBHs, the sphalerons are inactive and baryon number violation is suppressed by the exponential term, $\exp(-E_{\mathrm{sph}}(0)/T)$. However, even after the electroweak phase transition, baryon number violation can occur efficiently around PBHs, with a rate faster than the Hubble rate, see the right panel of Fig. 6.4, where for each BH mass we have taken the maximum between the temperature of the plasma and the Hawking temperature to evaluate the rate.

One may wonder if this picture represents a problem for those scenarios where the baryon asymmetry of the universe is generated before or at the electroweak phase transition. At the formation time, the fraction of PBHs per horizon is given by [83]

$$\beta(T_{\mathrm{f}}) \simeq 10^{-19} \left( \frac{M_{\mathrm{BH}}}{5 \cdot 10^{-22} M_\odot} \right)^{1/2} f_{\mathrm{PBH}}, \tag{6.66}$$

which is constrained by Big Bang nucleosynthesis limits to be $\beta(T_{\mathrm{f}}) \lesssim 10^{-23}$ for the mass range of interest [596]. The number $\mathcal{N}$ of causally independent regions with temperature $T$, which are currently within our horizon, is given by $\mathcal{N} \sim 10^{34} (T/\mathrm{GeV})^3$. This implies that the PBH number density at a given temperature $T$, normalized to the photon number density $n_\gamma$, is approximately given by

$$\frac{n_{\text{PBH}}}{n_\gamma}\frac{1}{\eta} \sim 10^{-34} \left(\frac{\eta}{10^{-9}}\right)^{-1} \left(\frac{M_{\text{BH}}}{5 \cdot 10^{-22} M_\odot}\right)^{-1} f_{\text{PBH}}, \tag{6.67}$$

in terms of the baryon asymmetry $\eta = n_{\text{b}}/n_\gamma$, normalised to the current constrained value [595]. Given that the PBH density is too small, one finally concludes that there is no impact on a pre-existing baryon asymmetry.

# Chapter 7

# Fundamental physics in gravitational waves: tidal Love number

In the previous chapter we have investigated some phenomena on the interplay between a population of black holes and fundamental particle physics, that may occur during the evolution of the universe. A crucial connection between these two topics is provided by gravitational wave observations. In particular, one may expect that the presence of a condensate of massive scalar fields around coalescing BHs could leave detectable imprints on the corresponding gravitational waveform through effects of tidal deformability. The latter is usually measured in terms of the tidal Love numbers (TLNs) [597, 598], which depend on the internal structure of the deformed body immersed in the external tides and impact the latest stages of the inspiral phase before the merger [195]. TLNs have already been computed for compact objects like neutron stars [599], for which they can be used to set stringent constraints on their equation of state using binary neutron-star coalescences [192, 196, 600–623], and BHs in isolation, for which they are found to be exactly zero [624–636].

Reporting the results of Ref. [4], we will show that the TLNs of dressed BHs, whose halos have grown due to secular effects like accretion or due to dramatic processes like superradiant instabilities [376, 637–641], are nonvanishing and provide therefore a departure from the standard "naked BH" paradigm.

## 7.1 Setup

We mainly focus on condensates born out of massive scalar fields growing around BHs either through a phase of accretion or due to superradiant instability if their Compton wavelength is comparable to the BH radius.

In the first case, accreting BHs are usually described in terms of generalisations of the Vaidya spacetime [442, 642–645], and the accretion flow strongly depends on the properties of the accreted particles. In the second case, ultralight bosonic degrees of freedom can efficiently extract rotational energy from a spinning BH over relatively short time scales [376, 646] and give rise to macroscopic quasi-stationary condensates. This happens if the superradiance condition for the mode's frequency $w < m\Omega$ is satisfied, in terms of the azimuthal mode quantum number $m$ and BH angular velocity $\Omega$. When the superradiance condition is saturated, the cloud will stop growing and the condensate may be slowly dissipated in GWs, depending on the nature of the scalar field. Specific features of the BH superradiant instability include gaps in the spin-mass distribution of astrophysical BHs [640, 641, 647], continuous GW signals emitted by the condensate [640, 648, 649], a stochastic GW background generated by unresolved sources [648, 649], and tidal effects [650–656] in binary inspirals. Negative searches for continuous GW signals [657, 658] and for the stochastic background from unresolved sources [648, 649, 659, 660] by the LVC put constraints on ultralight boson masses around $m_b \sim 10^{-13}\,\mathrm{eV}$. On the other hand, electromagnetic and GW observations together can potentially constrain a wider range $10^{-22} \lesssim m_b c^2 / \mathrm{eV} \lesssim 10^{-10}$, although the window around $10^{-14}\,\mathrm{eV}$ remains hard to probe (see Ref. [376] for a recent summary of the constraints).

In the following, we consider a spherically symmetric isolated body which absorbs matter from a source placed at very large distances. We assume that the flow of matter is quasi-stationary, with the infall of matter from infinity balancing the loss of matter into the BH. For simplicity we assume again the geometrical units $G = \hbar = 1$.

One can describe the flow of matter onto the naked BH in a perturbative scheme [661, 662], assuming that the matter fields are small perturbations. The naked BH is described by a Schwarzschild metric which, in terms of the ingoing Eddington-Finkelstein coordinates $v = t + r_* = r + r_{\mathrm{BH}}\ln(r/r_{\mathrm{BH}} - 1)$, is given by

$$\mathrm{d}s^2 = -f\mathrm{d}v^2 + 2\mathrm{d}v\mathrm{d}r + r^2\mathrm{d}\Omega^2, \qquad f(r) = 1 - \frac{2M_{\mathrm{BH}}}{r}, \tag{7.1}$$

in terms of the BH Schwarzschild radius $r_{\mathrm{BH}} = 2M_{\mathrm{BH}}$.

## 7.1.1 Massive scalar perturbations

Let us focus on the setup where the matter source is given by a real or complex scalar field $\Phi$ with mass $m_b = \mu\hbar$, which satisfies the Klein-Gordon equation $\Box\Phi = \mu^2\Phi$. We stress that the main difference between real and complex bosonic condensates is that the latter, in a nearly stationary regime, have a stress energy tensor which is constant-in-time, and therefore do not emit gravitational radiation. At linear order in the matter fields, the Klein-Gordon equation should be evaluated on the Schwarzschild metric.

One can decompose the background scalar field in terms of spin-0 spheroidal harmonics $S_{\ell m}(\theta)$

as [384]

$$\Phi = \int \mathrm{d}w \sum_{\ell m} e^{-iwv} e^{im\varphi} S_{\ell m}(\theta) \psi_{\ell m}(r). \tag{7.2}$$

In the limit of small gravitational coupling, $\alpha \equiv M_{\mathrm{BH}}\mu$, defined as the ratio between the reduced Compton wavelength of the scalar field and the horizon radius of the BH, the radial part of the wavefunction can be expressed as [384, 663]

$$\psi(r) = \Phi_0 \left( \frac{4\alpha^2}{\ell + 1} \frac{r}{r_{\mathrm{BH}}} \right)^\ell \exp\left( -\frac{2\alpha^2}{\ell + 1} \frac{r}{r_{\mathrm{BH}}} \right), \tag{7.3}$$

in terms of an arbitrary amplitude $\Phi_0$ and focusing on the fundamental ($n = 0$ node) mode. The corresponding eigenfrequency is complex, $w = w_R + iw_I$, with $w_R \sim \mu$ and $w_I \sim -w_R\alpha^{4\ell+5}$ [376]. One can appreciate that the radial eigenfunction peaks at $r_s = r_{\mathrm{BH}}(\ell + 1)^2/2\alpha^2$ [664], that is much larger than the BH horizon for small values of $\alpha$. Indeed, the eigenfunctions can be also calculated in the flat spacetime limit [384, 663].

We stress that a complex scalar field circumvents the hypotheses of the no-hair theorems [665, 666], possibly giving rise to stationary hairy BHs. On the other hand, in the limit of small gravitational coupling $\alpha \ll 1$, a real scalar field admits quasi-bound states that can slowly dissipate through gravitational radiation, even though the time scales for this emission are so long that we can assume the hairy BH to be nearly stationary.

## 7.1.2 Backreaction on the metric

**Spherical case**

In the spherically symmetric ($\ell = 0$) case, the perturbation to the metric due to a small spherically symmetric accretion flow can be expressed as [661, 662]

$$\mathrm{d}s^2 = -Fe^{2\delta\lambda(v,r)}\mathrm{d}v^2 + 2e^{\delta\lambda(v,r)}\mathrm{d}v\mathrm{d}r + r^2\mathrm{d}\Omega^2, \tag{7.4}$$

as a function of the metric perturbation $\delta\lambda(v, r)$ and

$$F(v, r) = f(r) - \frac{2\delta M(v, r)}{r} = 1 - \frac{2M_{\mathrm{BH}} + 2\delta M(v, r)}{r}. \tag{7.5}$$

The presence of accretion is captured in the mass perturbation $\delta M(v, r)$ which provides an additional contribution to the BH mass.

Assuming that the accretion flow comes from an asymptotically constant energy density $\rho = -T_v^v$, one can solve the perturbed Einstein's equations and get [661, 662]

$$\delta M(v, r) = \delta\tilde{M}\, v + \int_{r_{\mathrm{BH}}}^r 4\pi r'^2 \rho \mathrm{d}r',$$

$$\delta\lambda(v,r) = -\int_{r_{\mathrm{BH}}}^{r} 4\pi r' T_{rr} \mathrm{d}r',$$

(7.6)

as a function of the rate of increase in the BH mass, $\delta\tilde{M}$, that depends on the configuration of the accreting field. We highlight that this perturbative scheme holds if these corrections to the metric are small, meaning that the total flux of the infalling matter should be sufficiently small, such that the matter energy density is small with respect to $\sim 1/M_{\mathrm{BH}}^2$ and the mass of matter inside a radius $r$ is much smaller than the bare BH mass.

For the spherically symmetric (real or complex) massive scalar field described above, the energy density takes the form [662]

$$\rho = f|\partial_r \Phi|^2 + \mu^2 |\Phi|^2,$$

(7.7)

which has a regular behavior at the BH horizon, $\rho \to \mu^2 |\Phi_0|^2$. The metric backreaction then takes the form [662]

$$\delta M(v,r) = 8\pi (2M_{\mathrm{BH}} w)^2 |\Phi_0|^2 v + \int_{r_{\mathrm{BH}}}^{r} 4\pi r'^2 (f|\partial_r \Phi|^2 + \mu^2 |\Phi|^2) \mathrm{d}r',$$

$$\delta\lambda(r) = -2\int_{r_{\mathrm{BH}}}^{r} 4\pi r' |\partial_r \Phi|^2 \mathrm{d}r'.$$

(7.8)

At leading order in the matter perturbations, one can appreciate that the horizon of the accreting BH increases as

$$r_{\mathrm{BH}}^{\mathrm{acc}}(v) = r_{\mathrm{BH}} + 64\pi |\Phi_0|^2 M_{\mathrm{BH}}^2 \mu^2 v,$$

(7.9)

in terms of the advanced-time coordinate $v$.

The dependence of the metric perturbations $\delta M$ and $\delta\lambda$ with respect to the radial coordinate is shown in Fig. 7.1, for fixed values of the gravitational coupling $\alpha$ and for the $\ell = 0, 1, 2$ modes (only the mode $\ell = 0$ is relevant for accretion, see next section for the nonaxisymmetric ones). One can notice that both perturbations approach zero in the limit $r \to r_{\mathrm{BH}}$ (assuming $v = 0$ for the mass correction), and that higher multipoles of the scalar field results in larger metric perturbations (for the simplified choice of the same amplitude for the $\ell = 1, 2$ modes, which is typically not the case as discussed below). Furthermore, in the limit $r \gg r_s$ both perturbations reach a plateau, which is due to the exponential suppression of the massive scalar field at large radii.

The scalar field amplitude can be expressed in terms of the mass enclosed in the scalar cloud $M_s$ as [384]

$$M_s = 4\pi \int_{r_{\mathrm{BH}}}^{\infty} \mathrm{d}r\, r^2 \rho = 4\pi \int_{r_{\mathrm{BH}}}^{\infty} \mathrm{d}r\, r^2 \left( f|\partial_r \Phi|^2 + \mu^2 |\Phi|^2 \right),$$

(7.10)

such that, for the $\ell = 0$ mode, one gets

$$|\Phi_0^{\ell=0}|^2 = \frac{1}{\pi} \left( \frac{M_s}{M_{\mathrm{BH}}} \right) \alpha^4 e^{4\alpha^2}.$$

(7.11)

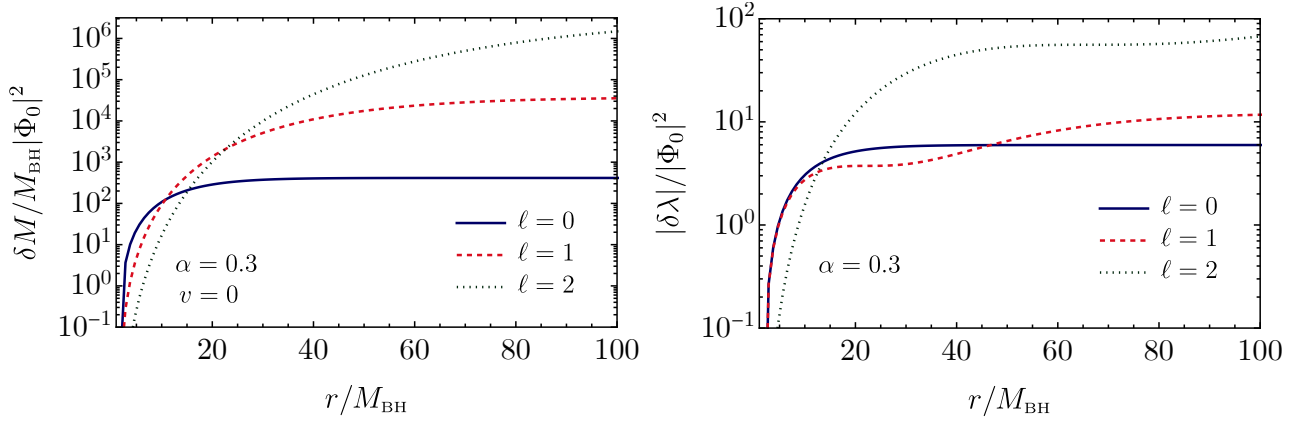

**Figure 7.1**: Radial profile of the metric corrections $\delta M$ and $\delta \lambda$, for fixed values of the gravitational coupling $\alpha = 0.3$ and for the modes $\ell = 0$, $\ell = 1$ and $\ell = 2$ of the complex scalar field background. Figure taken from Ref. [4].

In this case the energy density associated to the scalar field cloud is given by

$$\rho = \frac{\alpha^6}{\pi M_{\text{BH}}^2} \frac{M_s}{M_{\text{BH}}} \sim 2 \times 10^{-2} \left( \frac{\alpha}{0.1} \right)^6 \left( \frac{10^6 M_\odot}{M_{\text{BH}}} \right)^2 \frac{M_s}{0.1 M_{\text{BH}}} \, \text{g/cm}^3 \,. \quad (7.12)$$

One can appreciate that in the limit of small gravitational coupling $\alpha \ll 1$, the energy density is much smaller than the one typically associated to BHs, $\rho_{\text{BH}} M_{\text{BH}}^2 = 3/(32\pi) \sim \mathcal{O}(0.1)$, even when the scalar cloud mass $M_s$ is a sizeable fraction of the BH mass. To give some numerical estimates, one can consider the density associated to a thin accretion disk around a $10^6 M_\odot$ BH as $\rho_{\text{disk}} \approx 6 \times 10^{-3} f_{\text{Edd}}^{11/20} \, \text{g/cm}^3$ [667] (in terms of the Eddington fraction for mass accretion $f_{\text{Edd}}$), which is found to be of the same order of Eq. (7.12) with the chosen normalization.

For the interesting case of a binary system during its inspiral, one can appreciate that the characteristic time scale for accretion, $\tau_{\text{Salpeter}} \simeq 4.55 \times 10^7 \, \text{yr}$, is much longer than the orbital period of the binary,

$$T = \sqrt{\frac{4\pi^2 a^3}{2 M_{\text{BH}}}} \sim 6 \times 10^{-2} \, \text{s} \left( \frac{\beta}{0.1} \right)^{-3} \left( \frac{M_{\text{BH}}}{M_\odot} \right) \,, \quad (7.13)$$

in terms of the binary semi-major axis $a$ and orbital velocity $\beta$. This implies that, during the inspiral phase, accretion may be treated as an adiabatic process, and neglect the dependence on time of the metric perturbations. In other words, one can assume that $\delta M(v, r) \sim \delta M(\tilde{v}, r)$, with $\tilde{v}$ approximately constant during the inspiral time scale. As we will show, in this regime the time parameter $\tilde{v}$ does not impact on the TLNs.

**Superradiantly-triggered bosonic condensate**

In the previous section we have studied how accretion may give rise to a spherical condensate, corresponding to the mode $\ell = 0$. In this section we want to focus on the alternative possibility that macroscopic bosonic condensates may be formed around spinning BHs by the superradiant instability of nonaxisymmetric modes of ultralight fields [376].

The condition for modes to trigger superradiant instability around a BH is given by $w_R < m\Omega$, which therefore requires nonaxisymmetric modes with azimuthal number $m$ and a spinning BH with angular velocity $\Omega = \chi/(2M_{\rm BH}(1 + \sqrt{1 - \chi^2}))$, in terms of the dimensionless BH spin $\chi$.

The $\ell = m = 1$ unstable mode may experience an exponential growth by extracting rotational energy and angular momentum from a spinning BH with a characteristic time scale of [668]

$$\tau_I \sim \frac{1}{w_I} \sim 7 \times 10^{-3} \chi^{-1} \left(\frac{M_{\rm BH}}{M_\odot}\right) \left(\frac{\alpha}{0.1}\right)^{-9} {\rm yr}, \tag{7.14}$$

which is found to be shorter than the characteristic one for accretion, $\tau_{\rm Salpeter}$ (see Ref. [384] for a detailed study on the interplay between these two processes). The superradiant instability will proceed as long as the condition

$$\chi > \chi_{\rm crit} \simeq \frac{4m\alpha}{m^2 + 4\alpha^2}, \tag{7.15}$$

is satisfied. When the condition is saturated, the cloud will stop growing. The subsequent evolution of the bosonic cloud depends on the nature of the ultralight field. In particular, for a real bosonic field the cloud will start dissipating by emitting gravitational radiation. On the other hand, a complex massive bosonic field satisfying the condition $w = m\Omega < \mu$ does not emit GWs neither at the horizon nor at infinity, and can therefore form stationary Kerr BHs with bosonic hair [666, 669]. In the former case, the relevant time scale for GW emission is given by

$$\tau_{\rm GW} \sim 6 \times 10^3 \chi^{-1} \left(\frac{M_{\rm BH}}{M_\odot}\right) \left(\frac{\alpha}{0.1}\right)^{-15} {\rm yr}, \tag{7.16}$$

that is much longer than the instability time scale, $\tau_I$. This implies that, in the $\alpha \ll 1$ regime, the condensate first grows on a time scale $\tau_I$, until the superradiant condition is saturated, and then it dissipates in gravitational radiation over a time scale $\tau_{\rm GW} \gg \tau_I$.

The spin extraction from the BH during superradiance is accompanied by a decrease of the BH mass as

$$\Delta M_{\rm BH} \sim M_s \sim -\mu \left[\chi_f M_{\rm BH}^2 - \chi_i (M_{\rm BH}^i)^2\right], \tag{7.17}$$

as a function of the initial (final) BH mass and spin $M_{\rm BH}^i$, $\chi_i$ ($M_{\rm BH}$, $\chi_f$). The scalar cloud mass $M_s$ can then be determined as

$$M_s = \frac{1 - 2M_{\rm BH}\chi_i\mu - \sqrt{1 - 4\chi_i M_{\rm BH}\mu + 4\chi_i\chi_f M_{\rm BH}^2\mu^2}}{2\chi_i\mu}, \tag{7.18}$$

and using the condition for superradiance

$$\chi_f = \chi_{\text{crit}} \simeq \frac{4}{m}\alpha + \mathcal{O}(\alpha^3), \tag{7.19}$$

at leading order in $\alpha$, one finds

$$\frac{M_s}{M_{\text{BH}}} = \frac{1 - 2\chi_i\alpha - \sqrt{1 - 4\chi_i\alpha + 16\chi_i\alpha^3}}{2\chi_i\alpha} \simeq \chi_i\alpha + \mathcal{O}(\alpha^2). \tag{7.20}$$

For a BH which is initially highly spinning $\chi_i \simeq 1$, one gets an estimate for the final mass of the scalar condensate

$$\frac{M_s}{M_{\text{BH}}} \simeq \alpha. \tag{7.21}$$

In the following we assume the upper bound $M_s = 0.1 M_{\text{BH}}$, which roughly corresponds to the maximum energy extracted by superradiance [376, 670, 671], and also saturates the validity regime of the perturbative expansion [661, 662].

At the end of the superradiant process, the BH is (slowly) spinning and the bosonic condensate is nonaxisymmetric. For example, for the dominant $\ell = m = 1$ mode of a real scalar field, one has [384]

$$\Phi \propto \psi(r)\cos(\varphi - w_R t)\sin\theta. \tag{7.22}$$

In the limit of small gravitational coupling $\alpha \ll 1$, since the condensate peaks at a radius $r_s \gg r_{\text{BH}}$, the spin of the BH can be neglected and the scalar field can be decomposed in spherical harmonics, $\Phi \sim \psi(r)\sum_{\ell m} a_{\ell m} Y^{\ell m}$. When the scalar field is plugged back into the energy momentum tensor $T_{\mu\nu}$ in the Einstein's equations, the angular-momentum selection rules states that an $\ell = 1$ mode will source $\ell = 0, 2$ modes in the metric perturbations of a naked Schwarzschild BH [672]. The $\ell = 0$ mode can be treated as described in the previous section, while the $\ell = 2$ mode will produce a quadrupole moment, $Q \propto M_s^2$, on the background.

For condensates triggered by superradiance, we will make the assumption of neglecting the spin of the final BH, which is found to be small after the rotational energy is extracted from the compact object. This does not impact on our results, since the TLN of a slowly-spinning object can be written as [627, 628]

$$\text{TLN} = \text{TLN}^{\text{nonspinning}}\left(1 + a_1 m\chi_f + \mathcal{O}(\chi_f)\right), \qquad a_1 \sim \mathcal{O}(1), \tag{7.23}$$

with $m\chi_f \sim \alpha \ll 1$, and the dominant contribution comes from the spherically-symmetric part of the metric. We also make the working assumption of neglecting the nonspherical backreaction of the condensate onto the metric. In particular, even though in principle the $\ell = 1$ (say) mode will source $\ell = 0$ and $\ell = 2$ metric perturbations, we focus only on the $\ell = 0$ contribution, which is simpler to compute and gives an estimate for the order of magnitude of tidal effects. The $\ell = 2$ background perturbations will yield a TLN of the same order of magnitude, which should sum with the $\ell = 0$ contribution, and that we neglect in our results.

Regardless of the nonspherical contributions, one can compute the spherically-symmetric part of the metric as done for the accretion case. In Fig. 7.1 we show the corresponding metric perturbations for the $\ell = 1, 2$ modes, whose amplitude can be mapped into the condensate mass as

$$|\Phi_0^{\ell=1}|^2 = \frac{1}{96\pi} \left( \frac{M_s}{M_{\mathrm{BH}}} \right) \alpha^4 e^{2\alpha^2}, \tag{7.24}$$

$$|\Phi_0^{\ell=2}|^2 = \frac{1}{9720\pi} \left( \frac{M_s}{M_{\mathrm{BH}}} \right) \alpha^4 e^{\frac{4}{3}\alpha^2}, \tag{7.25}$$

where in the last step we have assumed that the scalar cloud mass is fully accounted by just one dominant mode.

This assumption may be justified by realising that the instability time scales for higher modes with $m > 1$ satisfying the superradiance condition are much longer than the one for the dominant mode $m = 1$, which has efficiently extracted angular momentum from the BH, and did not have enough time to grow sufficiently [671]. Higher-$m$ modes may be relevant if the superradiant condition is not satisfied for $m = 1$ but only for higher values of $m$. In this case the next dominant mode with the smallest value of $m$ becomes the superradiantly unstable one, producing a cascade process of ever longer time scales.

## 7.2   Scalar and vector Love numbers

After the description of the physical setup in the previous section, we can now compute the TLNs for a BH surrounded by a scalar field cloud under the presence of an external scalar ($\phi$) and vector ($A_\mu$) tidal field.

### 7.2.1   Definition of scalar and vector Love numbers

The multipole moments of a nonspinning compact object with mass $M$ can be extracted by asymptotically expanding the relevant fields. In particular, for a gravitational tidal field one can extract the (electric) multipole moments from [673, 674]

$$g_{tt} = 1 + \frac{2M}{r} + \sum_{l \geq 2} \left( \frac{2}{r^{l+1}} \left[ \sqrt{\frac{4\pi}{2l+1}} M_l Y^{l0} + (l' < l \, \mathrm{pole}) \right] - \frac{2}{l(l-1)} r^l \left[ \mathcal{E}_l Y^{l0} + (l' < l \, \mathrm{pole}) \right] \right), \tag{7.26}$$

in terms of the mass multipole moments $M_l$ ($l = 2, 3, ..$, different from the $\ell$ index describing the multipolar structure of the background scalar condensate $\Phi$) and the amplitude $\mathcal{E}_l$ of the multipole moment $l$ of the tidal field.

Similar asymptotic expansions can be performed for the (axial, or parity-odd) vector field $A_\varphi$ and for the scalar field $\phi$ as [674]

$$A_\varphi = \sum_{l \geq 1} \left( \frac{2}{r^l} \left[ \sqrt{\frac{4\pi}{2l+1}} J_l S_\varphi^{l0} + (l' < l \text{ pole}) \right] - \frac{2r^{l+1}}{l\,(l-1)} \left[ \mathfrak{B}_l S_\varphi^{l0} + (l' < l \text{ pole}) \right] \right), \qquad (7.27)$$

$$\phi = \phi_0 + \sum_{l \geq 1} \left( \frac{1}{r^{l+1}} \left[ \sqrt{\frac{4\pi}{2l+1}} \phi_l Y^{l0} + (l' < l \text{ pole}) \right] - \frac{1}{l(l-1)} r^l \left[ \mathcal{E}_l^{\mathrm{S}} + (l' < l \text{ pole}) \right] \right), \quad (7.28)$$

as a function of the magnetic $J_l$ and scalar $\phi_l$ multipole moments, in terms of the axial (parity odd) vector tidal field $\mathfrak{B}_l$ and the scalar one $\mathcal{E}_l^{\mathrm{S}}$.

The TLNs due to the presence of a gravitational ($s = 2$), axial vector ($s = 1$), and scalar ($s = 0$) tidal perturbation can be then defined to be

$$\begin{aligned}
k_l^{(s=2)} &\equiv -\frac{1}{2} \frac{l(l-1)}{M^{2l+1}} \sqrt{\frac{4\pi}{2l+1}} \frac{M_l}{\mathcal{E}_l}, \\
k_l^{(s=1)} &\equiv -\frac{1}{2} \frac{l(l-1)}{M^{2l+1}} \sqrt{\frac{4\pi}{2l+1}} \frac{J_l}{\mathfrak{B}_l}, \\
k_l^{(s=0)} &\equiv -\frac{1}{2} \frac{l(l-1)}{M^{2l+1}} \sqrt{\frac{4\pi}{2l+1}} \frac{\phi_l}{\mathcal{E}_l^{\mathrm{S}}}.
\end{aligned} \qquad (7.29)$$

We stress that for the spins $s = 0, 1$ the normalization used in the above definitions is arbitrary and was chosen to be the same of the $s = 2$ case.

The TLNs can be calculated by solving the corresponding perturbation equations, with the condition of regularity at the BH horizon, and finally matching the solution with the asymptotic expansion outlined above.

## 7.2.2 Scalar Love numbers

The equation of motion for the external scalar tidal field $\phi$ (different from the scalar field $\Phi$ that creates a cloud around the BH and that is part of the background) is given by the (real) Klein-Gordon equation

$$\Box \phi(v, r, \theta, \varphi) = 0. \qquad (7.30)$$

Using the spherical symmetry of the background and focusing on static perturbations (being the relevant ones to compute the TLNs, that impact on the GW waveform at leading-order in a PN expansion), one can perform a spherical-harmonic decomposition,

$$\phi = \sum_{lm} \frac{R(r)}{r} Y^{lm}, \qquad (7.31)$$

which gives the radial equation

$$F\frac{R''}{r} + \left(F' + F\delta\lambda'\right)\frac{R'}{r} - \left(\frac{l(l+1)}{r^2} + \frac{1}{r}F' + \frac{F}{r}\delta\lambda'\right)\frac{R}{r} = 0\,, \tag{7.32}$$

where differentiation with respect to $r$ has been denoted with a prime. Since we wish to use this computation as a proxy for the gravitational case (for which $l \geq 2$), we shall focus on the $l = 2$ scalar mode.

This equation can be solved in a perturbative expansion in the coupling strength of the source, which is expressed in terms of a dimensionless parameter $\epsilon = M_{\rm BH}^2\Phi_0^2$. At zeroth order in $\epsilon$, the solution has the asymptotic form at spatial infinity

$$\frac{R^{(0)}}{r} = 6c_1\left(\frac{r}{r_{\rm BH}}\right)^2\,, \tag{7.33}$$

after imposing the condition of regularity of the solution at the BH horizon[1]. Since the solution above does not have an induced quadrupolar term $\sim 1/r^3$, one concludes that the scalar TLNs of a naked Schwarzschild BH are zero [674, 675]. At first order in $\epsilon$, the solution has the asymptotic form at spatial infinity

$$\frac{R^{(1)}}{r} = -\mathcal{N}c_1\frac{\epsilon}{G^{10}M_{\rm BH}^{11}\mu^{12}}\frac{1}{r^3} = -\frac{\mathcal{N}}{8}c_1\frac{\epsilon}{M_{\rm BH}^{14}\mu^{12}}\left(\frac{r_{\rm BH}}{r}\right)^3\,, \tag{7.34}$$

in terms of a numerical coefficient $\mathcal{N}$, which depends on the considered background, such that the total solution reads

$$\frac{R}{r} = \frac{R^{(0)}}{r} + \frac{R^{(1)}}{r} = 6c_1\left(\frac{r}{r_{\rm BH}}\right)^2\left[1 - \frac{\mathcal{N}}{48}\frac{\epsilon}{M_{\rm BH}^{14}\mu^{12}}\left(\frac{r_{\rm BH}}{r}\right)^5\right]\,. \tag{7.35}$$

By comparing the behavior of the solution at spatial infinity to Eq. (7.28), one can extract the scalar TLN as[2]

$$k_2^{(\ell,s=0)} = -\frac{\mathcal{N}^{(\ell)}}{96}\frac{\epsilon^{(\ell)}}{M_{\rm BH}^{14}\mu^{12}}\,, \tag{7.36}$$

where we have stressed that the numerical coefficient $\mathcal{N}^{(\ell)}$ and the expansion parameter $\epsilon^{(\ell)}$ depend on the mode $\ell$ of the background complex scalar field $\Phi$. Note that, in the adiabatic regime, the scalar TLNs do not depend explicitly on the time coordinate $\tilde{v}$. At leading order, the scalar TLN for the $\ell = 0, 1, 2$ modes of the background condensate reads

$$k_2^{(\ell=0,s=0)} = -\frac{27}{192}\pi\frac{|\Phi_0^{\ell=0}|^2}{\alpha^{12}}e^{-4\alpha^2}\,,$$

---

[1]We stress that, since the background solution is perturbative in the matter field, regularity should be imposed on the horizon of the naked BH both to zeroth and to first order in the matter fields.

[2]In the following we use the symbol $k_l^{(\ell,s)}$ to indicate the $l$-th order TLN of a tidal field with spin $s$ for a BH dressed with a background scalar cloud with harmonic index $\ell$.

$$k_2^{(\ell=1,s=0)} = -1008\pi \frac{|\Phi_0^{\ell=1}|^2}{\alpha^{12}} e^{-2\alpha^2},$$

$$k_2^{(\ell=2,s=0)} = -\frac{6200145}{4}\pi \frac{|\Phi_0^{\ell=2}|^2}{\alpha^{12}} e^{-\frac{4}{3}\alpha^2}, \tag{7.37}$$

which can be recast in terms of the gravitational coupling $\alpha$ and the scalar cloud mass $M_s$ as

$$k_2^{(\ell=0,s=0)} \simeq -0.14 \frac{1}{\alpha^8} \left(\frac{M_s}{M_{\rm BH}}\right),$$

$$k_2^{(\ell=1,s=0)} \simeq -10 \frac{1}{\alpha^8} \left(\frac{M_s}{M_{\rm BH}}\right),$$

$$k_2^{(\ell=2,s=0)} \simeq -160 \frac{1}{\alpha^8} \left(\frac{M_s}{M_{\rm BH}}\right). \tag{7.38}$$

These expressions show that higher modes of the complex scalar background experience larger deformability, and hence larger TLNs, since boson clouds with larger values of $\ell$ are more diluted, and hence more deformable.

The behavior of the Love number as a function of the gravitational coupling $\alpha$ reproduces the one found in Ref. [650], using a dimensional analysis approach for a gravitational tidal perturbation.

### 7.2.3   Axial vector Love numbers

The equation of motion for a vector field $A_\mu$ in the spacetime of a BH surrounded by a bosonic scalar cloud is given by

$$\nabla_\nu F^{\mu\nu} = 0, \tag{7.39}$$

in terms of the electromagnetic tensor $F_{\mu\nu} = \partial_\mu A_\nu - \partial_\nu A_\mu$. Focusing on the axial sector, one can decompose the spin-1 field as [676]

$$A_\mu = \sum_{lm} \left(0, 0, a^{lm}(r)\frac{\partial_\varphi Y^{lm}}{\sin\theta}, -a^{lm}(r)\sin\theta\, \partial_\theta Y^{lm}\right), \tag{7.40}$$

and insert this decomposition in the equation of motion to find

$$F a'' + (F' + F\delta\lambda') a' - \frac{l(l+1)}{r^2} a = 0, \tag{7.41}$$

where $a \equiv a^{lm}$. This equation can be solved in a perturbative expansion in the coupling strength of the background matter source. For $l = 2$, after requiring regularity at the naked BH horizon, the solution at zeroth order has the following asymptotic form at spatial infinity

$$a^{(0)} = c_1 \left(\frac{r}{r_{\rm BH}}\right)^3. \tag{7.42}$$

The absence of any induced quadrupolar term implies that the axial vector TLNs of a naked Schwarzschild BH are zero. At first order, instead, the solution has the asymptotic form

$$a^{(1)} = -\mathcal{N}c_1 \frac{\epsilon}{M_{\text{BH}}^{11}\mu^{12}} \frac{1}{r^2} = -\frac{\mathcal{N}}{4} c_1 \frac{\epsilon}{M_{\text{BH}}^{14}\mu^{12}} \left(\frac{r_{\text{BH}}}{r}\right)^2, \tag{7.43}$$

such that the total solution reads

$$a = a^{(0)} + a^{(1)} = c_1 \left(\frac{r}{r_{\text{BH}}}\right)^3 \left[1 - \frac{\mathcal{N}}{4} \frac{\epsilon}{M_{\text{BH}}^{14}\mu^{12}} \left(\frac{r_{\text{BH}}}{r}\right)^5\right], \tag{7.44}$$

from which one can deduce the axial-vector TLN as

$$k_2^{(\ell,s=1)} = -\frac{\mathcal{N}^{(\ell)}}{8} \frac{\epsilon^{(\ell)}}{M_{\text{BH}}^{14}\mu^{12}}. \tag{7.45}$$

The leading-order term of the dimensionless TLN for the $\ell = 0, 1, 2$ modes of the background complex scalar field $\Phi$ then reads

$$k_2^{(\ell=0,s=1)} = -\frac{81}{256}\pi \frac{|\Phi_0^{\ell=0}|^2}{\alpha^{12}} e^{-4\alpha^2},$$

$$k_2^{(\ell=1,s=1)} = -2268\pi \frac{|\Phi_0^{\ell=1}|^2}{\alpha^{12}} e^{-2\alpha^2},$$

$$k_2^{(\ell=2,s=1)} = -\frac{55801305}{16}\pi \frac{|\Phi_0^{\ell=2}|^2}{\alpha^{12}} e^{-\frac{4}{3}\alpha^2}, \tag{7.46}$$

which can be recast in terms of the mass of the scalar cloud as

$$k_2^{(\ell=0,s=1)} \simeq -0.32 \frac{1}{\alpha^8} \left(\frac{M_s}{M_{\text{BH}}}\right),$$

$$k_2^{(\ell=1,s=1)} \simeq -24 \frac{1}{\alpha^8} \left(\frac{M_s}{M_{\text{BH}}}\right),$$

$$k_2^{(\ell=2,s=1)} \simeq -360 \frac{1}{\alpha^8} \left(\frac{M_s}{M_{\text{BH}}}\right). \tag{7.47}$$

Compatibly with the results for the scalar TLNs, higher modes of the complex field $\Phi$ experience larger responses to tidal perturbations, and the scaling with the gravitational coupling $\alpha$ matches the one obtained for a scalar or gravitational tidal perturbation. Indeed, we confirm the general scaling properties predicted in Ref. [650], that is

$$k_2 \sim \frac{1}{\alpha^8} \left(\frac{M_s}{M_{\text{BH}}}\right), \tag{7.48}$$

with the numerical prefactor which depends on the background solution and on the type of tidal perturbation. One can appreciate that $k_2$ can be very large in the $\alpha \ll 1$ regime. Adopting the same normalization used in the gravitational case, the axial vector Love numbers are found to be slightly bigger than the ones for scalar tides, for the same configuration of the background complex scalar field.

In the appendix of Ref. [4] we showed that, for a perfect-fluid neutron star, a hierarchy between the scalar, axial vector, and gravitational TLNs holds, with the TLNs growing for larger spins $s = 0, 1, 2$ of the external tidal field. Based on this (small) hierarchy, in the following we will adopt the result for the axial vector TLNs as a proxy for the real gravitational TLNs and investigate their detectability at future GW experiments. This assumption is conservative, since the gravitational TLNs are expected to be slightly larger.

## 7.3   Detectability of tidal effects on dressed BHs

### 7.3.1   Forecast

The detectability of TLNs through GW observations may be investigated with a Fisher matrix approach as described in Refs. [674, 677, 678] and summarized in Appendix C.

In the previous section we found that the TLNs $k_2$ depend on the BH mass and gravitational coupling $\alpha$. We can now focus on a binary system, whose BH components are characterised by a bosonic scalar cloud, and use the results for the TLNs due to vector tidal fields as a proxy for the gravitational ones induced by the binary companion. Then, a GW measurement of the tidal deformability will automatically translate into a measurement of the coupling $\alpha$ and, hence, on the scalar field mass.

In the Fisher matrix approach, the set of intrinsic hyperparameters which we consider are $\vec{\xi} = (\phi_c, t_c, \ln\mathcal{M}, \ln\nu, \ln\alpha, \chi_1, \chi_2)$ (phase, time of coalescence, chirp mass, symmetric mass ratio, gravitational coupling, and individual BH spins). We investigate the TLN detectability at ET and LISA, assuming for the latter an observation time of $T_{\rm obs} = 5\,{\rm yr}$.[3] For the physical system at hand, a detailed study of the frequency domain in the Fisher analysis is particularly relevant since the latter is dictated by different physical phenomena.

The maximum frequency in the integration range, $f_{\rm max}$, is given by the minimum between three characteristic frequencies:

(i)   the first frequency is dictated by tidal forces which may disrupt the scalar cloud around the BH as the inspiral proceeds. This occurs when the binary semi-major axis is comparable to

---

[3]We stress that for LISA the GW waveform amplitude has been multiplied by a factor $\sqrt{3}/2$ to account for the triangular shape of the detector [541, 679], and by a factor $\sqrt{3/20}$ to include the sky average [680] as done in Ref. [674].

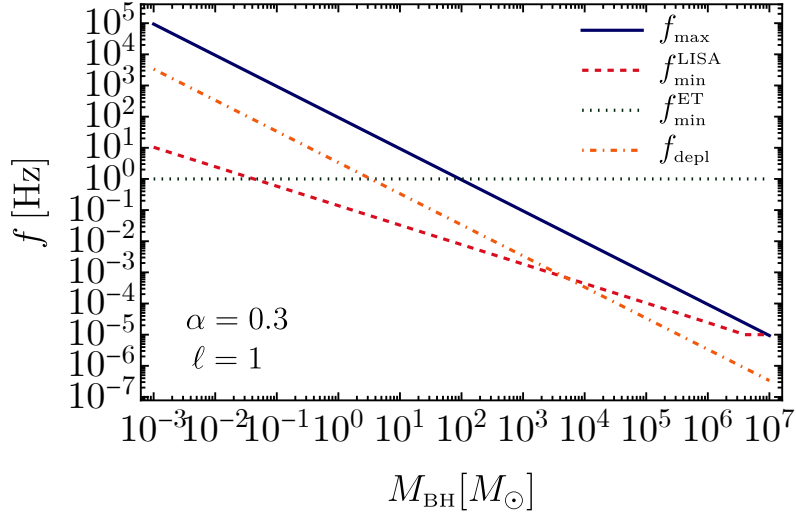

**Figure 7.2**: Comparison between the relevant frequencies as a function of the BH mass for an equal-mass binary system, fixing the value of the gravitational coupling $\alpha$ and for the mode $\ell = 1$ of the background complex scalar field. Figure taken from Ref. [4].

the Roche radius [369]

$$r_{\rm Roche} = \gamma r_{\rm BH} \left( \frac{\rho_{\rm BH}}{\rho} \right)^{1/3} = \gamma r_s \left( \frac{M_{\rm BH}}{M_s} \right)^{1/3} = \frac{\gamma(\ell+1)^2}{2} \frac{r_{\rm BH}}{\alpha^2} \left( \frac{M_{\rm BH}}{M_s} \right)^{1/3}, \tag{7.49}$$

in terms of the numerical coefficient $\gamma \sim \mathcal{O}(2)$, whose value ranges from 1.26 for rigid bodies to 2.44 for fluid ones. In this estimate we have assumed the primary binary component to be dominated by the BH itself (neglecting its halo) with density $\rho_{\rm BH} = 3M_{\rm BH}/4\pi r_{\rm BH}^3$, destroying the halo of the secondary component with corresponding density $\rho = 3M_s/4\pi r_s^3$. The corresponding frequency for a circular, equal-mass binary reads

$$f_{\rm Roche} = \frac{2^{3/2}}{(\ell+1)^3 \pi \gamma^{3/2}} \frac{\alpha^3}{r_{\rm BH}} \left( \frac{M_s}{M_{\rm BH}} \right)^{1/2}; \tag{7.50}$$

(ii) the second frequency is dictated by the applicability of the multipole expansion in the PN waveform, which holds as long as the semi-major axis is larger than the cloud radius. This translates into the condition

$$r \gtrsim \frac{(\ell+1)^2}{2} \frac{r_{\rm BH}}{\alpha^2}, \tag{7.51}$$

or into a maximum frequency

$$f_{\rm me} = \frac{2^{3/2}}{(\ell+1)^3 \pi} \frac{\alpha^3}{r_{\rm BH}}; \tag{7.52}$$

(iii) the third frequency is set at the ISCO, at which early-inspiral waveform approximants have usually a cut-off, and it is given by

$$f_{\text{ISCO}} = \frac{1}{6\sqrt{6}\pi r_{\text{BH}}}. \tag{7.53}$$

The resulting maximum frequency is therefore set by $f_{\text{max}} = \text{Min}\,[f_{\text{Roche}}, f_{\text{me}}, f_{\text{ISCO}}]$. For our scenario, the relevant frequency is always the Roche one, meaning that, well before the binary approaches the ISCO, the tidal field can destroy the condensate. This implies that at $f > f_{\text{Roche}}$ tidal effects can be neglected and the binary evolution proceeds with the merger of two naked BHs. One can appreciate that higher $\ell$-modes of the condensate result in smaller maximum frequencies. Furthermore, since $f_{\text{Roche}} \propto \alpha^3/M_{\text{BH}}$, the strong dependence on $\alpha$ limits our capability of detecting tidal effects in the waveform. In particular, even though $k_2 \propto \alpha^{-8}$, the 5PN correction is much smaller than unity at $f = f_{\text{Roche}}$, so the PN expansion holds in the relevant frequency range.

The minimum frequency, $f_{\text{min}}$, depends on other physical processes. For naked BHs one typically sets it to the minimum frequency detectable by the interferometer, which for ET is $f_{\text{min}}^{\text{ET}} = 1\,\text{Hz}$, whereas for LISA is $f_{\text{min}}^{\text{LISA}} = \text{Max}[10^{-5}\,\text{Hz}, f_{\text{5yr}}]$, where the second term describes the initial frequency of a binary system that spends $T_{\text{obs}} = 5\,\text{yr}$ to span the frequency band up to $f_{\text{max}}$ [680], and is given by

$$f_{\text{5yr}} = \left[\frac{1}{f_{\text{max}}^{8/3}} + 1.27 \times 10^{12}\,\text{Hz}^{-8/3}\left(\frac{\mathcal{M}}{10^6 M_\odot}\right)^{5/3}\left(\frac{T_{\text{obs}}}{5\,\text{yr}}\right)\right]^{-3/8}. \tag{7.54}$$

For BHs surrounded by superradiantly-triggered bosonic clouds one has also to consider the gravitational perturbation induced by the presence of the BH companion, which triggers a transition of the dominant growing mode to decaying modes and results into a depletion of the cloud. For nonaxisymmetric condensates this condition therefore translates into the requirement that the coalescence time has to be shorter than the lifetime of the cloud, which is valid for frequencies bigger than the critical value [650]

$$f_{\text{depl}} = 0.3\,\text{mHz}\left(\frac{3M_\odot}{M_{\text{BH}}}\right)\frac{(1+q)^{1/8}}{q^{3/8}}\left(\frac{\alpha}{0.07}\right)^{45/8}, \tag{7.55}$$

in terms of the binary's mass ratio $q$. This frequency is smaller than the Roche frequency for the relevant superradiant modes $\ell = 1, 2$, and is comparable with the sensitivity frequencies of the considered GW experiments in setting the minimum frequency. For the $\ell = 1, 2$ condensates we therefore set $f_{\text{min}} = \text{Max}[f_{\text{min}}^{\text{detector}}, f_{\text{depl}}]$, while for spherical condensates there are no level transitions and we simply set $f_{\text{min}} = f_{\text{min}}^{\text{detector}}$. The relevant frequencies are shown in Fig. 7.2, fixing the value of the gravitational coupling $\alpha$ and mode $\ell = 1$ of the scalar condensate.

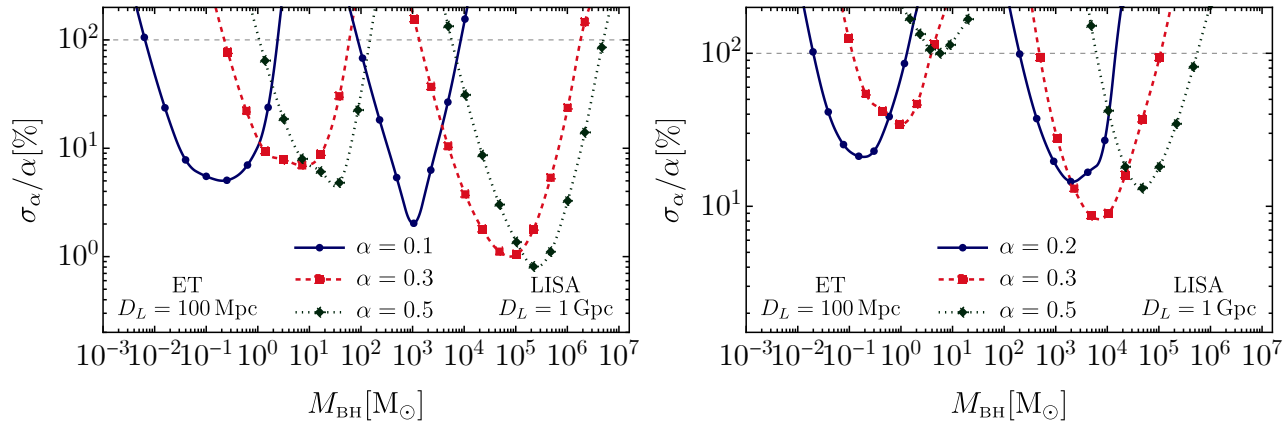

**Figure 7.3**: Relative percentage error on the gravitational coupling $\alpha = \mu M_{\rm BH}$ for equal-mass, nonspinning binaries for the modes $\ell = 0$ (left) and $\ell = 1$ (right) of the background complex scalar field, assuming an axial vector TLN in the GW waveform and scalar cloud mass $M_s(\alpha) = 0.1 M_{\rm BH}$. The left-most (right-most) curves are related to binaries detectable by ET (LISA) at luminosity distance $D_L = 100\,{\rm Mpc}$ ($D_L = 1\,{\rm Gpc}$). The horizontal dashed lines represent the value $\sigma_\alpha/\alpha = 1$ above which no constraint can be set on the gravitational coupling. Figure taken from Ref. [4].

## 7.3.2 Projected bounds on ultralight fields from tidal-deformability measurements

The result of the Fisher analysis is shown in Fig. 7.3, where we plot the relative percentage error on the gravitational coupling $\alpha$ for a nonspinning, equal-mass binary system, for the modes $\ell = 0$ (left panel) and $\ell = 1$ (right panel) of the background complex scalar field $\Phi$. As already anticipated before, we have assumed the value $M_s = 0.1 M_{\rm BH}$ for the scalar cloud mass, which represents the most optimistic scenario, since smaller values would make the TLN smaller and reduce the Roche frequency.

One can appreciate that both ET and LISA would be able to set a constraint on the gravitational coupling, and thus on the tidal deformability, in a wide and complementary range of BH masses. In particular, the determined bounds are found to extend to smaller masses than those usually probed by inspiral tests. This is due to the behaviour of the maximum frequency, which for inspiral tests is roughly given by $f_{\rm ISCO} \approx 0.01/M_{\rm BH}$ and matches the optimal frequency of ground-based (space-based) detectors – i.e., roughly $100\,{\rm Hz}$ ($1\,{\rm mHz}$) – for $M_{\rm BH} \sim 20\,M_\odot$ ($M_{\rm BH} \sim 2 \times 10^6\,M_\odot$), while in our case is given by the Roche frequency, that matches the optimal detector frequencies for much smaller BH masses, that is $M_{\rm BH} \approx 0.02 M_\odot$ ($M_{\rm BH} \approx 2 \times 10^3 M_\odot$). We stress that, even though these frequencies are much smaller than the characteristic ones for mergers, tidal deformations may still be detectable because of the large value of the TLNs in the limit of small $\alpha$. This implies that LISA could detect the tidal deformability of dressed

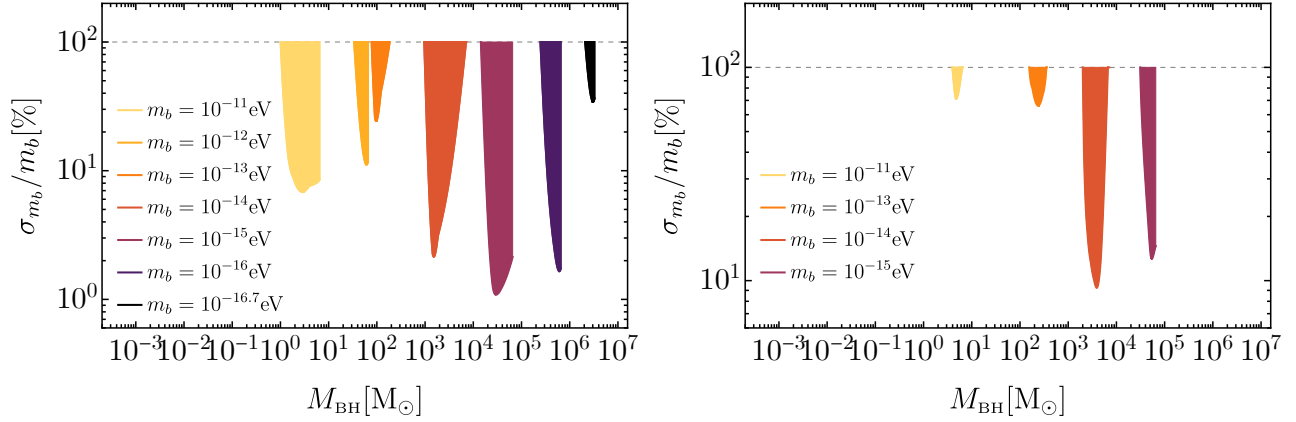

**Figure 7.4**: Relative percentage error on the mass of the scalar field $m_b = \mu \hbar$ for equal-mass, nonspinning binaries for the modes $\ell = 0$ (left) and $\ell = 1$ (right) of the background complex bosonic condensate, assuming an axial vector TLN and $M_s(\alpha) = 0.1 M_{\rm BH}$. The horizontal dashed lines identify the value $\sigma_{m_b}/m_b = 1$ above which no constraint can be set on the scalar field mass. Figure taken from Ref. [4].

BHs across the mass range from stellar-mass ($\approx 10^2 M_\odot$) to supermassive ($\approx 10^7 M_\odot$) compact objects, providing a measurement of $\alpha$ as accurate as a few percent for optimal configurations. Furthermore, we stress that the bounds coming from ET and LISA are well complementary to each other, leaving no gap for intermediate-mass BHs.

From the comparison between the left and right panels of Fig. 7.3 one deduces that condensates with smaller-$\ell$ can be better constrained than those with higher-$\ell$. This occurs due to the fact that, although the TLNs for the $\ell = 1$ condensate are bigger than those for the spherical one $\ell = 0$ by approximately an order of magnitude, the corresponding maximum frequency is much smaller (because of the behavior of $f_{\rm max} \sim (\ell+1)^{-3}$), resulting in a weaker bound on the tidal-deformability term. For this reason we chose not to show the bounds for $\ell = 2$ condensates, which are found to be not constraining ($\sigma_\alpha/\alpha > 1$).

The plot shows that the relative errors, at fixed value of $\alpha$, decrease for larger BH masses, reach a minimum, and then increase again. This trend can be understood by considering that tidal deformations grow for larger masses which, on the other hand, would also reduce the number of effective cycles in the detector bandwidth due to the dependence of the maximum frequency $f_{\rm max} \sim 1/M_{\rm BH}$. Furthermore, curves at different $\alpha$ overlap since larger values of $\alpha$, corresponding to smaller TLNs, imply a larger number of effective cycles ($f_{\rm max} \sim \alpha^3$), thus increasing their possible detection.

A measurement/bound on the gravitational coupling $\alpha = M_{\rm BH}\mu$ can be translated into a measurement/bound on the boson mass $m_b = \mu \hbar$. In particular, since the typical SNRs for the binary systems we consider are high enough, the corresponding chirp mass can be measured very well, and the binary mass ratio is also measured with a comparable, or even better, precision than $\alpha$. One can therefore apply the standard propagation of errors to determine the relative error on $m_b$, finding that they are comparable to those on $\alpha$. The corresponding result is shown in Fig. 7.4, where we have spanned the range $10^{-17} \lesssim m_b/\text{eV} \lesssim 10^{-11}$, for both the spherical $\ell = 0$ (left panel) and $\ell = 1$ (right panel) modes. In particular, for a fixed value of $m_b$, we perform a Fisher analysis for different values of the BH mass such that $\alpha < 0.5$ and translate the bounds on $\alpha$ to bounds on the scalar field mass. Our results suggest that the mass of the bosonic field can be measured with roughly 10% accuracy and down to 1% accuracy in some optimal configurations.

For a given GW experiment, the bounds on the boson mass extend across a certain range of BH masses. Indeed, since the gravitational coupling has values around $\alpha = \mathcal{O}(0.1)$ in the relevant parameter space, the lower (upper) boundary of this mass range is set by the heaviest (lightest) BH that gives a meaningful constraint, that is $\sigma_\alpha/\alpha < 1$. ET could probe the range $10^{-13} \lesssim m_b/\text{eV} \lesssim 10^{-11}$ for $\ell = 0$ condensates and a narrower range (especially at high BH masses) for $\ell = 1$ condensates, with the upper bound $m_b \approx 3 \times 10^{-13}\,\text{eV}$ reached for $\alpha \approx 0.5$ and $M_{\text{BH}} \approx 200 M_\odot$. Furthermore, it can probe values of the coupling slightly smaller than $\alpha = 0.1$ down to masses $M_{\text{BH}} = 10^{-3} M_\odot$, which could only be ascribed to PBHs. On the other hand, LISA could probe the range of ultralight bosons masses $10^{-17} \lesssim m_b/\text{eV} \lesssim 10^{-13}$ for $\ell = 0$ condensates, that becomes slightly narrower for $\ell = 1$ condensates, partially overlapping with the one of ET. Its range is however wider due to the larger SNR and its capability to detect BHs in the entire mass range available for this measurement. Furthermore, LISA could measure masses around $m_b \approx 10^{-14}\,\text{eV}$, which are difficult to probe with other GW-based tests [649, 681, 682]. Therefore, LISA measurements of the tidal deformability of light BHs would be complementary to other searches and would also provide a smoking gun for departure from the standard "naked BH" picture.

Finally, let us comment on the fact that, even though we focused on bosonic condensates, in principle any matter configuration like accretion disks around BHs could feature nonvanishing tidal deformations [683]. Quantifying this effect for coalescence GW signals could be of utmost importance, especially in the framework of PBHs, which could be characterised by accretion processes during their evolution and thus experience accretion disks around them.

# Part IV

# Conclusions

# Chapter 8

# Conclusions

The goal of this thesis has been to investigate some of the phenomena which could have occurred during the very early universe, focusing in particular on the formation of primordial black holes, their interplay with fundamental physics, and describe the many ways with which gravitational-wave observations provide fascinating insights related to the physics of the primordial universe.

In part I, we investigated the generation of PBHs from the gravitational collapse of cosmological perturbations produced during the inflationary era. We saw how the properties of the collapsing perturbations are crucial to determine the features of the resulting PBHs, that are their mass function, initial spin and abundance in the universe. Then, we studied possible GW signatures associated to the PBHs formation, focusing on the stochastic gravitational wave background generated from the same curvature perturbations which give rise to PBHs, due to the intrinsic non-linear nature of gravity. We showed that, depending on the characteristic frequencies and amplitude of the scalar inflationary perturbations, the corresponding GW spectrum may fall within the sensitivity curves of present experiments like NANOGrav or future experiments like LISA, providing insights on the connection between PBHs and the dark matter in the universe.

In part II, we studied the evolution of PBHs across the cosmic history. We show that PBHs may assemble in binary systems before the matter-radiation equality, and that those binaries may experience a phase of baryonic mass accretion. Indeed, the latter is responsible for increasing the mass of the binary components, along with their individual spins, giving rise to specific correlations between these parameters, that could be used to disentangle them from astrophysical BHs. Furthermore, we saw that, if PBHs comprise a large fraction of the dark matter in the universe, after the matter-radiation equality they will form clusters, whose main effect would be to modify the binary parameters and affect their corresponding merger rates. However, for the range of PBH masses that could be observed by the LIGO/Virgo/KAGRA experiments, there are stringent constraints on the PBH abundance which limit the effective role of clustering. In the same part of the thesis, we compared the PBH model with the current data detected by the LIGO/Virgo/KAGRA collaboration, finding that the scenario in which PBHs explain all the observed events is compatible with the current constraints on the PBH abundance. However, PBHs do not seem to be able to explain all the features measured in the GWTC-2 catalog, pointing towards the direction of multiple channels playing a role in the present data. Then, we studied the perspectives of future GW detectors like Einstein Telescope, Cosmic Explorer and LISA to discover these compact objects, for example through the observation of merger events at very high redshift or measuring a coalescence with a subsolar component.

Part III is dedicated to the interplay between PBHs and fundamental physics during the early stages of the universe. In particular, we studied some properties of first-order phase transitions, that can occur at very high energies, and showed that PBHs may arise from the collisions of bubbles of the new phase. We also proved the role of PBHs, as gravitational impurities, in catalysing baryon number violation processes through Standard Model sphaleron transitions. Finally, we investigated the capability of ET and LISA in providing insights on the properties of scalar fields condensed in halos around BHs through the tidal Love numbers, showing the powerful role of GW observations to shed light on the connection between particle physics and cosmology.

Along this direction, much work is still needed in order to understand the role of PBHs in the evolution of the universe and their connection to fundamental physics. However, the large number of GW events which would be detected in the near future by present and future GW experiments do represent a promising path towards a better understanding of the many phenomena and signals coming from the early universe.

# Appendices

## A Hierarchical Bayesian inference

In this appendix we provide an introduction to the hierarchical Bayesian inference formalism, which we have adopted in several analysis of the data found by the LVKC. More comprehensive descriptions are provided in the literature in Refs. [684–687].

GW interferometers record a time series of the GW-induced strain in the detectors. In order to compare these data with astrophysical models, which predict intrinsic properties of the individual binary sources, one has to convert these time series into astrophysically meaningful quantities through a process of parameter estimation [684]. The LVKC's Gravitational Wave Open Science Center provides the output of this parameter estimation analysis in terms of posteriors $p(\boldsymbol{\theta}|\boldsymbol{d}_i)$ characterizing the expectation value and uncertainty on the individual merger events properties, where $\boldsymbol{\theta}$ is a vector of source parameters and $\boldsymbol{d}_i$ denotes the time series of the $i$-th event in the catalog.

Given some data $\boldsymbol{d}$, one can use the Bayes' theorem to relate the posterior probability distribution of the signal from a given source with parameters $\boldsymbol{\theta}$,

$$p(\boldsymbol{\theta}|\boldsymbol{d}) \propto p(\boldsymbol{d}|\boldsymbol{\theta})p(\boldsymbol{\theta}), \tag{A.1}$$

in terms of the likelihood $p(\boldsymbol{d}|\boldsymbol{\theta})$ of observing the data given our model of the astrophysical signal and detector, and the assumed prior on the source parameters $p(\boldsymbol{\theta})$. The latter encodes the previous knowledge of the underlying physics and it is relevant in interpreting the results [28, 688, 689].

A hierarchical Bayesian analysis parametrizes the choice of priors by some hyperparameters vector $\boldsymbol{\lambda}$, whose posterior distribution can be deduced from the data as

$$p(\boldsymbol{\lambda}|\boldsymbol{d}) \propto p(\boldsymbol{\lambda}) \int p(\boldsymbol{d}|\boldsymbol{\theta})p_{\text{pop}}(\boldsymbol{\theta}|\boldsymbol{\lambda})\mathrm{d}\boldsymbol{\theta}, \tag{A.2}$$

in terms of the single-event likelihood $p(\boldsymbol{d}|\boldsymbol{\theta})$, a prior on the hyperparameters $p(\boldsymbol{\lambda})$, and the population likelihood $p_{\text{pop}}(\boldsymbol{\theta}|\boldsymbol{\lambda})$, which represents a prior parametrized by some hyperparameters.

In the following, we denote the single source parameters, like the BH masses, spins and redshifts, as the event parameters, while the hyperparameters describing the entire sample are identified as population parameters.

A given population model characterized by population parameters $\boldsymbol{\lambda}$ will predict some distribution of the event parameters $\boldsymbol{\theta}$ such that the differential rate $\frac{\mathrm{dr}}{\mathrm{d}\boldsymbol{\theta}}(\boldsymbol{\lambda})$ is given by

$$\frac{\mathrm{dr}}{\mathrm{d}\boldsymbol{\theta}}(\boldsymbol{\lambda}) = R(\boldsymbol{\lambda})\, p_{\mathrm{pop}}(\boldsymbol{\theta}|\boldsymbol{\lambda})\,, \tag{A.3}$$

where the population likelihood is normalised such that $\int p_{\mathrm{pop}}(\boldsymbol{\theta}|\boldsymbol{\lambda})\mathrm{d}\boldsymbol{\theta} = 1$, and the total rate $R(\boldsymbol{\lambda})$ is typically measured in $\mathrm{yr}^{-1}$. The number of predicted events is given by

$$N(\boldsymbol{\lambda}) = \int \frac{\mathrm{dr}}{\mathrm{d}\boldsymbol{\theta}}(\boldsymbol{\lambda})\mathrm{d}\boldsymbol{\theta} \times T_{\mathrm{obs}}, \tag{A.4}$$

in terms of the duration of the observing run(s) $T_{\mathrm{obs}}$.

The sensitivity of the detectors induce selection effects which can be taken into account by a function $0 \leq p_{\mathrm{det}}(\boldsymbol{\theta}) \leq 1$, describing the probability that an event with parameters $\boldsymbol{\theta}$ can be detected. A more detailed description of the detection probability is provided in the next appendix. The observable rate distribution is then given by

$$\frac{\mathrm{dr}_{\mathrm{det}}}{\mathrm{d}\boldsymbol{\theta}}(\boldsymbol{\lambda}) = R(\boldsymbol{\lambda})\, p_{\mathrm{pop}}(\boldsymbol{\theta}|\boldsymbol{\lambda})\, p_{\mathrm{det}}(\boldsymbol{\theta})\,, \tag{A.5}$$

such that the expected number of observations is

$$N_{\mathrm{det}}(\boldsymbol{\lambda}) = \int \mathrm{d}\boldsymbol{\theta}\frac{\mathrm{dr}_{\mathrm{det}}}{\mathrm{d}\boldsymbol{\theta}}(\boldsymbol{\lambda}) \times T_{\mathrm{obs}}. \tag{A.6}$$

The population posterior then takes the standard expression for an inhomogeneous Poisson process as [684, 687, 690, 691]

$$p(\boldsymbol{\lambda}|\boldsymbol{d}) \propto \pi(\boldsymbol{\lambda})\, e^{-N_{\mathrm{det}}(\boldsymbol{\lambda})} N(\boldsymbol{\lambda})^{N_{\mathrm{obs}}} \prod_{i=1}^{N_{\mathrm{obs}}} \int \frac{p(\boldsymbol{\theta}_i|\boldsymbol{d})}{\pi(\boldsymbol{\theta}_i)} p_{\mathrm{pop}}(\boldsymbol{\theta}_i|\boldsymbol{\lambda})\mathrm{d}\boldsymbol{\theta}_i\,, \tag{A.7}$$

as a function of the number of observations $N_{\mathrm{obs}}$ and the population prior $\pi(\boldsymbol{\lambda})$.

The information coming from the rate can be excluded by the parameter inference by marginalizing over $N(\boldsymbol{\lambda})$ with prior $\propto 1/N(\boldsymbol{\lambda})$, to get [692]

$$p(\boldsymbol{\lambda}|\boldsymbol{d}) \propto \pi(\boldsymbol{\lambda}) \prod_{i=1}^{N_{\mathrm{obs}}} \int \frac{p(\boldsymbol{\theta}_i|\boldsymbol{d})}{\pi(\boldsymbol{\theta}_i)} \frac{p_{\mathrm{pop}}(\boldsymbol{\theta}_i|\boldsymbol{\lambda})}{\alpha(\boldsymbol{\lambda})}\mathrm{d}\boldsymbol{\theta}_i\,, \tag{A.8}$$

in terms of the selection bias $\alpha(\boldsymbol{\lambda})$

$$\alpha(\boldsymbol{\lambda}) = \int p_{\mathrm{pop}}(\boldsymbol{\theta}'|\boldsymbol{\lambda}) p_{\mathrm{det}}(\boldsymbol{\theta}')\mathrm{d}\boldsymbol{\theta}' = \frac{N_{\mathrm{det}}(\boldsymbol{\lambda})}{N(\boldsymbol{\lambda})}, \tag{A.9}$$

which describes the fraction of events one would detect given a population.

The event posterior probability distribution function $p(\boldsymbol{\theta}_i|\boldsymbol{d})$ is usually provided in the form of $\mathcal{S}_i$ discrete samples by a process of parameter estimation [693, 694]. Making use of these posterior samples, one can avoid reevaluations of the likelihood $p(\boldsymbol{d}|\boldsymbol{\theta})$ and reduce the computation load for each population inference run. The integral in Eq. (A.7) can be therefore evaluated by using importance sampling, that is by computing the expectation value of the prior-reweighted population likelihood. The latter can be turned into a discrete sum over the event posterior samples, to finally get

$$p(\boldsymbol{\lambda}|\boldsymbol{d}) = \pi(\boldsymbol{\lambda})e^{-N_{\det}(\boldsymbol{\lambda})}N(\boldsymbol{\lambda})^{N_{\mathrm{obs}}}\prod_{i=1}^{N_{\mathrm{obs}}}\frac{1}{\mathcal{S}_i}\sum_{j=1}^{\mathcal{S}_i}\frac{p_{\mathrm{pop}}(^j\boldsymbol{\theta}_i|\boldsymbol{\lambda})}{\pi(^j\boldsymbol{\theta}_i)}, \tag{A.10}$$

where $j$ indicates the $j$-th sample of the $i$-th event. We stress that the priors on the event parameters $\pi(\boldsymbol{\theta}_i)$ are weighted out, such that they do not contribute to the resulting $p(\boldsymbol{\lambda}|\boldsymbol{d})$.

The last information is encoded in the population likelihood, which in turn depends on the predicted number of events. The latter can be obtained in a given population model from the differential merger rate $\mathrm{d}R/(\mathrm{d}m_1\mathrm{d}m_2)$ as

$$p_{\mathrm{pop}}(\boldsymbol{\theta}|\boldsymbol{\lambda}) \equiv \frac{1}{N(\boldsymbol{\lambda})}\left[T_{\mathrm{obs}}\frac{1}{1+z}\frac{\mathrm{d}V}{\mathrm{d}z}\frac{\mathrm{d}R}{\mathrm{d}m_1\mathrm{d}m_2}(\boldsymbol{\theta}|\boldsymbol{\lambda})\right], \tag{A.11}$$

and

$$N_{\det}(\boldsymbol{\lambda}) \equiv T_{\mathrm{obs}}\int \mathrm{d}m_1\mathrm{d}m_2\mathrm{d}z\, p_{\det}(m_1, m_2, z)\frac{1}{1+z}\frac{\mathrm{d}V_c}{\mathrm{d}z}\frac{\mathrm{d}R}{\mathrm{d}m_1\mathrm{d}m_2}(m_1, m_2, z|\boldsymbol{\lambda})\,. \tag{A.12}$$

The comparison between various models in this framework is usually performed, in a quantitative assessment, through the statistical evidence $Z$. In particular, given a model $\mathcal{M}$, the evidence is defined as the integral of the population posterior $p(\boldsymbol{\lambda}|\boldsymbol{d})$, i.e.

$$Z_{\mathcal{M}} \equiv \int \mathrm{d}\boldsymbol{\lambda}\, p(\boldsymbol{\lambda}|\boldsymbol{d}), \tag{A.13}$$

which provides a measure of the support for a model given the data $\boldsymbol{d}$. The comparison then occurs by computing the Bayes factors

$$\mathcal{B}_{\mathcal{M}_2}^{\mathcal{M}_1} \equiv \frac{Z_{\mathcal{M}_1}}{Z_{\mathcal{M}_2}}. \tag{A.14}$$

According to the Jeffreys' scale criterion [695], a Bayes factor larger than $(10, 10^{1.5}, 10^2)$ would imply a strong, very strong, or decisive evidence in favour of model $\mathcal{M}_1$ with respect to model $\mathcal{M}_2$.

# B   Binary detectability

In this appendix we review the main steps to compute the probability of binary detection following Ref. [12].

The event parameters which characterise a compact quasi-circular binary are the BHs masses $m_1$ and $m_2$, the dimensionless spins $\boldsymbol{\chi}_1$ and $\boldsymbol{\chi}_2$, the merger redshift $z$ and the position and orientation with respect to the detectors. The latter are usually defined in terms of right ascension $\alpha$, declination $\delta$, orbital-plane inclination $\iota$, and polarization angle $\psi$. One can thus designate the intrinsic and extrinsic parameters as $\theta_{\rm i} = \{m_1, m_2, \boldsymbol{\chi}_1, \boldsymbol{\chi}_2, z\}$ and $\theta_{\rm e} = \{\alpha, \delta, \iota, \psi\}$, respectively.

Given that spinning binaries have typically larger SNR than nonspinning binaries with same masses and redshift, and that the majority of events in GWTC-2 are compatible with small spins, in the computation of the BH binary detectability we have neglected the role of spins.

For each value of the intrinsic parameters $\theta_{\rm i}$, one can define the detection probability by averaging over the extrinsic ones $\theta_{\rm e}$ as

$$p_{\rm det}(\theta_{\rm i}) = \int p(\theta_{\rm e}) \, \Theta[\rho(\theta_{\rm i}, \theta_{\rm e}) - \rho_{\rm thr}] \, d\theta_{\rm e} \, , \tag{B.1}$$

in terms of the extrinsic parameters probability distribution function $p(\theta_e)$ and SNR $\rho$. Following the procedure frequently adopted for GWTC-1 also for the O3 run, one can compute $p_{\rm det}$ by assuming the single-detector semianalytic framework of Refs. [696, 697] and a SNR threshold $\rho_{\rm thr} = 8$. This approach has no significant departures from the large-scale injection campaigns.

One can factor out the dependency on the extrinsic parameters $\theta_{\rm e}$ in the SNR to obtain $\rho(\theta_{\rm i}, \theta_{\rm e}) = \omega(\theta_{\rm e})\rho_{\rm opt}(\theta_{\rm i})$, in terms of the SNR of an "optimal" source located overhead the detector with face-on inclination $\rho_{\rm opt}$. The marginalized distribution $p_{\rm det}(\theta_{\rm i})$ is then computed by evaluating the cumulative distribution function $P(\omega_{\rm thr}) = \int_{\omega_{\rm thr}}^1 p(\omega')d\omega'$ at $\omega_{\rm thr} = \rho_{\rm thr}/\rho_{\rm opt}(\theta_{\rm i})$. For the case of isotropic sources, where $\alpha, \cos\delta, \cos\iota$, and $\psi$ are uniformly distributed, and non-precessing binaries, and considering only the dominant quadrupole moment, the function $P(\omega_{\rm thr})$ is found in Ref. [504].

The optimal SNR $\rho_{\rm opt}$ of individual GW events for a source with masses $m_1$ and $m_2$ at redshift $z$ is given in terms of the GW waveform in Fourier space $h(f)$ by

$$\rho_{\rm opt}^2(m_1, m_2, z) \equiv \int_0^\infty \frac{4|h(f)|^2}{S_n(f)} df, \tag{B.2}$$

in terms of the strain noise $S_n$. One can easily notice that the optimal SNR scales like the inverse of the luminosity distance, i.e. $\rho \propto 1/D_{\rm L}(z)$.

The sensitivity curves for the LIGO/Virgo observation runs and 3G detectors are shown in Fig 1.3. In particular, for the O1-O2 (O3a) observation runs, we have adopted the `aLIGOEarlyHighSensitivityP1200087` (`aLIGOMidHighSensitivityP1200087`) noise power spectral densities, as implemented in the publicly available repository `pycbc` [698]. Furthermore, in order to be consistent with the framework described to compute the binary detectability, we have also adopted the non-precessing waveform model IMRPhenomD.

# C   Fisher analysis

The detectability of some parameters entering the GW waveform can be estimated using a Fisher matrix approach as described in Refs. [674, 677, 678]. Following Ref. [4], we review the basic notions of the formalism.

The output $s(t)$ of a generic GW interferometer can be expressed as the sum of the GW signal $h(t, \vec{\xi})$ and the stationary detector noise $n(t)$. The posterior on the hyperparameters $\vec{\xi}$ is given by

$$p(\vec{\xi}|s) \propto \pi(\vec{\xi}) e^{-\frac{1}{2}(h(\vec{\xi})-s|h(\vec{\xi})-s)}, \tag{C.1}$$

in terms of the prior on the hyperparameters $\pi(\vec{\xi})$, where the inner product is defined to be

$$(g|h) = 2 \int_{f_{\min}}^{f_{\max}} df \frac{h(f)g^*(f) + h^*(f)g(f)}{S_n(f)}. \tag{C.2}$$

The frequencies $f_{\min}$ and $f_{\max}$ identify the characteristic minimum and maximum frequencies of integration, which depends on the system under consideration.

According to the principle of the maximum-likelihood estimator, the values of the hyperparameters can be determined by maximising the posterior. In the limit of large signal-to-noise ratio (SNR$= \sqrt{(h|h)}$), for which the posterior is peaked around the true hyperparameters values $\vec{\zeta}$, one can perform a Taylor expansion to get

$$p(\vec{\xi}|s) \propto \pi(\vec{\xi}) e^{-\frac{1}{2}\Gamma_{ab}\Delta\xi^a\Delta\xi^b}, \tag{C.3}$$

with $\Delta\vec{\xi} = \vec{\zeta} - \vec{\xi}$, in terms of the Fisher matrix,

$$\Gamma_{ab} = \left( \frac{\partial h}{\partial \xi^a} \middle| \frac{\partial h}{\partial \xi^b} \right)_{\vec{\xi}=\vec{\zeta}}. \tag{C.4}$$

The errors on the hyperparameters are given by $\sigma_a = \sqrt{\Sigma^{aa}}$, where $\Sigma^{ab} = (\Gamma^{-1})^{ab}$ is the covariance matrix. In the limit of large SNR, the errors (and the inverse SNR) scale linearly with the luminosity distance of the source, and provide insights on the expected precision with which one can measure the parameters that enter the GW waveform.